\documentclass[12pt,PhD]{muthesis}

\usepackage{graphicx}
\usepackage{amsmath}
\usepackage{amssymb}
\usepackage{epsfig}
\usepackage{epstopdf}
\usepackage[numbers,sort&compress]{natbib}
\usepackage{subfigure}
\usepackage[hidelinks]{hyperref}
\newcommand{\mytilde}{\sim}

\newcommand{\Pe}{Pe}
\newcommand{\Le}{Le_F}
\newcommand{\pd}[2]{\frac{\partial #1}{\partial #2}}
\newcommand{\ed}[2]{\frac{d #1}{d #2}}

\newcommand{\sd}[2]{\frac{\partial^2 #1}{\partial #2^2}}
\DeclareGraphicsExtensions{.pdf,.eps,.png,.jpg,.mps} 
\begin{document}
\title{Propagation and stability of flames in inhomogeneous mixtures}
\author{Philip Pearce}
\school{Mathematics}
\faculty{Engineering and Physical Sciences}
\def\wordcount{31991}

\tablespagefalse



\beforeabstract

We investigate the effect of thermal expansion and gravity on the propagation and stability of flames in inhomogeneous mixtures. We focus on laminar flames in the simple configuration of an infinitely long channel with rigid porous walls in order to understand the effect of inhomogeneities on these fundamental structures.

The first part of the thesis is concerned with premixed flames propagating against a prescribed parallel (Poiseuille) flow and subject to thermal expansion. We show that in a narrow channel (corresponding to a relatively thick flame), if the Peclet number is fixed and of order unity, a premixed flame propagating against a parallel flow is governed by the equation for a planar premixed flame with an effective diffusion coefficient. The enhanced diffusion is shown to correspond to Taylor dispersion, or shear-enhanced diffusion. Several important applications of the results are discussed. One of the topics of relevance is the bending effect of turbulent combustion. The results of our analysis show that, for a large flow intensity, the effective propagation speed of the premixed flame for depends only on the Peclet number (which is equal to the Reynolds number if the Prandtl number is unity). This mimics the behaviour of the turbulent premixed flame when the effective propagation speed is plotted versus the turbulence intensity for fixed values of the Reynolds number.

The second part of the thesis is concerned with triple flames, subject to thermal expansion and buoyancy. A study is undertaken to investigate the stability of a diffusion flame subject to these effects, which gives rise to a problem analogous to the classical Rayleigh--B\'{e}nard convection problem. A linear stability analysis in the Boussinesq approximation is performed, which leads to analytical results showing that the Burke-Schumann flame is unstable if the Rayleigh number is above a critical value which is determined. Numerical results confirm and complement the analytical results. A full numerical investigation of the effects of gravity and thermal expansion on triple flames propagating in a direction perpendicular to the direction of gravity is then carried out. This configuration does not seem to have received dedicated attention in the literature. It is found that the well-known monotonic relationship between the propagation speed $U$ and the flame-front thickness $\epsilon$, which exists in the constant density case when
the Lewis numbers are of order unity or larger, persists for triple flames undergoing thermal expansion. Under strong enough gravitational effects, however, the relationship is no longer found to be monotonic, exhibiting hysteresis if the Rayleigh number is large enough. Finally, the initiation of triple flames from a hot two-dimensional ignition kernel is investigated. Particular attention is devoted to the energy required for ignition and the transient
evolution of triple flames after initiation. Steady, non-propagating, two-dimensional solutions representing
``flame tubes" are determined; their thermal energy is used to define a minimum
ignition energy for the two-dimensional triple flame in the mixing layer. The transient behaviour of triple flames following ``energy-increasing" or ``energy-decreasing" perturbations to the flame tube solutions is described in situations where the underlying diffusion flame is either stable or unstable.

\afterabstract

\prefacesection{Acknowledgements}
I would like to express my gratitude to my supervisor Dr. Joel Daou for introducing me to combustion theory and for his expert guidance and encouragement during this work. I would also like to acknowledge the EPSRC for providing me with the financial support to undertake my PhD. Finally, I would like to thank all of the friends and family who have made the time I've spent working on this thesis enjoyable.

\afterpreface


\chapter{Introduction}
\label{chapter:intro}
\section{Introduction}
In many practical situations involving a propagating flame, inhomogeneities are present in the mixture through which the flame propagates. These inhomogeneities can be caused by fluctuations or stratifications in the temperature, the composition or the flow field . Understanding the effects of such inhomogeneities on the propagation and stability of laminar flames in simple configurations is crucial to provide a platform for further investigations that take into account more complex aspects such as turbulence. Similarly, since in most applications combustion generates large amounts of heat and occurs in a gravitational field, it is vital to understand the combined effects of thermal expansion and buoyancy on flame propagation and stability in such simple situations.

The aim of this thesis is to investigate the combined effect of thermal expansion and buoyancy on the propagation and stability of flames propagating through inhomogeneous mixtures. The inhomogeneity is prescribed in the unburned gas, into which the flame propagates, in one of two ways: either a) as a non-uniform flow field against which a premixed flame propagates, or b) as a stratification in the concentrations of the fuel and oxidiser, which leads to the propagation of a triple flame. The simple configuration considered throughout the thesis is a channel with rigid walls that are impermeable to the fluid. We begin with a brief literature review that summarises the work done in the areas relevant to each chapter of the thesis. More thorough reviews and descriptions of further areas of relevance to each chapter are contained within the chapters themselves and the papers by \citet{pearce2013effect,pearce2013rayleigh,pearce2014taylor}, on which much of this thesis is based.

The coupling between the flame and the flow is modelled with the Navier--Stokes equations, coupled to equations for the temperature and the mass fractions of fuel and oxidiser, with a one-step Arrenhius reaction. A short derivation of the governing equations is given in \S \ref{intro:sec:background} of this chapter. Before discussing work that has been undertaken using these governing equations, it is instructive to first note a simplification that has been used in many studies, known as the \emph{constant density approximation}. This approximation neglects the effect of the flame on the flow by assuming that the density of the fluid is constant. The effect of the flow on the flame is taken into account through the advection term in the temperature and mass fraction equations, where the flow can be prescribed. The approximation has been justified asymptotically from the governing equations of combustion theory in the limit of weak heat release in \cite{matkowsky1979asymptotic}. Decoupling the temperature and mass fraction equations from the Navier--Stokes equations considerably simplifies combustion problems and has been useful for investigating, for example, the so-called \emph{thermo-diffusive} \cite{barenblatt1962diffusive} instabilities in combustion, which arise due to differences in the rate of transport of fuel and oxidiser.  Although we are not concerned with such instabilities in this thesis, we occasionally utilise the constant density approximation, either for comparison of results to help understand the effects of thermal expansion, or in order to investigate an effect that arises from combustion without having to account for the complex interactions brought about by the effect of the flame on the fluid through which it propagates.

Another significant simplification of the governing equations of combustion was utilised by Darrieus and Landau in their early studies on the stability of a planar premixed flame \cite{landau1944theory,darrieus1938propagation}. These studies took the effect of the flame on the flow into account through thermal expansion but ignored the transport of temperature and mass fractions. Darrieus and Landau found using this approximation that a planar flame should always be unstable due to the difference in density across the flame. Planar flames can, however, be observed in the laboratory; the analysis of Darrieus and Landau fails at short wavelengths, where transport processes inside the flame influence the flame structure and velocity \cite{clavin1985dynamic}.  The \emph{hydrodynamic} or \emph{Darrieus--Landau} instability of premixed flames has been the focus of several studies, as reviewed in \cite{bychkov2000dynamics}.

Later studies investigated the effects of a full coupling between the Navier--Stokes and the transport equations on the propagation speed and stability of premixed flames in the limit of infinite activation energy and an infinitely thin flame front \cite{sivashinsky1976distorted,frankel1982effect,clavin1982effects,pelce1982influence,matalon1982flames}. These studies provided the necessary correction terms to the dispersion relation derived by Darrieus and Landau, finding that planar premixed flames can indeed be stable. Since the aforementioned studies, there has been a significant amount of work investigating the effects of thermal expansion on the propagation and stability of thin flames in both laminar and turbulent regimes (see e.g. the reviews given in \cite{clavin1985dynamic,bychkov2000dynamics,peters2000turbulent}).

There has been significantly less work, however, on thick flames, which correspond to flames in relatively narrow channels. These are the focus of \textbf{Chapter \ref{chapter:premixed1}} of this thesis.  Since the development of a suitable analytical methodology based on a \emph{thick flame asymptotic limit} by \citet{daou2002thick},  studies on thick flames in the constant density approximation have addressed the effect of heat loss \cite{daou2002thick,daou2002influence}, the effect of nonunity Lewis numbers \cite{kurdyumov2002lewis,cui2004effects,kurdyumov2011lewis} and the influence of oscillatory flow \cite{daou2007flame}. More recently, the influence of thermal expansion on thick flames has been investigated \cite{short2009asymptotic, kurdyumov2013flame} under different distinguished limits of the governing parameters. In Chapter \ref{chapter:premixed1}, which is based on work by \citet{pearce2014taylor}, we extend the knowledge of premixed flame propagation by investigating thick premixed flames subject to the effects of thermal expansion in cases where the prescribed flow against which the flame propagates has infinitely large amplitude. The results of the analysis in this distinguished limit are relevant to several important topics of research, as will be discussed in more detail in Chapter \ref{chapter:premixed1}.

One such topic of interest is the so-called \emph{bending effect} of turbulent combustion. The bending effect is observed experimentally when the effective flame speed $U_T$ of a premixed flame is plotted versus the turbulence intensity \cite{ronney1995some}. In \textbf{Chapter \ref{chapter:premixed2}} we provide a discussion of the bending effect in laminar premixed flames and explain how this relates to turbulent combustion. The main motivation in this chapter is to describe the relevance of asymptotic results in the \emph{thick flame} asymptotic limit to the bending effect.

As well as investigating the effect of inhomogeneities in the flow on premixed flames, in this thesis we also investigate combustion in inhomogeneous mixtures of fuel and oxidiser, in the same channel configuration as the one utilised in Chapters \ref{chapter:premixed1} and \ref{chapter:premixed2}. Premixed flames can still propagate through mixtures with small fluctuations in the temperature or the concentrations of fuel and oxidiser. There have been several studies investigating how such fluctuations affect the propagation and stability of premixed flames (see \cite{wu2010large} and the references therein). However, if the fuel and oxidiser are stratified, a different structure known as a \emph{triple flame} propagates through the mixture. Triple flames consist of a fuel-rich premixed branch, a fuel-lean premixed branch and a trailing diffusion flame.

Triple flames were first identified experimentally by \citet{phillips}. Initial theoretical investigations were carried out by \citet{2}, followed by Dold and collaborators \cite{3,4}. Much research has focused on triple flames and their properties since, mostly concerning triple flames in the constant density approximation (see the review papers \cite{buckmaster2002edge} and \cite{chung2007stabilization}). The first work to investigate the effects of thermal expansion on triple flames was a mainly numerical study by \citet{5}. There have been several studies since that have investigated the effects of thermal expansion on triple flames \cite{6,8,9,10}, with a key result being the increase in triple flame speed due to thermal expansion above that of the planar premixed flame, when the flame-front is thin (corresponding to a wide channel).

One aspect of triple flames that is not very well understood is the effect of buoyancy on their propagation and stability. Triple flames propagating in a direction parallel to the direction of gravity have been investigated numerically and experimentally in \cite{11,12,chen2001numerical,echekki2004numerical}. It was found that the propagation speed of a triple flame propagating downwards is decreased in comparison to that of a triple flame in the absence of gravity. The change in the propagation speed was explained in \cite{chen2001numerical} as being due to an increase in the acceleration of the gas ahead of the triple flame leading edge, caused by buoyancy. Conversely, upward propagation leads to an increase in the propagation speed. It seems, however, that no prior dedicated studies have been undertaken on triple flames propagating in a direction perpendicular to gravity. In this thesis we provide such a study.

In order to investigate the effect of buoyancy on a triple flame, it is imperative to first understand this effect on the ‘‘strongly burning’’ diffusion flame, which forms one of the triple flame’s branches; steadily propagating triple flames are only expected for parameter values where a planar diffusion flame exists and is stable. For this reason \textbf{Chapter \ref{chapter:diffusion}} of this thesis contains a stability analysis of a horizontal planar diffusion flame, subject to the combined effect of thermal expansion and gravity. The chapter is based on work by \citet{pearce2013rayleigh}, which seems to be the first study in the literature to investigate the instability of a planar diffusion flame due to buoyancy-driven convection.

In \textbf{Chapter \ref{chapter:triple}} we move on to investigate the combined effect of thermal expansion and gravity on triple flames steadily propagating perpendicular to the direction of gravity, using the results of Chapter \ref{chapter:diffusion} to concentrate on areas in parameter space where the planar diffusion flame is stable. The original published work by \citet{pearce2013effect} seems to be the first dedicated study of this aspect of triple flame behaviour.

To complete our investigation of the combined effect of thermal expansion and gravity on triple flames, in  \textbf{Chapter \ref{chapter:ignition}} we study the transient behaviour of triple flames from their initiation in contexts where the underlying planar diffusion flame is either stable or unstable. Included in this study is an investigation of the energy required for initiation of triple flames from a two-dimensional ignition kernel. Steady, non-propagating, two-dimensional solutions representing ``flame tubes" are determined; their thermal energy is used to define a minimum ignition energy for the two-dimensional triple flame in the mixing layer. Similar axisymmetric structures representing inhomogeneous flame balls \cite{daou2014flame,daou2015flame}, flame disks \cite{buckmaster2000holes} and flame isolas \cite{lu2004flame} have previously been identified and linked to the ignition of axisymmetric flames in the mixing layer, but to our understanding, no such study has yet been performed for two-dimensional triple flames in the mixing layer.

The thesis is structured as follows. The remainder of the current chapter is given to a theoretical background of mathematical combustion, including the derivation of the governing equations and some quantities used for scaling throughout the thesis. Chapter \ref{chapter:premixed1} contains a study of the effects of thermal expansion on premixed flames propagating through a narrow channel against a parallel flow of large intensity. Chapter \ref{chapter:premixed2} focuses on the bending effect of laminar premixed flames. Chapter \ref{chapter:diffusion} consists of an investigation of the instability of a planar diffusion flame, caused by buoyancy-driven convection. Chapter \ref{chapter:triple} discusses the behaviour of steadily propagating triple flames under the combined effect of buoyancy and thermal expansion. Chapter \ref{chapter:ignition} presents a study of two-dimensional triple flame initiation in mixing layers. Finally, we end the thesis with conclusions and recommendations for further study in Chapter \ref{chapter:conc}. 

\section{Theoretical background}
\label{intro:sec:background}
\subsection{Governing equations}
\begin{figure}
\centering
\includegraphics{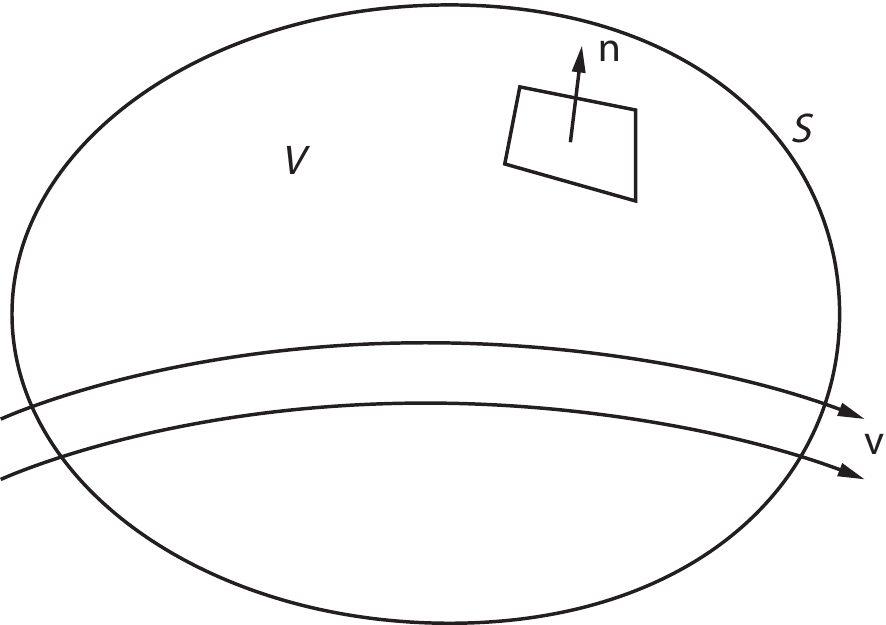}
\caption{Fixed control volume $V$, bounded by a surface $S$, with unit normal $\mathbf{n}$ and bulk velocity $\mathbf{v}$ passing through.}
\label{fig:control}
\end{figure}
In this section we provide the governing equations of combustion theory, with an overview of their derivation from a continuum mechanics perspective. Expanded versions of the derivation (including the derivation of the transport equations using the kinetic theory of gases) can be found in \cite{williams1985combustion} and \cite{law2006combustion}. The derivation of the general governing equations of continuum mechanics can be found in \cite{spencer2004continuum}. A general introduction to fluid dynamics, including a derivation of the Navier--Stokes equations, is given in \cite{batchelor2000introduction}.

Consider a fixed control volume $V$, depicted in figure \ref{fig:control}. The volume is bounded by a control surface $S$ with outer unit normal $\mathbf{n}$. A bulk velocity $\mathbf{v}$ passes through the volume. Note that the mass-weighted $\mathbf{v}$ is the resultant of the individual velocities $\mathbf{v}_i$ of the $N$ individual species, so that
\begin{gather}
\sum_{i=1}^{N}\rho_i \mathbf{v}_i=\rho \mathbf{v},\label{deriv:individualdensity}
\end{gather}
where $\rho_i$ is the density of each species. Then the \emph{molecular diffusion velocity} $\mathbf{V}_i$ is given by
\begin{gather}
\mathbf{V}_i=\mathbf{v}_i-\mathbf{v}.\label{deriv:individualvelocity}
\end{gather}
Equations \eqref{deriv:individualdensity} and \eqref{deriv:individualvelocity} lead to
\begin{gather}
\sum_{i=1}^{N}\rho_i\mathbf{V}_i=0.\label{deriv:sumdensities}
\end{gather}

Now, consider some extensive property $\Psi$, whose magnitude depends on the size of the control volume $V$. Then a quantity $\psi$, whose magnitude does not depend on the size of $V$, can be considered to be the ``density" of $\Psi$ per unit volume of the fluid. The two quantities can be related by the formula
\begin{gather}
\Psi=\int_V \psi ~\mathrm{d}V.
\end{gather}
Suppose the amount of $\Psi$ changes due to external influences at a rate given by
\begin{gather}
\int_V Q ~\mathrm{d}V, \label{deriv:materialtime}
\end{gather}
where $Q$ is an effective density of source strength. The rate of change of $\Psi$ is given by the amount of $\Psi$ lost through the surface $S$ plus the increase of $\Psi$ associated with external influences \cite{batchelor2000introduction}. This statement may be written mathematically as
\begin{gather}
\frac{d}{d t}\int_V \psi ~\mathrm{d}V=-\int_S \psi \left(\mathbf{v}\cdot\mathbf{n}\right)~\mathrm{d}S+\int_V Q ~\mathrm{d}V.\label{deriv:reynoldstransporta}
\end{gather}
Rearranging and applying the divergence theorem and Leibniz's rule to \eqref{deriv:reynoldstransporta} gives
\begin{gather}
\frac{D\Psi}{D t}=\int_V\left(\frac{\partial \psi}{\partial t} +\nabla \cdot \psi \mathbf{v}\right)~\mathrm{d}V=\int_V Q ~\mathrm{d}V,\label{deriv:generalcontinuity}
\end{gather}
where we have used the fact that $V$ is fixed. The left hand side of \eqref{deriv:generalcontinuity} is referred to as the \emph{material time derivative} of $\Psi$ \cite{spencer2004continuum}. Equation \eqref{deriv:generalcontinuity} will now be used to derive the conservation equations for mass, momentum, energy and the concentration of species.

\subsubsection{Conservation of mass}
Let $\Psi$ be the total mass $m$ of the fluid. Then $\psi$ is the mass density $\rho$ of the fluid. Since matter can neither be created nor destroyed, the quantity \eqref{deriv:materialtime} can be set to zero in this case. Then equation \eqref{deriv:generalcontinuity} becomes
\begin{gather*}
\int_V\left(\frac{\partial \rho}{\partial t} +\nabla \cdot \rho \mathbf{v}\right)~\mathrm{d}V=0,
\end{gather*}
or, since the control volume $V$ is arbitrary,
\begin{gather}
\frac{\partial \rho}{\partial t} +\nabla \cdot \rho \mathbf{v}=0. \label{deriv:continuity}
\end{gather}
Equation \eqref{deriv:continuity} is the \emph{continuity equation}.
\subsubsection{Conservation of momentum}
Let $\Psi$ be the momentum of the flow, $\mathbf{M}$. Then $\psi$ is the momentum flux $\rho \mathbf{v}$. In this case the material time derivative \eqref{deriv:generalcontinuity} is given by
\begin{gather}
\frac{D \mathbf{M}}{D t}=\mathbf{p},
\end{gather}
where $\mathbf{p}$ is the resultant force acting on the fluid. This follows from Newton's second law, which states that the force acting on the fluid is equal to the rate of change of the fluid's momentum. We can split the resultant force $\mathbf{p}$ into two parts: the force acting on the surface of $V$, represented by the stress tensor $\mathbf{P}$, and the resultant of the body forces $\mathbf{f}_i$ per unit mass, acting on the $i$th species \cite{williams1985combustion,spencer2004continuum,law2006combustion,batchelor2000introduction}. The surface force is given by
\begin{gather*}
\int_S\left(\mathbf{P}\cdot\mathbf{n}\right)~\mathrm{d}S=\int_V\nabla\cdot \mathbf{P}~\mathrm{d}V,
\end{gather*}
using the divergence theorem. Thus, equation \eqref{deriv:generalcontinuity} becomes
\begin{gather}
\int_V\left(\frac{\partial \rho \mathbf{v}}{\partial t} +\nabla \cdot \rho \mathbf{v}\mathbf{v}\right)~\mathrm{d}V=\int_V\nabla\cdot \mathbf{P}~\mathrm{d}V+\sum_{i=1}^{N}\int_V\rho_i \mathbf{f}_i ~\mathrm{d}V, \label{deriv:momentum1}
\end{gather}
where $\mathbf{v}\mathbf{v}$ is a dyadic tensor. Using equation \eqref{deriv:continuity} and the fact that $V$ is arbitrary, we can rewrite equation \eqref{deriv:momentum1} as
\begin{gather}
\rho\left(\frac{\partial\mathbf{v}}{\partial t}+\mathbf{v}\cdot\nabla\mathbf{v}\right)=\nabla \cdot \mathbf{P} +\sum_{i=1}^{N}\rho_i \mathbf{f}_i. \label{deriv:momentum2}
\end{gather}
For a viscous Newtonian fluid the stress tensor $\mathbf{P}$ can be written as
\begin{gather}
\mathbf{P}=-p\mathbf{I}+\mu\left(\left(\nabla\mathbf{v}\right)+\left(\nabla\mathbf{v}\right)^T-\frac{2}{3}\left(\nabla \cdot\mathbf{v}\right)\mathbf{I}\right)=-p\mathbf{I}+\tau, \label{deriv:stresstensor}
\end{gather}
where $p$ is the pressure, $\mu$ is the dynamic viscosity and $\mathbf{I}$ is the unit tensor. This form of the stress tensor for fluids, along with physical descriptions of the meaning of each term, is discussed in \cite{spencer2004continuum} and \cite{batchelor2000introduction}.

We assume the only external body force acting on the fluid is gravity. Since gravitational acceleration is the same for all species this leads to
\begin{gather}
\mathbf{f}_i=\mathbf{g}.\label{deriv:gravity}
\end{gather}
Finally, if the viscosity $\mu$ does not depend on the temperature of the fluid, which we assume for simplicity throughout this thesis, we have, using \eqref{deriv:momentum2}, \eqref{deriv:stresstensor} and \eqref{deriv:gravity},
\begin{gather}
\rho\left(\frac{\partial\mathbf{v}}{\partial t}+\mathbf{v}\cdot\nabla\mathbf{v}\right)=-\nabla p+\mu\left(\nabla^2 \mathbf{u}+\frac{1}{3}\nabla\left(\nabla \cdot \mathbf{u}\right)\right)+\rho \mathbf{g}. \label{deriv:navierstokes}
\end{gather}
Equation \eqref{deriv:navierstokes} is referred to as the \emph{Navier--Stokes equation}.
\subsubsection{Conservation of species}
Let $\Psi$ be the mass $m_i$ of the $i$th species. Then $\psi$ is the density $\rho_i$ of the $i$th species. The mass of the $i$th species inside $V$ can be changed either as a result of a chemical reaction, with rate of production $\omega_i$ of the $i$th species per unit volume, or due to diffusion across $S$. The magnitude of this diffusive transport is proportional to the mass flux $\rho_i \mathbf{V}_i$ of the molecular random motion \cite{law2006combustion} and can be written
\begin{gather*}
\int_S \left(\rho_i \mathbf{V}_i \cdot \mathbf{n}\right)~\mathrm{d}S=\int_V\nabla \cdot \rho_i \mathbf{V}_i~\mathrm{d}V,
\end{gather*}
using the divergence theorem. In this case the quantity \eqref{deriv:materialtime} is given by
\begin{gather}
\int_V Q \mathrm~{d}V=\int_V\left( \omega_i-\nabla \cdot \rho_i \mathbf{V}_i \right)~\mathrm{d}V,
\end{gather}
so that, using the fact that $V$ is arbitrary, equation \eqref{deriv:generalcontinuity} becomes
\begin{gather}
\frac{\partial \rho_i}{\partial t} +\nabla \cdot \left(\rho_i\left( \mathbf{v}+\mathbf{V}_i\right)\right)=\omega_i.\label{deriv:species1}
\end{gather}
Now, defining the \emph{mass fraction} of the $i$th species as
\begin{gather}
Y_i=\frac{\rho_i}{\rho} \label{deriv:massfrac}
\end{gather}
and using equation \eqref{deriv:continuity}, we can write equation \eqref{deriv:species1} as
\begin{gather}
\rho\left(\frac{\partial Y_i}{\partial t}+\mathbf{v}\cdot\nabla Y_i\right)=-\nabla \cdot \left(\rho \mathbf{V}_i Y_i\right)+\omega_i. \label{deriv:species2}
\end{gather}
Finally, assuming Fick's law (see \cite{williams1985combustion} or \cite{law2006combustion} for a derivation in this context), which states
\begin{gather}
Y_i \mathbf{V}_i=-D_i \nabla Y_i, \label{deriv:fickslaw}
\end{gather}
where $D_i$ is the diffusion coefficient of the $i$th species, we can write equation \eqref{deriv:species2} as
\begin{gather}
\rho\left(\frac{\partial Y_i}{\partial t}+\mathbf{v}\cdot\nabla Y_i\right)=\rho D_i \nabla^2 Y_i+\omega_i. \label{deriv:species3}
\end{gather}
Here we have assumed $\rho D_i$ is constant. We will discuss the form of the reaction term $\omega_i$ later.
\subsubsection{Conservation of energy}
Let $\Psi$ be the total energy of the material inside $V$. The total energy can be written as $K+E$, where $K$ is the kinetic energy and $E$ is the internal energy \cite{spencer2004continuum}. In this case $\psi$ can be written as
\begin{gather*}
\psi=\rho e + \frac{1}{2}\rho |\mathbf{v}|^2,
\end{gather*}
where $e$ is the internal energy density and the term on the right hand side defines the kinetic energy of the fluid.  The energy of the fluid inside $V$ can be changed by work done by the surface or body forces, or by energy entering $V$ through the boundary $S$; we ignore radiative heat transfer. Thus equation \eqref{deriv:generalcontinuity} becomes
\begin{gather}
\int_V\left(\frac{\partial \rho\left(e+\frac{1}{2}|\mathbf{v}|^2\right)}{\partial t} +\nabla \cdot \rho \left(e+\frac{1}{2}|\mathbf{v}|^2\right)\right)~\mathrm{d}V=\tilde{Q}+W_S+W_V,\label{deriv:energy1}
\end{gather}
where
\begin{gather}
\tilde{Q}=-\int_S\left(\mathbf{q}\cdot\mathbf{n}\right)~\mathrm{d}S=-\int_V\nabla \cdot \mathbf{q}~\mathrm{d}V \label{deriv:heatflux}
\end{gather}
is due to the energy flux $\mathbf{q}$ across $S$,
\begin{gather}
W_S=-\int_S \mathbf{v}\cdot\left(\mathbf{P}\cdot\mathbf{n}\right)~\mathrm{d}S=-\int_V\nabla\cdot\left(\mathbf{v}\cdot\mathbf{P}\right)~\mathrm{d}V \label{deriv:surfacework}
\end{gather}
is the work done by the surface forces acting on $V$, and
\begin{gather}
W_V=\sum_{i=1}^{N}\int_V\mathbf{v}_i\cdot\left(\rho_i \mathbf{f}_i\right)~\mathrm{d}V\label{deriv:bodywork}
\end{gather}
is the work done by the body forces acting on each species in $V$, which are moving at $\mathbf{v}_i$. Note that the divergence theorem was used in rewriting the above equations. Using equations \eqref{deriv:energy1}--\eqref{deriv:bodywork} and the fact that $V$ is arbitrary leads to
\begin{gather}
\frac{\partial \rho\left(e+\frac{1}{2}|\mathbf{v}|^2\right)}{\partial t} +\nabla \cdot \rho \left(e+\frac{1}{2}|\mathbf{v}|^2\right)=-\nabla \cdot \mathbf{q}-\nabla\cdot\left(\mathbf{v}\cdot\mathbf{P}\right)+\sum_{i=1}^{N}\mathbf{v}_i\cdot\left(\rho_i \mathbf{f}_i\right). \label{deriv:energy2}
\end{gather}
Now, taking the scalar product of equation \eqref{deriv:momentum2} with $\mathbf{v}$ and subtracting from \eqref{deriv:energy2}, we obtain a simpler form of the energy conservation equation, given by
\begin{gather}
\frac{\partial \rho e}{\partial t} +\nabla \cdot \rho \mathbf{v} e=-\nabla \cdot \mathbf{q}-\mathbf{P}:\nabla\mathbf{v}+\sum_{i=1}^{N}\mathbf{V}_i\cdot\left(\rho_i \mathbf{f}_i\right), \label{deriv:energy3}
\end{gather}
where the symbol $:$ indicates that the two tensors are to be contracted twice to form a scalar \cite{williams1985combustion}. We now make several assumptions to simplify the energy conservation equation \eqref{deriv:energy3}. Firstly, we define the enthalpy $h$ by
\begin{gather}
e=h-\frac{p}{\rho}. \label{deriv:caloric}
\end{gather}
Secondly, from \eqref{deriv:stresstensor} we have
\begin{gather}
\mathbf{P}:\nabla\mathbf{v}=p\nabla \cdot \mathbf{v}+ \Phi, \label{deriv:simplifystress}
\end{gather}
where we assume the \emph{viscous dissipation} \cite{batchelor2000introduction} term $\Phi=\tau:\nabla\mathbf{v}$, with $\tau$ defined in \eqref{deriv:stresstensor}, takes the value
\begin{gather}
\Phi=0,
\end{gather}
which is justifiable for low speed, subsonic flows \cite{law2006combustion}. Thirdly, we assume the energy flux $\mathbf{q}$ takes the form
\begin{gather}
\mathbf{q}= -\lambda \nabla T+\rho\sum_{i=1}^{N}h_iY_i\mathbf{V}_i\label{deriv:simplifyflux},
\end{gather}
where the first term on the right hand side results from Fourier's law of heat conduction, and the second term is due to partial enthalpy transport by diffusion \cite{law2006combustion}; for simplicity, we assume the thermal conductivity $\lambda$ is constant. In equation \eqref{deriv:simplifyflux} the quantities $h_i$ relate to the enthalpy $h$ by
\begin{gather}
h=\sum_{i=1}^{N}Y_i h_i. \label{deriv:enthalpies}
\end{gather}
Now, if the body force is gravity, using \eqref{deriv:sumdensities} and \eqref{deriv:gravity}, we have 
\begin{gather}
\sum_{i=1}^{N}\mathbf{V}_i\cdot\left(\rho_i \mathbf{f}_i\right)=0.\label{deriv:simplifybody}
\end{gather}
Using equations \eqref{deriv:continuity} and \eqref{deriv:energy3}, the assumptions \eqref{deriv:caloric}--\eqref{deriv:simplifybody}, together with the assumption that the process is isobaric, lead to the enthalpy equation
\begin{gather}
\rho \left(\frac{\partial h}{\partial t}+ \mathbf{v}\cdot\nabla h\right)-\frac{\partial p}{\partial t}=\lambda\nabla^2 T+\nabla\cdot\left(\rho\sum_{i=1}^{N}Y_i h_i \mathbf{V}_i\right).\label{deriv:enthalpyequation}
\end{gather}
Finally, we assume the \emph{caloric equation of state} \cite{borgnakke2009fundamentals}
\begin{gather}
h_i=h_i^0+c_p\left(T-T^0\right),\label{deriv:caloric2}
\end{gather}
where $T$ is the temperature and $T^0$ and $h_i^0$ are the reference temperature and enthalpies, respectively. Here we have assumed that each species has the same constant specific heat capacity $c_p$. Using \eqref{deriv:species3}, \eqref{deriv:enthalpies} and \eqref{deriv:caloric2}, equation \eqref{deriv:enthalpyequation} can be written in terms of the temperature as
\begin{gather}
\rho \left(\frac{\partial T}{\partial t} +\mathbf{v} \cdot \nabla T\right) - \frac{\partial p}{\partial t}=\frac{\lambda}{c_p} \nabla^2 T -\frac{1}{c_P}\omega_T. \label{deriv:temperature}
\end{gather}
Here $\omega_T$ is the temperature change due to chemical reactions given by
\begin{gather}
\omega_T=\sum_{i=1}^{N} \omega_i h_i. \label{deriv:tempreact}
\end{gather}
Note that $\lambda \left/ c_p\right.$ can also be written in terms of the thermal diffusivity $D_T$ as
\begin{gather}
\frac{\lambda}{c_p}=\rho D_T, \label{deriv:thermaldiff}
\end{gather}
where we have assumed $\rho D_T$ is constant.
\subsubsection{Chemical reactions}
In this section we prescribe the form of the chemical reaction terms $\omega_i$ in equations \eqref{deriv:species3} and \eqref{deriv:tempreact}. Reactions in combustion applications can be extremely complicated, consisting of multi-step reactions of many different species. A summary of common reaction mechanisms used in mathematical modelling of combustion is given in \cite{law2006combustion}. Here we assume a simple, one-step reaction between fuel F and oxidiser O
\begin{gather}
\nu_1 \text{F} + \nu_2 \text{O} \rightarrow \left(\nu_1+\nu_2\right)\text{Products}+\tilde{q}, \label{deriv:reaction1}
\end{gather}
where $\nu_1$ and $\nu_2$ denote the amount of fuel and oxidiser in the reaction, respectively, and $\tilde{q}$ denotes the heat released in the reaction. Then
\begin{gather}
\omega_i=-m_i \nu_i \omega,\label{deriv:massreactions}
\end{gather}
with the \emph{Arrenhius law} \cite{buckmaster1983lectures} assumed for the global reaction rate $\omega$, given by
\begin{gather}
\omega=\rho B Y_F Y_O \exp\left(-E\left/ RT \right.\right). \label{deriv:arrenhius}
\end{gather}
Here $Y_F$, $Y_O$, $R$, $T$, $B$ and $E$ are the fuel mass fraction, the oxidiser mass fraction, the universal gas constant, the temperature, the pre-exponential factor and the activation energy of the reaction, respectively. Then the temperature change due to chemical reactions is
\begin{gather}
\omega_T =\tilde{q} \omega,
\end{gather}
where $\tilde{q}$ is given by
\begin{gather}
\tilde{q}=-\sum_{i=1}^{N}m_i\nu_i h_i. \label{deriv:heatrelease}
\end{gather}
Now, substituting the relations \eqref{deriv:massreactions}--\eqref{deriv:heatrelease} into the equations \eqref{deriv:species3} and \eqref{deriv:temperature} and rescaling the mass fractions by $m_1\nu_1$, we obtain
\begin{gather}
\rho \left(\frac{\partial T}{\partial t} +\mathbf{v} \cdot \nabla T\right) - \frac{\partial p}{\partial t}=\rho D_T \nabla^2 T +\frac{q}{c_P}\omega, \label{deriv:temperaturefinal}\\
\rho\left(\frac{\partial Y_F}{\partial t}+\mathbf{v}\cdot\nabla Y_F\right)=\rho D_F \nabla^2 Y_F-\omega, \label{deriv:fuelfinal}\\
\rho\left(\frac{\partial Y_O}{\partial t}+\mathbf{v}\cdot\nabla Y_O\right)=\rho D_O \nabla^2 Y_O-s\omega, \label{deriv:oxidiserfinal}
\end{gather}
where $q=\tilde{q}\left/ m_1\nu_1\right.$ is the heat released per unit mass of fuel and $s$ is the amount of oxidiser consumed per unit mass of fuel, given by
\begin{gather*}
s=\frac{m_2\nu_2}{m_1\nu_1}.
\end{gather*}
\subsubsection{Equation of state}
We complete the set of governing equations by specifying the \emph{ideal gas law} equation of state \cite{batchelor2000introduction,williams1985combustion,law2006combustion}, which gives the pressure in terms of the density, the temperature and the universal gas constant $R$ as
\begin{gather}
p=\rho R T. \label{deriv:idealgas}
\end{gather}
\subsubsection{Low Mach number approximation}
To finish the formulation of the governing equations we adopt the low Mach number approximation, common in flame theory and more rigorously justified using asymptotic analyses in several studies, such as those by \citet{rehm1978equations} and \citet{majda1985derivation}.  If we assume low Mach number, the spatial variations in pressure are small. The total pressure can therefore be split into a background term consisting of thermodynamic pressure and a perturbational term consisting of  hydrostatic pressure and hydrodynamic pressure (see \citep[][]{rehm1978equations}). We define the hydrodynamic pressure as
\begin{gather*}
p\left(\mathbf{x},t\right)=P\left(\mathbf{x},t\right)-P_0-P_s\left(\mathbf{x}\right),
\end{gather*}
where $P_0$ is the thermodynamic pressure, which we assume to be constant (see \citep[][p. 14]{buckmaster1983lectures}) and given by the equation of state (\ref{deriv:idealgas}) as
\begin{gather}
P_0=\rho_u RT_u. \label{diff:state3}
\end{gather}
Here $\rho_u$ is the density in the absence of combustion.
$P_s(\mathbf{x})$ is the hydrostatic pressure which satisfies the equation
\begin{gather}
\nabla P_s=\rho_u \mathbf{g} \label{diff:ambient}.
\end{gather}
This is found by considering (\ref{deriv:navierstokes}) in the frozen limit with no flow (i.e. in hydrostatic equilibrium) and noting that, following from equation (\ref{diff:state3}), the ambient atmosphere in the absence of heating must be taken to have constant density $\rho_u$. Subtracting (\ref{diff:ambient}) from (\ref{deriv:navierstokes}) then gives
\begin{gather}
\rho \frac{\partial \mathbf{u}}{\partial t} + \rho \mathbf{u} \cdot \nabla \mathbf{u}+ \nabla p=\mu\left(\nabla^2\mathbf{u} + \frac{1}{3}\nabla \left(\nabla \cdot \mathbf{u}\right)\right) + \left(\rho-\rho_u\right)\mathbf{g}.\label{deriv:navierstokesfinal}
\end{gather}
Now, since
\begin{gather*}
\frac{p\left(\mathbf{x},t\right)+P_s\left(\mathbf{x}\right)}{P_0}=O\left(M^2\right),
\end{gather*}
where $M$ is the Mach number (see \citep[][]{paillere2000comparison}), the perturbational pressure term can be neglected in the ideal gas equation (\ref{diff:state}), which can then be written $P_0=\rho R T$ or, after considering (\ref{diff:state3}),
\begin{gather}
\rho T=\rho_u T_u \label{diff:state2}.
\end{gather}
\subsubsection{Summary of governing equations}
The governing equations \eqref{deriv:continuity}, \eqref{deriv:temperaturefinal}--\eqref{deriv:oxidiserfinal}, \eqref{deriv:navierstokesfinal} and \eqref{diff:state2} can now be written together in full:
\begin{gather}
\frac{\partial \rho}{\partial t} +\nabla \cdot \rho \mathbf{v}=0, \label{govern:continuity}\\
\rho\left(\frac{\partial\mathbf{v}}{\partial t}+\mathbf{v}\cdot\nabla\mathbf{v}\right)=-\nabla p+\mu\left(\nabla^2 \mathbf{u}+\frac{1}{3}\nabla\left(\nabla \cdot \mathbf{u}\right)\right)+\left(\rho-\rho_u\right) \mathbf{g}. \label{govern:navierstokes}\\
\rho \left(\frac{\partial T}{\partial t} +\mathbf{v} \cdot \nabla T\right) - \frac{\partial p}{\partial t}=\rho D_T \nabla^2 T +\frac{q}{c_P}\omega, \label{govern:temperature}\\
\rho\left(\frac{\partial Y_F}{\partial t}+\mathbf{v}\cdot\nabla Y_F\right)=\rho D_F \nabla^2 Y_F-\omega, \label{govern:fuel}\\
\rho\left(\frac{\partial Y_O}{\partial t}+\mathbf{v}\cdot\nabla Y_O\right)=\rho D_O \nabla^2 Y_O-s\omega, \label{govern:oxidiser}\\
\rho T=\rho_u T_u, \label{govern:idealgas}
\end{gather}
where the reaction term $\omega$ is given by \eqref{deriv:arrenhius}. These equations must be supplemented by suitable initial conditions and boundary conditions, which depend on the configuration and will be specified in future chapters.
\subsection{Planar premixed flame}
\begin{figure}
\centering
\includegraphics[scale=0.7]{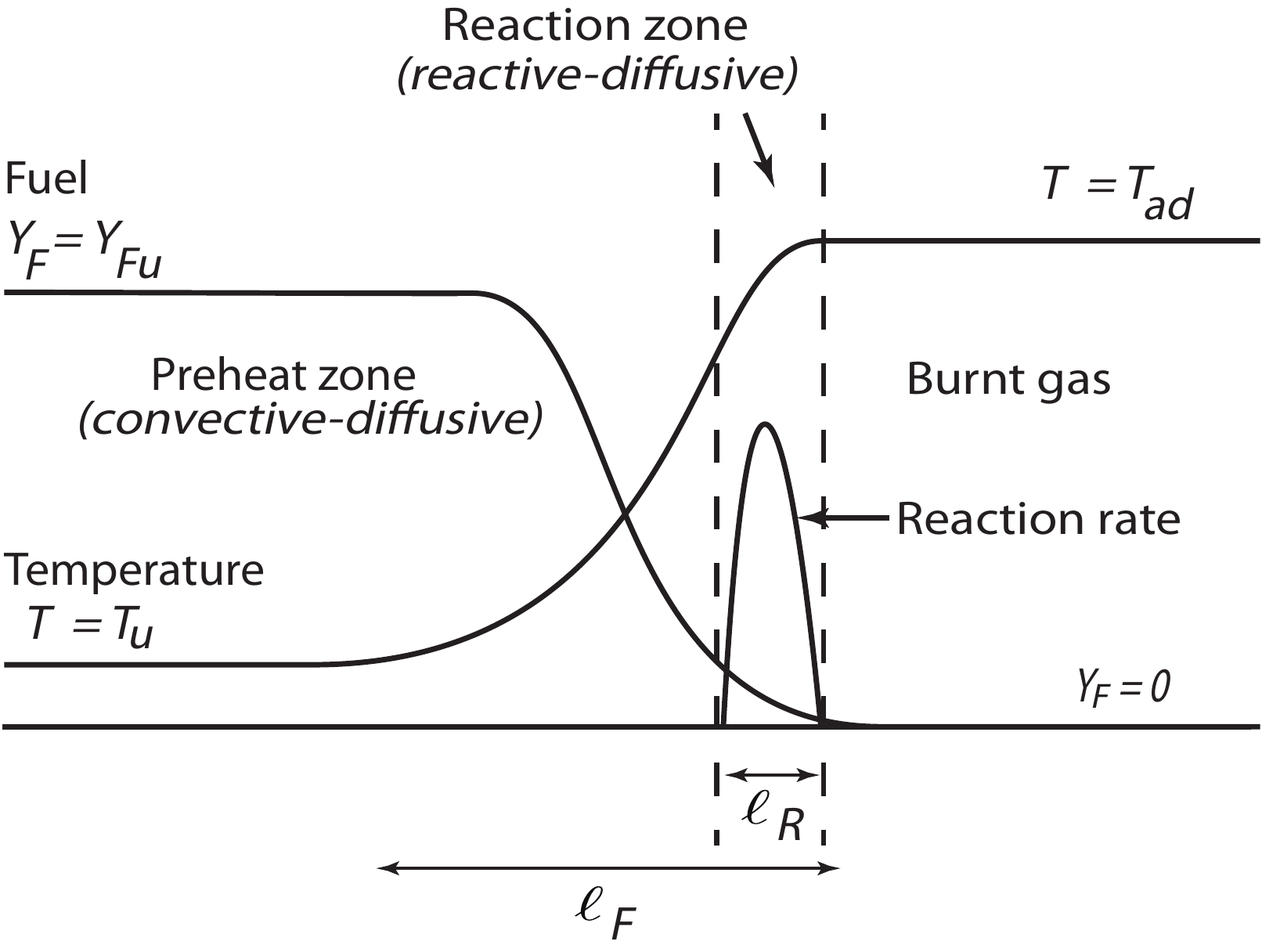}
\caption[The structure of a planar premixed flame, propagating at speed $S_L$ through an unbounded premixed gas. The flame thickness is given by $l_F$ and the reaction zone thickness is $l_R$. For large activation energies, the reaction is negligible except in a thin \emph{reaction zone}. Upstream is the \emph{preheat zone}, where a convective-diffusive balance holds, while in the reaction zone a reactive-diffusive balance holds. Behind the flame is the \emph{burnt gas}, which if the mixture is assumed to be stoichiometric has the adiabatic temperature $T_{ad}$.]{The structure of a planar premixed flame, propagating at speed $S_L$ through an unbounded premixed gas. The flame thickness is given by $l_F$ and the reaction zone thickness is $l_R$. For large activation energies, the reaction is negligible except in a thin \emph{reaction zone} \cite{linan1993fundamental}. Upstream is the \emph{preheat zone}, where a convective-diffusive balance holds, while in the reaction zone a reactive-diffusive balance holds. Behind the flame is the \emph{burnt gas}, which if the mixture is assumed to be stoichiometric has the adiabatic temperature $T_{ad}$.}
\label{intro:fig:planarflame}
\end{figure}
A fundamental problem in combustion is the propagation of the planar premixed flame through an unbounded premixed gas. The asymptotic structure of the planar premixed flame for large activation energies $E$ is shown in figure \ref{intro:fig:planarflame}. Throughout this thesis we will use several properties of the planar premixed flame as reference quantities, namely the adiabatic flame temperature $T_{ad}$, the flame thickness $l_F$ and the propagation velocity $S_L$. In this section we derive expressions for these quantities.

In order to derive such expressions we utilise the well-known technique of \emph{activation energy asymptotics}. This technique has been used effectively in many combustion studies, including generalisations of the following analysis to more complex situations (see e.g. \cite{sivashinsky1976distorted,frankel1982effect,clavin1982effects,pelce1982influence,matalon1982flames}). The method relies on the fact that the activation energy of the reaction is large, which is true in most combustion applications.

The governing equations of a planar premixed flame are given by the steady, one-dimensional form of equations \eqref{govern:continuity}--\eqref{govern:idealgas}. In a frame of reference attached to the flame-front, these equations become
\begin{gather}
\frac{d}{dx}\left(\rho u \right)=0, \label{planar:cont}\\
\rho u \frac{dT}{dx}=\rho D_T \frac{d^2T}{dx^2}+\frac{q}{c_p}\omega, \label{planar:temp}\\
\rho u \frac{dY_F}{dx}=\rho D_F \frac{d^2Y_F}{dx^2}-\omega,\label{planar:fuel}\\
\rho u \frac{dY_O}{dx}=\rho D_O \frac{d^2Y_O}{dx^2}-s\omega. \label{planar:ox}
\end{gather}
The conditions far upstream as $x\to -\infty$ correspond to a fresh mixture with unburnt temperature $T_u$ and mass fractions $Y_{Fu}$ and $Y_{Ou}$. Far downstream as $x \to \infty$ the conditions correspond to burnt gas where if we assume a stoichiometric mixture the temperature is adiabatic and the fuel and oxidiser have been fully consumed. The planar flame speed $S_L$ is defined as the speed of the incoming flow at $x=-\infty$ in the current frame of reference. The boundary conditions can be written
\begin{gather}
T=T_u \quad Y_F=Y_{Fu}, \quad Y_O=Y_{Ou}, \quad \rho=\rho_u,\quad u=S_L\quad \text{as } x \to -\infty,\label{planar:bc1}\\
T=T_{ad}, \quad Y_F=0, \quad Y_O=0, \quad \rho=\rho_b,\quad u=S_b \quad \text{as } x \to \infty, \label{planar:bc2}
\end{gather}
where $\rho_u$ is the density in the fresh mixture and $\rho_b$ is the density under adiabatic conditions. Note that \eqref{planar:cont} can be integrated to give $\rho u=\text{constant}$, so that
\begin{gather}
\rho_u S_L=\rho_b S_b. \label{planar:constantmassflux}
\end{gather}
\subsubsection{Planar flame temperature}
Multipling equation \eqref{planar:fuel} by $q \left/ c_P \right.$, then integrating from $x=-\infty$ to $x=+\infty$ and adding to the integral of equation \eqref{planar:temp} from $x=-\infty$ to $x=+\infty$, gives the adiabatic flame temperature as
\begin{gather}
T_{ad}=T_u+\frac{q Y_{Fu}}{c_P}, \label{planar:adiabatic}
\end{gather}
having used the boundary conditions \eqref{planar:bc1}--\eqref{planar:bc2}.
\subsubsection{Planar flame speed}
Using equations \eqref{planar:temp}--\eqref{planar:ox} and \eqref{planar:constantmassflux} with the fact that $\rho D_T$, $\rho D_F$ and $\rho D_O$ are assumed constant, the governing equations for a planar premixed flame can be written
\begin{gather}
S_L\frac{dT}{dx}=D_{Tu} \frac{d^2T}{dx^2}+\frac{q}{c_p} \frac{\rho}{\rho_u}B Y_F Y_O \exp(-E/RT),\label{planarspeed:temp}\\
S_L\frac{dY_F}{dx}=D_{Fu} \frac{d^2Y_F}{dx^2}-\frac{\rho}{\rho_u}B Y_F Y_O \exp(-E/RT), \label{planarspeed:fuel}\\
S_L\frac{dY_O}{dx}= D_{Ou} \frac{d^2Y_O}{dx^2}-s \frac{\rho}{\rho_u}B Y_F Y_O \exp(-E/RT)\label{planarspeed:ox},
\end{gather}
where the subscript $u$ denotes diffusivities in the unburnt gas at $x=-\infty$. These equations can be non-dimensionalised by writing
\begin{gather}
x^*=\frac{x}{D_{Tu} \left/ S_L \right.}, \quad y_F=\frac{Y_F}{Y_{Fu}}, \quad y_O=\frac{Y_O}{Y_{Ou}},\quad \theta=\frac{T-T_u}{T_{ad}-T_u},\label{planarspeed:scalings}
\end{gather}
which after substitution into equations \eqref{planarspeed:temp}-\eqref{planarspeed:ox} and dropping the superscript gives the non-dimensional governing equations
\begin{gather}
\frac{d \theta}{dx}=\frac{d^2 \theta}{dx^2}+ \lambda \frac{\rho}{\rho_u}y_F y_O \exp\left(\frac{\beta(\theta-1)}{1+\alpha(\theta-1)}\right),\label{planarspeed:nondimtemp}\\
\frac{d y_F}{dx}=\frac{1}{Le_F}\frac{d^2 y_F}{dx^2}-\lambda \frac{\rho}{\rho_u}y_F y_O \exp\left(\frac{\beta(\theta-1)}{1+\alpha(\theta-1)}\right),\label{planarspeed:nondimfuel}\\
\frac{d y_O}{dx}=\frac{1}{Le_O}\frac{d^2 y_O}{dx^2}-\lambda \frac{\rho}{\rho_u}y_F y_O \exp\left(\frac{\beta(\theta-1)}{1+\alpha(\theta-1)}\right),\label{planarspeed:nondimox}
\end{gather}
where 
\begin{gather}
\lambda=\frac{D_{Tu}BY_{Ou}\exp(-E/RT_{ad})}{S_L^2},\quad Le_F=\frac{D_T}{D_F}, \quad Le_O=\frac{D_T}{D_O},\label{planarspeed:parameters1} \\
\beta=E(T_{ad}-T_u) / RT_{ad}^2, \quad \alpha=(T_{ad}-T_{u})/T_{ad}.\label{planarspeed:parameters1}
\end{gather}
Here we have noted that $sY_{Fu}=Y_{Ou}$, based on our assumption that the mixture is stoichiometric. Substituting \eqref{planarspeed:scalings} into the boundary conditions \eqref{planar:bc1}--\eqref{planar:bc2} leads to the non-dimensional boundary conditions
\begin{gather}
\theta=0, \quad y_F=1, \quad y_O=1 \quad \text{as } x \to -\infty,\label{planarspeed:bc1}\\
\theta=1, \quad y_F=0, \quad y_O=0 \quad \text{as } x \to \infty. \label{planarspeed:bc2}
\end{gather}

Now we consider the commonly utilised limit of infinite activation energy, $\beta \to \infty$.  In this limit the reaction term is negligible to all orders in $\beta^{-1}$, except in a thin layer corresponding to $\theta-1=O(\beta^{-1})$. Thus we can split the domain into three zones: two outer zones consisting of the preheat zone and the burnt gas and a thin inner zone consisting of the reaction layer (see figure \ref{intro:fig:planarflame}). In the outer zones the reaction rate is zero so that the equations \eqref{planarspeed:nondimtemp}--\eqref{planarspeed:nondimox} become
\begin{gather*}
\frac{d\theta}{dx}=\frac{d^2 \theta}{dx^2},\\
\frac{dy_F}{dx}=\frac{1}{Le_F}\frac{d^2y_F}{dx^2},\\
\frac{dy_O}{dx}=\frac{1}{Le_O}\frac{d^2y_O}{dx^2}.
\end{gather*}
These equations can be solved to give the outer solutions
\begin{gather}
\theta^{\text{outer}}=A\exp(x)+B, \quad y_F^{\text{outer}}=C\exp(Le_F x)+D, \quad y_O^{\text{outer}}=E\exp(Le_O x)+F.
\end{gather}
In the burnt gas we have $A=C=E=0$ to prevent unboundedness as $x \to \infty$; thus from the boundary conditions \eqref{planarspeed:bc2} we have
\begin{gather*}
\theta^{\text{outer}}=1, \quad y_F^{\text{outer}}=0, \quad y_O^{\text{outer}}=0 \quad \text{ in the burnt gas}.
\end{gather*}
In the preheat zone we have $B=0$ and $D=F=1$ from boundary conditions (\ref{planarspeed:bc1}), which gives
\begin{align*}
\theta^{\text{outer}}=A\exp(x),  \quad y_F^{\text{outer}}=C\exp(Le_F x)+1, \quad &y_O^{\text{outer}}=E\exp(Le_O x)+1.
\end{align*}
The outer fuel mass fraction profiles intersect at the point $x_0$ where $C\exp(Le_F x_0)+1=0$. Since the problem is translationally invariant, we can choose $x_0=0$ to be the origin so that $C=-1$. Thus the outer profiles are given by
\begin{align}
\theta^{\text{outer}}=A\exp(x), \quad y_F^{\text{outer}}=1-\exp(Le_F x), \quad y_O^{\text{outer}}=1+E\exp(Le_O x), \quad &x<0,\label{planarspeed:unburnt}\\
\theta^{\text{outer}}=1, \quad y_F^{\text{outer}}=0, \quad y_O^{\text{outer}}=0, \quad &x>0. \label{planarspeed:burnt}
\end{align}
The constants $A$ and $E$ in \eqref{planarspeed:unburnt} can be determined by matching with the inner solution. We begin by expanding both in terms of $\beta^{-1}$
\begin{gather*}
A=A^0+\frac{A^1}{\beta}+O\left(\frac{1}{\beta^{2}}\right), \quad E=E^0+\frac{E^1}{\beta}+O\left(\frac{1}{\beta^{2}}\right),
\end{gather*}
where superscripts are used to denote successive terms in inner expansions in terms of $\beta^{-1}$. Now, since the reaction zone thickness is $O(\beta^{-1})$, we let
\begin{gather*}
X=\frac{x}{\beta^{-1}}, \quad \theta^{\text{inner}}(x)=\Theta(X), \quad y_F^{\text{inner}}(x)=F(X), \quad y_O^{\text{inner}}(x)=R(X),
\end{gather*}
where
\begin{gather*}
\Theta(X)=1+\frac{\Theta^1(X)}{\beta}+...,\quad F(X)=\frac{F^1(X)}{\beta}+...\quad R(X)=\frac{R^1(X)}{\beta}+...
\end{gather*}
Here we have anticipated that to leading order $y_F=y_O=0$ and $\theta=1$ inside the reaction zone. Now, for a balance between the reaction and diffusion terms inside the reaction zone in equation (\ref{planarspeed:nondimtemp}), it is clear that $\lambda=O(\beta^3)$. Also, since $\Theta=1$ to leading order, we have $\rho \sim \rho_b$ to leading order. Thus we let
\begin{gather}
\lambda \frac{\rho_b}{\rho_u}=\beta^3 \Lambda_0. \label{planarspeed:lambda}
\end{gather}
Hence the governing equations for the inner solution are given by
\begin{gather}
\Theta^1_{XX}+\Lambda_0 F^1 R^1 \exp(\Theta^1)=0,\label{planarspeed:inner1}\\
\frac{1}{Le_F}F^1_{XX}-\Lambda_0 F^1 R^1 \exp(\Theta^1)=0,\label{planarspeed:inner2}\\
\frac{1}{Le_O}R^1_{XX}-\Lambda_0 F^1 R^1 \exp(\Theta^1)=0.\label{planarspeed:inner3}
\end{gather}
The boundary conditions on equations \eqref{planarspeed:inner1}--\eqref{planarspeed:inner3} are now found by matching with the outer solutions using the formula
\begin{gather}
\phi^{\text{inner}}(X \to \pm \infty)=\phi^{\text{outer}}(x \to 0^{\pm})
\end{gather}
for each dependent variable $\phi$. Matching with the solution for the burnt gas, given in (\ref{planarspeed:burnt}), gives
\begin{gather}
\Theta^1=F^1=R^1=0 \quad \text{as }X \to \infty. \label{planarspeed:burntbc}
\end{gather}
Expanding the solution for the temperature in the unburnt gas, given by \eqref{planarspeed:unburnt}, as $x \to 0^-$ leads to
\begin{gather*}
\theta^{\text{outer}}(x \to 0^-)=(A_0+\frac{A_1}{\beta}+...)(1+\frac{X}{\beta}+...).
\end{gather*}
Matching with the inner solution as $X \to -\infty$ gives $A_0=1$ and thus
\begin{gather}
\Theta^1=X+A_1 \quad \text{as}\quad X \to -\infty. \label{planarspeed:unburntbc1}
\end{gather}
Similarly, it can be shown that $E_0=-1$, so that
\begin{gather}
F^1=-Le_F X, \quad R^1=-Le_O X+E_1 \quad \text{as}\quad X \to -\infty.\label{planarspeed:unburntbc2}
\end{gather}
Now, adding equations \eqref{planarspeed:inner1} and \eqref{planarspeed:inner2}, we obtain
\begin{gather*}
\frac{d^2}{dX^2}\left(\Theta^1+\frac{F^1}{Le_F}\right)=0,
\end{gather*}
which can be integrated twice, using boundary condition \eqref{planarspeed:burntbc}, to find
\begin{gather}
\Theta^1+\frac{F^1}{Le_F}=0.\label{planarspeed:innera}
\end{gather}
Since this is also valid as $X \to - \infty$, we have $A_1=0$ from boundary conditions \eqref{planarspeed:unburntbc1} and \eqref{planarspeed:unburntbc2}. Similarly, adding equations \eqref{planarspeed:inner1} and \eqref{planarspeed:inner3} and integrating with the use of \eqref{planarspeed:burntbc} shows that 
\begin{gather}
\Theta^1+\frac{R^1}{Le_O}=0,\label{planarspeed:innerb}
\end{gather}
and thus $E_1=0$ from \eqref{planarspeed:unburntbc1} and \eqref{planarspeed:unburntbc2}. Using \eqref{planarspeed:innera} and \eqref{planarspeed:innerb} we can now write the inner problem as
\begin{gather}
\Theta^1_{XX}+\Lambda_0Le_F Le_O\left(\Theta^1\right)^2 \exp(\Theta^1)=0,\label{planarspeed:innerfinal1}
\end{gather}
with
\begin{gather}
\Theta^1=X\quad \text{as} \quad X\to -\infty,\label{planarspeed:innerfinal2}\\
\Theta^1=0 \quad \text{as} \quad X \to \infty.\label{planarspeed:innerfinal3}
\end{gather}
The problem \eqref{planarspeed:innerfinal1}--\eqref{planarspeed:innerfinal3} can be solved by multiplying equation \eqref{planarspeed:innerfinal1} by $\Theta^1_X$ and integrating from $X=-\infty$ to $X=+\infty$, using conditions \eqref{planarspeed:innerfinal2}-\eqref{planarspeed:innerfinal3}, to find
\begin{gather*}
\Lambda_0=\frac{1}{4Le_FLe_O},
\end{gather*}
so that, using \eqref{planarspeed:lambda},
\begin{gather}
\lambda=\frac{\rho_u}{\rho_b}\frac{\beta^3}{4Le_FLe_O} \label{planarspeed:lambda2}
\end{gather}
and finally, inserting \eqref{planarspeed:lambda2} into the definition of $\lambda$ on the left hand side of \eqref{planarspeed:parameters1},
\begin{gather}
S_L=\left(\frac{\rho_b}{\rho_u}\frac{4 Le_F Le_O}{\beta^3}Y_{O,st}D_{Tu}B \exp(-E/RT_{ad})\right)^{1/2}
\end{gather}
to leading order in $\beta^{-1}$. This is the required result, giving the planar premixed flame speed, with thermal expansion taken into account.
\subsubsection{Planar flame thickness}
The planar flame thickness is defined by the diffusive length scale \cite{sivashinsky1976distorted,clavin1982effects,matalon1982flames}
\begin{gather}
l_F=\frac{D_T}{S_L}.
\end{gather}

\chapter{Taylor dispersion and thermal expansion effects on flame propagation in a narrow channel}
\chaptermark{Taylor dispersion effects on flames}
\label{chapter:premixed1}
\section{Introduction}
\label{premixed1:sec:intro}
In this chapter, which is based on a paper by \citet{pearce2014taylor}, we provide a theoretical study of a variable density premixed flame propagating through a narrow channel against a Poiseuille flow of large amplitude. Under these conditions, the dependence of the propagation speed of the premixed flame on the Peclet number is investigated. The essential governing parameters are the flame-front thickness $\epsilon$ and the amplitude of the flow $A$ (which together define the Peclet number $\Pe=A \left / \epsilon \right.$), as well as the activation energy of the reaction $\beta$. The problem studied has relevance to several important areas of research.

The first area concerns premixed flames propagating through narrow channels, which have been the subject of considerable renewed interest in recent years. In addition to traditional applications such as fire safety in mine shafts \citep[][p. 271]{kanury1975introduction}, recent applications are concerned with emerging technologies that utilise microscale combustion \citep[][]{fernandez2002micropower}. Related investigations have addressed the development of a suitable analytical methodology, based on a \emph{thick flame asymptotic limit} \citep{daou2002thick}, the effect of heat loss \citep{daou2002thick,daou2002influence}, the effect of non-unity Lewis numbers \citep{kurdyumov2002lewis,cui2004effects,kurdyumov2011lewis}, the influence of oscillatory flow \citep{daou2007flame} and the influence of thermal expansion \citep{short2009asymptotic, kurdyumov2013flame} under different distinguished limits of the governing parameters. The asymptotic results in the current chapter can be considered to be an extension of the results of \citet{daou2002thick} and \citet{short2009asymptotic}, who studied the same configuration but in the limit of small Peclet number in the constant density and variable density cases, respectively. A low value of $\Pe$ is not the case, however, in many practical applications (see, for example, the experimental results given in the review article \citep[][]{bradley1992fast}, which were obtained for a fixed value of $\Pe$). For this reason the asymptotic analysis in the current study is performed in the limit $\epsilon \to \infty$ with $\Pe=O(1)$ and numerical results are obtained for moderately large Peclet numbers.

The second area of research is related, albeit indirectly, to turbulent combustion. At high values of $\Pe$ the flame could become turbulent, an aspect of the problem not addressed here. Nevertheless, the results are still useful as a first step towards an understanding of the effects of the small scales in the flow on a turbulent premixed flame; at present there seems to be no analytical description of even laminar premixed flames for arbitrary values of $\Pe$ in situations where the flame is thick compared to the length scale of the flow. This latter situation is fundamental to a proper evaluation of Damk\"{o}hler's second hypothesis \citep[][]{damkohler1940influence} concerning the effect of small scale flow on turbulent premixed flames, which has received little attention in the literature. Conversely, there have been many studies on turbulent premixed flames in the flamelet regime of large flow scales compared to the flame thickness \citep[][]{clavin1979theory,kerstein1988field,aldredge1992propagation,yakhot1988propagation,sivashinsky1988cascade}, which was the subject of Damk\"{o}hler's first hypothesis. A detailed discussion of the relevance of Damk\"{o}hler's second hypothesis to turbulent premixed flames can be found in the paper by \citet{daou2002thick}. A thorough review of turbulent combustion in general can be found in the monograph by \citet{peters2000turbulent}.

The third area of relevant research is Taylor dispersion, a well-studied topic that began with Taylor's seminal paper discussing the enhanced dispersion of a solute due to a parallel flow in a channel \cite[][]{taylor1953dispersion}. Taylor investigated a distinguished limit characterised by a small diffusion time in comparison to the advective time; in this limit the depth-averaged concentration of the solute was shown to be governed by a one-dimensional equation with an effective diffusion coefficient $D_{\text{eff}}$, which was found to be larger than the diffusion coefficient $D$ and dependent upon the profile of the parallel flow. Specifically, in the case of a cylindrical channel of radius $a$ and an imposed Poiseuille flow of cross-sectional average $\bar{w}$, Taylor found the effective diffusion coefficient to be given by
\begin{gather}
D_{\text{eff}}=D\left(1+\frac{a^2\bar{w}^2}{48 D^2}\right),\label{eq:taylor}
\end{gather}
for a solute with diffusion coefficient $D$.

A comprehensive review of the subject of Taylor dispersion can be found in the book by \citet{brenner1993macrotransport}. Here we simply note that there seem to be relatively few analytical studies in the literature that investigate Taylor dispersion with a variable density flow (see \cite[][]{oltean2004transport,felder2004dispersion,dentz2006variable}). In these studies the effective diffusion coefficient has been found to be a function of the density. Although there has been a small number of studies on Taylor dispersion in reaction-diffusion equations (e.g. \cite[][]{leconte2008taylor}), this study is the first to discuss Taylor dispersion in the context of combustion. One of the limits taken in the current chapter can be considered to characterise the Taylor regime of a premixed flame, whereby the flame is described by the one-dimensional planar premixed flame equation with an effective diffusion coefficient. The determination of the propagation speed (an eigenvalue representing the speed of the travelling premixed flame) is intimately linked to the effective diffusion coefficient in the limit of infinite activation energy. It is surprising that despite this direct link, Taylor dispersion does not yet seem to have been investigated in the context of premixed (laminar or turbulent) combustion.

The main aims of the investigation are: 1) to quantify the effect of a small-scale parallel (Poiseuille) flow on the propagation speed of a premixed flame for fixed values of the Peclet number, taking gas expansion into account (see formula \eqref{result:infinitebeta} later); 2) to demonstrate that the enhancement of the propagation speed coincides exactly with the Taylor dispersion formula \eqref{eq:taylor}; 3) to provide an analytical confirmation of Damk\"{o}hler's second hypothesis in our particular case corresponding to a laminar flow with a single scale which is small compared to the flame thickness (see the discussion in \S \ref{premixed1:sec:conc}). We believe that achieving these aims, albeit in a simplified adiabatic context (such as in \cite[][]{daou2001flame}), as is carried out in this study,  is a contribution of a fundamental nature that will provide a solid basis for future studies accounting for additional realistic effects. These include more complex multi-scale flows and the influence of heat losses, which are not accounted for here to concentrate on the pure interaction between the flow and the flame and to ensure that the analysis is tractable. The practical aspects of heat losses are known to be important in real micro-combustion applications; indeed, to minimise the influence of such heat losses, it is well known that thermal management is required experimentally, such as external wall heating \cite[][]{fan2007experimental} or heat recirculation \cite[][]{sitzki2001combustion,ahn2005gas}; see also the review by \citet{fernandez2002micropower}.

The chapter is structured as follows. In \S \ref{premixed1:sec:formulation} we formulate the problem. \S \ref{premixed1:sec:asymptotics} consists of an asymptotic analysis in the limit $\epsilon \to \infty$, with $\Pe=O(1)$. In \S \ref{premixed1:sec:activation} we consider the limit of infinite activation energy, $\beta \to \infty$, in order to provide an analytical description of the propagation speed in terms of $\Pe$. In \S \ref{premixed1:sec:results} we expand and discuss the results of the preceding asymptotic analyses and compare with numerical solutions of the governing equations, with particular emphasis on describing the relationship between the effective propagation speed $U_T$ and Peclet number $\Pe$ for several values of the flame front thickness $\epsilon$ and activation energy $\beta$. Finally, a summary of the main findings is given in \S \ref{premixed1:sec:conc}.
\section{Formulation}
\label{premixed1:sec:formulation}
\begin{figure}
\centering
\includegraphics[scale=0.75,trim=0 15 0 0, clip=true]{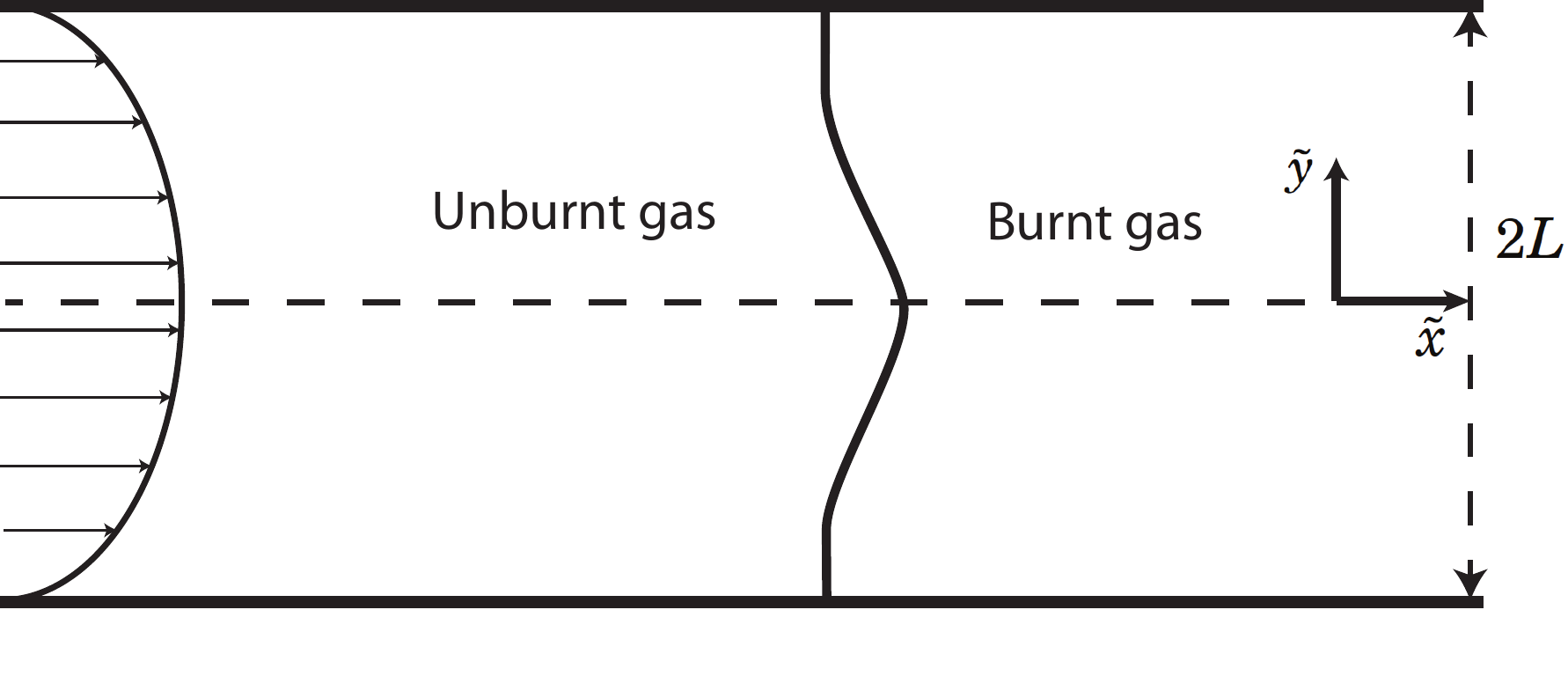}
\caption{An illustration of a premixed flame propagating against a Poiseuille flow in a channel of height $2L$.}
\label{fig:diagram}
\end{figure}
Consider a premixed flame propagating through a channel of height $2L$. Far upstream of the flame a fully developed Poiseuille flow, defined by 
\begin{gather*}
\tilde{u}=\tilde{A}\left(1-\frac{\tilde{y}^2}{L^2}\right),
\end{gather*}
is prescribed (see figure \ref{fig:diagram}). The governing
equations at low Mach number are given by the Navier--Stokes equations coupled
to equations for temperature and mass fractions, along with the ideal gas equation of state. The fluid velocity is given by $\left(\tilde{u},\tilde{v}\right)$. The combustion is modelled as a
single irreversible one-step reaction of the form
\begin{align*}
\text{F} \to \text{Products}+q,
\end{align*}
where F (assumed to be the deficient reactant) denotes the fuel and $q$ the heat released per unit mass of fuel.\\
The overall reaction rate $\tilde{\omega}$ is taken to follow an Arrhenius law of the form
\begin{align*}
\tilde{\omega}=\tilde{\rho} B Y_F \exp{(-
E/R\tilde{T})}.
\end{align*}
Here $\tilde{\rho}$, $Y_F$, $R$, $\tilde{T}$, $B$ and $E$ are the density, the fuel mass fraction, the
universal gas constant, the temperature, the pre-exponential factor and the activation
energy of the reaction, respectively. The flame propagates through the channel in the $\tilde{x}$-direction
with velocity $-\tilde{U}\mathbf{i}$, where $\tilde{U}$ is an eigenvalue to be determined as part of the
solution to the problem.\\
With tilda denoting dimensional quantities, scaled non-dimensional variables are introduced using
\begin{gather*}
x=\frac{\tilde{x}}{ L},\quad y=\frac{\tilde{y}}{ L},\quad u=\frac{\tilde{u}}{S_L^0},\quad v=\frac{\tilde{v}}{ S_L^0 },\\ t=\frac{\tilde{t}}{L\left/ S_L^0 \right.},\quad \theta=\frac{\tilde{T}-\tilde{T}_u}{\tilde{T}_{ad}-\tilde{T}_u},\quad y_F=\frac{Y_F}{ Y_{Fu}},\quad  p=\frac{\tilde{p}}{ \tilde{\rho}_u \left(S_L^0\right)^2}.
\end{gather*}
The unit speed is taken to be
\begin{gather*}
S_L^0=\left(2 \Le \beta^{-2}\left(1-\alpha\right)D_T B \exp\left(E\left/ R \tilde{T}_{ad}\right.\right)\right)^{1/2},
\end{gather*}
which is the laminar burning speed of the planar flame for $\beta \gg 1$. Here $\tilde{T}_{ad}\equiv \tilde{T}_u+q Y_{Fu}/c_P$ is the adiabatic flame temperature, $\beta \equiv E\left(\tilde{T}_{ad}-\tilde{T}_u\right)\left/ R\tilde{T}_{ad}^2 \right. $ is the Zeldovich number or non-dimensional activation energy and $\alpha \equiv \left(\tilde{T}_{ad}-\tilde{T}_u\right)\left/ \tilde{T}_{ad}\right.$ is the thermal expansion coefficient. The quantities $\tilde{T}_u$, $Y_{Fu}$, and $\tilde{\rho}_u$ denote the values of the temperature, fuel mass fraction and density in the unburnt gas as $\tilde{x} \to -\infty$, respectively.\\
In non-dimensional form the governing equations in a coordinate system attached to the flame front, which is travelling in the negative $x$-direction at speed $U=\tilde{U}\left/ S_L^0 \right.$, are given by
\begin{gather}
\frac{\partial \rho}{\partial t}+\nabla \cdot \rho \hat{\mathbf{u}}=0,\label{nondim1}\\
\rho \frac{\partial \mathbf{u}}{\partial t}+\rho\hat{\mathbf{u}}\cdot\nabla\mathbf{u} + \nabla p = \epsilon Pr\left(\nabla^2 \mathbf{u}+\frac{1}{3}\nabla\left(\nabla\cdot\mathbf{u}\right)\right),\\
\rho \frac{\partial \theta}{\partial t}+ \rho\hat{\mathbf{u}}\cdot\nabla\theta =\epsilon \nabla^2 \theta + \frac{\epsilon^{-1} \omega}{1-\alpha},\label{nondim4}\\
\rho \frac{\partial y_F}{\partial t}+\rho\hat{\mathbf{u}}\cdot\nabla y_F = \frac{\epsilon}{ \Le} \nabla^2 y_F - \frac{\epsilon^{-1} \omega}{1-\alpha},\label{nondim7}\\
\rho=\left(1+\frac{\alpha}{1-\alpha}\theta\right)^{-1},\label{nondim8}
\end{gather}
assuming that the thermal diffusivity $D_T$ and the fuel mass diffusion coefficient $D_F$ satisfy $\tilde{\rho}D_T=\tilde{\rho}D_F=\text{constant}$. Here $\hat{\mathbf{u}}=\mathbf{u}+U\mathbf{i}$ and $p$ is the hydrodynamic pressure.\\
The walls located at $y=-1$ and $y=1$ are considered to be rigid and adiabatic. Symmetry conditions are applied at $y=0$. The boundary conditions are therefore given by
\begin{gather}
\pd{\theta}{y}=\pd{y_F}{y}=\pd{u}{y}=v=\pd{p}{y}=0 \quad \text{at } y=0, \label{bc:1}\\
\pd{\theta}{y}=\pd{y_F}{y}=u=v=0 \quad \text{at } y=1,\label{bc:2}\\
\theta=0,\quad y_F=1,\quad u=A\left(1-y^2\right)=\epsilon \Pe \left(1-y^2\right),\nonumber\\ \quad v=0\quad \text{at } x=-\infty, \label{bc:minusinfty}\\
\pd{\theta}{x}=\pd{y_F}{x}=\pd{u}{x}=\pd{v}{x}=p=0 \quad \text{at } x=+\infty \label{bc:last},
\end{gather}
along with suitable initial conditions.
The non-dimensional parameters are defined as
\begin{gather*}
\epsilon=\frac{\delta_{L}}{L}=\frac{D_T \left/ S_L^0 \right.}{L}, \quad \Pe=\frac{A}{\epsilon},\\
\Le=\frac{D_T}{D_F}, \quad \text{and}\quad Pr=\frac{\nu}{D_T},
\end{gather*}
which are the non-dimensional flame-front thickness, the Peclet number, the fuel Lewis number and the Prandtl number, respectively. Here $\nu$ is the kinematic viscosity $\nu=\mu \left/ \tilde{\rho}_u \right.$. Note that $\delta_L$ is the dimensional flame-front thickness given by $\delta_L=D_T\left/ S^0_L\right.$ and $A$ is the non-dimensional amplitude of the imposed Poiseuille flow, $A=\tilde{A}\left/ S_L^0 \right. $. Finally, the non-dimensional reaction rate $\omega$ is defined as
\begin{gather}
\omega=\frac{\beta^2}{2 \Le}\rho y_F\exp\left(\frac{\beta\left(\theta-1\right)}{1+\alpha\left(\theta-1\right)}\right).\label{eq:reaction2}
\end{gather}
The problem is now fully formulated and is given by equations (\ref{nondim1})-(\ref{nondim8}), with boundary conditions (\ref{bc:1})-(\ref{bc:last}). The non-dimensional parameters in this problem are $\Pe$, $\epsilon$, $\beta$, $\alpha$, $Pr$ and $\Le$.

Note that by integrating the steady form of equation (\ref{nondim4}) over the whole domain, using the continuity equation (\ref{nondim1}) with the boundary conditions (\ref{bc:1})-(\ref{bc:last}) and assuming total fuel consumption downstream, we find
\begin{gather}
U+\bar{u}=\int^1_0 \int^\infty_{-\infty} \frac{\epsilon^{-1} \omega}{1-\alpha} \mathrm{d}x \mathrm{d}y,
\end{gather}
where $\bar{u}$ is the mean speed of the parallel inflow at $x=-\infty$. Therefore
\begin{gather}
U_T \equiv U+\bar{u}  \label{effective}
\end{gather}
appears as an effective propagation speed, as commonly defined in turbulent combustion. In the current study of a Poiseuille flow in a rectangular channel, using the boundary condition (\ref{bc:minusinfty}), the effective propagation speed is given by
\begin{gather*}
U_T\equiv U+\frac{2}{3}\epsilon \Pe.
\end{gather*}
\section{Asymptotic analysis in the limit $\epsilon \to \infty$}
\label{premixed1:sec:asymptotics}
To simplify the problem we consider the steady form of equations \eqref{nondim1}-\eqref{nondim8} with unity Lewis number
\begin{gather}
\Le=1.
\end{gather}
In this case only the equation for temperature needs to be considered, since $y_F=1-\theta$. This follows from adding equations (\ref{nondim4}) and (\ref{nondim7}) and using boundary conditions (\ref{bc:minusinfty}).

We now consider the limit $\epsilon \to \infty$ with $\Pe=O(1)$, $Pr=O(1)$ and $\beta=O(1)$. The flow amplitude $A=O\left(\epsilon\right)$ for $\Pe=O(1)$. We introduce a rescaled coordinate
\begin{gather}
\xi=\frac{x}{\epsilon},\label{eq:scaling}
\end{gather}
so that the governing equations (\ref{nondim1})-(\ref{nondim8}) become
\begin{gather}
\frac{\partial}{\partial \xi}\left(\rho(u+U)\right)+\epsilon\frac{\partial}{\partial y}\left( \rho v\right)=0,\label{2:nondim1}\\
\rho(u+U)\frac{\partial u}{\partial \xi} +\epsilon \rho v \frac{\partial u}{\partial y} +\frac{\partial p}{\partial \xi} = Pr\left( \frac{4}{3}\sd{u}{\xi}+\epsilon^2\sd{u}{y}+\frac{\epsilon}{3}\frac{\partial^2v}{\partial\xi\partial y}\right),\\
\rho(u+U)\frac{\partial v}{\partial \xi} +\epsilon \rho v \frac{\partial v}{\partial y} +\epsilon\frac{\partial p}{\partial y} = Pr\left( \sd{v}{\xi}+\frac{4\epsilon^2}{3}\sd{v}{y}+\frac{\epsilon}{3}\frac{\partial^2u}{\partial\xi\partial y}\right),\\
\rho(u+U)\frac{\partial \theta}{\partial \xi} + \epsilon\rho v \frac{\partial \theta}{\partial y} = \sd{\theta}{\xi}+\epsilon^2 \sd{\theta}{y} + \frac{ \omega}{1-\alpha},\label{2:nondim4}\\
\rho=\left(1+\frac{\alpha}{1-\alpha}\theta\right)^{-1},\label{2:idealgas}
\end{gather}
where
\begin{gather*}
\omega=\frac{\beta^2}{2}\rho \left(1-\theta \right)\exp\left(\frac{\beta\left(\theta-1\right)}{1+\alpha\left(\theta-1\right)}\right).
\end{gather*}
These equations are subject to the boundary conditions (\ref{bc:1}) and (\ref{bc:2}), with
\begin{gather}
\theta=0,\quad u=A\left(1-y^2\right)=\epsilon \Pe \left(1-y^2\right), \quad v=0\quad \text{at } \xi=-\infty,\label{bc2:3}\\
\pd{\theta}{\xi}=\pd{u}{\xi}=\pd{v}{\xi}=p=0 \quad \text{at } \xi=+\infty. \label{bc2:last}
\end{gather}
We now introduce expansions for $\epsilon \to \infty$ in the form
\begin{equation}
\left.
\begin{aligned}
&U=-\frac{2}{3}\epsilon \Pe+U_0+\epsilon^{-1}U_1+...\\  &u=\epsilon u_0 +u_1+...\quad v=v_0+\epsilon^{-1}v_1+...\\&\theta=\theta_0+\epsilon^{-1}\theta_1+...\quad p=\epsilon^3 p_0+\epsilon^2 p_1+...
\end{aligned}
\right\}
\label{premixed1:eq:expansions}
\end{equation}
Note that here $U_0$ is the leading order approximation to the effective flame speed $U_T$, defined in (\ref{effective}). Note also that the horizontal velocity component $u$ is $O\left(\epsilon\right)$, due to the imposed Poiseuille flow at $\xi=-\infty$ given by (\ref{bc2:3}), while the vertical component of the velocity $v$ is taken to be $O(1)$ to balance the two terms in the continuity equation (\ref{2:nondim1}).

Substituting (\ref{premixed1:eq:expansions}) into equations (\ref{2:nondim1})-(\ref{2:nondim4}), we obtain to leading order
\begin{gather}
\pd{}{\xi}\left(\rho_0\left(u_0-\frac{2}{3}\Pe\right)\right)+\pd{}{y}\left(\rho_0v_0\right)=0\label{leading1},\\
\pd{p_0}{\xi}=Pr\sd{u_0}{y},\label{leading2} \\
\pd{p_0}{y}=0,\label{leading3}\\
\sd{\theta_0}{y}=0.\label{leading4}
\end{gather}
Equations (\ref{leading3}) and (\ref{leading4}) can be integrated with respect to $y$ to give $p_0=p_0(\xi)$ and $\theta_0=\theta_0(\xi)$, after using the boundary condition (\ref{bc:1}) on $\theta_0$, so that $\rho_0=\rho_0(\xi)$ from (\ref{2:idealgas}).\\
Now, using a similar method to \citet{short2009asymptotic}, we look for a separable solution for  $u_0(\xi,y)$ in the form
\begin{gather}
u_0(\xi,y)=\hat{u}_0(y)\check{u}_0(\xi).\label{splitveloc}
\end{gather}
Substitution of (\ref{splitveloc}) into equation (\ref{leading2}) gives
\begin{gather}
\sd{\hat{u}_0}{y}=\frac{1}{\check{u}_0 Pr}\pd{p_0}{\xi}=-2C, \label{splitveloc2}
\end{gather}
where $C$ is a constant. Equation (\ref{splitveloc2}) can be integrated twice with respect to $y$, using the boundary conditions (\ref{bc:1}) and (\ref{bc:2}), to yield
\begin{gather*}
\hat{u}_0(y)=C(1-y^2),
\end{gather*}
so that
\begin{gather*}
u_0(\xi,y)=\check{u}_0(\xi)(1-y^2),
\end{gather*}
where $C$ has been absorbed into $\check{u}_0(\xi)$.\\
Integrating equation (\ref{leading1}) with respect to $y$ from $y=0$ to $y=1$, we obtain
\begin{gather}
\pd{}{\xi}\left(\rho_0(\xi)\left(\frac{2}{3}\check{u}_0(\xi)-\frac{2}{3}\Pe\right)\right)=0,\label{leadcont}
\end{gather}
after using boundary conditions (\ref{bc:1})-(\ref{bc:2}) on $v_0$. Equation (\ref{leadcont}) implies that
\begin{gather*}
\frac{2}{3}\rho_0(\xi)\left(\check{u}_0(\xi)-\Pe\right)=\frac{2}{3}\left(\check{u}_0\left(\xi \to -\infty\right)-\Pe\right)=0,
\end{gather*}
using the fact that $\rho_0\left(\xi \to -\infty\right)=1$ from equation (\ref{2:idealgas}) and boundary condition (\ref{bc2:3}). Thus $\check{u}_0(\xi)=\Pe$, so that
\begin{gather}
u_0=\Pe(1-y^2).\label{leadingu}
\end{gather}
The continuity equation (\ref{leading1}) can then be integrated with respect to $y$, using (\ref{leadingu}) and condition (\ref{bc:1}), to yield
\begin{gather}
v_0=-\frac{1}{\rho_0}\pd{\rho_0}{\xi}\frac{\Pe}{3}\left(y-y^3\right).\label{leadingv}
\end{gather}
Now, at $O\left(\epsilon\right)$ in equation (\ref{2:nondim4}) we have
\begin{gather}
\rho_0\left(u_0-\frac{2}{3}\Pe\right)\pd{\theta_0}{\xi}=\sd{\theta_1}{y},
\end{gather}
which, after using (\ref{leadingu}) and condition (\ref{bc:1}), can be integrated twice with respect to $y$ to give
\begin{gather}
\theta_1=\rho_0\pd{\theta_0}{\xi}\Pe\left(\frac{y^2}{6}-\frac{y^4}{12}\right)+\check{\theta}_1(\xi).\label{theta_1}
\end{gather}
Next we look to $O\left(1\right)$ in equation (\ref{2:nondim1}) to find
\begin{gather}
\pd{}{\xi}\left(\rho_1\left(u_0-\frac{2}{3}\Pe\right)\right)+\pd{}{\xi}\left(\rho_0\left(u_1+U_0\right)\right)+\pd{}{y}\left(\rho_0v_1\right)+\pd{}{y}\left(\rho_1v_0\right)=0.\label{order1}
\end{gather}
Equation (\ref{order1}) can be integrated first with respect to $y$ from $y=0$ to $y=1$, utilising the boundary conditions (\ref{bc:1})-(\ref{bc:2}) on $v_0$, and then with respect to $\xi$ to give
\begin{gather}
\int_0^1\left(\rho_1\left(u_0-\frac{2}{3}\Pe\right)\right)\mathrm{d}y+\int_0^1\left(\rho_0\left(u_1+U_0\right)\right)\mathrm{d}y=K.\label{cont1}
\end{gather}
To evaluate $K$, we use boundary conditions (\ref{bc2:3}) to obtain
\begin{gather}
K=\int_0^1\left(\rho_1(\xi \to -\infty)\left(u_0(\xi \to -\infty)-\frac{2}{3}\Pe\right)\right)\mathrm{d}y~+\nonumber\\\int_0^1\left(\rho_0(\xi \to -\infty)\left(u_1(\xi \to -\infty)+U_0\right)\right)\mathrm{d}y=U_0.\label{cont2}
\end{gather}
Finally, at $O\left(1\right)$ of equation (\ref{2:nondim4}) we have
\begin{gather}
\rho_0\left(u_1+U_0\right)\pd{\theta_0}{\xi}+\rho_1\left(u_0-\frac{2}{3}\Pe\right)\pd{\theta_0}{\xi}+\rho_0\left(u_0-\frac{2}{3}\Pe\right)\pd{\theta_1}{\xi}+\rho_0v_0\pd{\theta_1}{y}=\nonumber\\\sd{\theta_0}{\xi}+\sd{\theta_2}{y}+\frac{\omega_0}{1-\alpha},\label{theta_order1}
\end{gather}
where $\omega_0(\xi)=\omega\left(\theta_0,\rho_0\right)$. Integrating (\ref{theta_order1}) with respect to $y$ from $y=0$ to $y=1$, taking into account the boundary conditions (\ref{bc:1})-(\ref{bc:2}) on $\theta$ and substituting (\ref{leadingu}), (\ref{leadingv}), (\ref{theta_1}), (\ref{cont1}) and (\ref{cont2}), we obtain
\begin{gather}
U_0\pd{\theta_0}{\xi}-\pd{}{\xi}\left(\left(1+\frac{8}{945}\Pe^2\rho_0^2\right)\pd{\theta_0}{\xi}\right)=\frac{\omega_0}{1-\alpha} \label{result:finitebeta},
\end{gather}
with
\begin{equation}
\left.
\begin{aligned}
&\rho_0=\left(1+\frac{\alpha}{1-\alpha}\theta_0\right)^{-1}, \\ 
&\omega_0=\frac{\beta^2}{2}\rho_0\left(1-\theta_0\right)\exp\left(\frac{\beta \left(\theta_0-1\right)}{1+\alpha\left(\theta_0-1\right)}\right),
\end{aligned}
\right\}
\label{res}
\end{equation}
subject to the boundary conditions
\begin{gather}
\theta_0(-\infty)=0, \quad \theta_{0\xi}(+\infty)=0. \label{res:bc}
\end{gather}
This shows that in the limit $\epsilon \to \infty$, with $\Pe=O(1)$, the problem of a variable density premixed flame in a two dimensional channel can be reduced to a one dimensional boundary value problem. Equation (\ref{result:finitebeta}) is the equation that would describe a premixed flame propagating through a one-dimensional channel with an effective diffusion coefficient
\begin{gather}
D_{\text{eff}}=D_T\left(1+\frac{8}{945}\Pe^2\frac{\tilde{\rho}^2}{\tilde{\rho}^2_u}\right). \label{equation:deffective}
\end{gather}
This is an important result because it corresponds to a generalised form (accounting for variable density effects) of the effective diffusion coefficient found when studying the effect of a Poiseuille flow on mixing in the non-reactive Taylor dispersion problem, originally investigated by \citet{taylor1953dispersion}. A premixed flame in the limit $\epsilon\to\infty$, with $\Pe=O(1)$ can be therefore considered to be in the Taylor dispersion regime.

The boundary value problem (\ref{result:finitebeta})-(\ref{res:bc}) will be solved numerically in \S \ref{premixed1:sec:results} to provide a description of the relationship between $U_T$ and $\Pe$. These results will also be compared to those of numerical solutions of the full problem. Firstly, however, we will proceed to study the limit $\beta\to\infty$ in order to find a leading order asymptotic solution for $U_T$.

\section{Explicit solution for large activation energy $\beta \to \infty$}
\label{premixed1:sec:activation}
Here we consider the solution to the problem (\ref{result:finitebeta})-(\ref{res:bc}) in the limit of infinite activation energy $\beta \to \infty$. Following a well known approach in this limit, the reaction is confined to a thin layer of thickness $O\left(\beta^{-1}\right)$. The domain can therefore be split into two outer zones (which we refer to as the preheat zone and the burnt gas) and an inner zone (the reaction zone). 
We use the condition $\theta_0\left(\xi \to \infty\right)=1$, which follows from the total completion of the reaction far downstream.

In the outer zones the reaction rate is set to zero so that, from (\ref{result:finitebeta}),
\begin{gather}
U_0\ed{\theta_0}{\xi}-\ed{}{\xi}\left(\left(1+\frac{8}{945}\Pe^2\rho_0^2\right)\ed{\theta_0}{\xi}\right)=0, \label{outerode},
\end{gather}
where $\rho_0=\rho\left(\theta_0\right)$. The solution to this equation in the burnt gas is found to be
\begin{gather}
\theta_0=1, \label{burntsol}
\end{gather}
while in the preheat zone we have, denoting $P=8\Pe^2\left/ 945\right.$ and $\tilde{\alpha}=\alpha\left/ \left(1-\alpha\right)\right.$,
\begin{gather}
\frac{\left(U_0\theta_0\right)^{\left(P+1\right)/U_0}}{\left(1+\tilde{\alpha}\theta_0\right)^{P/U_0}}\exp\left(\frac{P}{U_0\left(1+\tilde{\alpha}\theta_0\right)}\right)=C_1 \exp\left(\xi\right), \label{unburntsol1}
\end{gather}
on using the condition (\ref{bc2:3}). The constant $C_1$ can be determined by choosing the origin where the outer profiles intersect, since the problem is translationally invariant in the $\xi$-direction. This gives
\begin{gather*}
C_1=(1+\tilde{\alpha})^{-P/U_0} \exp\left({\frac{P}{(1+\tilde{\alpha}) U_0}}\right) U_0^{\left(1+P\right)/U_0}.
\end{gather*}
The propagation speed $U_0$  can now be determined by matching with the inner solution. Since the reaction layer is of thickness $O\left(\beta^{-1}\right)$, in the inner region we let
\begin{gather}
X=\frac{\xi}{\beta^{-1}},\quad \theta_0^{\text{inner}}=\Theta\left(X\right)=1+\beta^{-1}\Theta^1+O\left(\beta^{-2}\right). \label{innerexpansion}
\end{gather}
Then to leading order we have
\begin{gather}
\Theta^1_{XX}-\Lambda \Theta^1\exp\left(\Theta^1\right)=0,\label{theta1govern}
\end{gather}
where
\begin{gather*}
\Lambda=\left(2 \left(1+\tilde{\alpha}\right)\left(1+\frac{P}{\left(1+\tilde{\alpha}\right)^2}\right) \sqrt{1-\alpha}\right)^{-1}.
\end{gather*}
The boundary conditions to equation (\ref{theta1govern}) are found by matching with the outer solutions using the formula
\begin{gather}
\theta_0^{\text{inner}}\left(X \to \pm \infty\right)=\theta_0^{\text{outer}}\left(\xi \to 0^{\pm}\right). \label{matching1}
\end{gather}
Matching with the solution in the burnt gas, given by (\ref{burntsol}), yields
\begin{gather}
\Theta^1\left(X \to +\infty\right)=0. \label{theta1bc1}
\end{gather}
Now, noting from (\ref{innerexpansion}) and (\ref{matching1}) that
\begin{gather*}
\theta_0\left(\xi \to 0^-\right)=1+\beta^{-1}\Theta^1\left(X\to-\infty\right)+O\left(\beta^{-2}\right),
\end{gather*}
we expand equation (\ref{unburntsol1}) for the temperature in the unburnt gas as $\xi \to 0$ to find
\begin{gather}
\Theta^1\left(X\to-\infty\right)=\frac{\left(1+\tilde{\alpha}\right)^2U_0}{\left(1+\tilde{\alpha}\right)^2+P}X.\label{theta1bc2}
\end{gather}
Now $U_0$ can be found by integrating equation (\ref{theta1govern}) subject to the boundary conditions (\ref{theta1bc1}) and (\ref{theta1bc2}).
Multiplying (\ref{theta1govern}) by $\Theta^1_X$ and integrating with respect to $X$ from $X=-\infty$ to $X=+\infty$ yields
\begin{gather}
\left[\frac{\left(\Theta^1_X\right)^2}{2}\right]^{X=+\infty}_{X=-\infty}=\int^{\Theta^1\left(+\infty\right)}_{\Theta^1\left(-\infty\right)}\Lambda \Theta^1\exp\left(\Theta^1\right) \mathrm{d}\Theta^1.\label{eq:innertheta1}
\end{gather}
Thus, using (\ref{theta1bc1}) and (\ref{theta1bc2}),
\begin{gather}
\left(\frac{\left(1+\tilde{\alpha}\right)^2U_0}{\left(1+\tilde{\alpha}\right)^2+P}\right)^2=2 \Lambda,
\end{gather}
so that
\begin{gather}
U_0=\sqrt{1+\frac{8}{945}\Pe^2\left(1-\alpha\right)^2} \label{result:infinitebeta}.
\end{gather}
This equation gives the leading order approximation $U_0$ to the effective flame speed $U_T$ for a given value of $\Pe$ in the limit $\epsilon\to\infty$, $\beta\to\infty$, with $\Pe=O(1)$.
\subsection{Constant density results}
The asymptotic results found for a variable density premixed flame can be similarly derived in the simpler case of a constant density premixed flame. The constant density form of the boundary value problem (\ref{result:finitebeta})-(\ref{res:bc}), derived in the limit $\epsilon\to\infty$ with $\Pe=O(1)$, is
\begin{gather}
U_0\pd{\theta_0}{\xi}-\left(1+\frac{8}{945}\Pe^2\right)\sd{\theta_0}{\xi}=\omega_0 \label{const:finitebeta},
\end{gather}
with
\begin{gather}
\omega_0=\frac{\beta^2}{2}\left(1-\theta_0\right)\exp\left(\frac{\beta \left(\theta_0-1\right)}{1+\alpha\left(\theta_0-1\right)}\right),\label{const:res}
\end{gather}
subject to the boundary conditions
\begin{gather}
\theta_0(-\infty)=0, \quad \theta_{0\xi}(+\infty)=0. \label{const:bc}
\end{gather}
In the limit $\beta\to\infty$, this problem has the solution
\begin{gather}
U_0=\sqrt{1+\frac{8}{945}\Pe^2} \label{const:infinitebeta}.
\end{gather}
\subsection{Cylindrical channel results}
\label{premixed1:sec:cylindrical}
A similar asymptotic analysis to the one above can be performed (see Appendix \ref{appendix:premixed1}) for a premixed flame propagating through a cylindrical channel of diameter $2L$ with an imposed Poiseuille flow. As in the case of a rectangular channel it is found that the flame is governed by the equation for a planar premixed flame with an effective diffusion coefficient, in this case given by
\begin{gather*}
D_{\text{eff}}=D_T\left(1+\frac{1}{192}\Pe^2\frac{\tilde{\rho}^2}{\tilde{\rho}^2_u}\right),
\end{gather*}
in the variable density case and
\begin{gather*}
D_{\text{eff}}=D_T\left(1+\frac{1}{192}\Pe^2\right).
\end{gather*}
in the constant density case. Using the definition of the Peclet number the constant density result can be written in dimensional form as
\begin{gather}
D_{\text{eff}}=D_T\left(1+\frac{L^2 \bar{u}^2}{48 D_T^2}\right), \label{cylind}
\end{gather}
where $\bar{u}=\tilde{A}\left/ 2 \right.$ is the cross-sectional average of the imposed Poiseuille flow. The result (\ref{cylind}) is exactly the result (\ref{eq:taylor}) found by \citet{taylor1953dispersion} in his original paper.
\begin{figure}
\subfigure[Constant density results]{
\includegraphics[scale=0.7,trim=20 0 10 10, clip=true]{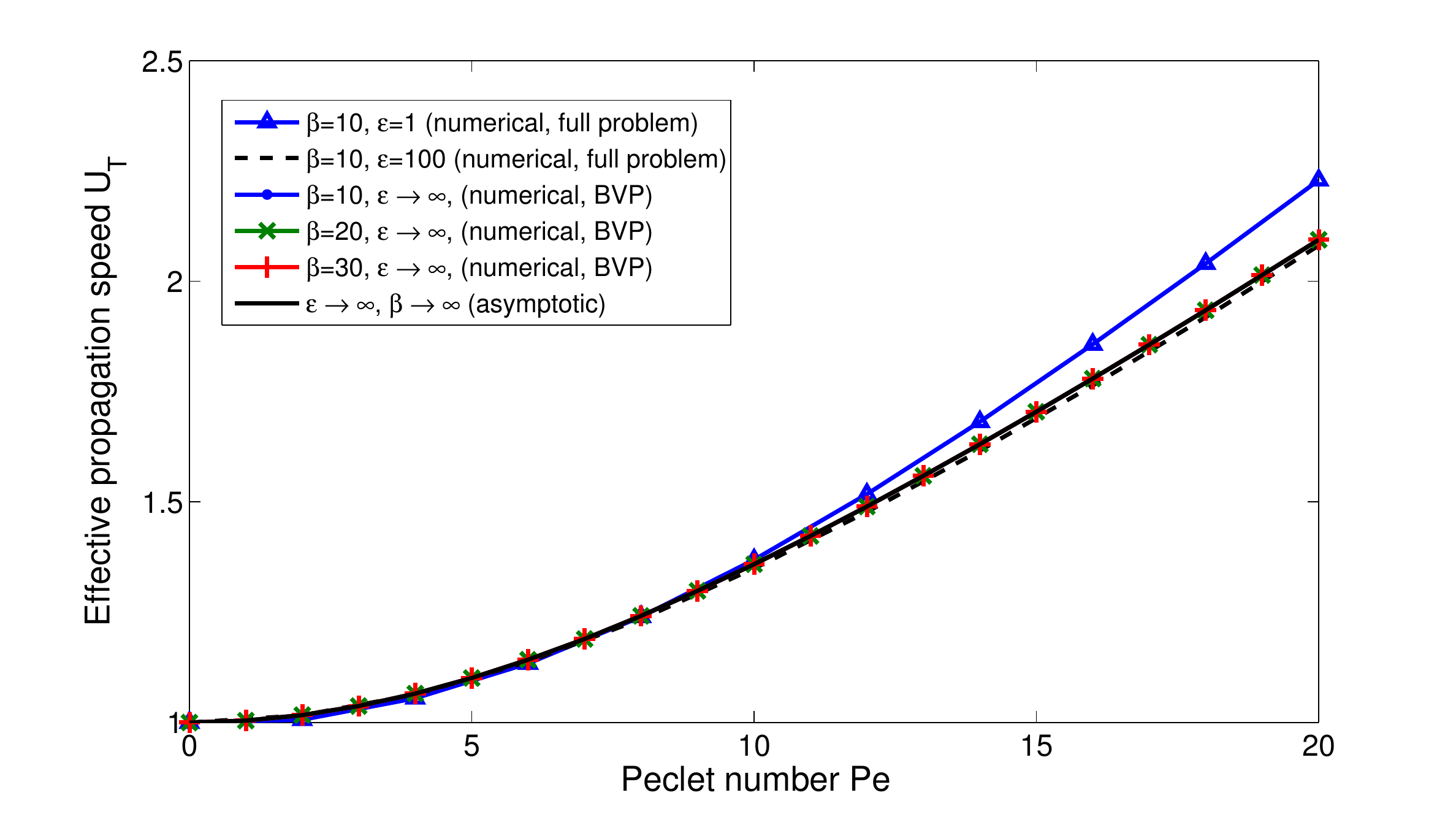}\label{fig:figure2}}
\subfigure[Variable density results]{
\includegraphics[scale=0.7,trim=20 0 10 10, clip=true]{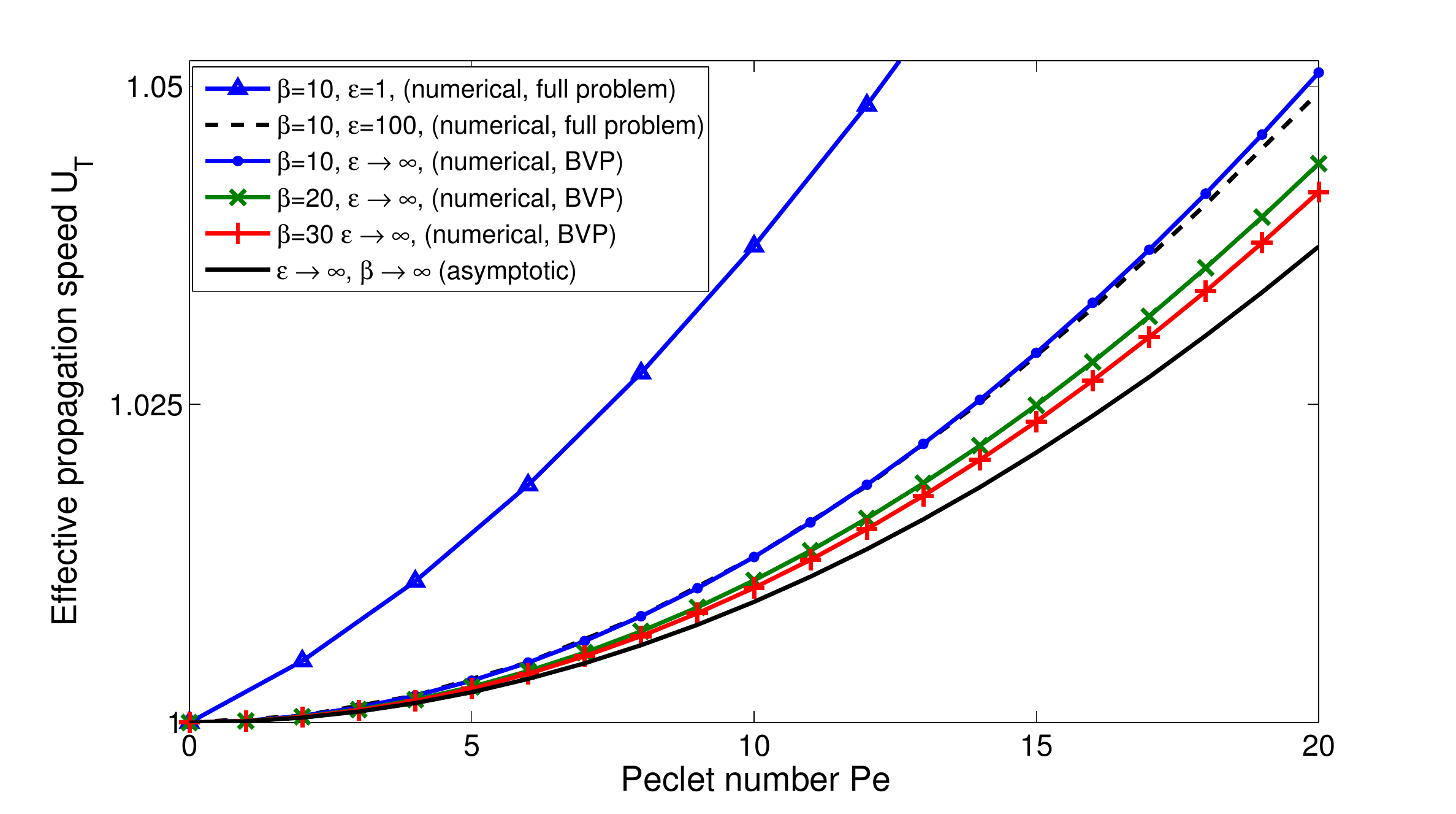}\label{fig:figure3}}
\caption{Summary of asymptotic and numerical results in a) constant density case and b) variable density case. Numerical simulations of the full system (\ref{nondim1})-(\ref{bc:last}) are performed for $\alpha=0.85$, $\beta=10$ and $Pr=1$. Numerical solutions of the boundary value problem  (\ref{result:finitebeta})-(\ref{res:bc}) are calculated for $\alpha=0.85$ and $\beta=10$.}
\end{figure}
\section{Further results and discussion}
\label{premixed1:sec:results}
In this section we compare the results of the asymptotic analyses undertaken in previous sections with the results of numerical computations. The main aim is to examine the relationship between the effective propagation speed $U_T$, defined in (\ref{effective}), and the Peclet number $\Pe$ for several values of the flame-front thickness $\epsilon$ and activation energy $\beta$, in both the variable density and constant density cases.

\subsection{Numerical procedure}
The numerical results are obtained by solving the steady form of equations (\ref{nondim1})-(\ref{nondim8}) with boundary conditions (\ref{bc:1})-(\ref{bc:last}) using the software package Comsol Multiphysics. This software has been extensively tested in combustion applications including our previous publications \cite[][]{pearce2013effect,pearce2013rayleigh}. The problem is entered into the partial differential equation (PDE) interface in Comsol, which uses a finite element discretisation to transform the set of non-linear PDEs into a set of non-linear algebraic equations in which the propagation speed $U_T$ appears as an additional unknown (eigenvalue). These equations, augmented by the requirement that the temperature is prescribed at the origin (an additional equation needed to determine the eigenvalue $U_T$), are then solved using an affine invariant form of the damped Newton method, as described by \citet{deuflhard1974modified}. In the constant density case we solve (\ref{nondim4}) with $u=\epsilon \Pe\left(1-y^2\right)$, $v=0$, $\rho=1$ and the reaction term replaced by $\epsilon^{-1}\omega$. The domain is covered by a non-uniform grid of approximately 200,000 triangular elements, with local refinement around the reaction zone. Various tests are performed to ensure the results are independent of the mesh. A channel of length $x=30\epsilon$ is taken to approximate an infinitely long channel. Throughout this section we let $\Le=1$, $Pr=1$ and $\alpha=0.85$ unless otherwise stated. The numerical calculations are performed for a fixed value of the activation energy, $\beta=10$, unless otherwise stated. For each $\Pe$ the value of $U_T$ is scaled by the value of $U$ calculated numerically for $\Pe=0$ and $\epsilon=1$. Finally the boundary value problems (\ref{result:finitebeta})-(\ref{res:bc}) and (\ref{const:finitebeta})-(\ref{const:bc}), derived in the limit $\epsilon\to\infty$, are solved using the BVP4C solver in Matlab, which uses a Lobatto IIIa method \cite[][]{shampine2000solving}.

\subsection{Comparison of the asymptotic and numerical results}
\begin{figure}
\subfigure[]{
\includegraphics[scale=0.8,trim=0 0 0 0, clip=true]{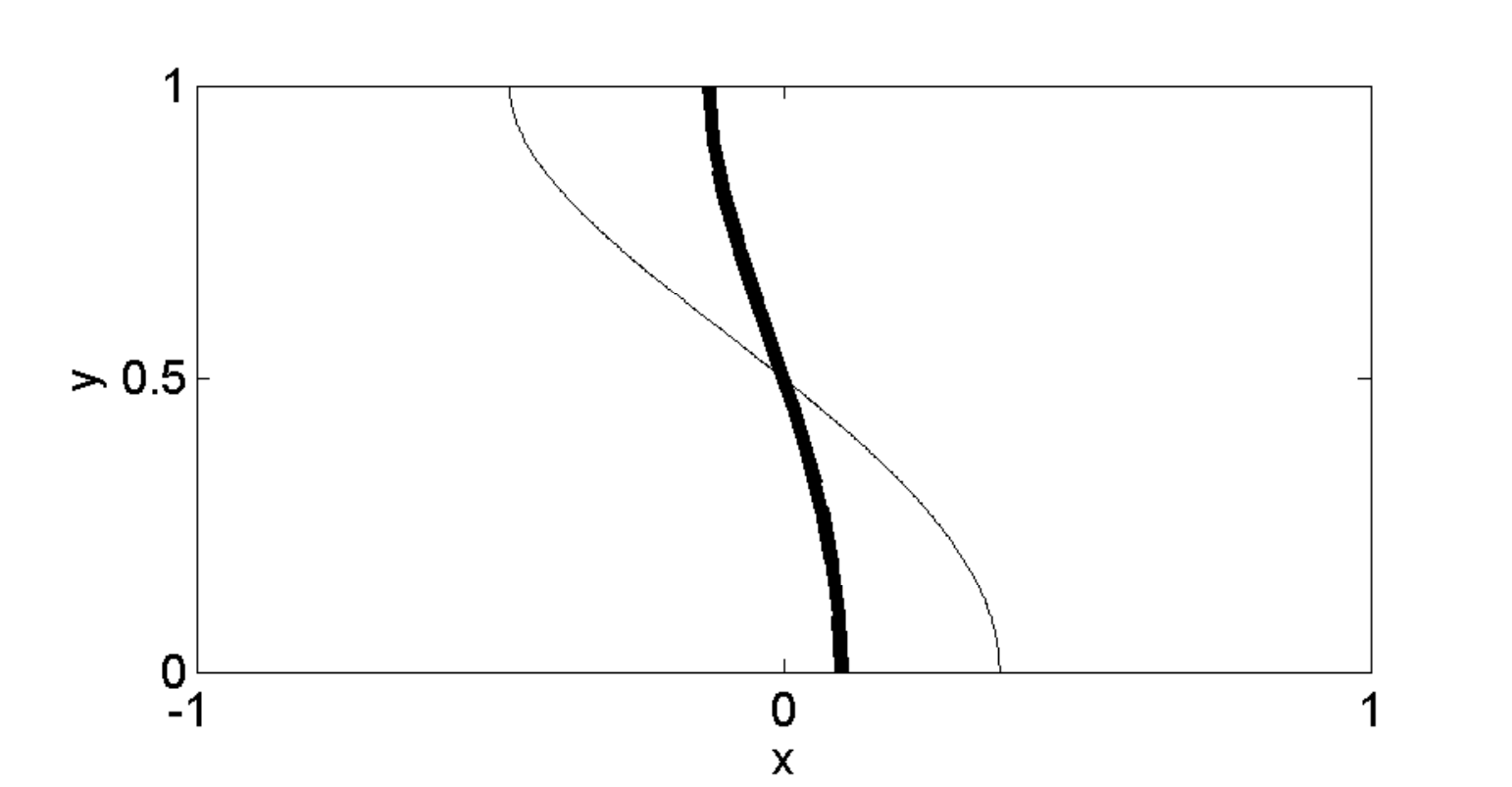}\label{fig:flameshape10}}
\subfigure[]{
\includegraphics[scale=0.8,trim=0 0 0 0, clip=true]{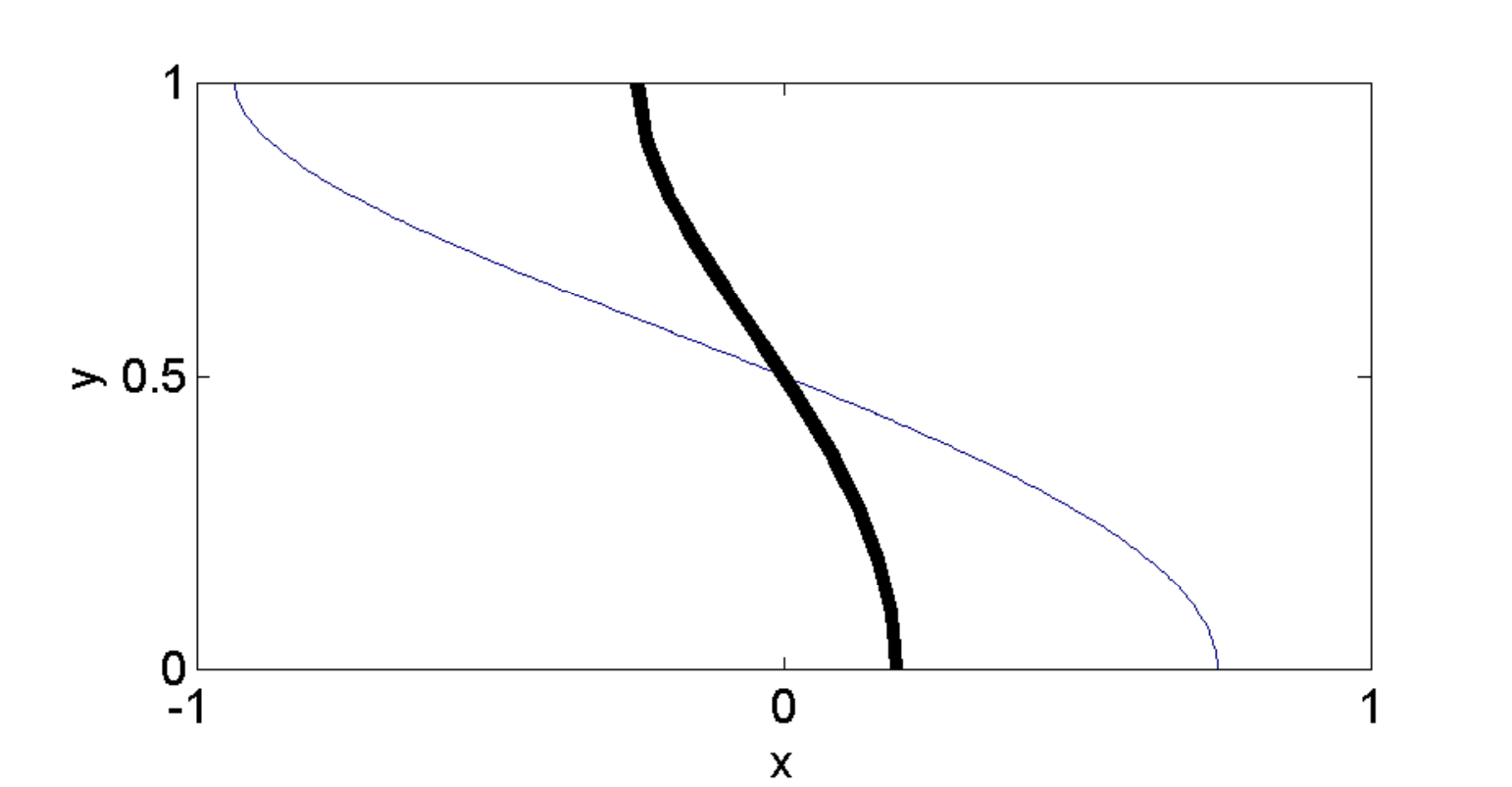}\label{fig:flameshape20}}
\caption{Flame shape (represented by the line $\theta=0.5$) for $\epsilon=10$, $\beta=10$  and a) $\Pe=10$, b) $\Pe=20$ in the constant density (thin line) and variable density (thick line) cases.}
\label{fig:shape}
\end{figure}
Figures \ref{fig:figure2} and \ref{fig:figure3} summarise both the asymptotic and numerical results in the constant density case and variable density case, respectively. Plotted are 1) the solutions to the boundary value problems (\ref{result:finitebeta})-(\ref{res:bc}) and (\ref{const:finitebeta})-(\ref{const:bc}), derived in the limit $\epsilon \to \infty$, $\Pe=O(1)$, for several values of $\beta$; 2) the asymptotic results (\ref{result:infinitebeta}) and (\ref{const:infinitebeta}), derived in the limit $\epsilon \to \infty$, $\beta \to \infty$, $\Pe=O(1)$; 3) the results of numerical solutions of the full problem, given by equations (\ref{nondim1})-(\ref{nondim8}) and boundary conditions (\ref{bc:1})-(\ref{bc:last}), for large values of both $\epsilon$ and $\beta$.

\begin{figure}
\subfigure[]{
\includegraphics[width=\textwidth,trim=40 0 0 5, clip=true]{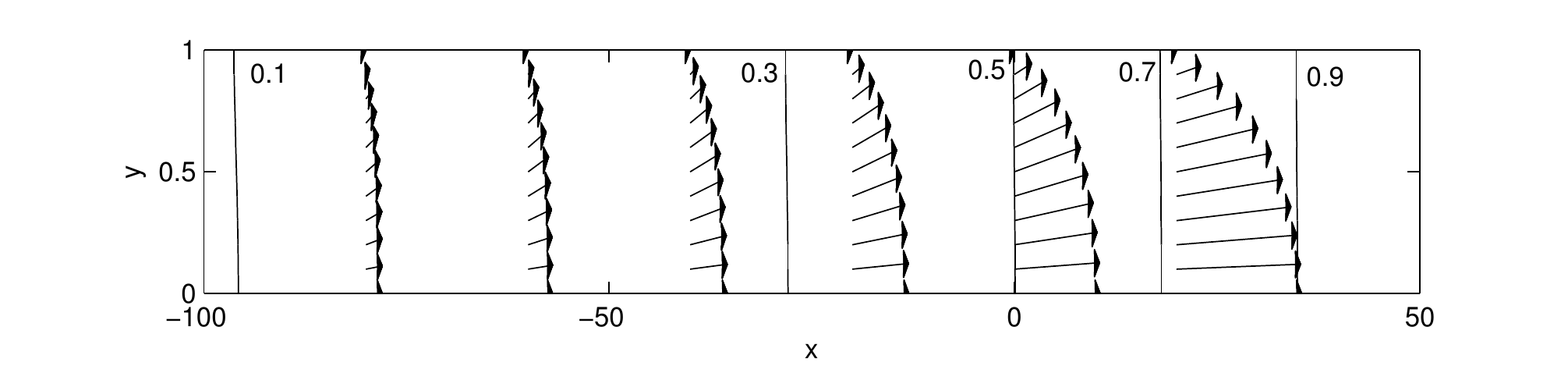}\label{fig:tempsurf}}\\ 
\subfigure[]{
\includegraphics[width=\textwidth,trim=40 0 0 5 clip=true]{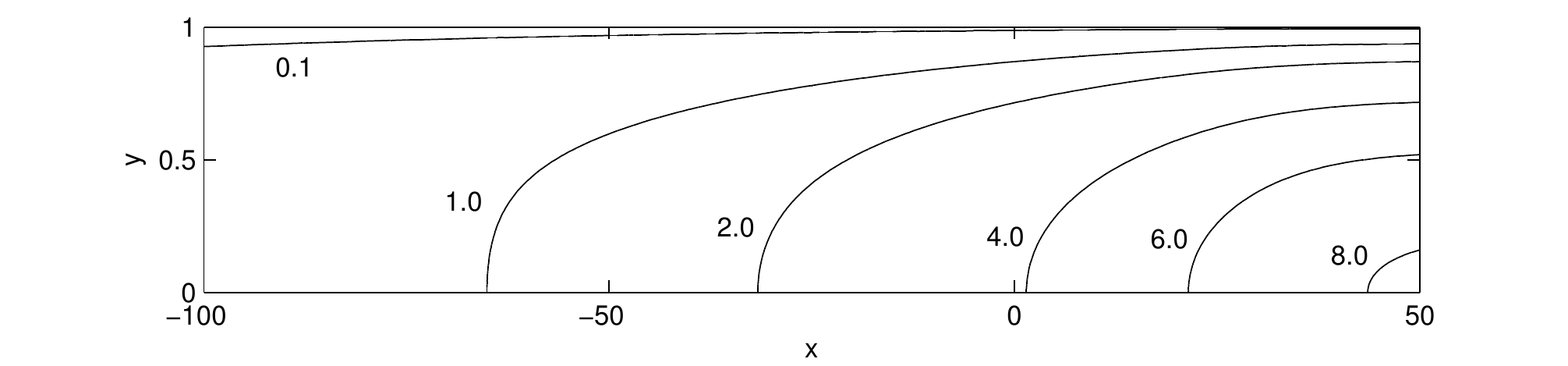}\label{fig:velocsurf}}
\subfigure[]{
\includegraphics[width=\textwidth,trim=40 0 0 5, clip=true]{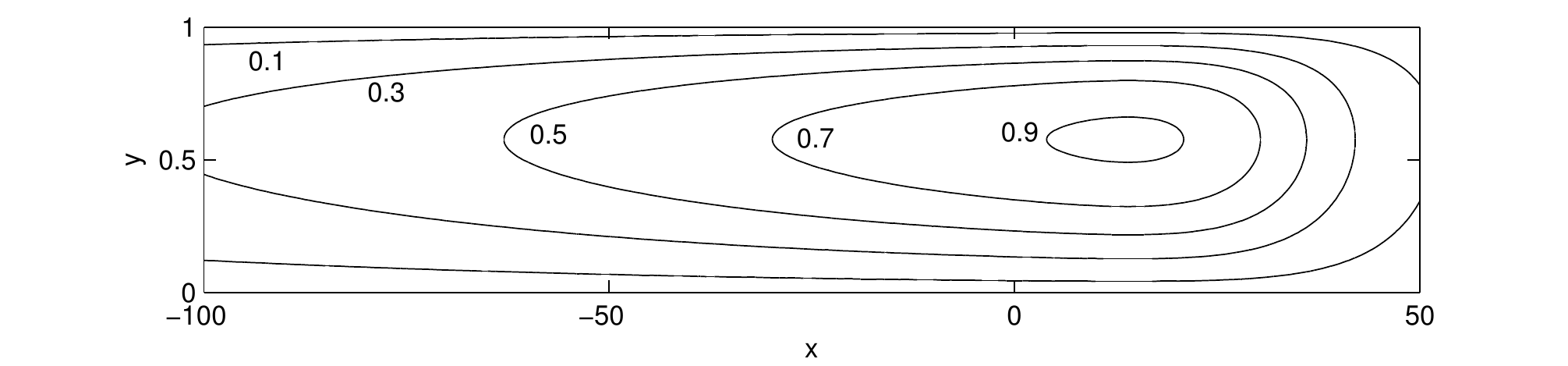}\label{fig:velocvsurf}}
\caption{a) Contour plot of the temperature $\theta$, with the velocity field induced by thermal expansion, which is given by $\left(u-\epsilon\Pe\left(1-y^2\right),v\right)$; b) contour plot of the horizontal velocity component due to thermal expansion, which is given by $u-\epsilon \Pe \left(1-y^2\right)$; c) contour plot of the vertical velocity component $v$. The plots correspond to $\epsilon=50$, $\Pe=10$ and $\beta=10$. The values of the quantities along each contour are indicated. The figures are plotted in the unscaled coordinates $(x,y)$.}
\label{fig:surfs}
\end{figure}

\begin{figure}
\subfigure[]{
\includegraphics[width=0.49\textwidth,trim=0 0 0 0, clip=true]{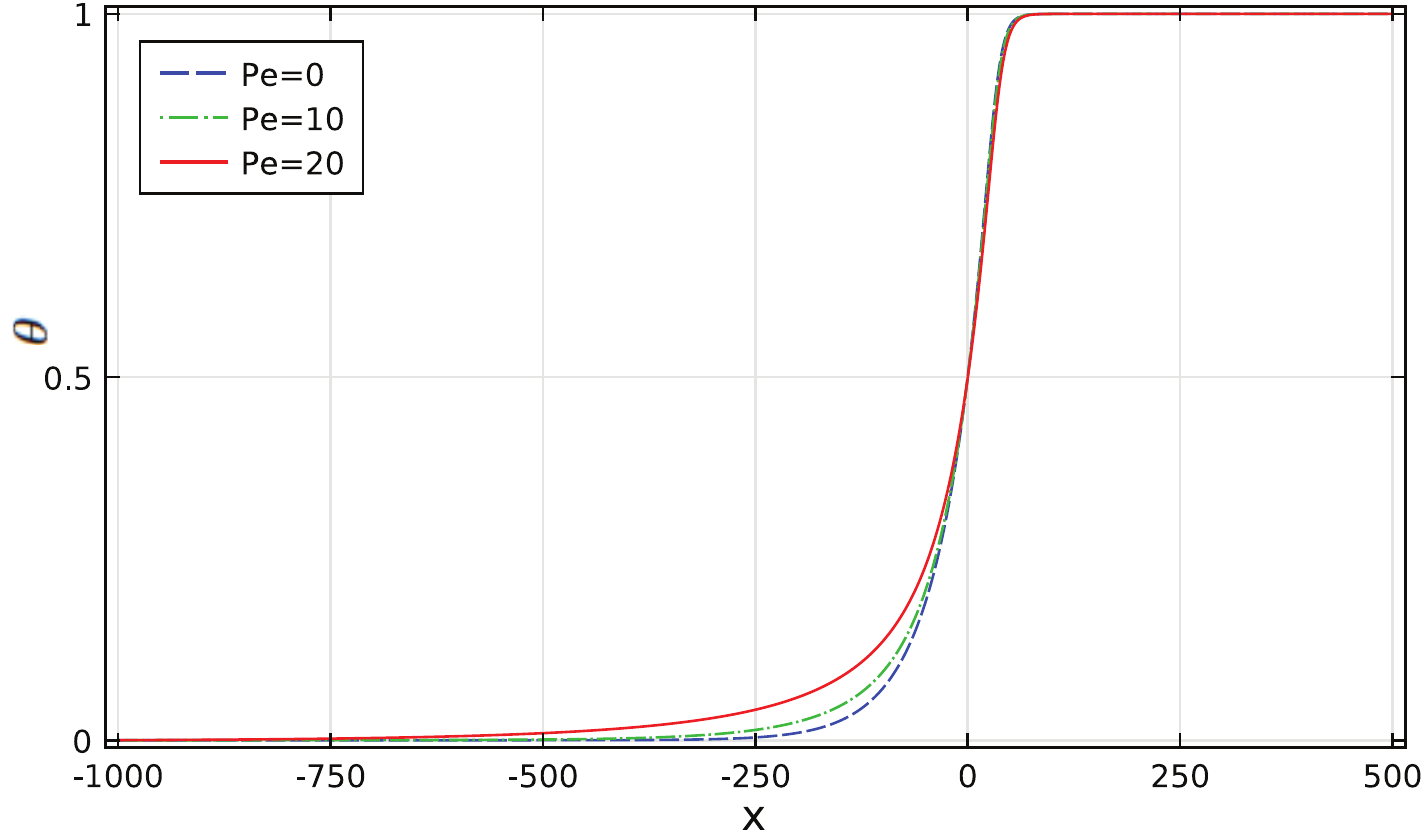}\label{fig:temp}}
\subfigure[]{
\includegraphics[width=0.49\textwidth,trim=0 0 0 0, clip=true]{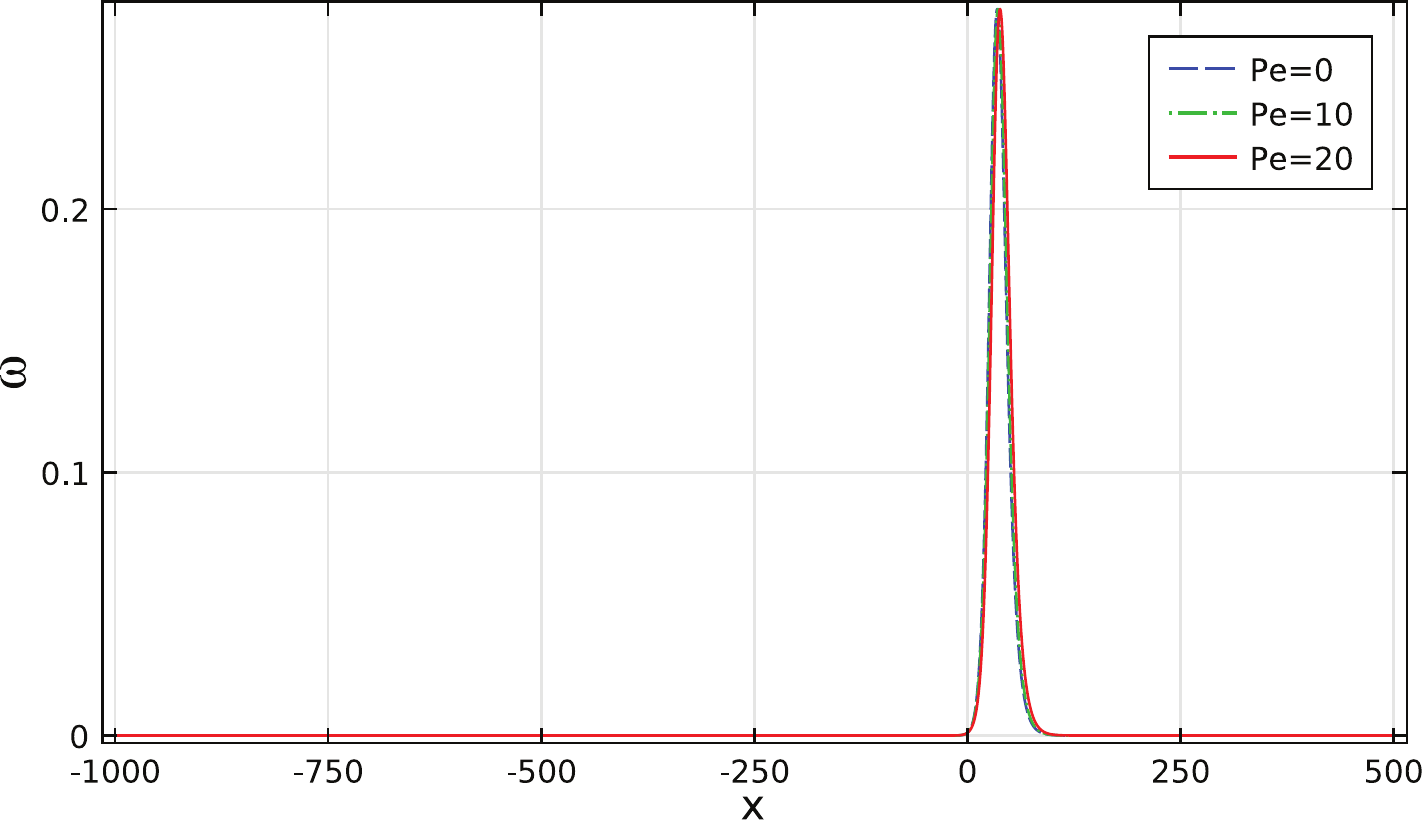}\label{fig:omega}}
\subfigure[]{
\includegraphics[width=0.49\textwidth,trim=0 0 0 0, clip=true]{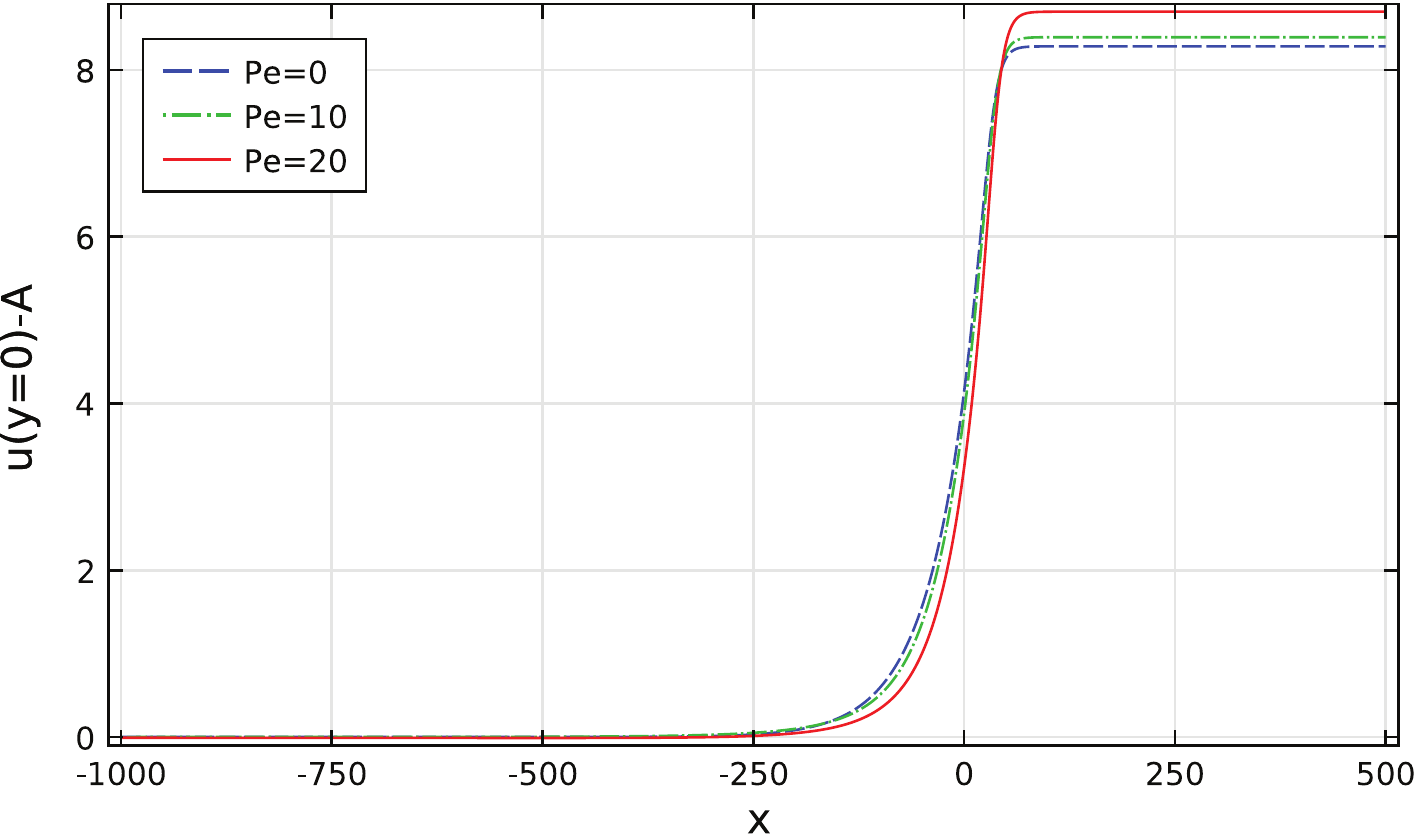}\label{fig:veloc}}
\subfigure[]{
\includegraphics[width=0.49\textwidth,trim=0 0 0 0, clip=true]{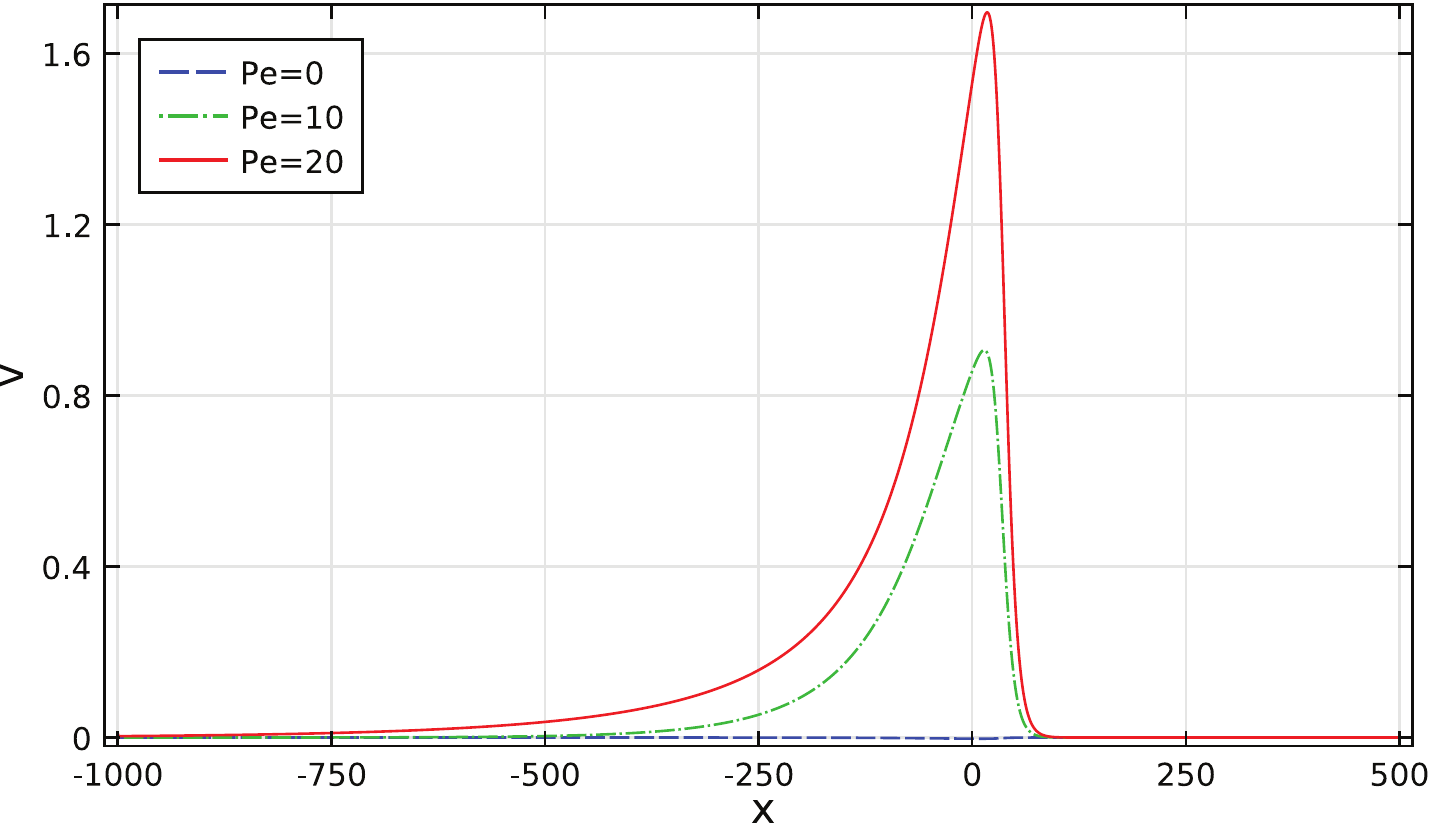}\label{fig:velocv}}
\caption{Plots of a) temperature $\theta$ along $y=0$; b) reaction rate $\omega$ along $y=0$; c) increase in horizontal velocity $u$ along $y=0$ due to thermal expansion, which is given by $u(y=0)-A$; d) vertical velocity component $v$ along $y=0.5$. The plots correspond to selected values of the Peclet number $\Pe$, with $\epsilon=50$ and $\beta=10$. The figures are plotted in the unscaled coordinates $(x,y)$.}
\label{fig:lines}
\end{figure}

It can be seen that in both the constant and variable density cases there is strong agreement between the numerical results calculated for high values of $\epsilon$ with $\beta=10$ and the asymptotic results derived in the limit $\epsilon \to \infty$ for $\beta=10$. It can also be seen that in both cases the asymptotic results derived in the limit $\epsilon \to \infty$ (for a chosen value of $\beta$) approach the results derived in the limit $\epsilon \to \infty$, $\beta \to \infty$ when $\beta$ is increased, as expected. Comparing the figures shows that a finite value of the activation energy $\beta$ has a larger effect on the propagation speed in the variable density case than in the constant density case.

A further comparison of figures \ref{fig:figure2} and \ref{fig:figure3} shows that $\Pe$ has a significantly larger effect on the propagation speed in the constant density case than when thermal expansion is taken into account. This can be explained by considering the perturbation to the flame shape using the method of \citet{daou2002influence} and \citet{short2009asymptotic}. Using the fact that $\theta_0=\theta_0\left(\xi\right)$ and equation (\ref{theta_1}) we have
\begin{gather*}
\theta = \theta_0\left(\xi\right)+\epsilon^{-1}\left(\rho_0\pd{\theta_0}{\xi}\Pe\left(\frac{y^2}{6}-\frac{y^4}{12}\right)+\check{\theta}_1(\xi)\right) +O\left(\epsilon^{-2}\right).
\end{gather*}
Now, let $\xi^*$ be the location at which the leading order temperature takes the constant value $\theta_0^*$. Defining the perturbation $\xi=\xi^*+\epsilon^{-1}\xi'$, and letting $^*$ denote the value of a variable at $\xi^*$, so that
\begin{gather*}
\theta\left(\xi^*+\epsilon^{-1}\xi'\right)=\theta_0^*+\epsilon^{-1}\left(\theta_1^*(y)+\xi'\pd{\theta_0^*}{\xi}\right)+O\left(\epsilon^{-2}\right),
\end{gather*}
we obtain the value of $\xi'$ for which $\theta\left(\xi^*+\epsilon^{-1}\xi'\right)=\theta_0^*$ as
\begin{gather*}
\xi'=-\rho_0^*\Pe\left(\frac{y^2}{6}-\frac{y^4}{12}\right)-\frac{\check{\theta}_1^*}{\theta_{0\xi}^*}.
\end{gather*}
Finally, the relative distance between the temperature reaching $\theta_0^*$ at $y=0$ and reaching the same value at $y=1$ is given by
\begin{gather}
\xi_r'=\frac{\rho_0^* \Pe}{12},\label{shapevar}
\end{gather}
which gives a measure of the deformation to the flame due to the flow. The equivalent of (\ref{shapevar}) in the constant density case is given by
\begin{gather}
\xi'_{r,\text{const}}=\frac{\Pe}{12}.\label{shapeconst}
\end{gather}
Thus since $\rho_0<1$, we have $\xi'_r<\xi'_{r,\text{const}}$  from (\ref{shapevar})-(\ref{shapeconst}) and therefore the deformation to the flame shape is smaller in the variable density case. This means that the effective propagation speed $U_T$, which gives a measure of the burning rate of the flame, is expected to be less in the variable density case. Note that in the $\Pe \to 0$ analysis of \citet{short2009asymptotic}, the flame deformation was found to be larger in the variable density case than the constant density case for values of $\Pe$ giving a propagation speed $U>0$, and smaller in the variable density case when $U<0$. Since $U=U_T-\frac{2}{3}\epsilon\Pe$, where $U_T=O\left(1\right)$ and $\epsilon \to \infty$, the propagation speed $U$ is always expected to be negative in our study and so the results (\ref{shapevar})-(\ref{shapeconst}) agree with those of \citet{short2009asymptotic}.

An illustration of the numerically calculated flame shape for selected values of the Peclet number in both the constant density and variable density cases is given in figure \ref{fig:shape}, which shows that the deformation to the flame shape is indeed larger in the variable density case, as found in (\ref{shapevar})-(\ref{shapeconst}).

The flame behaviour and its interaction with the flow is further illustrated in figures \ref{fig:surfs} and \ref{fig:lines}, corresponding to numerical simulations. Figure \ref{fig:tempsurf} shows a contour plot of the temperature $\theta$; also shown is the velocity field induced by thermal expansion, namely $\left(u-\epsilon\Pe\left(1-y^2\right),v\right)$. This is plotted rather than the full velocity field $(u,v)$ for clarity since the imposed Poiseuille flow is large compared to the induced flow, which is consistent with the asymptotic findings (see equation \eqref{leadingu}). Figures \ref{fig:velocsurf} and \ref{fig:velocvsurf} provide contour plots of the horizontal and vertical components of the induced flow, respectively. It is seen that for a fixed value of $x$, the maximum of the horizontal component of the induced flow is located at the centreline $y=0$, and the maximum of the vertical component is located around $y=0.5$. Figures \ref{fig:veloc} and \ref{fig:velocv} plot these horizontal and vertical components at $y=0$ and $y=0.5$, respectively for selected values of $\Pe$. Corresponding plots of the temperature $\theta$ and the reaction rate $\omega$ along the centreline $y=0$ are shown in figures \ref{fig:temp} and \ref{fig:omega}. Figure \ref{fig:lines} illustrates the gas expansion through the flame.  Furthermore the figure demonstrates that the effective flame thickness increases with increasing $\Pe$; this is in line with the asymptotic results (see the asymptotic formula \eqref{equation:deffective} and also the discussion in \S \ref{premixed1:sec:conc} related to equation \eqref{eq:effectiveflamespeed}).

Returning now to the effect of the Peclet number on the propagation speed, comparing the asymptotic result (\ref{const:infinitebeta}), derived in the constant density approximation, with (\ref{result:infinitebeta}), derived in the variable density case, provides a further reason for the larger effect of $\Pe$ on $U_T$ in the constant density case. The constant density results are the same as the variable density results, but with the term $\Pe\left(1-\alpha\right)$ replaced by $\Pe$ in the leading order term for the effective propagation speed\footnote{Note that the constant density asymptotic results (\ref{const:finitebeta})-(\ref{const:bc}) are not recovered by simply setting $\alpha=0$ in equations (\ref{result:finitebeta})-(\ref{res:bc}) due to the presence of $\alpha$ in the reaction term, which throughout this study is set to $\alpha=0.85$ in both cases.}. This suggests that replacing the Peclet number in the variable density case by a scaled Peclet number, given by
\begin{gather*}
Pe_\text{scaled}=\Pe\left(1-\alpha\right),
\end{gather*}
\begin{figure}
\centering
\includegraphics[scale=0.7,trim=20 0 0 10, clip=true]{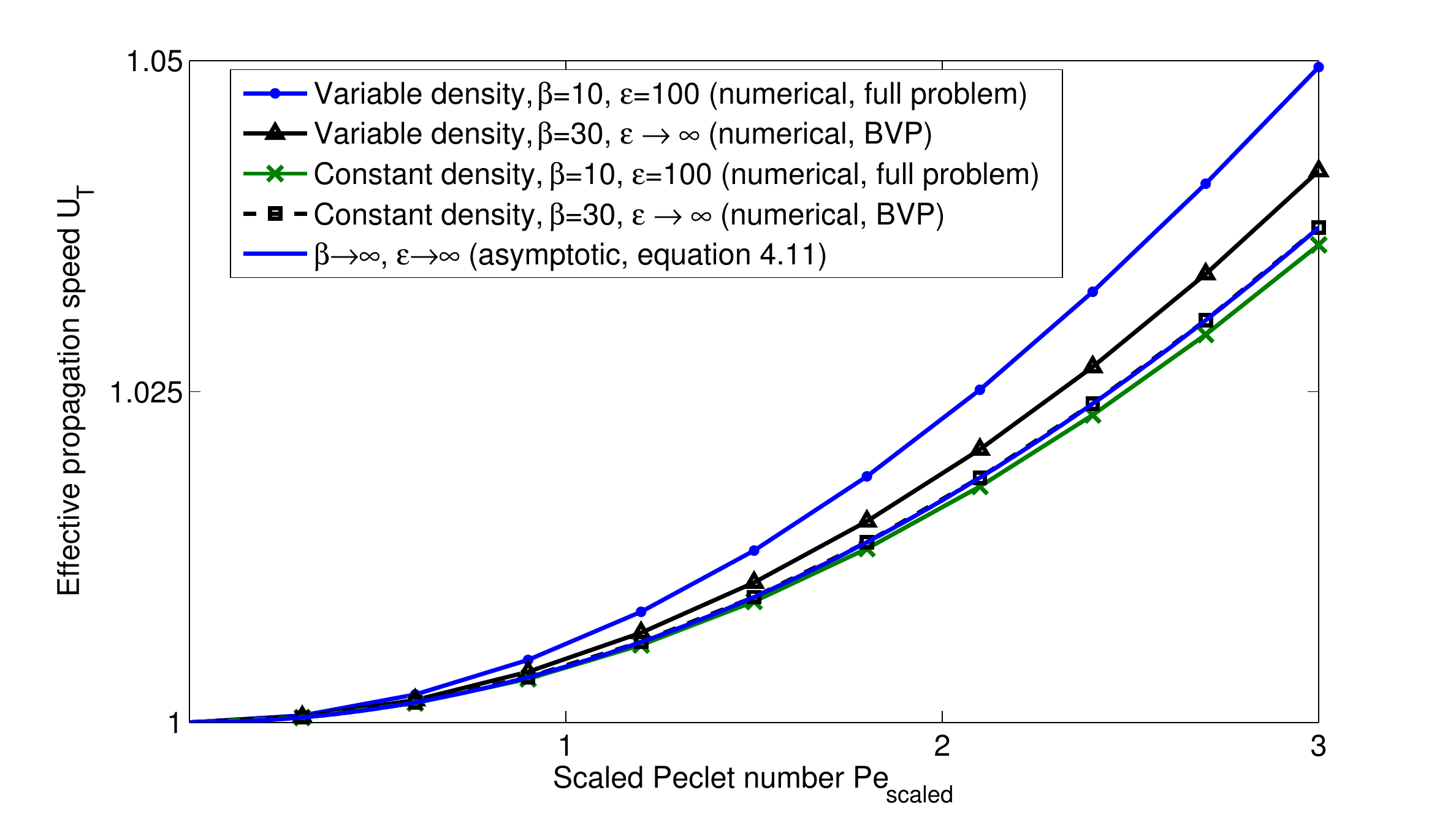}
\caption{Comparison of numerical results for the effective propagation speed $U_T$ versus the scaled Peclet number $\Pe_{\text{scaled}}$ in the variable density and constant density cases. Numerical simulations of the full system (\ref{nondim1})-(\ref{bc:last}) are performed for the parameter values $Pr=1$, $\alpha=0.85$ and $\beta=10$. Also included are numerical results of the solutions to (\ref{const:finitebeta})-(\ref{const:bc}) and (\ref{result:finitebeta})-(\ref{res:bc}) for $\alpha=0.85$ and $\beta=30$, to illustrate that for higher values of $\beta$ the lines of $U_T$ versus $\Pe_{\text{scaled}}$ in the constant density and variable density cases collapse onto the theoretical asymptotic curve.}
\label{fig:scaledcomp}
\end{figure}
would lead to strong agreement between the variable density and constant density numerical results.

A plot of the numerically calculated value of $U_T$ versus $\Pe_{\text{scaled}}$ in the constant density and variable density cases is given in figure \ref{fig:scaledcomp}. As expected, the relationship between $U_T$ and $\Pe$ in the two cases is much more similar than in figures \ref{fig:figure2} and \ref{fig:figure3}, but there is still a quantitative difference. This can be attributed to the fact that the finite activation energy has a more significant effect in the variable density case than in the constant density case, as described above. It is therefore expected that for larger values of $\beta$ the agreement between the numerically calculated value of $U_T$ and $\text{Pe}_\text{scaled}$ would be closer between the two cases. To illustrate this, included in figure \ref{fig:scaledcomp} is a comparison of $U_T$ versus $\Pe_{\text{scaled}}$ from the numerical solution to (\ref{const:finitebeta})-(\ref{const:bc}) and (\ref{result:finitebeta})-(\ref{res:bc}), valid as $\epsilon \to \infty$, with $\beta=30$. The figure shows that in this case the values of $U_T$ in the constant and variable density cases are indeed closer together. However, performing numerical calculations of the full system with a larger value of $\beta$ involves a significant amount of extra computation and is beyond the scope of this study.

Finally, it should be noted that the asymptotic results found in this study agree with results obtained previously in the limit $\Pe \to 0$. Expanding the constant density result (\ref{const:infinitebeta}) as $\Pe \to 0$ gives
\begin{gather*}
U_T=1+\frac{4}{945}\Pe^2+O\left(\Pe^3\right),
\end{gather*}
which agrees with \citet{daou2002thick}. Expanding the variable density result (\ref{result:infinitebeta}) as $\Pe \to 0$ gives
\begin{gather*}
U_T=1+\frac{4}{945}\Pe^2\left(1-\alpha\right)^2+O\left(\Pe^3\right).
\end{gather*}
This agrees with the results found by \citet{short2009asymptotic}, which found $U_T \mytilde 1$ to leading order.
\section{Conclusion}
\label{premixed1:sec:conc}
In this study we have investigated the propagation of a premixed flame through a narrow channel against a flow of large amplitude, taking the effect of the flame on the flow into account through the action of thermal expansion. This is the first study to consider a variable density premixed flame in a narrow channel with Peclet number $\Pe=O(1)$, which characterises the large amplitude flow. It is also the first to investigate Taylor dispersion in the context of combustion. The problem has been studied analytically to determine the effective propagation speed $U_T$ for $\Pe=O(1)$, in the limit $\epsilon \to \infty$ with both finite and infinite values of the activation energy $\beta$. The asymptotic studies are complemented by a numerical study whose results have been compared to the analytical results to test their effectiveness.

It has been found that, in the limit $\epsilon \to \infty$, a two-dimensional premixed flame propagating through a rectangular channel against a Poiseuille flow can be described by a boundary value problem that corresponds to a one-dimensional premixed flame with an effective diffusion coefficient, given by
\begin{gather*}
D_{\text{eff}}=D_T\left(1+\frac{8}{945} \Pe^2 \right),
\end{gather*}
in the constant density case and
\begin{gather*}
D_{\text{eff}}=D_T\left(1+\frac{8}{945} \Pe^2 \frac{\tilde{\rho}^2}{\tilde{\rho}^2_u} \right),
\end{gather*}
in the variable density case. These values correspond to those found in studies of enhanced dispersion due to a Poiseuille flow in non-reactive fluids, known as Taylor dispersion. A premixed flame propagating through a channel in the limit $\epsilon \to \infty$, with $\Pe=O(1)$ can therefore be considered to be in the Taylor regime.

Further, analytical solutions to the derived one-dimensional boundary value problems have been obtained in the limit $\beta \to \infty$ in both the constant density and variable density cases. The asymptotic results have been found to show strong agreement with the numerical results in both cases, as well as with results derived in previous studies in the limit $\Pe\to 0$. Physical reasons for the differences between the constant and variable density cases in the relationship of the propagation speed versus the Peclet number have been discussed.

The analytical results (\ref{result:infinitebeta}) and (\ref{const:infinitebeta}) can provide some insight towards understanding the effect of small-scale eddies on the propagation of a turbulent premixed flame, when the flow amplitude $A$ in our study is identified with the turbulence intensity and the channel height $L$ is identified with the turbulent flow (integral) scale. The situation where the flame is thick compared to the scale of the flow is described by Damk\"{o}hler's second hypothesis, which may be stated in the form of a relationship between the dimensional effective propagation speed and the effective thermal diffusivity, given by $\tilde{U}_\text{eff}=\sqrt{D_\text{eff} \left/ \tau \right.}$. Here $\tau$ is the chemical time related to the planar premixed flame speed $S_L^0$ (used in this study as unit speed to non-dimensionalise velocities) by $S_L^0=\sqrt{D_T \left/ \tau \right.}$. Therefore on dividing these two equations,  Damk\"{o}hler's second hypothesis is recovered, to leading order, in equation (\ref{const:infinitebeta}), as can be seen by noting that
\begin{gather}
U_T\equiv\frac{\tilde{U}_\text{eff}}{S^0_L}=\sqrt{1+\frac{8}{945}\Pe^2}\equiv\sqrt{\frac{D_\text{eff}}{D_T}}.\label{eq:effectiveflamespeed}
\end{gather}

The results (\ref{result:infinitebeta}) and (\ref{const:infinitebeta}) may also be used to provide a possible explanation of the so-called bending effect of the turbulent premixed flame speed when plotted in terms of the turbulence intensity for fixed values of the Reynolds number (see e.g. \cite[][]{bradley1992fast,ronney1995some}). Therefore our distinguished limit, namely $\epsilon \to \infty$ with $\Pe$ fixed (note that the Reynolds number and Peclet number are equal if $Pr=1$), mimics the experimental conditions of \citet{bradley1992fast} and can be used to shed some light on the experimental findings. An initial discussion of the relevance of the asymptotic results in this chapter to the bending effect, along with further asymptotic results, is provided in Chapter \ref{chapter:premixed2} of this thesis.

Finally, it has been shown that, in the limit $\epsilon \to \infty$, $\beta \to \infty$, the graphs of $U_T$ in the constant and variable density cases are identical when plotted against a scaled Peclet number
\begin{gather*}
Pe_\text{scaled}=\Pe\left(1-\alpha\right),
\end{gather*}
and graphs of the numerically calculated propagation speed against this scaled Peclet number have also been provided.

\chapter{The thick flame asymptotic limit and the bending effect of premixed flames}
\chaptermark{The bending effect of premixed flames}
\label{chapter:premixed2}
\section{Introduction}
\begin{figure}
\centering
\includegraphics[scale=0.7, trim=10 0 0 0,clip=true]{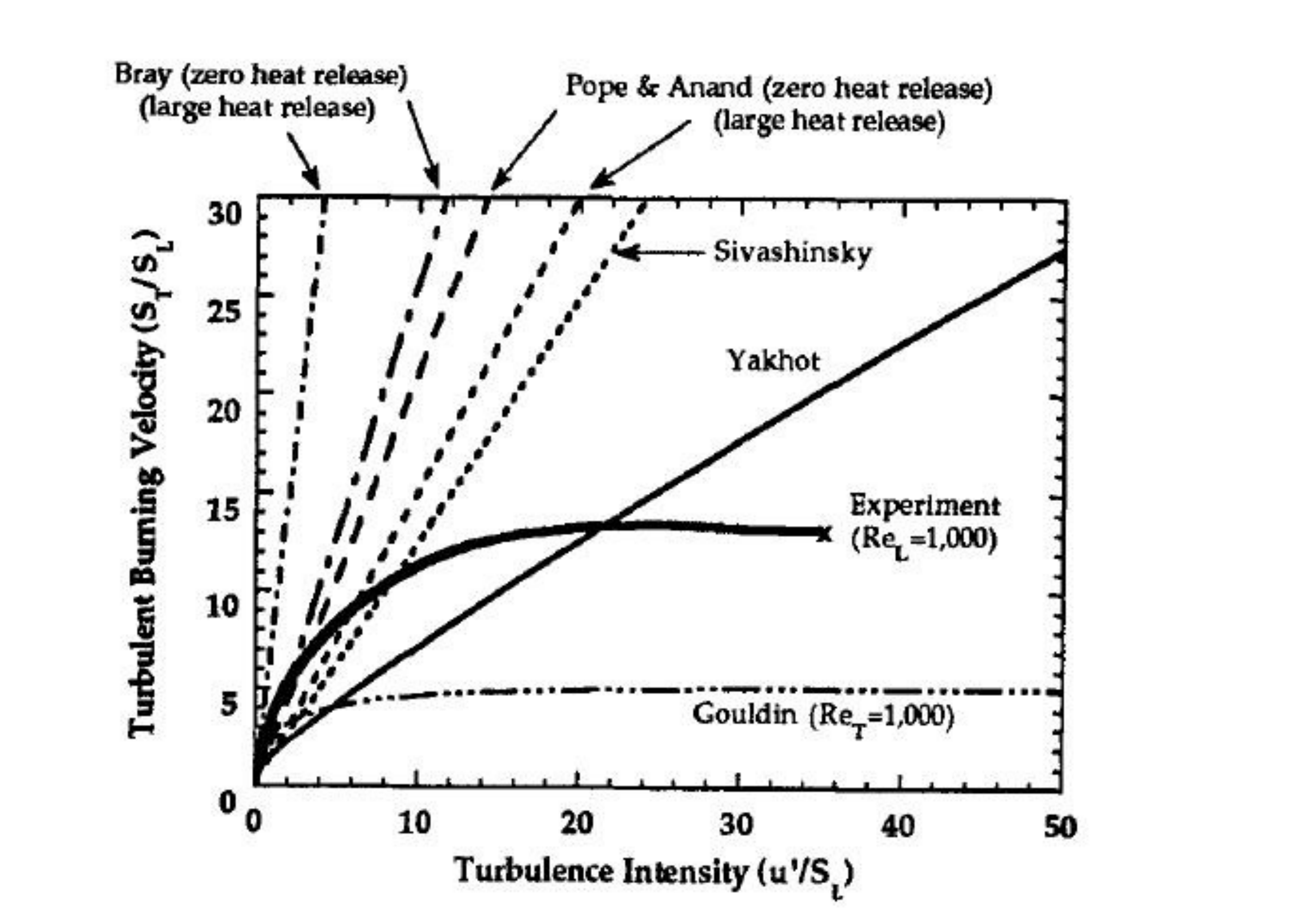}
\caption{Comparison between theoretical and experimental predictions of the turbulent burning velocity $S_T \left/ S_L\right.$, or $U_T$ in our notation, as a function of turbulent intensity $u' \left/S_L \right.$, or $A$ in our notation \cite{ronney1995some}. Theoretical studies (carried out in the ``thin-flame" limit) include: Bray \cite{bray1990studies} with zero
heat release and large (density ratio = 7) heat release; Anand and Pope \cite{anand1987calculations}
with zero and infinite heat release; Yakhot \cite{yakhot1988propagation}; Sivashinsky \cite{sivashinsky1990cascade}; Gouldin \cite{gouldin1987application} with $Re_L$ = 1,000; experimental values from Bradley \cite{bradley1992fast} for $Re_L = 1,000$. Where $Re_L$  is not specified, predictions are independent of $Re_L$. Here, $S_T$ is the turbulent flame speed, $S_L$ laminar flame speed, $u'$ turbulent intensity and $Re_L$ is the turbulent Reynolds number.}
\label{fig:ronney_premixed}
\end{figure}

In this chapter, which provides a continuation of the work undertaken in Chapter \ref{chapter:premixed1}, we are again mainly concerned with premixed flame propagation against a steady parallel flow of large amplitude, for fixed values of the Peclet number. The main motivation in this chapter is to describe the relevance of the asymptotic results in the \emph{thick flame} asymptotic limit to the so-called \emph{bending effect} of turbulent combustion, as mentioned in the conclusion of Chapter \ref{chapter:premixed1}. The bending effect is reported to be observed experimentally when the effective flame speed $U_T$ is plotted in terms of the flow intensity $A$, as seen in figure \ref{fig:ronney_premixed}, adapted from \cite{ronney1995some}. As discussed in \cite{almalkithesis}, this is a fundamental problem in turbulent combustion which has received considerable attention in the literature. Previous theoretical studies have predicted linear, sub-linear, or exponential dependence of $U_T$ on $A$ \cite{ronney1995some}. Experiments, on the other hand, have shown that the effective flame speed increases with the flow intensity $A$ at lower values of $A$, but levels off at higher values of $A$ \cite{abdel1987turbulent}, which is widely known as the bending effect. The constant value to which $U_T$ seems to converge is found to depend on the Reynolds number \cite{abdel1987turbulent}, which in our case is equivalent to the Peclet number since we assume unity Prandtl number. Taking into account the complexity associated with the multi-scale nature of turbulence, it is of great interest to examine this relationship for cases such as parallel or vortical flows, because of their relative simplicity.

In this chapter, we investigate the relationship between the effective propagation speed of a laminar premixed flame and the amplitude of a prescribed Poiseuille flow. The amplitude of the flow can be thought of as analagous to the turbulence intensity, as plotted in figure \ref{fig:ronney_premixed}. We are motivated by the asymptotic result \eqref{result:infinitebeta}, derived in Chapter \ref{chapter:premixed1}, which shows a constant dependence of the effective propagation speed on the Peclet number in the thick flame asymptotic limit. Furthermore, to complement the investigation other distinguished limits are considered to help understand the discrepencies between theory and experiments. Finally, we will provide numerical results to aid in this discussion.

The chapter is structured as follows. In \S \ref{premixed2:sec:infinite} we perform an asymptotic analysis in the infinite activation energy limit, to provide a complementary asymptotic result to equation \eqref{result:infinitebeta}, derived in Chapter \ref{chapter:premixed1}. In \S \ref{premixed2:sec:results}, we provide some further asymptotic results in the constant density case, as well as numerical results for a full range of values of the flow amplitude $A$ and the Peclet number $\Pe$. This section also contains a discussion of the relevance of the results to the bending effect of turbulent combustion. We end the chapter with conclusions in \S \ref{premixed2:sec:conc}.

\section{Infinite activation energy asymptotic analysis}
\label{premixed2:sec:infinite}
The result \eqref{result:infinitebeta} in Chapter \ref{chapter:premixed1} provided an asymptotic result for the effect of a Poiseuille flow on the propagation speed of a variable density premixed flame. The limits taken were $\epsilon \to \infty$, followed by $\beta \to \infty$, with $\Pe=O(1)$. In this chapter we provide a complementary asymptotic analysis to the one performed in Chapter \ref{chapter:premixed1}, this time in the limit $\beta \to \infty$, followed by $\epsilon \to \infty$, with $\Pe=O(1)$, in order to assess the effect of taking the limits in this order. Note that, since the three main non-dimensional parameters in the problem can be related to each other by
\begin{gather}
A=\epsilon \Pe, \label{parameterrelationship}
\end{gather}
taking $\epsilon\to\infty$ with $\Pe=O(1)$ is equivalent to taking $A\to\infty$ with $\Pe=O(1)$. All parameters have the same definitions as in Chapter \ref{chapter:premixed1}. In this chapter, attention will be restricted to steady flame propagation with unity Lewis number
\begin{gather}
\Le=1.
\end{gather}

In the limit of infinite activation energy, $\beta \to \infty$, the reaction zone is confined to a thin sheet, $x=F(y)$, say. On either side of the thin sheet, the reaction rate can be set to zero. We introduce the change of variables
\begin{gather}
\xi=x-F(y), \quad \eta=y, \quad f(\eta)=F(y).\label{flamecoordinates}
\end{gather}
Then the steady form of the governing equations \eqref{nondim1}-\eqref{nondim8} can be written
\begin{gather}
\pd{}{\xi}\left(\rho\left(U+u-f'(\eta)v\right)\right)+ \pd{}{\eta}\left(\rho v\right)=0,\label{b:1}\\
\rho\left(u+U-f'(\eta)v\right)\pd{u}{\xi}+\rho v \pd{u}{\eta}+\pd{p}{\xi}=
\epsilon Pr \left(\nabla^2 u + \frac{1}{3}\pd{}{\xi}\nabla \cdot \mathbf{u}\right),\\
\rho\left(u+U-f'(\eta)v\right)\pd{v}{\xi}+\rho v \pd{v}{\eta}+\pd{p}{\eta}-f'(\eta) \pd{p}{\xi}=\nonumber\\
\epsilon Pr \left(\nabla^2 v+\frac{1}{3}\left(\pd{}{\eta}-f'\left(\eta\right)\pd{}{\xi}\right)\left(\nabla \cdot \mathbf{u}\right)\right),\\
\rho\left(u+U-f'(\eta)v\right)\pd{\theta}{\xi}+\rho v \pd{\theta}{\eta}=\epsilon\nabla^2 \theta,\label{b:4}\\
\rho=\left(1+\frac{\alpha}{1-\alpha}\theta\right)^{-1},\label{b:last}
\end{gather}
where 
\begin{gather*}
\nabla^2u=\left(1+f'(\eta)^2\right)\sd{u}{\xi}+\sd{u}{\eta}-2 f'(\eta)\frac{\partial^2 u}{\partial \eta \partial \xi}- f''(\eta) \pd{u}{\xi},
\end{gather*}
and
\begin{gather*}
\nabla \cdot \mathbf{u}=\pd{u}{\xi}+\pd{v}{\eta}-f'\left(\eta\right)\pd{v}{\xi}.
\end{gather*}
These equations are subject to boundary conditions, which using \eqref{flamecoordinates} in \eqref{bc:1}-\eqref{bc:last} are given by
\begin{gather}
\pd{\theta}{\eta}=f'\left(\eta\right)=\pd{u}{\eta}=v=\pd{p}{\eta}=0 \quad \text{at } \eta=0, \label{b:bc:1}\\
\pd{\theta}{\eta}=f'\left(\eta\right)=u=v=0 \quad \text{at } \eta=1,\label{b:bc:2}\\
\theta=0,\quad u=A\left(1-\eta^2\right)=\epsilon\Pe\left(1-\eta^2\right), \quad v=0\quad \text{at } \xi=-\infty, \label{b:bc:minusinfty}\\
\pd{\theta}{\xi}=\pd{u}{\xi}=\pd{v}{\xi}=p=0 \quad \text{at } \xi=+\infty \label{b:bc:last}.
\end{gather}

Equations \eqref{b:1}-\eqref{b:last} are also subject to jump conditions across the flame sheet, which is located at $\xi=0$. The derivation of these jump conditions is not given here, but can be found in e.g. \cite{matalon1982flames}, where the full set of conditions are listed; here we simply give the conditions used in the following asymptotic analysis:
\begin{align}
&\theta=1, \quad \theta_\xi=\epsilon^{-1}\left(1+f'\left(\eta\right)^2\right)^{-1\left/ 2\right.}, \quad \text{at}\quad \xi=0\label{jump:last}.
\end{align}
\subsection{Solution in the limit $\epsilon \to \infty$, with $\Pe=O(1)$}
\label{premixed2:sec:infinit2}
In this section we provide the leading order asymptotic solution to \eqref{b:1}-\eqref{jump:last} in the limit $\epsilon \to \infty$, with $\Pe=O(1)$. Note that this limit is equivalent to $A\to\infty$, with $\Pe=O(1)$, due to \eqref{parameterrelationship}. A similar asymptotic analysis was performed in \cite{almalkithesis} in the constant density approximation. Using a similar method to the one used in \S \ref{premixed1:sec:asymptotics} in Chapter \ref{chapter:premixed1}, we introduce a scaled coordinate
\begin{gather}
\xi'=\frac{\xi}{\epsilon},\label{eq:scaling}
\end{gather}
and expand each variable in powers of $\epsilon^{-1}$, so that
\begin{equation}
\left.
\begin{aligned}
&U=-\frac{2}{3}\epsilon \Pe+U_0+\epsilon^{-1}U_1+...,\quad f=F_0+\epsilon^{-1}F_1+...\\  &u=\epsilon u_0 +u_1+...\quad v=v_0+\epsilon^{-1}v_1+...\\&\theta=\theta_0+\epsilon^{-1}\theta_1+...\quad p=\epsilon^3 p_0+\epsilon^2 p_1+...
\end{aligned}
\right\}
\label{eq:expansions}
\end{equation}
where $U_0$ is the leading order term for the effective propagation speed $U_T$, defined in \eqref{effective}.

\begin{figure}
\subfigure[]{
\includegraphics[width=0.49\textwidth,trim=0 0 0 0, clip=true]{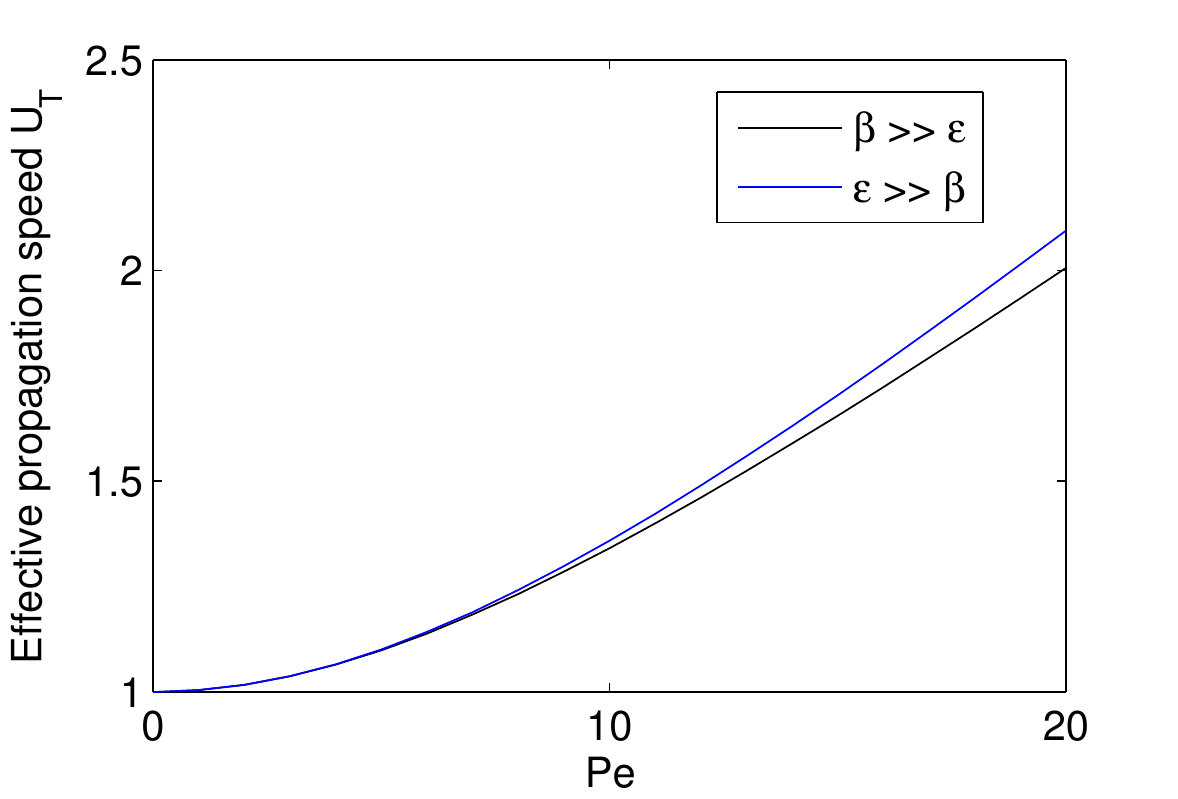}}
\subfigure[]{
\includegraphics[width=0.49\textwidth,trim=0 0 0 0, clip=true]{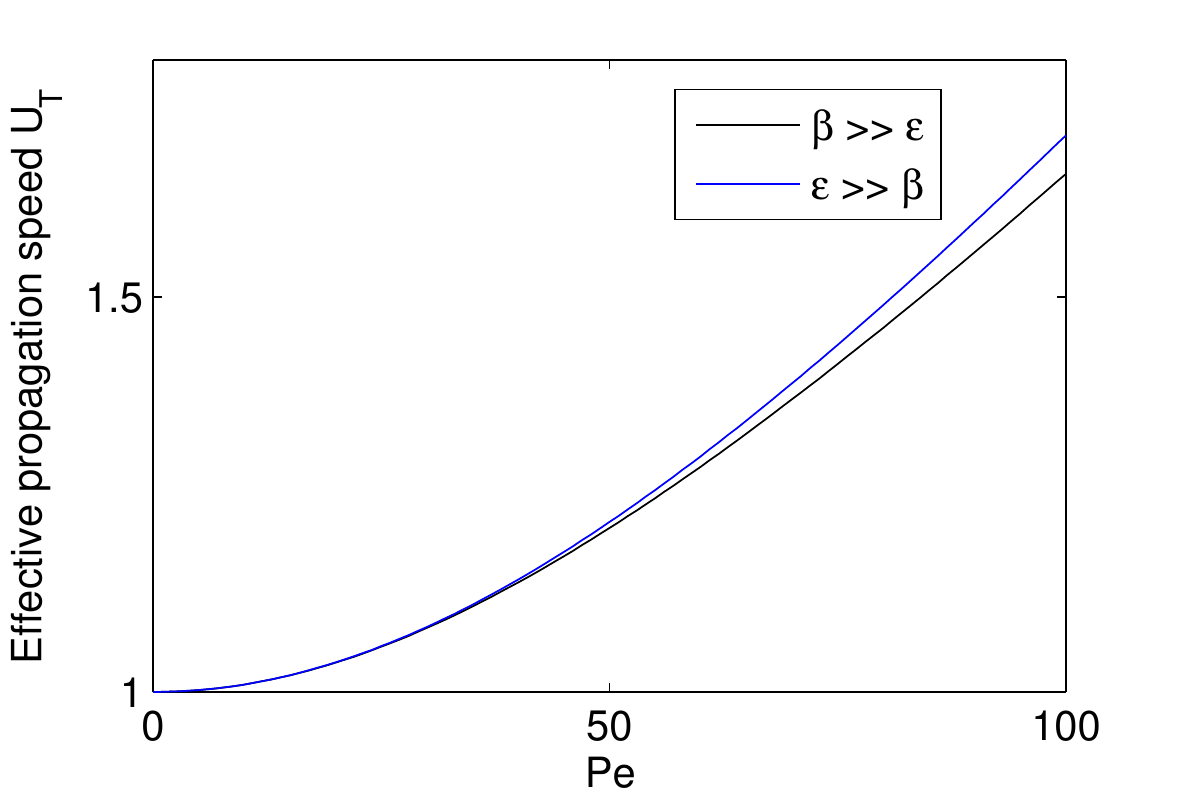}}
\caption{Comparison of asymptotic results in the limit $\Pe=O(1)$, $\epsilon \to \infty$ and $\beta \to \infty$. Included are results for $\epsilon \to \infty$ followed by $\beta \to \infty$, or $\epsilon \gg \beta$, given by \eqref{result:infinitebeta}, and $\beta \to \infty$ followed by $\epsilon \to \infty$, or $\beta \gg \epsilon$, given by \eqref{premixed2:res:2}, for a) $\alpha=0$ (constant density) and b) $\alpha=0.85$ (variable density).}
\label{fig:zonecomparison}
\end{figure}

Inserting the scaling \eqref{eq:scaling} into \eqref{b:1}-\eqref{jump:last} and dropping the prime notation, we find that the leading order solutions are the same as those found in \S 2.3 of Chapter \ref{chapter:premixed1}, and are given by
\begin{gather}
u_0=\Pe \left(1-\eta^2 \right), \label{b:u}\\
p_0=p_0(\xi), \quad \theta_0=\theta_0(\xi), \quad \rho_0=\rho_0(\xi).
\end{gather}
Then to $O\left(\epsilon\right)$ in equation (\ref{b:4}) we have
\begin{gather}
\rho_0 \Pe \left(u_0-\frac{2}{3} \Pe \right)\pd{\theta_0}{\xi}=\sd{\theta_1}{\eta}-F_0''(\eta)\pd{\theta_0}{\xi},
\end{gather}
which can be integrated twice with respect to $\eta$, using the boundary conditions \eqref{b:bc:1}-\eqref{b:bc:2} on $\theta_0$ and $F_0$, and substituting in (\ref{b:u}), to give
\begin{gather}
\left( \rho_0 \Pe \left(\frac{\eta^2}{6}-\frac{\eta^4}{12}\right)+F_0(\eta)\right)\pd{\theta_0}{\xi}=\theta_1+\bar{\theta}_1\left(\xi\right),
\end{gather}
where $\bar{\theta}_1$ is an arbitrary function of integration. Now, using the conditions \eqref{b:bc:minusinfty} and \eqref{jump:last} on $\theta$ and setting $F(0)=0$, a condition which can be prescribed due to the translational invariance of the problem, we obtain
\begin{gather}
F_0(\eta)=-\rho_0(0) \Pe \left(\frac{\eta^2}{6}-\frac{\eta^4}{12}\right)=-\left(1-\alpha\right) \Pe \left(\frac{\eta^2}{6}-\frac{\eta^4}{12}\right).\label{premixed2:flameshapeleading}
\end{gather}
The total burning rate is proportional to the flame surface area \cite{daou2001flame} by the relation 
\begin{gather}
U_T=\int_0^1\sqrt{1+F_{0\eta}^2}~\mathrm{d}\eta,\label{eq:flamesurfacearea}
\end{gather}
a relationship which can be derived by integrating the steady form of equation \eqref{nondim4} up to the reaction sheet and utilising the continuity equation \eqref{nondim1} with the boundary conditions \eqref{bc:1}-\eqref{bc:last}. Thus we can find the first approximation $U_0$ to the effective flame speed $U_T$, which using \eqref{premixed2:flameshapeleading} and \eqref{eq:flamesurfacearea} is given by
\begin{gather}
U_0=\int_0^1\sqrt{1+\left(\frac{\Pe \left( 1- \alpha\right)}{3}\left(1-\eta^2\right)\eta\right)^2}~\mathrm{d}\eta. \label{premixed2:res:2}
\end{gather}
This result provides a complementary asymptotic expression to equation \eqref{result:infinitebeta} in Chapter \ref{chapter:premixed1}. Note that equation \eqref{result:infinitebeta} in Chapter \ref{chapter:premixed1} is expected to be valid for $\epsilon \gg \beta$, while equation \eqref{premixed2:res:2} is expected to be valid for $\beta \gg \epsilon$.  It can be seen that both expressions predict a constant dependence of the effective propagation speed $U_T$ on the Peclet number in the thick flame asymptotic limit. A comparison of the two expressions is given in figure \ref{fig:zonecomparison}, where it can be seen that there is good agreement between the two results for low and moderate values of the Peclet number (approximately $\Pe<10$ in the constant density case and $\Pe<50$ in the variable density case). For high values of $\Pe$, the results start to diverge, although this divergence is more marked in the constant density case than the variable density case. The relevance of the results to the bending effect will be discussed in the following section.
\section{Further results and discussion}
\label{premixed2:sec:results}
In this section we present the results of the asymptotic analysis of \S\ref{premixed2:sec:infinite} and compare with the results of numerical simulations of the full system of governing equations, given by the steady form of \eqref{nondim1}-\eqref{bc:last}, with $\Le=1$. We also compare with some further asymptotic results that are available in the constant density case. All asymptotic results in this section have been obtained by first taking the infinite activation energy limit $\beta \to \infty$. The aim in this section is to provide a discussion of the relevance of these asymptotic and numerical results to the bending effect of premixed flames.

Note that throughout this section the numerical method used is the same as the one described in Chapter \ref{chapter:premixed1}. Numerical results in the constant density case are obtained by solving \eqref{nondim4} with $u=\epsilon \Pe\left(1-y^2\right)$, $v=0$, $\rho=1$ and the reaction term replaced by $\epsilon^{-1}\omega$, as described in Chapter \ref{chapter:premixed1}. Finally, note that the effect of thermal expansion is investigated by varying the thermal expansion coefficient $\alpha$, while taking the reaction term $\omega$ to be given by
\begin{gather}
\omega=\frac{\beta^2}{2}\rho y_F\exp\left(\frac{\beta\left(\theta-1\right)}{1+\alpha_h\left(\theta-1\right)}\right),\label{premixed2:reaction2}
\end{gather}
with $\alpha_h$ given the fixed value $\alpha_h=0.85$, as is done in \cite{pearce2013effect}.

\subsection{Constant density results}
In this section we provide asymptotic and numerical results in the constant density case. The  asymptotic result \eqref{premixed2:res:2} obtained in \S\ref{premixed2:sec:infinit2} in the limit $A\to\infty$, with $\Pe=O(1)$, can be written in the constant density case simply by setting $\alpha=0$, so that the leading order term for the effective propagation speed is given by
\begin{gather}
U_0=\int_0^1\sqrt{1+\left(\frac{\Pe }{3}\left(1-\eta^2\right)\eta\right)^2}\mathrm{d}\eta. \label{premixed2:constantresulta}
\end{gather}
Further asymptotic results are available in the constant density case, in the separate limits a) $A \to 0$, with $\Pe=O(1)$ and b) $A=O(1)$, with $\Pe\to\infty$. Note that these limits both correspond to the thin flame limit $\epsilon \to 0$. In both of these limits, the propagation speed $U_T$ can be written asymptotically as
\begin{gather}
U_T=1+\frac{2}{3}A-2K\Pe^{-2/3}A^{4/3}+...,\label{premixed2:constantresultb}
\end{gather}
where with $K\approx1.02$. A derivation of \eqref{premixed2:constantresultb} in the limit $A=O(1)$, with $Pe\to\infty$ can be found in \cite{daou2001flame}. The derivation of \eqref{premixed2:constantresultb} in the limit $A \to 0$, with $\Pe=O(1)$ is slightly different and can be found in Appendix \ref{appendix:premixed2}. Equation \eqref{premixed2:constantresultb} predicts a linear dependence of $U_T$ on $A$ when $\epsilon=A \left/ \Pe \right.$ is small, which is in agreement with several of the theoretical studies shown in figure \ref{fig:ronney_premixed}.
\begin{figure}
\centering
\includegraphics[width=.49\textwidth,trim=0 0 15 15, clip=true]{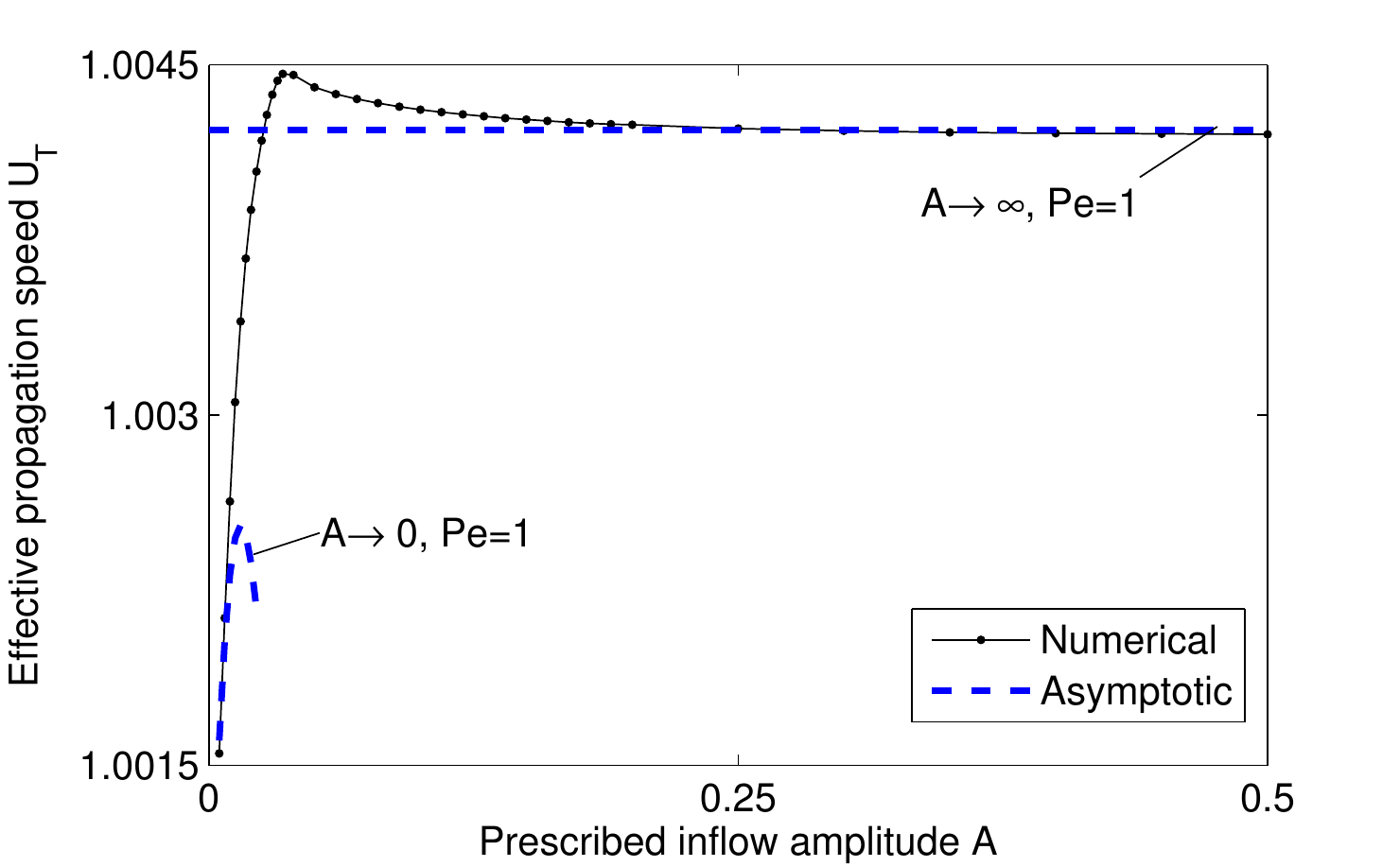}
\includegraphics[width=.49\textwidth,trim=0 0 15 15, clip=true]{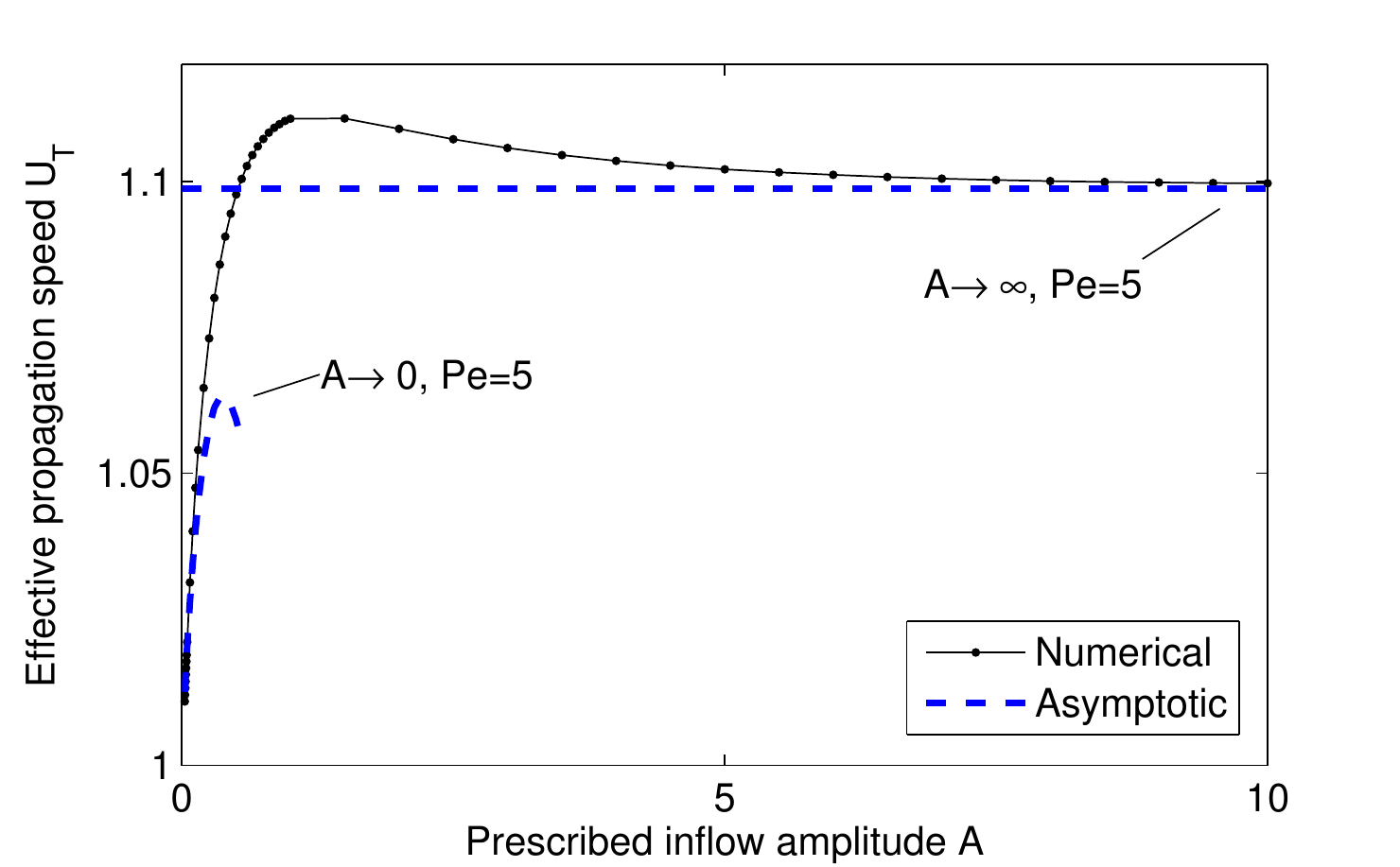}
\includegraphics[width=.49\textwidth,trim=0 0 15 15, clip=true]{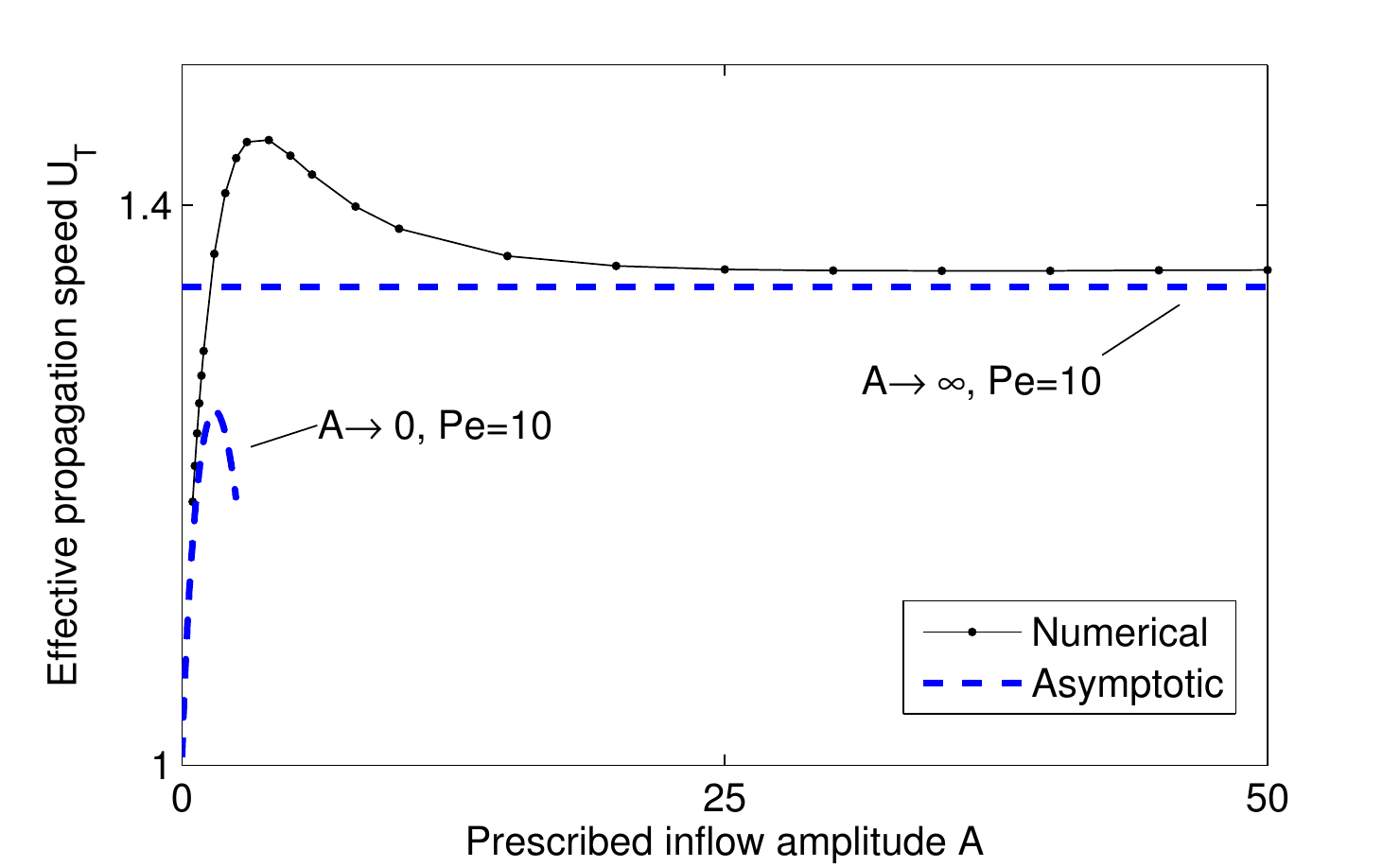}
\includegraphics[width=.49\textwidth,trim=0 0 15 15, clip=true]{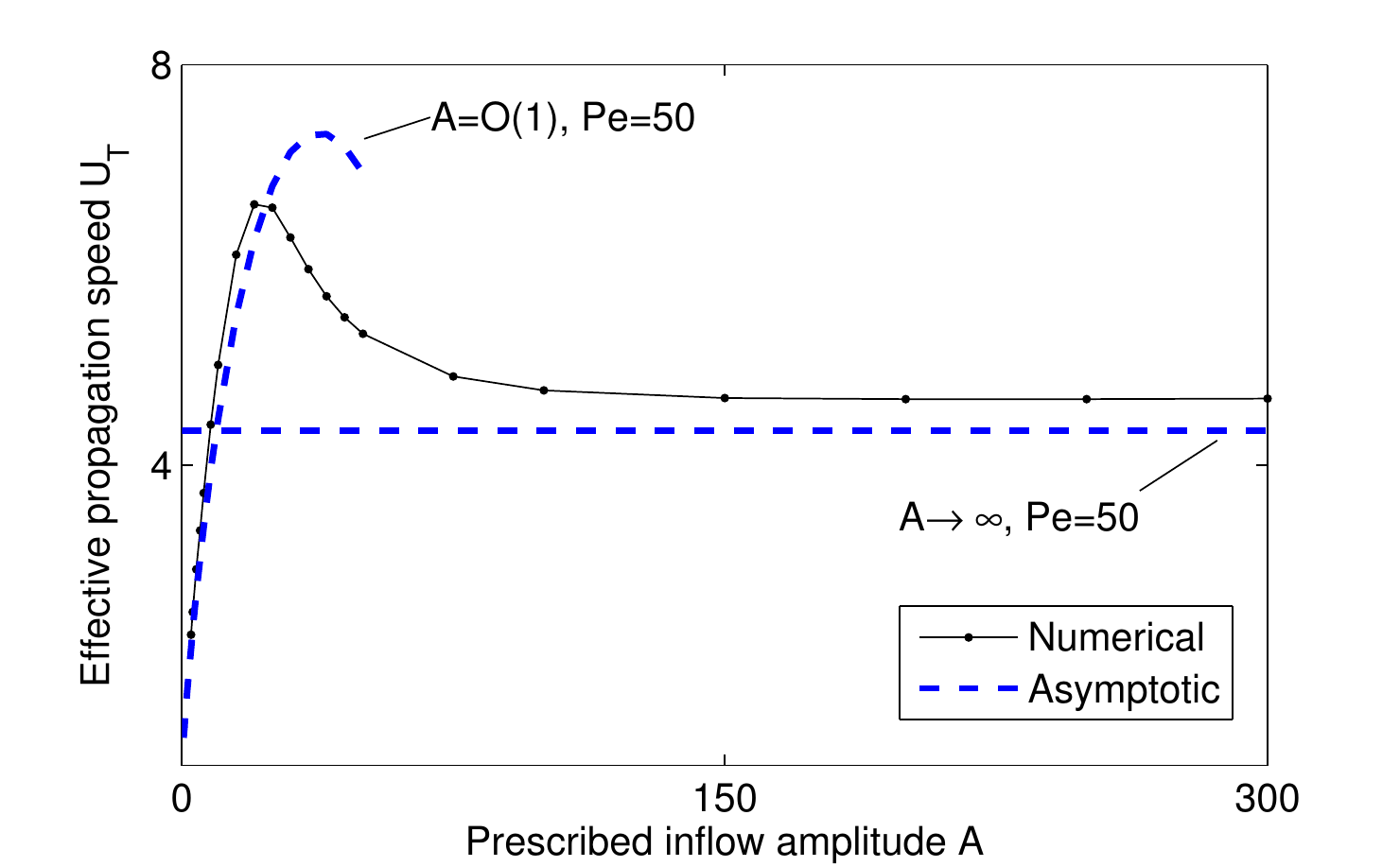}
\caption{Effective propagation speed $U_T$ versus prescribed flow amplitude $A$ for selected values of the Peclet number $\Pe$. Included are the asymptotic results \eqref{premixed2:constantresulta} and \eqref{premixed2:constantresultb}. Equation \eqref{premixed2:constantresulta} is derived in the limit $A\to \infty$, $\Pe=O(1)$ and equation \eqref{premixed2:constantresultb} is derived in the separate limits a) $A\to 0$, $\Pe=O(1)$ and b) $A=O(1)$, $\Pe \to \infty$. The numerical results are for the constant density case $\alpha=0$, with $\beta=10$ and $\alpha_h=0.85$.}
\label{fig:pecletcomparison}
\end{figure}

We now have a fairly complete asymptotic picture of the problem. The asymptotic results \eqref{premixed2:constantresulta} and \eqref{premixed2:constantresultb} are shown in figure \ref{fig:pecletcomparison}, where they are compared with and complemented by numerical results in the constant density case, obtained for a full range of values of $\Pe$ and $A$. It is clear that in all cases the numerically calculated propagation speed approaches a constant value, which depends on the Peclet number, as the flow amplitude $A$ increases to large values; this is in line with the the asymptotic results \eqref{premixed2:constantresulta} and shows a similar behaviour to the available experimental results summarised in figure \ref{fig:ronney_premixed}. It can also be seen in figure \ref{fig:pecletcomparison} that there is a maximum in the curve of $U_T$ versus $A$, which occurs at a higher value of $A$ as $\Pe$ increases, as predicted by the asymptotic results \eqref{premixed2:constantresultb}. Finally, it can be seen that the numerical results for small values of $A$ agree very well with the formula \eqref{premixed2:constantresultb}, as is to expected.

\subsection{Variable density results}
\begin{figure}
\centering
\includegraphics[scale=0.8, trim=0 0 0 0,clip=true]{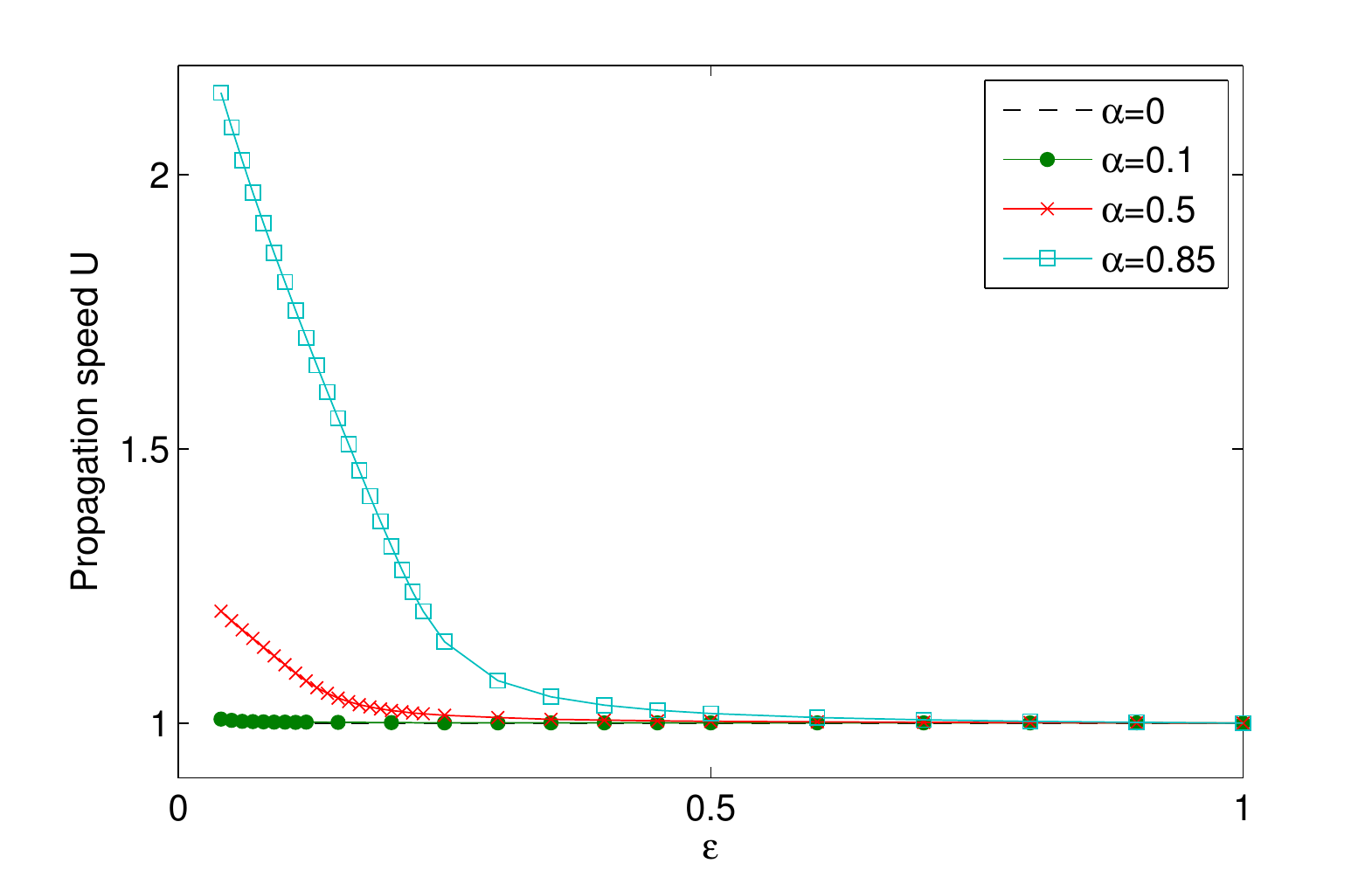}
\caption{Effective propagation speed $U_T$ versus $\epsilon$ for a premixed flame in a channel with no prescribed inflow ($A=0$). For small values of $\epsilon$, $U>1$ due to frictional effects from the no-slip walls. For each value of $\alpha$, the propagation speed $U$ is scaled by the corresponding numerically calculated planar flame speed. The dashed line denotes the planar flame propagation speed found in the constant density case, $\alpha=0$, and in the case of ``free-slip" sidewalls with $u_y=0$. For all simulations we take $Pr=1$, $\beta=10$ and $\alpha_h=0.85$.}
\label{fig:no_flow}
\end{figure}
\begin{figure}
\centering
\includegraphics[scale=0.8, trim=0 0 0 0,clip=true]{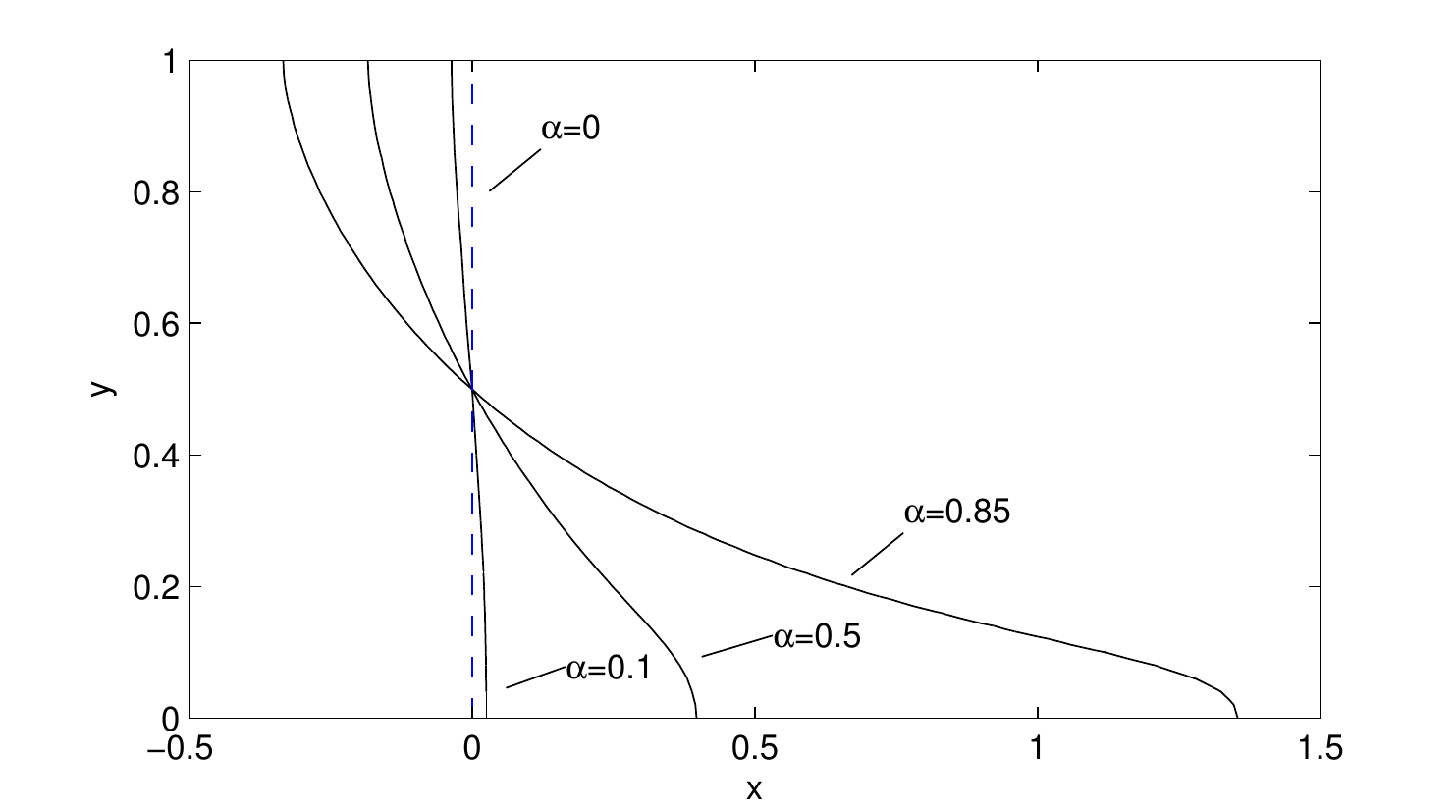}
\caption{Flame shape, defined by the line $\theta=0.5$, for a premixed flame in a channel with no prescribed inflow ($A=0$), for several values of the thermal expansion coefficient $\alpha$. The dashed line denotes the planar flame found in the constant density case, $\alpha=0$, and in the case of ``free-slip" sidewalls. For all simulations we take $Pr=1$, $\beta=10$ and $\alpha_h=0.85$.}
\label{fig:no_flow_shapes}
\end{figure}
\begin{figure}
\centering
\includegraphics[scale=0.9, trim=0 0 0 0,clip=true]{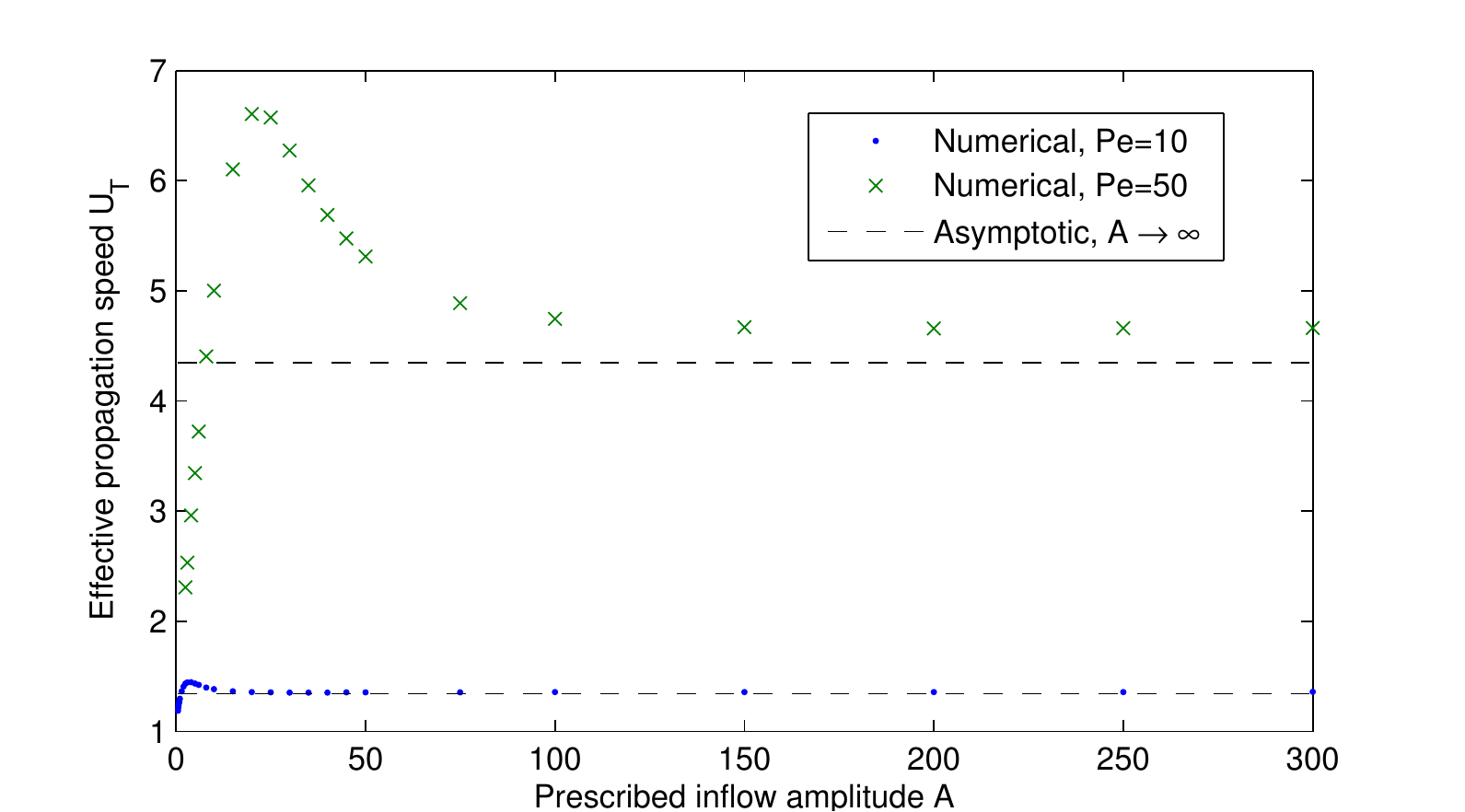}
\includegraphics[scale=0.9, trim=0 0 0 0,clip=true]{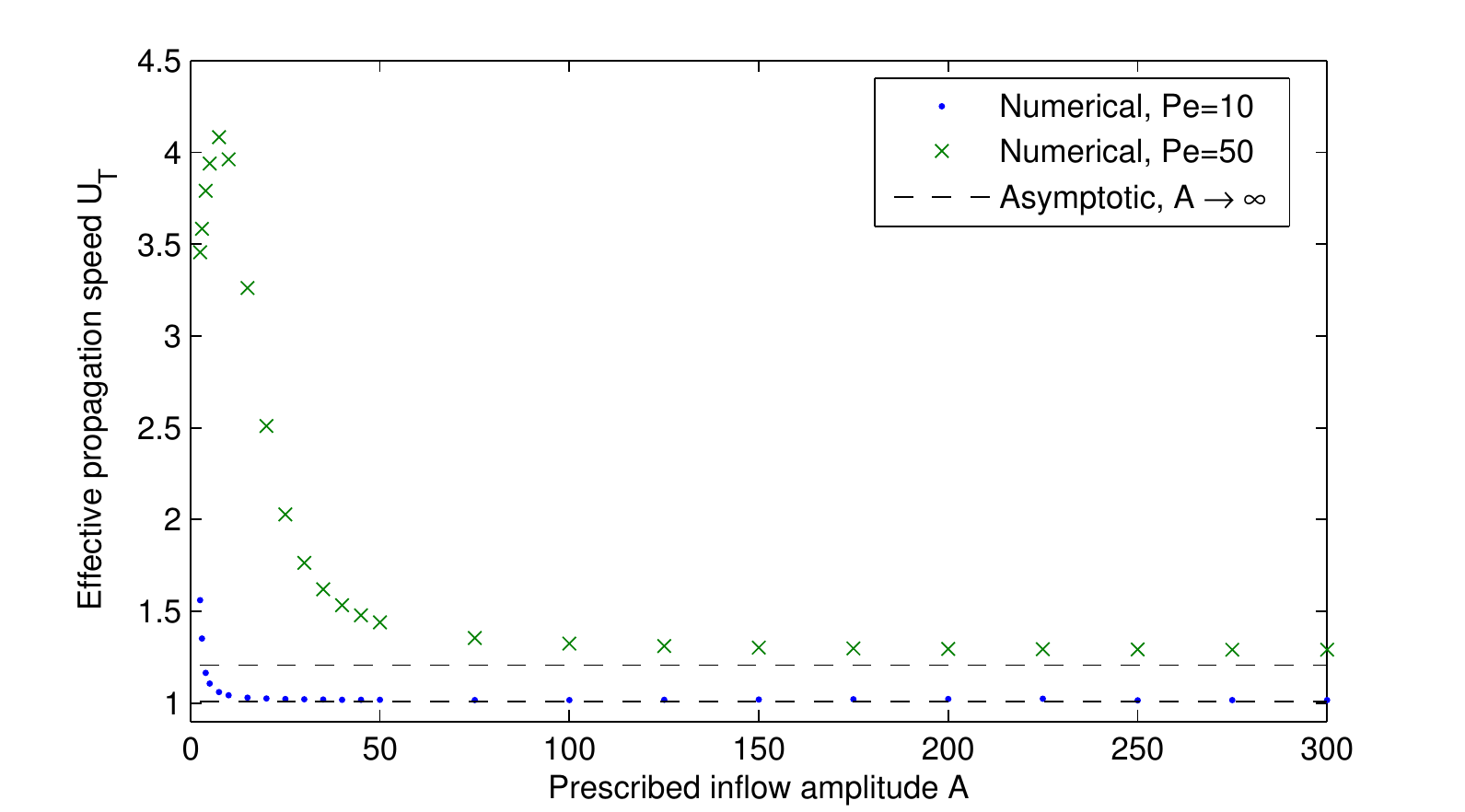}
\caption{Comparison of numerical results in a) constant density case $\alpha=0$ and b) variable density case $\alpha=0.85$. The effective propagation speed $U_T$ is scaled in each case by the numerical value of the planar premixed flame with no prescribed inflow ($A=0$) and ``free-slip" sidewalls. Also included are asymptotic results in the limit $A\to\infty$, $\Pe=O(1)$, from equation \eqref{premixed2:res:2}. For all simulations we take $Pr=1$, $\beta=10$ and $\alpha_h=0.85$.}
\label{fig:comb_comp}
\end{figure}
In this section we investigate the effect of thermal expansion on premixed flames propagating against a prescribed parallel inflow. We begin with a discussion of the behaviour of a premixed flame in a channel subject to thermal expansion when no flow is prescribed (i.e. the case $A=0$).

If no flow is prescribed in the constant density case, the flame is planar and propagates at the laminar flame speed, so that $U_T=1$. However, this is not the case when thermal expansion is present. As can be seen in figure \ref{fig:no_flow}, the propagation speed of variable density flames is larger than the laminar flame speed if $\epsilon$ is small enough, even though no flow is prescribed at $x=-\infty$ in these calculations. For larger values of $\alpha$, the propagation speed is increased by a larger amount.

The increase in the propagation speed as $\epsilon \to 0$ is due to the flame being necessarily curved. This is attributable to the effect of the no-slip sidewalls. Thermal expansion across the flame causes a shear (Poiseuille) flow to be induced downstream, and the flame becomes curved (or ``tulip" shaped), as can be seen in figure \ref{fig:no_flow_shapes}. Figures \ref{fig:no_flow} and \ref{fig:no_flow_shapes} also show results for a flame with $\alpha=0.85$, but $u_y$=0 at the sidewalls (instead of the no-slip condition).  As can be seen, in this case the flame is planar and propagates at the laminar flame speed, which shows that the no-slip sidewalls are the cause of the increase in propagation speed due to thermal expansion as $\epsilon \to 0$. A similar effect was found in \cite{kagan2003transition} and is numerically studied in detail in \cite{marra1996numerical}.

At this point it should be noted that various transient behaviours of premixed flames in closed channels that arise as a result of the frictional effects of sidewalls, combined with other effects of thermal expansion such as the Darrieus--Landau instability, have been investigated in the literature (for a review, see e.g. \cite{bychkov2000dynamics}). However, in the infinitely long channel configuration studied here, it is sufficient to note that steady solutions exist and that the steady solutions calculated here are all found to be stable to small perturbations.

Now, we plot in figure \ref{fig:comb_comp} the effective propagation speed $U_T$ versus the prescribed inflow amplitude $A$ for selected values of the Peclet number $\Pe$ in both the constant and variable density cases. As expected, it can be seen that as $A \to 0$, the propagation speed in the constant density case approaches $U_T=1$ and the propagation speed in the variable density case approaches a value $U_T>1$. It can also be seen that the prescribed inflow has a larger effect on the propagation speed in the constant density case than in the variable density case. Thus for the same value of the Peclet number, the propagation speed in the constant density case is larger than the propagation speed in the variable density case, if $A$ is large enough. This is to be expected as the asymptotic results \eqref{premixed2:res:2}, valid in the limit $A \to \infty$, show this behaviour. The results \eqref{premixed2:res:2} are included in figure \ref{fig:comb_comp}.

To summarise, despite the effect of wall friction and the lessened effect of a prescribed parallel flow on a variable density premixed flame, the qualitative shape of the curve of $U_T$ versus $A$ remains the same in the constant and variable density cases. This shows that in a laminar flame at least, the bending effect cannot be attributed to the effects of thermal expansion.
\section{Conclusion}
\label{premixed2:sec:conc}
In this chapter we have discussed the relevance of asymptotic results obtained in the thick flame asymptotic limit to the bending effect of turbulent combustion. The asymptotic result \eqref{result:infinitebeta}, obtained in Chapter \ref{chapter:premixed1} in the limit $\epsilon \to \infty$ followed by $\beta \to \infty$, with $\Pe=O(1)$, has been compared to the further asymptotic result \eqref{premixed2:res:2}, which has been derived in the limit $\beta \to \infty$ followed by $\epsilon \to \infty$, with $\Pe=O(1)$. It has been found that the results agree for moderately small values of the Peclet number, with closer agreement in the variable density case than in the constant density case. Numerical results have been provided, showing that as the prescribed inflow amplitude $A$ increases to large values, the effective propagation speed $U_T$ approaches a constant value, which depends on the Peclet number. This result is in line with both the asymptotic results \eqref{result:infinitebeta} and \eqref{premixed2:res:2} and with the available experimental results on turbulent combustion, summarised in figure \ref{fig:ronney_premixed}. We have therefore shown that for laminar flames in the context of a parallel flow, premixed flames in the thick flame asymptotic limit mimic the behaviour of turbulent premixed flames in the relationship between the effective propagation speed and the flow intensity $A$, which is analagous to the turbulence intensity.

We have also provided some asymptotic results for small values of $A$ in the constant density case, given by equation \eqref{premixed2:constantresultb}, in order to provide a full asymptotic picture of the relationship between the effective propagation speed and $A$. These results show that for small values of $A$, the relationship between $U_T$ and $A$ is linear. Finally, we have investigated the effect of thermal expansion on the problem; it has been found that in the limit $A\to 0$, variable density premixed flames in channels propagate at a speed larger than the laminar flame speed, due to the fact that the flame curves as a result of the effect of wall friction. Despite this difference between constant and variable density premixed flames, it has been found that for variable density premixed flames the effective propagation speed still approaches a constant value which depends on the Peclet number for large values of $A$, which again mimics the behaviour of the turbulent premixed flames shown in figure \ref{fig:ronney_premixed}.

\chapter{Rayleigh--B\'{e}nard Instability Generated by a Diffusion Flame}
\chaptermark{Diffusion flame instability}
\label{chapter:diffusion}
\section{Introduction}
The presence of flames in chemically reacting systems, whether premixed or non-premixed, naturally generates temperature gradients. Such systems are therefore prone to buoyancy-driven instabilities which have a paradigm in Rayleigh--B\'{e}nard convection. In this chapter, which is based on a paper by \citet{pearce2013rayleigh}, we revisit the Rayleigh--B\'{e}nard problem in the specific context of a diffusion or non-premixed flame, a fundamental problem which seems to have received no attention in the literature. The aim of the work is to complement the available knowledge on flame stability by determining the critical conditions which define the threshold of instability of a planar diffusion flame under gravitational effects.

The Rayleigh--B\'{e}nard problem itself has been important in work on the stability of physical systems since the early studies at the beginning of the 20th century by \citet{benard1} and \citet{rayleigh1916convection} on the natural convection of a fluid layer heated from below. The temperature gradient in the system causes the fluid at the bottom of the layer to be lighter than the fluid at the top, an arrangement which becomes unstable if the temperature gradient is strong enough. The instability is opposed by the viscous forces of the fluid. It was in Lord Rayleigh's seminal paper that it was first demonstrated that a non-dimensional parameter which later became known as the Rayleigh number must exceed a critical value in order for the aforementioned instability to manifest itself. This parameter was defined in terms of the gravitational acceleration $g$, the height of the fluid layer $L$, the temperature gradient $\hat{\beta}$ and the coefficients of thermal expansion $\hat{\alpha}$, thermal diffusivity $\kappa$ and kinematic viscosity $\nu$ as
\begin{gather*}
Ra=\frac{g \hat{\alpha} \hat{\beta} L^4}{\kappa \nu}.
\end{gather*}

If the Rayleigh number exceeds its critical value, which may be denoted $Ra_c$, the resulting instability resolves itself into a steady state of 'convection rolls', observed by B\'{e}nard in his original experiments as a regular structure of hexagonal cells. The problem of determining the critical Rayleigh number and characterising the resulting instabilities has been studied experimentally, numerically and analytically in a huge amount of research, which we do not review here. For a complete overview of the canonical problem see \citet[pp. 1--75]{chandrasekhar1961hydrodynamic} and literature reviews by \citet[pp.1--26]{getling1998rayleigh} and \citet{bodenschatz2000recent}; more recent reviews concerning turbulence in Rayleigh--B\'{e}nard convection have been performed by \citet{ahlers2009heat} and \citet{lohse2010small}.

The stability of steady states has also formed a crucial aspect of the study of both premixed and non-premixed flames. A thorough review of recent literature concerning flame instabilities has been performed by \citet{matalon2009flame}. A further review concentrating on instabilities in premixed flames can be found in \cite{bychkov2007rayleigh}; here we concentrate on instabilities in non-premixed combustion.

There have been a large number of studies on the diffusive-thermal instability of planar diffusion flames. Early studies focused on oscillatory instabilities of diffusion flames for fuel and oxidiser Lewis numbers greater than 1 \citep{kirkbey1966analytical,cheatham1996heat,cheatham1996near} before further studies including cellular instabilities for values of the Lewis number below 1 \citep{kim1996diffusional,kim1997linear,cheatham2000general,kukuck2001onset,vance2001stability,miklavvcivc2005oscillations,matalon2006diffusive}. The common factor in these studies is the use of the \emph{constant density approximation} which simplifies the study of combustion phenomena by separating the hydrodynamics from the equations for temperature and mass fractions. Recently this approximation was dispensed with by \citet{matalon2010effect} to investigate the effect of thermal expansion on the stability of diffusion flames, again using the planar diffusion flame as the unperturbed state. It was found that, although thermal expansion does not play as crucial a role as it does for premixed flames, it influences the regions of parameter space for which the diffusive-thermal instability occurs.

A key point in the stability analyses previously performed on the planar diffusion flame in the absence of gravity is that of the unconditional stability of the Burke--Schumann diffusion flame, which arises in the limit of infinite reaction rate. Our study is motivated by the idea that the mechanism for the Rayleigh--B\'{e}nard instability described above could be expected to have a similar effect on a layer of fluid heated from below by a horizontal diffusion flame if buoyancy is taken into account, leading to instabilities of a system previously considered unconditionally stable. There seems to our knowledge to be very little work addressing the effect of gravity on planar diffusion flames in the literature. Stability studies that have taken gravity into account have been focused on the flickering motion of diffusion flames \citep{buckmaster1988infinite,arai1999gravity,jiang2000combustion}, diffusion flames over a solid fuel \citep{wu2003numerical} or triple flames aligned with the gravity vector \citep{echekki2004numerical}.

In the present study we consider the stability of a horizontal planar diffusion flame in an infinitely long channel of given height under the effects of gravity. The main aim is to find the critical conditions for instability in the form of the critical value of a suitably defined Rayleigh number which defines the threshold of instability. An investigation into the instabilities in this problem will provide a crucial step towards a full understanding of the interaction between diffusion flames and hydrodynamics. This will also provide a basis for investigating the combined effect of thermal expansion and gravity on the propagation of triple flames, which leave a trailing diffusion flame behind them.

We formulate the problem in the low Mach number approximation, as derived in Chapter \ref{chapter:intro}. Further, in order to treat the problem analytically as far as possible we employ the Boussinesq approximation commonly used to study non-reactive Rayleigh--B\'{e}nard convection, which was derived by \citet{boussinesq1903theorie}. A rigorous derivation of this model from the general equations of combustion theory has been performed by \citet{matkowsky1979asymptotic}. The results of a linear stability analysis in this formulation will be compared to a numerical treatment of the non-Boussinesq equations in order to test the effect of compressibility on the system.

The chapter is structured as follows. We begin in \S \ref{diff:sec:form} by providing the governing equations, which we then non-dimensionalise and simplify in the Boussinesq approximation. We proceed in \S \ref{diff:sec:aa} with an asymptotic analysis in the Burke--Schumann limit of infinite Damk\"{o}hler number, followed by a linear stability analysis in \S \ref{diff:sec:stab} using the planar Burke--Schumann flame as the base state. In \S \ref{diff:sec:sol} we solve the linear stability problem for the growth rate eigenvalue numerically and analytically as far as possible to investigate the Rayleigh number at which the planar Burke--Schumann flame becomes unstable. In \S \ref{diff:sec:res} we present the results of the linear stability analysis, including how the critical Rayleigh number depends on the position of the flame in the channel. Finally, we compare these results in \S \ref{diff:section:numerical} with a full numerical treatment to investigate the accuracy of the Boussinesq approximation and the effects of compressibility upon the stability of the system, followed by a short investigation of the effect of finite chemistry. The chapter is closed with a discussion of the main findings and recommendations for future related studies.
\section{Formulation}\label{diff:sec:form}
\begin{figure}
\includegraphics[scale=0.8]{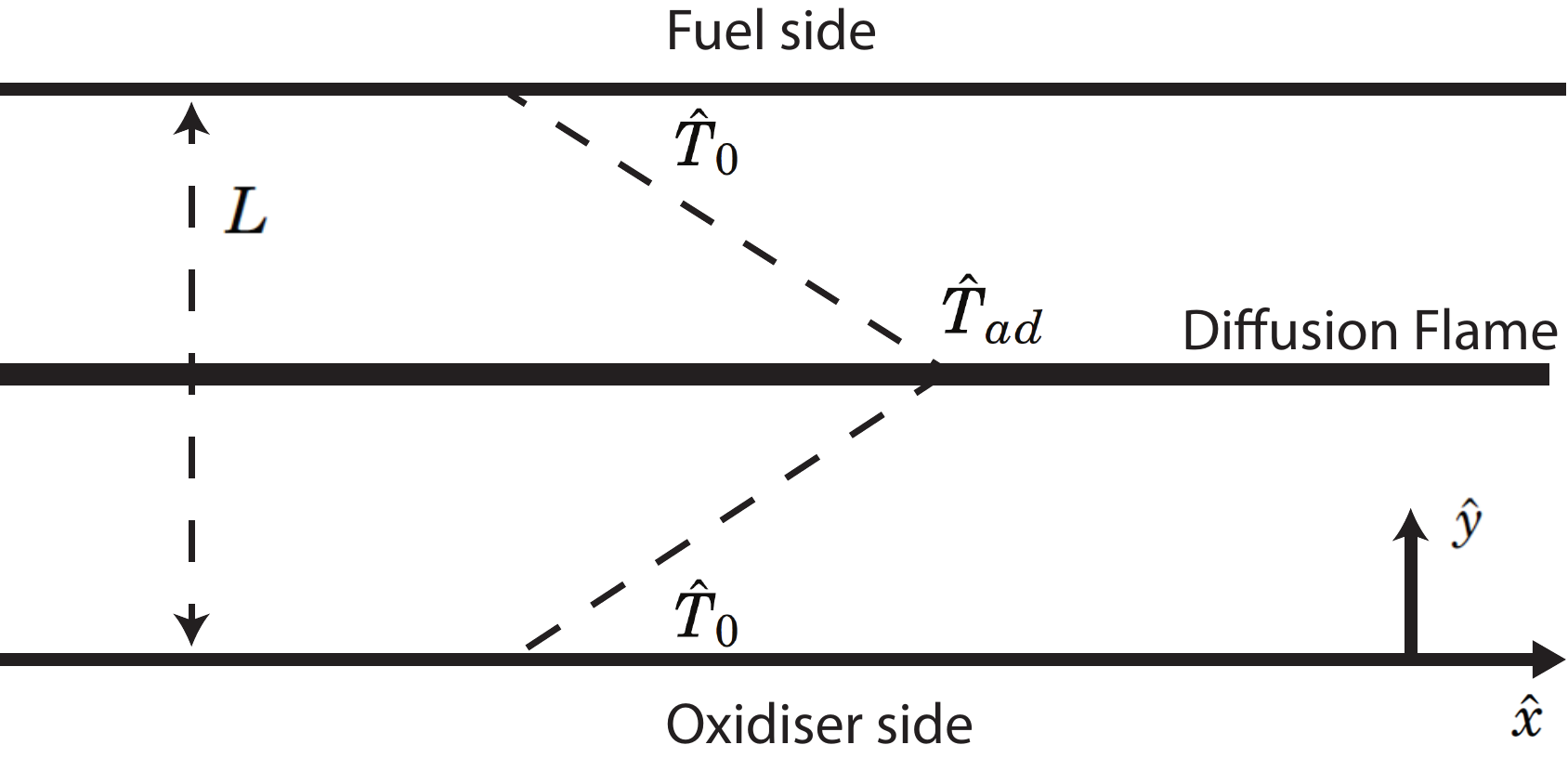}
\caption{An illustration of a planar diffusion flame in a channel of height $L$ with the fuel provided at the upper wall where $\hat{Y}_F=\hat{Y}_{Fu}$, $\hat{Y}_O=0$ and the oxidiser provided at the lower wall where $\hat{Y}_F=0$, $\hat{Y}_O=\hat{Y}_{Ou}$. Both walls are assumed to be rigid and to have equal temperatures $\hat{T}=\hat{T}_u$.}
\label{diff:fig:diagram}
\end{figure}
We investigate the problem of the stability of a planar diffusion flame in an infinitely long channel of height $L$. The channel walls are assumed to be porous with the fuel being provided at the upper wall and the oxidiser provided at the lower wall (see figure \ref{diff:fig:diagram}); for simplicity, the temperatures of the walls are assumed to be equal. Although this setup may be difficult to achieve experimentally, it is adopted here as a simple theoretical model to aid understanding of the effect of gravity on the planar diffusion flame. A similar setup has been used in several previous theoretical investigations (see e.g. \citep[][]{sohn1999instability,buckmaster2000holes,daou2010triple}). Since we are taking the effects of density changes and gravity into account, the governing equations will consist of the Navier--Stokes equations, coupled to equations for temperature and mass fractions of fuel and oxidiser. The combustion is modelled as a single irreversible one-step reaction of the form
\begin{align*}
\text{F}+s\text{O} \to (1+s)\text{Products}+q,
\end{align*}
where F denotes the fuel and O the oxidiser. The quantity $s$ denotes
the mass of oxidiser consumed and $q$ the heat released, both per unit mass of fuel.\\
The overall reaction rate $\hat{\omega}$ is taken to follow an Arrhenius law of the form
\begin{align*}
\hat{\omega}=\hat{\rho} B \hat{Y}_F \hat{Y}_O \exp{\left(-
E\left/R\hat{T}\right.\right)}.
\end{align*}
Here $\hat{\rho}$, $\hat{Y}_F$, $\hat{Y}_O$, $R$, $\hat{T}$, $B$ and $E$ are the density, the fuel mass fraction, the oxidiser mass fraction, the
universal gas constant, the temperature, the pre-exponential factor and the activation
energy of the reaction, respectively.
\subsection{Governing Equations and Boundary Conditions}
In the low Mach number approximation, the governing equations are
\begin{gather}
\frac{\partial \hat{\rho}}{\partial \hat{t}} + \nabla\cdot \left(\hat{\rho} \hat{\bf{u}}\right) = 0, \label{diff:1}\\
\hat{\rho} \frac{\partial \hat{\bf{u}}}{\partial \hat{t}} + \hat{\rho} \hat{\bf{u}} \cdot \nabla \hat{\bf{u}}+ \nabla \hat{p}=\mu\left(\nabla^2\hat{\bf{u}} + \frac{1}{3}\nabla \left(\nabla \cdot \hat{\bf{u}}\right)\right) + \hat{\rho}\bf{\hat{g}},\label{diff:eq:lowmachgravity}\\
\hat{\rho} \frac{\partial \hat{T}}{\partial \hat{t}} + \hat{\rho} \hat{\bf{u}} \cdot \nabla \hat{T}=\hat{\rho}D_T\nabla^2\hat{T} + \frac{q}{c_P}\hat{\omega}, \label{diff:3}\\
\hat{\rho} \frac{\partial \hat{Y}_F}{\partial \hat{t}} + \hat{\rho} \hat{\bf{u}} \cdot \nabla \hat{Y}_F=\hat{\rho}D_F\nabla^2\hat{Y}_F -\hat{\omega},\\
\hat{\rho} \frac{\partial \hat{Y}_O}{\partial \hat{t}} + \hat{\rho} \hat{\bf{u}} \cdot \nabla \hat{Y}_O=\hat{\rho}D_O\nabla^2\hat{Y}_O -s\hat{\omega},\label{diff:5}\\
\hat{\rho}\hat{T}=\hat{\rho}_u  \hat{T}_u, \label{diff:state}
\end{gather}
where the $\hat{}$ notation denotes dimensional terms and $D_T$, $D_F$, and $D_O$ denote the diffusion coefficients of heat, fuel and oxidiser respectively.
Here $\hat{\rho} D_T$, $\hat{\rho} D_F$ and $\hat{\rho} D_O$ are all assumed constant, as are the specific heat capacity $c_P$, thermal conductivity $\lambda$ and dynamic viscosity $\mu$.

As shown in figure \ref{diff:fig:diagram} the walls are assumed to be rigid, porous and at the same temperature, giving the boundary conditions
\begin{gather}
\hat{T}=\hat{T}_u,\text{ }\hat{Y}_F=0,\text{ }\hat{Y}_O=\hat{Y}_{Ou},\text{ }\hat{u}=\hat{v}=0,\text{ at }\hat{y}=0,\label{diff:bc:dim1}\\
\hat{T}=\hat{T}_u,\text{ }\hat{Y}_F=\hat{Y}_{Fu},\text{ }\hat{Y}_O=0,\text{ }\hat{u}=\hat{v}=0,\text{ at }\hat{y}=L.\label{diff:bc:dim2}
\end{gather}

Before continuing we note that equations (\ref{diff:3})-(\ref{diff:5}) have a steady, one dimensional solution with no flow which, for large activation energy $E$, as typically encountered in combustion, is very close to the frozen solution (with zero reaction rate $\hat{\omega}$) given by
\begin{gather}
\hat{T}=\hat{T}_u, \label{diff:frozen1}\\
\hat{Y}_F=\hat{Y}_{Fu}\frac{\hat{y}}{L},\\
\hat{Y}_O=\hat{Y}_{Ou}\left(1-\frac{\hat{y}}{L}\right) \label{diff:frozen3}.
\end{gather}
Equations (\ref{diff:frozen1})-(\ref{diff:frozen3}) determine the location of the stoichiometric surface $\hat{y}=\hat{y}_{st}$, where $\hat{Y}_{O}=s\hat{Y}_F$, and the values of the mass fractions there in the absence of combustion as
\begin{gather}
\frac{\hat{y}_{st}}{L}=\frac{1}{1+S}\quad \hat{Y}_{F,st}=\frac{\hat{Y}_{Fu}}{1+S},\quad \hat{Y}_{O,st}=\frac{S\hat{Y}_{Ou}}{1+S} \label{diff:eq:frozen4},
\end{gather}
where $S \equiv s\hat{Y}_{Fu} / \hat{Y}_{Ou}$ is a normalised stoichiometric coefficient.

We now introduce the non-dimensional variables
\begin{align*}
x=\frac{\hat{x}}{L},\text{ }y=\frac{\hat{y}}{L},\text{ }u=\frac{\hat{u}}{D_T\left/L\right.},\text{ }v=\frac{\hat{v}}{D_T\left/L\right.},\\
t=\frac{\hat{t}}{L^2\left/D_T\right.},\text{ }\theta=\frac{\hat{T}-\hat{T}_u}{\hat{T}_{ad}-\hat{T}_u},\text{ }y_F=\frac{\hat{Y}_F}{\hat{Y}_{F,st}},\text{ }\\ y_O=\frac{\hat{Y}_O}{\hat{Y}_{O,st}}, \text{ }p=\frac{\hat{p}}{\hat{p}_0};
\end{align*}
note that $p$ is the hydrodynamic pressure with reference unit $\hat{p}_0$=$\left(D_T\left/L\right.\right)^2 \hat{\rho}_u$. Here the reference length $L$ is the height of the channel. $\hat{T}_{ad}$ is the adiabatic flame temperature given by $\hat{T}_{ad}=\hat{T}_u+q\hat{Y}_{F,st}\left/ c_P\right.$.
The non-dimensional governing equations are then
\begin{gather}
\frac{\partial \rho}{\partial t} + \nabla\cdot \left(\rho {\bf{u}}\right) = 0,\label{diff:eq:machnondim1}\\
\rho \frac{\partial {\bf{u}}}{\partial t} + \rho {\bf{u}} \cdot \nabla {\bf{u}}+ \nabla P^*=Pr \nabla^2{\bf{u}} +\frac{Pr Ra}{\alpha} \left(\rho-1 \right) \frac{{\bf{g}}}{|{\bf{g}}|} \label{diff:eq:machgrav},\\
\rho \frac{\partial \theta}{\partial t} + \rho {\bf{u}} \cdot \nabla \theta= \nabla^2 \theta + Da\omega,\label{diff:eq:machtemp1}\\
\rho \frac{\partial y_F}{\partial t} + \rho {\bf{u}} \cdot \nabla y_F= \frac{1}{Le_F}\nabla^2 y_F - Da\omega,\label{diff:eq:machfuel}\\
\rho \frac{\partial y_O}{\partial t} + \rho {\bf{u}} \cdot \nabla y_O= \frac{1}{Le_O}\nabla^2 y_O - Da\omega,\label{diff:eq:machox}
\end{gather}
where $\alpha$ is the thermal expansion coefficient $\alpha=\left(\hat{T}_{ad}-\hat{T}_u\right)\left/\hat{T}_{ad}\right.$. Note that $P^*$ is a modified pressure given by $P^*=p-\frac{Pr}{3}\left(\nabla \cdot \bf{u}\right)$. The non-dimensional parameters are
\begin{align*}
Ra &=\frac{g\left(\hat{T}_{ad}-\hat{T}_{u}\right)L^3}{\nu \hat{T}_{ad} D_T}, \quad Da =\frac{4L^2}{\beta^3 D_T} Le_F Le_O B\hat{Y}_{O,st}\exp{\left(-E\left/R\hat{T}_{ad}\right.\right)},\\
Pr &=\frac{\mu c_P} {\lambda},\quad
Le_F =\frac{D_T}{D_F},\quad
Le_O =\frac{D_T}{D_O},
\end{align*}
which are the Rayleigh number, the Damk\"{o}hler number, the Prandtl number and the fuel and oxidiser Lewis numbers, respectively. Here, $\nu$ is the kinematic viscosity $\nu=\mu \left/ \hat{\rho}_u \right.$ and $\beta$ is the Zeldovich number or non-dimensional activation energy defined as $\beta=E\left(\hat{T}_{ad}-\hat{T}_u\right) \left/ R\hat{T}_{ad}^2\right.$. It is worth mentioning that we have defined the Rayleigh number to be of the form $Ra=g \Delta T L^3 \left/ \nu T_r D_T \right.$, where $\Delta T$ measures the temperature difference and $T_r$ is a reference temperature taken here to be $\hat{T}_{ad}$. This form has been adopted previously in the literature, for example by \citet{frohlich1992large}, where the reference temperature was taken to be an average value.\\
The non dimensional reaction rate is given by
\begin{gather}
\omega=\frac{\beta^3}{4Le_F Le_O} \rho y_F y_O \exp{\left(\frac{\beta ( \theta-1)}{1+\alpha(\theta-1)}\right)} \label{diff:eq:rrate}
\end{gather}
and the ideal gas equation (\ref{diff:state}) takes the non-dimensional form
\begin{gather}
\rho=\left(1+\frac{\alpha}{1-\alpha}\theta\right)^{-1}. \label{diff:eq:idealgas}
\end{gather}
Finally, (\ref{diff:bc:dim1})-(\ref{diff:bc:dim2}) and (\ref{diff:eq:frozen4}) imply that the boundary conditions are
\begin{gather}
\theta=0,\text{ }y_F=0,\text{ }y_O=\frac{S+1}{S},\text{ }u=v=0,\text{ at }y=0,\label{diff:eq:bc1}\\
\theta=0,\text{ }y_F=1+S,\text{ }y_O=0,\text{ }u=v=0,\text{ at }y=1.\label{diff:eq:bc2}
\end{gather}
The non-dimensional problem is now fully formulated and is given by equations (\ref{diff:eq:machnondim1})-(\ref{diff:eq:idealgas}) with boundary conditions (\ref{diff:eq:bc1}) and (\ref{diff:eq:bc2}). The non-dimensional parameters in this problem are $\alpha$, $\beta$, $Pr$, $Ra$, $Da$, $S$, $Le_F$ and $Le_O$.

\subsection{Mixture fraction formulation}
\label{diffusion:sec:mixfracform}
\subsubsection{Formulation}
We can simplify the problem by making the assumption of unity Lewis numbers,
\begin{gather}
 Le_F=Le_O=1.
\end{gather}
In this case we note that, from equations (\ref{diff:eq:machtemp1})-(\ref{diff:eq:machox}) and the boundary conditions (\ref{diff:eq:bc1})-(\ref{diff:eq:bc2}), the quantity
\begin{gather}
\Phi=y_F+Sy_O+\left(S+1\right)\left(\theta-1\right)\label{diff:eq:phidefinition}
\end{gather}
satisfies the equation
\begin{gather}
\rho  \frac{\partial \Phi}{\partial t} +\rho {\bf{u}} \cdot \nabla \Phi= \nabla^2 \Phi,\label{diff:eq:machscalar}
\end{gather}
subject to the boundary conditions
\begin{gather}
\Phi=0 \quad \text{at } y=0\text{ and }y=1.\label{diff:eq:machscalarbc}
\end{gather}
Clearly $\Phi=0$ is a stationary solution of (\ref{diff:eq:machscalar})-(\ref{diff:eq:machscalarbc}). This will be the only solution we retain, in view of the focus of our linear stability analysis on the base state with no flow and $\Phi=0$. A justification for retaining only this solution is presented below.

For solutions that are periodic in $x$ of period $L$, define the $L^2$ scalar product
\begin{gather*}
\left<\Phi,\Psi\right>=\int \limits_{x=0}^L \int\limits_{y=0}^1 \Phi \Psi \, \mathrm{d}x\, \mathrm{d}y,
\end{gather*}
with $||\Phi||^2=\left<\Phi,\Phi\right>$. Then equation (\ref{diff:eq:machscalar}) leads to
\begin{gather*}
\left<\rho \frac{\partial \Phi}{\partial t}, \Phi\right> + \left<\rho {\bf{u}} \cdot \nabla \Phi,\Phi\right>=\left<\nabla^2 \Phi,\Phi\right>.
\end{gather*}
Using integration by parts and equation (\ref{diff:eq:machnondim1}), this can be written
\begin{gather*}
\frac{1}{2}\frac{d}{dt}\left<\rho\Phi,\Phi\right>=-\left(||\Phi_x ||^2+||\Phi_y||^2\right),
\end{gather*}
or, using the Poincar\'{e} inequality and the fact that $0<\rho\leq 1$,
\begin{gather*}
\frac{1}{2}\frac{d}{dt}\left<\rho\Phi,\Phi\right> \leq -c^2||\Phi||^2 \leq -c^2\left<\rho \Phi,\Phi\right>,
\end{gather*}
for some constant $c$. Hence,
\begin{gather*}
\left<\rho \Phi,\Phi\right> \leq \Phi_0 e^{-2c^2t},
\end{gather*}
where $\Phi_0=\left<\rho\Phi,\Phi\right>\left(t=0\right)$. We therefore conclude that \[\lim_{t\to\infty}\,\Phi=0,\] which justifies retaining only the stationary solution $\Phi=0$.

It follows on using (\ref{diff:eq:phidefinition}) that
\begin{gather}
y_F+Sy_O+\left(S+1\right) \theta = S+1.
\end{gather}
We now observe, by adding equations (\ref{diff:eq:machtemp1}) and (\ref{diff:eq:machfuel}), that the quantity $y_F+\theta$ satisfies the equation
\begin{gather}
\rho  \frac{\partial \left(y_F+\theta\right)}{\partial t} +\rho {\bf{u}} \cdot \nabla \left(y_F+\theta\right)= \nabla^2 \left(y_F+\theta\right),
\end{gather}
subject to the boundary conditions
\begin{gather}
y_F+\theta=0 \quad \text{at } y=0,\\
y_F+\theta=1+S \quad \text{at } y=1.
\end{gather}
This suggests defining the \emph{mixture fraction} $Z$ by
\begin{gather}
y_F+\theta=\left(1+S\right)Z,\label{diff:eq:fuelmixfrac}
\end{gather}
which implies that
\begin{gather}
y_O+\theta=\frac{S+1}{S}\left(1-Z\right).\label{diff:eq:oxmixfrac}
\end{gather}
The governing equations (\ref{diff:eq:machnondim1})-(\ref{diff:eq:machox}) then become
\begin{gather}
\frac{\partial \rho}{\partial t} + \nabla\cdot \left(\rho {\bf{u}}\right) = 0,\label{diff:eq:2machnondim1}\\
\rho \frac{\partial {\bf{u}}}{\partial t} + \rho {\bf{u}} \cdot \nabla {\bf{u}}+ \nabla P^*=Pr \nabla^2{\bf{u}} +\frac{Pr Ra}{\alpha} \left(\rho-1 \right) \frac{{\bf{g}}}{|{\bf{g}}|},\label{diff:eq:2machnondim2}\\
\rho \frac{\partial \theta}{\partial t} + \rho {\bf{u}} \cdot \nabla \theta= \nabla^2 \theta + Da\omega,\label{diff:eq:mixtemp}\\
\rho \frac{\partial Z}{\partial t} +  \rho {\bf{u}} \cdot \nabla Z= \nabla^2 Z,\label{diff:eq:machmixfrac}
\end{gather}
where
\begin{gather}
\omega=\frac{\beta^3}{4} \rho  \left(\left(1+S\right)Z-\theta\right)\left(\frac{1+S}{S}\left(1-Z\right)-\theta\right)\exp{\left(\frac{\beta\left(\theta-1\right)}{1+\alpha\left(\theta-1\right)}\right)}, \label{diff:eq:mixfracreaction}
\end{gather}
and $\rho$ is given by equation (\ref{diff:eq:idealgas}).
These equations are subject to the boundary conditions
\begin{gather}
\theta=0,\text{ }Z=0,\text{ }u=v=0,\text{ at }y=0,\label{diff:eq:mixbc1}\\
\theta=0,\text{ }Z=1,\text{ }u=v=0,\text{ at }y=1.\label{diff:eq:mixbc2}
\end{gather}

\subsubsection{Stationary planar diffusion flame}
The problem defined by (\ref{diff:eq:2machnondim1})-(\ref{diff:eq:mixbc2}) admits a stationary planar solution with no flow given by $Z=y$. The temperature can then be determined, using (\ref{diff:eq:mixtemp}), by the numerical solution to the equation
\begin{gather}
\frac{d^2\theta}{dy^2}+Da \frac{\beta^3}{4} \left(1+\frac{\alpha}{1-\alpha}\theta\right)^{-1}  \left(\left(1+S\right)y-\theta\right)\left(\frac{1+S}{S}(1-y)-\theta\right)\exp{\left(\frac{\beta\left(\theta-1\right)}{1+\alpha\left(\theta-1\right)}\right)}\nonumber\\=0,\label{diff:eq:scurve}
\end{gather}
with
\begin{gather}
\theta=0 \quad \text{at }y=0,\label{diff:bc:scurve1}\\
\theta=0 \quad \text{at }y=1.\label{diff:bc:scurve2}
\end{gather}
Note that the Rayleigh number merely affects the pressure (as dictated by equation (\ref{diff:eq:2machnondim2})) in the one-dimensional stationary system and does not affect the solution for temperature.

\begin{figure}
\includegraphics[scale=0.8]{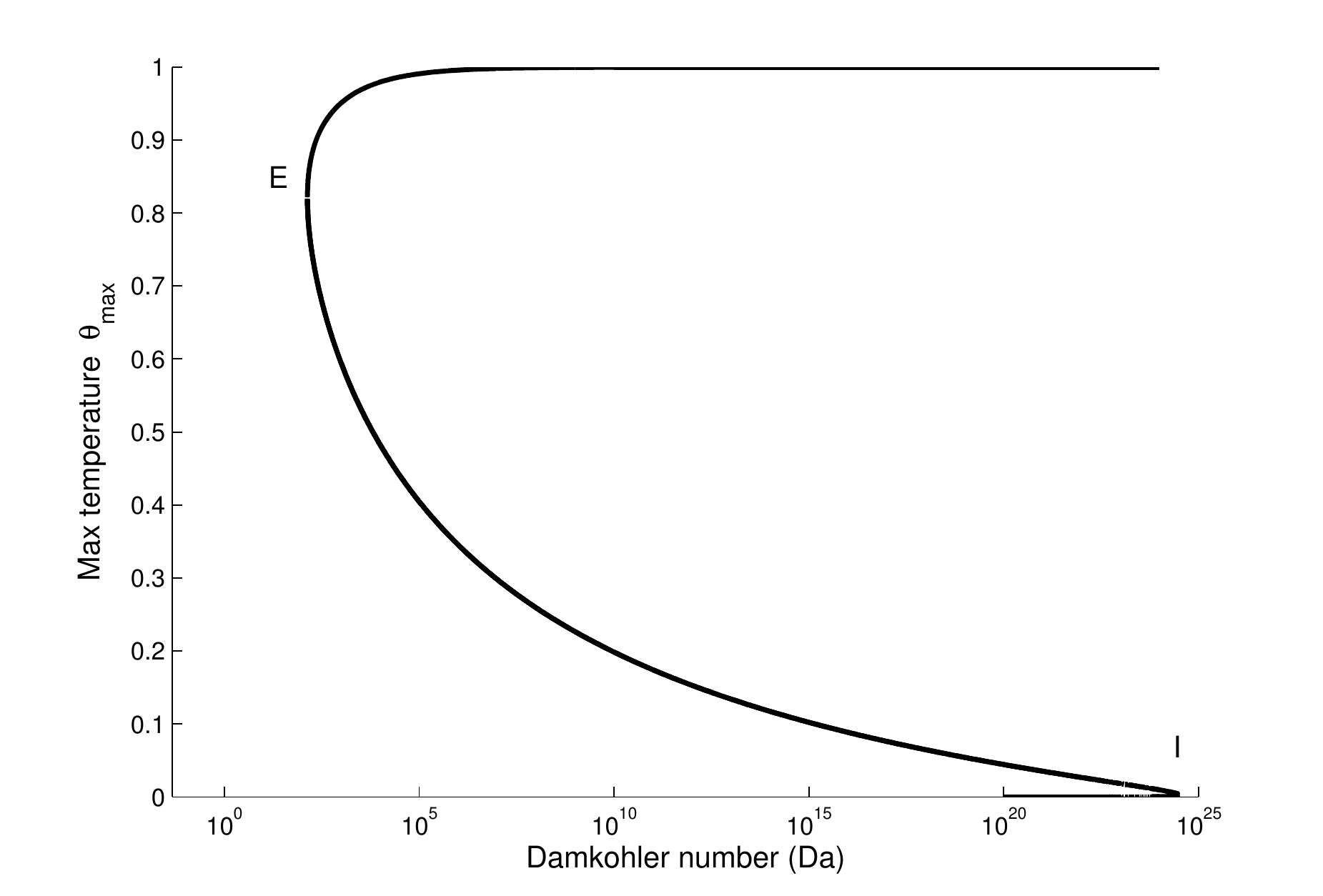}
\caption{S-shaped curve generated by plotting the maximum temperature of the planar diffusion flame against the Damk\"{o}hler number with labelled extinction (E) and ignition (I) points.}
\label{diff:fig:scurve}
\end{figure}

We now let $\beta,\alpha$ and $S$ take typical values of 10, 0.85 and 1 respectively and solve equation (\ref{diff:eq:scurve}) with conditions (\ref{diff:bc:scurve1}) and (\ref{diff:bc:scurve2}) for selected values of the Damk\"{o}hler number $Da$. We use the boundary value problem solver BVP4C, a finite difference code that implements the three-stage Lobatto IIIa formula  in Matlab \citep[][]{shampine2000solving}. We then plot the maximum temperature of the solution against $Da$. Figure \ref{diff:fig:scurve} shows the S-shaped curve that is generated; this is a classical curve characterising diffusion flames and has been comprehensively studied in the context of the constant density approximation \citep[][]{linan1974asymptotic}. The upper and lower branches are known as the strongly burning and weakly burning branches respectively, and both have been shown to be stable in the context of constant density diffusion flames, while the middle branch has been shown to be unstable \citep[][]{buckmaster1983fast}.

Notable in figure \ref{diff:fig:scurve} is the presence of an extinction and an ignition value of the Damk\"{o}hler number represented by points $E$ and $I$ respectively; for values of $Da$ below the extinction value the strongly burning solution cannot exist and for $Da$ above the ignition value there is no weakly burning solution.

We are interested in instabilities caused by the effect of the hydrodynamics on the upper branch of the S-shaped curve, in particular in the Burke--Schumann limit of infinite Damk\"{o}hler number. In \S \ref{diff:section:numerical}, we will numerically solve the governing equations in the mixture fraction formulation to investigate these instabilities; however, in order to treat the problem analytically as far as possible we now proceed with a reformulation of the problem in the Boussinesq approximation.
 \subsection{Boussinesq approximation}
For a detailed derivation of the governing equations of combustion theory in the Boussinesq approximation in the context of a premixed flame see \cite{sivashinsky1977nonlinear}. Here we use a similar approach and assume that the thermal expansion parameter $\alpha$ is small. Thus we expand equation (\ref{diff:eq:idealgas}) as $\alpha \to 0$ to obtain
\begin{gather*}
\rho=1-\alpha \theta +O\left(\alpha^2\right).
\end{gather*}
Using this result, and expanding all variables in successive powers of $\alpha$ in equations (\ref{diff:eq:2machnondim1})-(\ref{diff:eq:mixbc2}), yields to leading order
\begin{gather}
 \nabla\cdot  {\bf{u}} = 0,\label{diff:eq:nondim1}\\
 \frac{\partial {\bf{u}}}{\partial t} + {\bf{u}} \cdot \nabla {\bf{u}}+ \nabla p=Pr \nabla^2{\bf{u}} -Pr Ra \theta \frac{{\bf{g}}}{|{\bf{g}}|} \label{diff:eq:grav},\\
\frac{\partial \theta}{\partial t} + {\bf{u}} \cdot \nabla \theta= \nabla^2 \theta + Da \frac{\beta^3}{4} \left(\left(1+S\right)Z-\theta\right)\left(\frac{1+S}{S}\left(1-Z\right)-\theta\right)\exp{\left(\beta\left(\theta-1\right)\right)}\label{diff:eq:temp2},\\
 \frac{\partial Z}{\partial t} +  {\bf{u}} \cdot \nabla Z= \nabla^2 Z,\label{diff:eq:mixfrac}
\end{gather}
which are subject to the boundary conditions (\ref{diff:eq:mixbc1})-(\ref{diff:eq:mixbc2}).

We have thus reduced the number of equations in the unity Lewis number case so that the problem is now given by equations (\ref{diff:eq:nondim1})-(\ref{diff:eq:mixfrac}) with boundary conditions (\ref{diff:eq:mixbc1})-(\ref{diff:eq:mixbc2}). We proceed with an asymptotic analysis of the problem in the infinitely fast chemistry limit $Da \to \infty$, which will reduce the problem to a form comparable to the classic non-reactive case studied in \cite{chandrasekhar1961hydrodynamic} and others.

\section{Asymptotic analysis}\label{diff:sec:aa}
We now study the problem of the Burke--Schumann diffusion flame, which arises in the limit of infinite Damk\"{o}hler number. In this case $y_Fy_O=0$ throughout the domain to prevent an unbounded reaction rate, except at an (infinitely thin) reaction sheet located at $y=y_{st}\left( t,x\right)$, say, where the temperature is equal to its adiabatic value. Hence, using (\ref{diff:eq:fuelmixfrac})-(\ref{diff:eq:oxmixfrac}),
\begin{align}
y_F=\left(1+S\right)Z-\theta & =0, \quad   y<y_{st},\label{diff:eq:mixtemp1}\\
y_O=\frac{1+S}{S}\left(1-Z\right)-\theta & =0, \quad y>y_{st},\label{diff:eq:mixtemp2}\\
\lim_{y\to y^\pm_{st}}\theta & =1. \label{diff:eq:adiabatic}
\end{align}
Thus the domain is split into two parts: the region above the reaction sheet and the region below it, with
\begin{gather}
Z=\frac{1}{S+1}\theta, \quad y<y_{st},\label{diff:eq:stoichdef1}\\
Z=1-\frac{S}{S+1}\theta, \quad y>y_{st}.\label{diff:eq:stoichdef2}
\end{gather}
Note that (\ref{diff:eq:adiabatic}) and (\ref{diff:eq:stoichdef1})-(\ref{diff:eq:stoichdef2}) imply that
\begin{gather}
\lim_{y\to y^\pm_{st}}Z=\frac{1}{S+1}=Z_{st},
\end{gather}
which defines the stoichiometric mixture fraction $Z_{st}$.

We expand all variables outside the thin reaction zone, as $Da \to \infty$, in terms of the thickness of the reaction zone $\delta$, where
\begin{gather*}
\delta \mytilde Da^{-1/3} \ll 1.
\end{gather*}
This scaling follows from a reactive-diffusive balance in the reaction sheet. Indeed, writing $y_F \mytilde \delta y_F'$, $y_O \mytilde \delta y_O'$ and $\theta \mytilde 1 + \delta \theta'$ with $n= \delta n'$ inside the reaction sheet gives, from the leading order of equation (\ref{diff:eq:machtemp1}),
\begin{gather*}
\frac{\partial^2 \theta'}{\partial n'^2} = Da \delta^3 \omega',
\end{gather*}
where $\omega'=O(1)$. Thus, since the diffusion term and the reaction term (which are given by the left hand side and the right hand side of the equation above, respectively) must balance, we have the required scaling. We therefore write the outer expansions as
\begin{gather*}
\mathbf{u}=\mathbf{u}^0+\frac{\mathbf{u}^1}{Da^{1/3}}+..., \quad p=p^0+\frac{p^1}{Da^{1/3}}+...,\\
\theta=\theta^0+\frac{\theta^1}{Da^{1/3}}+...,\quad Z=Z^0+\frac{Z^1}{Da^{1/3}}+...
\end{gather*}
We now substitute into the governing equations (\ref{diff:eq:nondim1})-(\ref{diff:eq:mixfrac}) which become, to leading order
\begin{gather}
\frac{\partial u^0}{\partial x} +\frac{\partial v^0}{\partial y} = 0,\label{diff:eq:bouss1}\\
 \frac{\partial u^0}{\partial t} +  u^0 \frac{\partial u^0}{\partial x}+v^0 \frac{\partial u^0}{\partial y}+\frac{\partial p^0}{\partial x}=Pr\left(\frac{\partial^2 u^0}{\partial x^2} +\frac{\partial^2 u^0}{\partial y^2}\right) \label{diff:eq:bouss2},\\
 \frac{\partial v^0}{\partial t} +  u^0 \frac{\partial v^0}{\partial x}+ v^0 \frac{\partial v^0}{\partial y}+\frac{\partial p^0}{\partial y}=Pr\left(\frac{\partial^2 v^0}{\partial x^2} +\frac{\partial^2 v^0}{\partial y^2}\right) +PrRa\theta^0,\label{diff:eq:bouss3}\\
 \frac{\partial \theta^0}{\partial t}+  u^0 \frac{\partial \theta^0}{\partial x} +  v^0\frac{\partial \theta^0}{\partial y}=\frac{\partial^2 \theta^0}{\partial x^2} +\frac{\partial^2 \theta^0}{\partial y^2},\label{diff:eq:bouss4}
\end{gather}
where we have
\begin{gather}
\theta^0=\left(1+S\right)Z^0 \text{ for }y<y_{st}\quad\text{ and }\quad\theta^0=\frac{1+S}{S}\left(1-Z^0\right)\text{ for }y>y_{st} \label{diff:eq:relations}
\end{gather}
from equations (\ref{diff:eq:stoichdef1})-(\ref{diff:eq:stoichdef2}), so that the equation for $Z^0$ is not necessary. These equations are subject to the boundary conditions
\begin{gather}
u^0=v^0=\theta^0=0 \text{ at y=0},\label{diff:eq:boussbc1}\\
u^0=v^0=\theta^0=0\text{ at y=1},\label{diff:eq:boussbc2}
\end{gather}
from (\ref{diff:eq:mixbc1})-(\ref{diff:eq:mixbc2}).
\begin{figure}
\centering
\includegraphics[scale=1]{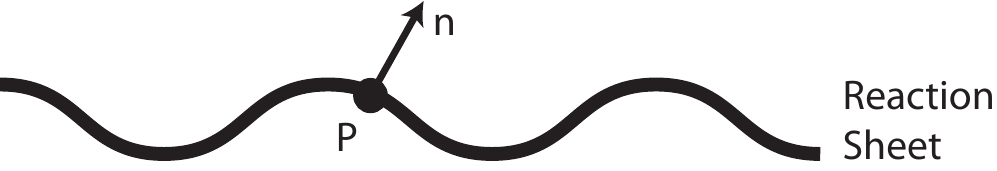}
\caption{A point $P$ on the reaction sheet and its normal coordinate.}
\label{diff:fig:sheet}
\end{figure}
To close the problem we need to provide jump conditions across the reaction sheet located at $y=y_{st}(t,x)$. These are given by
\begin{align}
\left[\theta^0 \right]=\left[u^0\right]=\left[v^0\right]=0 &,\label{diff:jc1}\\
\left[\frac{\partial u^0}{\partial n}\right]=\left[p^0\right]- Pr \left[\frac{\partial v^0}{\partial n}\right]=0 &,\\
\lim_{y\to y^\pm_{st}}\theta^0 =1, \quad \frac{\partial\theta^0\left(y_{st}^-\right)}{\partial n}+S\frac{\partial\theta^0\left(y_{st}^+\right)}{\partial n}=0 &,\label{diff:jc3}
\end{align}
where $[f]=f\left(y_{st}^+\right)-f\left(y_{st}^-\right)$ and $n$ is a coordinate normal to the reaction sheet. A derivation of these in a general context has been presented by \citet{cheatham2000general}, who assumed a large activation energy parameter $\beta \gg 1$, with leakages of the components through the reaction sheet. Since this is not the case here, for the convenience of the reader a short explanation of the derivation of conditions (\ref{diff:jc1})-(\ref{diff:jc3}) is given in the remainder of this section, which can be skipped by those familiar with such an approach.

Consider a small neighbourhood of a point $P$ on the reaction sheet, which has normal $n$ as shown in figure \ref{diff:fig:sheet}. We temporarily consider a coordinate system with origin $P$ and the $y$-axis directed along $n$. Thus in a small neighbourhood of $P$ the reaction sheet is located at $y=0$. We begin by assuming all state variables are continuous across the reaction sheet:
\begin{gather*}
\left[\theta^0 \right]=\left[Z^0\right]=0,
\end{gather*}
with the notation $\left[f\right]=f\left(y=0^+\right)-f\left(y=0^-\right)$.

Now we introduce a stretched variable $\eta$ given by
\begin{gather*}
y=\frac{\eta}{Da^{1/3}}
\end{gather*}
and inner expansions in the thin reaction layer for all variables in terms of the thickness of the reaction sheet $\delta \mytilde Da^{-1/3}$,
\begin{equation}
\left.
\begin{aligned}
&\mathbf{u}=\mathbf{u}_0+\frac{\mathbf{u}_1}{Da^{1/3}}+..., \quad p=p_0+\frac{p_1}{Da^{1/3}}+...,\\
&\theta=\theta_0+\frac{\theta_1}{Da^{1/3}}+...,\quad Z=Z_0+\frac{Z_1}{Da^{1/3}}+...
\end{aligned}
\right\}
\label{diff:eq:inner1}
\end{equation}
Note that a subscript denotes successive terms in the inner expansion and a superscript denotes successive terms in the outer expansion of a variable. The inner and outer variables must satisfy matching conditions given by
\begin{gather*}
\mathbf{u}_{\text{inner}}\left(\eta \to \pm \infty\right)=\mathbf{u}_{\text{outer}}\left(y \to 0^\pm\right),
\end{gather*}
for the velocity (and similar conditions on the other variables). Note that by expanding the outer solution around $y=0$  we find
\begin{gather}
\mathbf{u}_0\left(\eta \to \pm \infty\right)=\mathbf{u}^0\left(y \to 0^\pm\right), \label{diff:eq:match1}\\
\mathbf{u}_1\left(\eta \to \pm \infty\right)=\eta\frac{\partial \mathbf{u}^0\left(y \to 0^\pm\right)}{\partial y}+\mathbf{u}^1\left(y \to 0^\pm\right) \label{diff:eq:match2}
\end{gather}
for the velocity (and similar conditions on the other variables). We substitute the inner variables into equations (\ref{diff:eq:nondim1})-(\ref{diff:eq:mixfrac}) and apply the matching conditions in order to derive jump conditions for the outer variables across the reaction sheet located at $y=0$.

To leading order in equations (\ref{diff:eq:nondim1})-(\ref{diff:eq:grav}), after substituting in the inner expansions (\ref{diff:eq:inner1}), we find
\begin{gather*}
\frac{\partial^2 u_0}{\partial \eta^2}=\frac{\partial v_0}{\partial \eta}=0,
\end{gather*}
which can be integrated, using matching condition (\ref{diff:eq:match1}), to find
\begin{gather}
u_0=\text{const}, \quad v_0=\text{const},\label{diff:eq:uvconst}
\end{gather}
so that
\begin{gather*}
\left[u^0\right]=\left[v^0\right]=0. \label{diff:eq:uvjumps}
\end{gather*}

At $O\left(Da^{1/3}\right)$ in the $u$ momentum equation (\ref{diff:eq:grav}), after noting (\ref{diff:eq:uvconst}), we have
\begin{gather}
\frac{\partial^2 u_1}{\partial \eta^2}=0.\label{diff:eq:unextorder}
\end{gather}
Differentiating matching condition (\ref{diff:eq:match2}) with respect to $\eta$ and applying to an integration of (\ref{diff:eq:unextorder}) then leads to
\begin{gather*}
\left[\frac{\partial u^0}{\partial y}\right]=0.\label{diff:eq:dujump}
\end{gather*}

Similarly, we look to $O\left(Da^{1/3}\right)$ in the $v$ momentum equation (\ref{diff:eq:grav}) and apply the differentiated form of matching condition (\ref{diff:eq:match2}) to find
\begin{gather*}
\left[p^0\right]- Pr \left[\frac{\partial v^0}{\partial y}\right]=0.
\end{gather*}

Since the temperature is adiabatic at the flame we already have condition (\ref{diff:eq:adiabatic}) on the outer temperature profile, which gives
\begin{gather*}
\lim_{y\to 0^\pm}\theta^0 =1.
\end{gather*}
Finally, we have the jump condition on the mixture fraction, given by
\begin{gather*}
\left[ \frac{\partial Z^0}{\partial y} \right]=0,\label{diff:eq:zjump}
\end{gather*}
which is found by substituting (\ref{diff:eq:inner1}) into (\ref{diff:eq:mixfrac}) and integrating across the reaction sheet. From the relations (\ref{diff:eq:relations}) between $\theta^0$ and $Z^0$, this gives the final condition
\begin{gather*}
\frac{\partial\theta^0\left(y=0^-\right)}{\partial y}+S\frac{\partial\theta^0\left(y=0^+\right)}{\partial y}=0.
\end{gather*}

The jump conditions above are valid at $y=0$ in the coordinate system chosen. We can now generalise this to all points lying on the reaction sheet instead of just a neighbourhood of $P$ (since $P$ is arbitrary) and thus, by substitution of $y$ with the normal coordinate $n$, the conditions at the reaction sheet are given by conditions (\ref{diff:jc1})-(\ref{diff:jc3}) across $y=y_{st}$.

\section{Linear stability analysis}\label{diff:sec:stab}
To summarise, dropping the superscript notation from equations (\ref{diff:eq:bouss1})-(\ref{diff:eq:relations}) gives the governing equations as
\begin{gather}
\frac{\partial u}{\partial x} +\frac{\partial v}{\partial y} = 0,\label{diff:2eq:1}\\
 \frac{\partial u}{\partial t} +  u \frac{\partial u}{\partial x}+v \frac{\partial u}{\partial y}+\frac{\partial p}{\partial x}=Pr\left(\frac{\partial^2 u}{\partial x^2} +\frac{\partial^2 u}{\partial y^2}\right),\label{diff:2eq:2}\\
 \frac{\partial v}{\partial t} +  u \frac{\partial v}{\partial x}+ v \frac{\partial v}{\partial y}+\frac{\partial p}{\partial y}=Pr\left(\frac{\partial^2 v}{\partial x^2} +\frac{\partial^2 v}{\partial y^2}\right) +PrRa\theta,\label{diff:2eq:3}\\
 \frac{\partial \theta}{\partial t}+  u \frac{\partial \theta}{\partial x} +  v\frac{\partial \theta}{\partial y}=\frac{\partial^2 \theta}{\partial x^2} +\frac{\partial^2 \theta}{\partial y^2},\label{diff:2eq:4}\\
\end{gather}
where
\begin{gather}
\theta=(1+S)Z \quad   y<y_{st}, \quad
\theta=\frac{1+S}{S}(1-Z)\quad y>y_{st}\label{diff:2eq:5}.
\end{gather}
These are to be solved on both sides of the reaction sheet located at $y=y_{st}$, with the boundary conditions
\begin{gather}
\theta=u=v=0,\text{ at }y=0,\label{diff:2bc:1}\\
\theta=u=v=0,\text{ at }y=1,\label{diff:2bc:2}
\end{gather}
and the jump conditions
\begin{align}
\left[\theta \right]=\left[u\right]=\left[v\right]=0 &, \label{diff:2jc:1}\\
\left[\frac{\partial u}{\partial n}\right]=\left[p\right]- Pr \left[\frac{\partial v}{\partial n}\right]=0 &,\label{diff:2jc:2}\\
\lim_{y\to y^\pm_{st}}\theta =1, \quad \frac{\partial \theta\left(y_{st}^-\right)}{\partial n}+S\frac{\partial\theta\left(y_{st}^+\right)}{\partial n}=0 &\label{diff:2jc:3}
\end{align}
across $y=y_{st}$ (where $n$ denotes a coordinate normal to the reaction sheet).
\subsection{Base State}
Equations (\ref{diff:2eq:1})-(\ref{diff:2eq:4}) admit a stationary planar solution with no flow given by, using bars to denote the base state,
\begin{gather}
\frac{d \bar{p}}{d y}=PrRa\bar{\theta}, \quad
\frac{d^2 \bar{\theta}}{d y^2}=0,\label{diff:eq:basetemppressure}
\end{gather}
subject to boundary conditions (\ref{diff:2bc:1}) and (\ref{diff:2bc:2}). Jump conditions (\ref{diff:2jc:1})-(\ref{diff:2jc:3}) become
\begin{align}
\left[\bar{\theta}\right]=\left[\bar{p}\right]=0 &, \label{diff:basejc:1}\\
\lim_{y\to \bar{y}^\pm_{st}}\bar{\theta} =1, \quad \frac{d \bar{\theta}\left(\bar{y}_{st}^-\right)}{d y}+S\frac{d\bar{\theta}\left(\bar{y}_{st}^+\right)}{d y}=0, &\label{diff:basejc:3}
\end{align}
where $[f]=f\left(\bar{y}_{st}^+\right)-f\left(\bar{y}_{st}^-\right)$. It follows from (\ref{diff:eq:basetemppressure})-(\ref{diff:basejc:3}) and conditions (\ref{diff:2bc:1}) and (\ref{diff:2bc:2}) that
\begin{gather}
\bar{\theta}=\frac{y}{\bar{y}_{st}},\quad y<\bar{y}_{st},\\
\bar{\theta}=\frac{1-y}{1-\bar{y}_{st}},\quad y>\bar{y}_{st}.
\end{gather}
Thus, using the condition on the right of (\ref{diff:basejc:3}),
\begin{gather}
\bar{y}_{st}=\frac{1}{1+S},\label{diff:base3}
\end{gather}
so that
\begin{gather}
\bar{\theta}=(1+S)y, \quad  y<\bar{y}_{st}, \label{diff:eq:base1}\\
\bar{\theta}=\frac{1+S}{S}(1-y), \quad y>\bar{y}_{st}.\label{diff:eq:base2}
\end{gather}
Finally, the base state pressure profile $\bar{p}$ can be found by integrating the equation on the left of (\ref{diff:eq:basetemppressure}) with respect to $y$.

\subsection{Linear stability problem}
We now perturb the base state by writing
\begin{gather}
u=\epsilon \tilde{u},\text{ }v=\epsilon \tilde{v},\text{ }p=\bar{p}+ \epsilon \tilde{p},\text{ }\theta=\bar{\theta}+\epsilon \tilde{\theta},\text{ } y_{st}=\bar{y}_{st}+\epsilon\tilde{y}_{st}, \label{diff:eq:perturb}
\end{gather}
where $\epsilon <<1$ is a small parameter measuring the magnitude of the perturbations (denoted by tilde).
To $O(\epsilon)$ in equations (\ref{diff:2eq:1})-(\ref{diff:2eq:4}) we find
\begin{gather}
\frac{\partial \tilde{u}}{\partial x} +\frac{\partial \tilde{v}}{\partial y} = 0, \label{diff:eq:tilde1}\\
 \frac{\partial \tilde{u}}{\partial t} +\frac{\partial \tilde{p}}{\partial x}=Pr \nabla^2\tilde{u}, \label{diff:eq:tilde2}\\
 \frac{\partial \tilde{v}}{\partial t} + \frac{\partial \tilde{p}}{\partial y}=Pr\nabla^2 \tilde{v} +PrRa\tilde{\theta}, \label{diff:eq:tilde3}\\
\frac{\partial \tilde{\theta}}{\partial t} +\tilde{v}\frac{d \bar{\theta}}{d y}=\nabla^2 \tilde{\theta}.\label{diff:eq:tilde4}
\end{gather}

The wall boundary conditions on the perturbed variables are given by, using (\ref{diff:2bc:1})-(\ref{diff:2bc:2}),
\begin{gather}
\tilde{\theta}=\tilde{u}=\tilde{v}=0,\text{ at }y=0,\label{diff:linearbc:1}\\
\tilde{\theta}=\tilde{u}=\tilde{v}=0,\text{ at }y=1.\label{diff:linearbc:2}
\end{gather}
Finally, we transfer the conditions (\ref{diff:2jc:1})-(\ref{diff:2jc:3}) at the reaction sheet $y=y_{st}(t,x)$ to $y=\bar{y}_{st}$ by using a Taylor expansion around $y_{st}=\bar{y}_{st}+\epsilon \tilde{y}_{st}$. For example,
\begin{gather*}
\theta(y_{st})=\theta(\bar{y}_{st}+\epsilon\tilde{y}_{st})
=\theta\left(\bar{y}_{st}\right)+\left(y_{st}-\bar{y}_{st}\right)\left(\theta_y\left(\bar{y}_{st}\right)\right)+...
\end{gather*}
which, using the perturbation to $\theta$ in (\ref{diff:eq:perturb}), becomes
\begin{gather*}
\theta(y_{st})=\bar{\theta}(\bar{y}_{st})+\epsilon\tilde{\theta}(\bar{y}_{st})+...+\epsilon \tilde{y}_{st}\bar{\theta}_y(\bar{y}_{st})+...
\end{gather*}
and therefore, at $O\left(\epsilon\right)$ of the condition to the left of (\ref{diff:2jc:3}),
\begin{gather*}
\tilde{\theta}=-\tilde{y}_{st}\bar{\theta}_{y}\text{ at } y=\bar{y}_{st}^\pm.
\end{gather*}
Hence, using the base state solution (\ref{diff:eq:base1})-(\ref{diff:eq:base2}),
\begin{gather*}
\tilde{\theta}=-\tilde{y}_{st}(1+S) \quad \text{at }y=\bar{y}_{st}^-,\\
\tilde{\theta}=\tilde{y}_{st}\frac{1+S}{S} \quad \text{at }y=\bar{y}_{st}^+.
\end{gather*}
The other reaction sheet conditions can be derived similarly, noting that $u$, $v$, their derivatives and $p$ are continuous across the reaction sheet in the base state. This leads to
\begin{gather}
\left[\tilde{u}\right]=\left[\tilde{v}\right]=\left[\frac{\partial \tilde{u}}{\partial n}\right]=\left[\tilde{p}\right]- Pr \left[\frac{\partial \tilde{v}}{\partial n}\right]=0\quad\text{ across }y=\bar{y}_{st} \label{diff:linearjc:1},\\
\tilde{\theta}\left(y=\bar{y}_{st}^-\right)=-\tilde{y}_{st}\left(1+S\right),\quad\tilde{\theta}\left(y=\bar{y}_{st}^+\right)=\tilde{y}_{st}\frac{1+S}{S},\nonumber\\ S\tilde{\theta}_y\left(y=\bar{y}_{st}^+\right)+\tilde{\theta}_y\left(y=\bar{y}_{st}^-\right)=0.\label{diff:linearjc:2}
\end{gather}

The linear stability problem has now been derived and is given by equations (\ref{diff:eq:tilde1})-(\ref{diff:eq:tilde4}) for $y\neq \bar{y}_{st}$, subject to the boundary conditions (\ref{diff:linearbc:1})-(\ref{diff:linearbc:2}) and the jump conditions (\ref{diff:linearjc:1})-(\ref{diff:linearjc:2}) across $y= \bar{y}_{st}$. It is worth noting that $d\bar{\theta} / dy$ takes different values below and above the reaction sheet, given by (\ref{diff:eq:base1}) and  (\ref{diff:eq:base2}), respectively.

Before continuing, it can be noted that equations (\ref{diff:eq:tilde1})-(\ref{diff:eq:tilde4}), which are four equations in four variables, can be simplified into two equations in two variables. To this end, we first take $\frac{\partial}{\partial x}$(\ref{diff:eq:tilde2})+$\frac{\partial}{\partial y}$(\ref{diff:eq:tilde3}), which with the use of (\ref{diff:eq:tilde1}) gives
\begin{gather}
\nabla^2 \tilde{p}=Pr Ra \frac{\partial \tilde{\theta}}{\partial y}. \label{diff:eq:pressurepert}
\end{gather}
We then take $\nabla^2$(\ref{diff:eq:tilde3}) to find
\begin{gather}
\frac{\partial}{\partial t}\nabla^2 \tilde{v}+\frac{\partial}{\partial y}\nabla^2 \tilde{p}=Pr\nabla^4 \tilde{v}+PrRa\nabla^2 \tilde{\theta}. \label{diff:eq:velocitypert}
\end{gather}
Finally, substitution of (\ref{diff:eq:pressurepert}) into (\ref{diff:eq:velocitypert}) yields
\begin{gather}
\frac{\partial}{\partial t}\nabla^2 \tilde{v}=Pr\nabla^4\tilde{v}+PrRa\frac{\partial^2 \tilde{\theta}}{\partial x^2}. \label{diff:eq:pertveloctheta}
\end{gather}
Thus the perturbations $\tilde{v}$ and $\tilde{\theta}$ are governed by equations (\ref{diff:eq:tilde4}) and (\ref{diff:eq:pertveloctheta}).

\subsection{Fourier analysis}
We consider normal mode solutions by setting
\begin{equation}
\left.
\begin{aligned}
&\tilde{u}=U(y)e^{\sigma t+i a x},\quad \tilde{v}=V(y)e^{\sigma t+i a x}, \quad \tilde{p}=\tilde{P}(y)e^{\sigma t+i a x},\\
&\tilde{\theta}=\phi(y)e^{\sigma t+i a x},\quad \tilde{y}_{st}=e^{\sigma t+i a x}.
\end{aligned}
\right\}
\label{diff:eq:normalmodes}
\end{equation}
At this point it should be noted that three-dimensional perturbations of the form $\tilde{u}=U(y)\exp\left(\sigma t+i \left(a_1 x+a_2 z\right)\right)$, say, do not need to be considered because they lead to exactly the same problem, as derived below, if the substitution $a^2= a_1^2 +a_2^2$ is made (known as Squire's transformation).
\subsubsection{Governing equations}
Note that for the derivation of the governing equations of the linear stability problem we do not need to consider $\tilde{u}$ or $\tilde{p}$, but we shall here nevertheless write the continuity equation and the momentum equations in terms of the Fourier variables, which follows from substituting (\ref{diff:eq:normalmodes}) into (\ref{diff:eq:tilde1})-(\ref{diff:eq:tilde3}), for future use:
\begin{gather}
iaU+V'=0,\label{diff:eq:mode1}\\
\sigma U +ia\tilde{P}=Pr(-a^2U+U'')\label{diff:eq:mode2},\\
\sigma V + \tilde{P}'=Pr(-a^2V + V'') +Pr Ra  \phi \label{diff:eq:mode3}.
\end{gather}
Continuing with the equations for $V(y)$ and $\phi(y)$, which follow from substituting (\ref{diff:eq:normalmodes}) into (\ref{diff:eq:tilde4}) and (\ref{diff:eq:pertveloctheta}), we have
\begin{gather}
\left(D^2-a^2-\sigma\right)\phi=V\frac{d \bar{\theta}}{d y} \label{diff:eq:phiv},\\
\sigma\left(D^2-a^2\right)V=Pr\left(D^2-a^2\right)^2V-a^2PrRa\phi, \label{diff:eq:mode5}
\end{gather}
where $D\equiv d / d y$.\\
Substitution of (\ref{diff:eq:phiv}) into (\ref{diff:eq:mode5}) gives a single equation for $V$ in each region
\begin{gather}
\left(D^2-a^2-\sigma\right)\left(D^2-a^2\right)\left(D^2-a^2-\frac{\sigma}{Pr}\right)V=a^2 Ra \frac{d \bar\theta}{d y} V.
\end{gather}
In other words, we have derived equations in the regions above and below the reaction sheet, which depend on the derivative of the base temperature in the respective region, given by (\ref{diff:eq:base1}) for $y<\bar{y}_{st}$ and (\ref{diff:eq:base2}) for $y>\bar{y}_{st}$.

Finally note that (\ref{diff:eq:mode5}) implies that
\begin{gather}
\left(D^2-a^2\right)\left(D^2-a^2-\frac{\sigma}{Pr}\right)V=a^2Ra \phi,\label{diff:vtemperature}
\end{gather}
which will be useful when deriving the boundary conditions in the next section.
\subsubsection{Boundary conditions}
On using equations (\ref{diff:eq:mode1}) and (\ref{diff:vtemperature}), which give $V$ in terms of $U$ and $\phi$ respectively, boundary conditions (\ref{diff:linearbc:1})-(\ref{diff:linearbc:2}) become, after substituting in (\ref{diff:eq:normalmodes}),
\begin{gather}
V=DV=\left(D^2-a^2\right)\left(D^2-a^2-\frac{\sigma}{Pr}\right)V=0 \quad \text{ at }y=0,1.
\end{gather}
\subsubsection{Jump conditions}
Conditions (\ref{diff:linearjc:2}) can be written, after substituting in (\ref{diff:eq:normalmodes}),
\begin{gather}
\phi\left(y=\bar{y}_{st}^-\right)=-\left(1+S\right),\quad\phi\left(y=\bar{y}_{st}^+\right)=\frac{1+S}{S},\nonumber\\ S\phi_y\left(y=\bar{y}_{st}^+\right)+\phi_y\left(y=\bar{y}_{st}^-\right)=0.\label{diff:reactionsheettemp}
\end{gather}
The velocity jump conditions (\ref{diff:linearjc:1}) convert to
\begin{gather}
\left[V\right]=\left[U\right]=\left[DU\right]=0\quad \text{across }y=\bar{y}_{st}. \label{diff:jc:uv}
\end{gather}
Also on using (\ref{diff:linearjc:1}) we have
\begin{gather}
\left[\tilde{P}\right]=Pr\left[DV\right] \quad \text{across }y=\bar{y}_{st}. \label{diff:jc:p}
\end{gather}
Now, considering equation (\ref{diff:eq:mode1}) and its successive differentiations we can convert (\ref{diff:jc:uv}) to conditions on $V$, namely,
\begin{gather*}
\left[V\right]=\left[DV\right]=\left[D^2V\right]=0\quad\text{ across }y=\bar{y}_{st},
\end{gather*}
and thus, using (\ref{diff:jc:p}),
\begin{gather}
\left[\tilde{P}\right]=0 \quad \text{across }y=\bar{y}_{st}.\label{diff:eq:pjump}
\end{gather}
Finally, substitution of (\ref{diff:eq:mode1}) into (\ref{diff:eq:mode2}) gives
\begin{gather*}
-a^2 \tilde{P}=\sigma DV + Pr\left(a^2V-D^3V\right)
\end{gather*}
and hence (\ref{diff:eq:pjump}) implies that
\begin{gather*}
\left[D^3V\right]=0 \quad \text{across }y=\bar{y}_{st}.
\end{gather*}
Thus the linear stability problem for the system is fully formulated. It is given by
\begin{gather}
\left(D^2-a^2-\sigma\right)\left(D^2-a^2\right)\left(D^2-a^2-\frac{\sigma}{Pr}\right)V=a^2 Ra (1+S) V, \quad y<\bar{y}_{st}, \label{diff:linearstabeq1}\\
\left(D^2-a^2-\sigma\right)\left(D^2-a^2\right)\left(D^2-a^2-\frac{\sigma}{Pr}\right)V=-a^2 Ra \frac{(1+S)}{S} V, \quad y>\bar{y}_{st},\label{diff:linearstabeq2}
\end{gather}
with the wall boundary conditions
\begin{gather}
V=DV=\left(D^2-a^2\right)\left(D^2-a^2-\frac{\sigma}{Pr}\right)V=0\text{ at }y=0,1, \label{diff:linearstabbc}
\end{gather}
the mass/momentum jump conditions
\begin{gather}
[V]=[DV]=[D^2V]=[D^3V]=0 \quad\text{ across }y=\bar{y}_{st}
\end{gather}
and the reaction sheet conditions, which follow from using \eqref{diff:vtemperature} in \eqref{diff:reactionsheettemp},
\begin{gather}
\left(D^2-a^2\right)\left(D^2-a^2-\frac{\sigma}{Pr}\right)V=-a^2 Ra(1+S)\text{ at }y=\bar{y}_{st}^-,\\
\left(D^2-a^2\right)\left(D^2-a^2-\frac{\sigma}{Pr}\right)V=a^2 Ra\frac{1+S}{S}\text{ at }y=\bar{y}_{st}^+,\\
S\left(D\left(D^2-a^2\right)\left(D^2-a^2-\frac{\sigma}{Pr}\right)V\left(\bar{y}_{st}^+\right)\right)+\nonumber\\D\left(D^2-a^2\right)\left(D^2-a^2-\frac{\sigma}{Pr}\right)V\left(\bar{y}_{st}^-\right)=0 \label{diff:linearstabjc3},
\end{gather}
where the averaged flame position is given by $\bar{y}_{st}=1/(1+S)$.

We now have a sixth order ordinary differential equation for the velocity perturbation $V$ in each region, with 13  auxiliary conditions (six boundary conditions at the wall and seven conditions at the averaged reaction sheet). These conditions are sufficient to determine $V$ along with the eigenvalue $\sigma=\sigma(a;Ra,S)$, which will determine the linear stability of the Burke--Schumann diffusion flame. For given values of $Ra$ and $S$, if the real part of the growth rate $\sigma$ is greater than zero for any value of the wavenumber $a$, the system is unstable. If the real part of $\sigma$ is negative for all values of $a$ the system is stable.
\section{Solution of the linear stability problem}\label{diff:sec:sol}
It is worth noting at this point that in the non-reactive case it can be shown that the growth rate $\sigma$ is real and the marginal state is characterised by $\sigma=0$; this is called the principle of the exchange of stabilities \citep[][pp. 24--26]{chandrasekhar1961hydrodynamic}. Since this is not straightforward in our case, we instead begin by solving the linear stability problem numerically using the BVP4C solver in Matlab to find the eigenvalue $\sigma$ and investigate whether its imaginary part is zero at marginal stability. If so, we can characterise the marginal state as the state where $\sigma=0$ and proceed to solve the problem in a similar approach to the non-reactive case.
\begin{figure}
\subfigure[Growth rate versus Rayleigh number for $S=1$ with wavenumber $a=3.95$.]{
\includegraphics[scale=0.8]{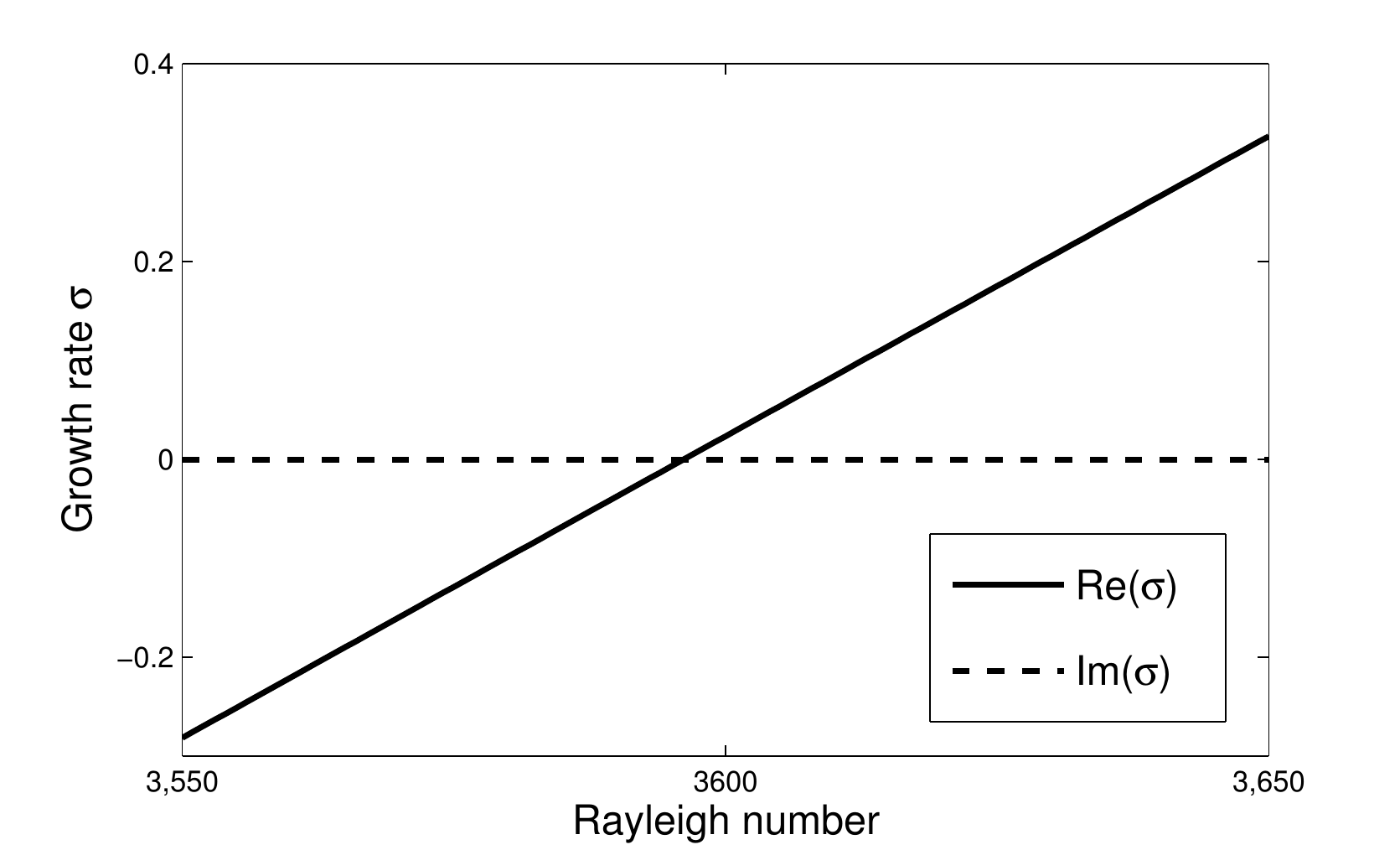}
}
\subfigure[Growth rate versus Rayleigh number for $S=5$ with wavenumber $a=3.13$.]{
\includegraphics[scale=0.8]{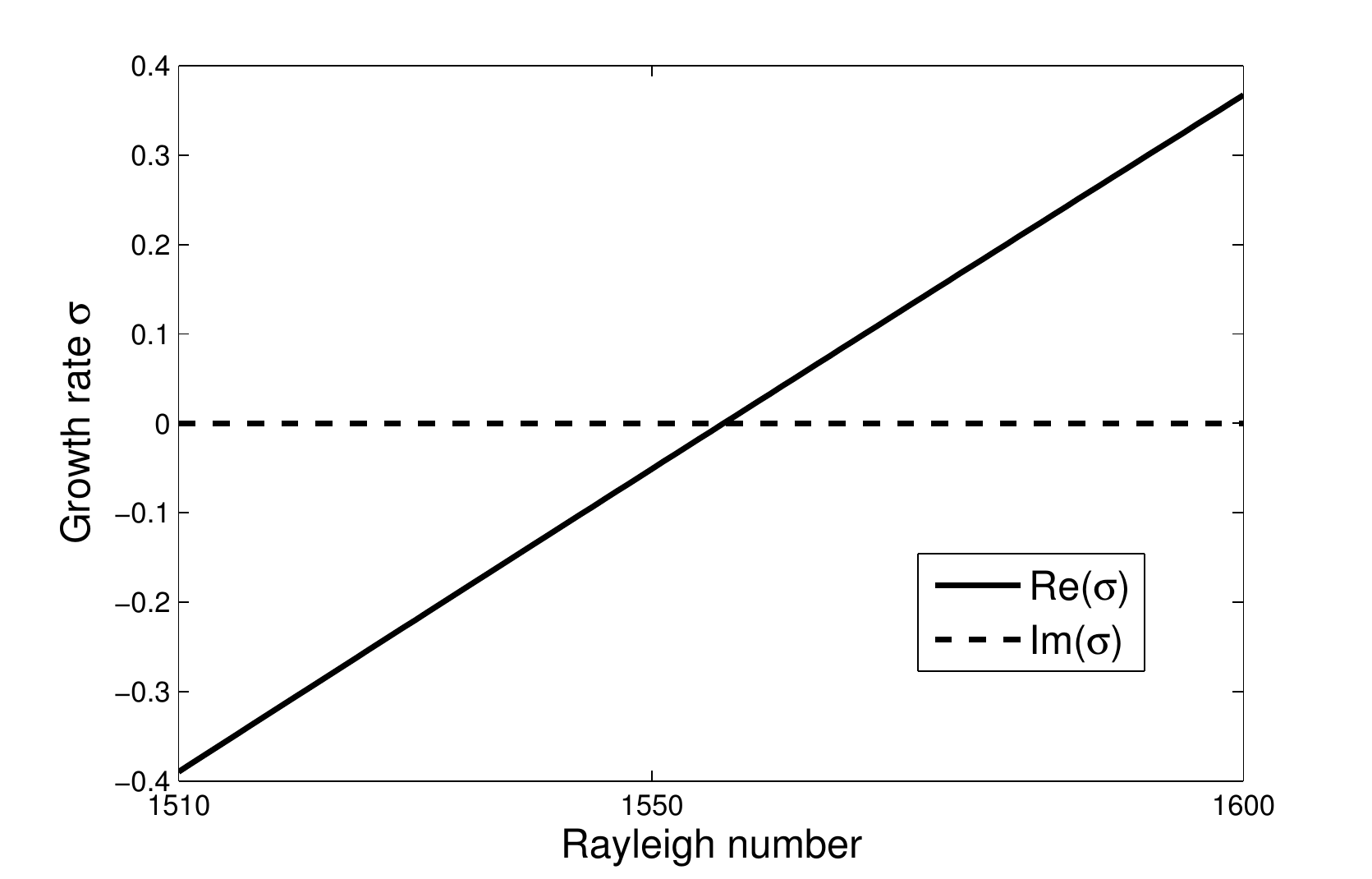}
}
\caption{Graphs of real and imaginary parts of the growth rate $\sigma$ versus the Rayleigh number $Ra$ for two selected values of the stoichiometric coefficient $S$ (with $Pr=1$). Note that the imaginary part of the growth rate is found to be zero for all values of $Ra$.}
\label{diff:fig:growthrate}
\end{figure}
\subsection{Numerical solution for $\sigma$}
\begin{figure}
\includegraphics[scale=0.8, trim=20 0 0 0, clip=true]{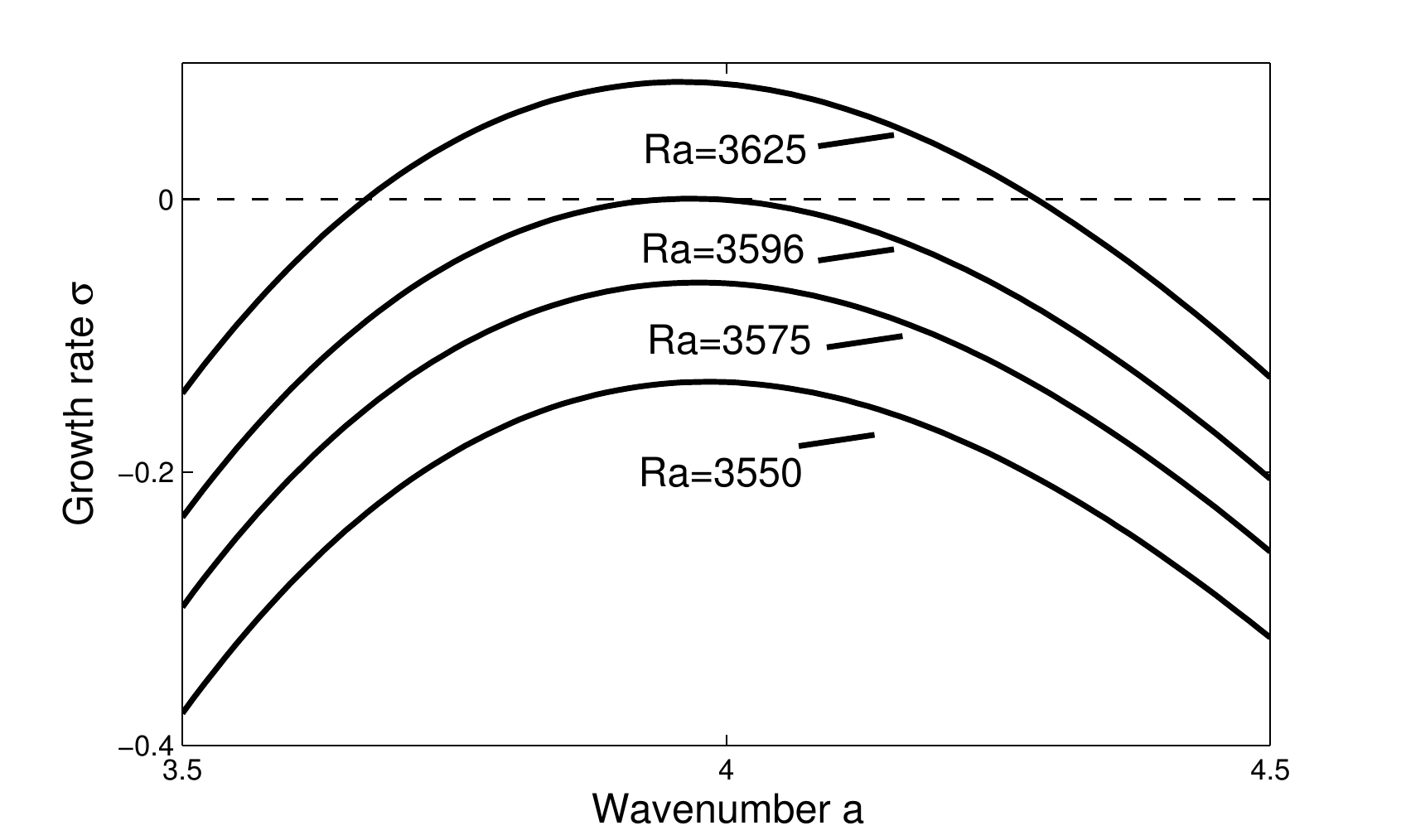}
\caption{Growth rate versus wavenumber for selected values of the Rayleigh number, with $S=1$ and $Pr=1$. Note that in this case the critical Rayleigh number is found to be $Ra_c=3596$.}
\label{diff:fig:growth_v_a}
\end{figure}

In this section, we numerically solve equations (\ref{diff:linearstabeq1}) and (\ref{diff:linearstabeq2}) with conditions (\ref{diff:linearstabbc})-(\ref{diff:linearstabjc3}). We use the eigen-boundary-value-problem Matlab solver BVP4C \citep[][]{shampine2000solving} to find the value of the growth rate $\sigma$ for given values of the wavenumber $a$, the Rayleigh number $Ra$ and the stoichiometric coefficient $S$. The key result is that $\sigma$ is always found to be real. Figure \ref{diff:fig:growthrate} shows that as $Ra$ increases for selected values of $a$ and $S$, $\sigma$ passes from negative values to positive values and there is a marginal value of $Ra$ at which the system changes from stability to instability and $\sigma=0$. We also plot the effect of the wavenumber on the growth rate for several values of the Rayleigh number in figure \ref{diff:fig:growth_v_a}, which clearly shows the existence of a critical Rayleigh number. If the Rayleigh number takes a value higher than its critical value, there is a band of wavenumbers for which $\sigma>0$ and the system is unstable.

At this point we could produce many more plots investigating the stability of the system, however to avoid repetition we merely note that the growth rate is always found to be real and thus $\sigma=0$ characterises the marginal state. We can therefore simplify the problem by setting $\sigma=0$ in the governing equations and then solve to find the marginal Rayleigh number. In the next section we will solve the marginal problem and then discuss the stability of the system in more detail.

\subsection{Marginal state}
Motivated by the conclusions of the previous section, here we set $\sigma=0$ in equations (\ref{diff:linearstabeq1}) and (\ref{diff:linearstabeq2}) with conditions (\ref{diff:linearstabbc})-(\ref{diff:linearstabjc3}), characterising the marginal state. This is the state at which the system passes from being stable to being unstable, and thus the Rayleigh number for which this state exists for a given wavenumber $a$ is the marginal Rayleigh number at that wavenumber. Hence
\begin{gather}
\left(D^2-a^2\right)^3V=a^2 Ra (1+S)V,\quad y<\bar{y}_{st},\label{diff:4eq:1}\\
\left(D^2-a^2\right)^3V=-a^2 Ra \frac{(1+S)}{S}V,\quad y>\bar{y}_{st}\label{diff:4eq:2},
\end{gather}
with the wall boundary conditions
\begin{gather}
V=DV=\left(D^2-a^2\right)^2V=0\text{ at }y=0,1,\label{diff:4bc:1}
\end{gather}
the mass/momentum jump conditions
\begin{gather}
\left[V\right]=\left[DV\right]=\left[D^2V\right]=\left[D^3V\right]=0 \quad\text{ across }y=\bar{y}_{st}\label{diff:4jc:1}
\end{gather}
and the reaction sheet conditions
\begin{gather}
\left(D^2-a^2\right)^2V=-a^2Ra(1+S)\text{ at }y=\bar{y}_{st}^-,\label{diff:4jc:2a}\\
\left(D^2-a^2\right)^2V=a^2Ra\frac{1+S}{S}\text{ at }y=\bar{y}_{st}^+,\label{diff:4jc:2b}\\
S\left(D\left(D^2-a^2\right)^2V\left(\bar{y}_{st}^+\right)\right)+D\left(D^2-a^2\right)^2V\left(\bar{y}_{st}^-\right)=0,\label{diff:4jc:3}
\end{gather}
where the averaged flame position is given by $\bar{y}_{st}=1/(1+S)$.

Following a similar approach to that of \citet[pp. 36--42]{chandrasekhar1961hydrodynamic}, we let the velocity perturbation $V$ take the form
\begin{gather*}
V=e^{\pm py}\quad y<\bar{y}_{st},\\
V=e^{\pm qy}\quad y>\bar{y}_{st}.
\end{gather*}
Then we have
\begin{gather*}
\left(p^2-a^2\right)^3=a^2 Ra (1+S)\quad y<\bar{y}_{st},\\
\left(q^2-a^2\right)^3=-a^2 Ra \frac{(1+S)}{S}\quad y>\bar{y}_{st}.
\end{gather*}
If we let
\begin{gather}
\frac{1+S}{S}a^2 Ra=\tau^3 a^6 \label{diff:definitionoftau}
\end{gather}
we find that the roots of these equations are given by
\begin{gather*}
p^2=a^2\left(S^{1/3}\tau+1\right)\quad \text{and} \quad p^2=a^2\left[1+S^{1/3}\frac{\tau}{2} \left(-1\pm i \sqrt{3}\right) \right],\\
q^2=-a^2\left(\tau-1\right)\quad \text{and} \quad q^2=a^2\left[1+\frac{\tau}{2} \left(1 \pm i\sqrt{3}\right)\right].
\end{gather*}
Thus we have six roots for the solution below the reaction sheet and six roots for the solution above it, given by
\begin{gather*}
\pm p_0,\quad \pm p\quad \text{and}\quad \pm p^*,\\
\pm iq_0,\quad \pm q\quad \text{and}\quad \pm q^*,
\end{gather*}
where * denotes the complex conjugate. Here
\begin{gather*}
p_0=a\sqrt{\left(S^{1/3}\tau+1\right)},\\
\mathbb{R}(p)=a\left[\frac{1}{2}\left(\sqrt{1-S^{1/3}\tau+S^{2/3}\tau^2}+\left(1-S^{1/3}\frac{\tau}{2} \right)\right)\right]^{1/2},\\
\mathbb{I}(p)=a\left[\frac{1}{2}\left(\sqrt{1-S^{1/3}\tau+S^{2/3}\tau^2}-\left(1-S^{1/3}\frac{\tau}{2}\right)\right)\right]^{1/2},
\end{gather*}
and
\begin{gather*}
q_0=a\sqrt{(\tau-1)},\\
\mathbb{R}(q)=a\left[\frac{1}{2}\left(\sqrt{1+\tau+\tau^2}+\left(1+\frac{\tau}{2}\right)\right)\right]^{1/2},\\
\mathbb{I}(q)=a\left[\frac{1}{2}\left(\sqrt{1+\tau+\tau^2}-\left(1+\frac{\tau}{2}\right)\right)\right]^{1/2}.
\end{gather*}
Thus the solution for the velocity perturbation can be written as
\begin{gather}
V^-=A_0\cosh{p_0 y}+A\cosh{p y} + A^* \cosh{p^* y}+\nonumber\\B_0\sinh{p_0y}+B\sinh{py}+B^*\sinh{p^*y},\label{diff:eq:vminus}\\
\left(y<\bar{y}_{st}\right), \nonumber \\
V^+=C_0\cos{q_0 y}+C\cosh{q y} + C^* \cosh{q^* y}+\nonumber\\D_0\sin{q_0y}+D\sinh{qy}+D^*\sinh{q^*y}, \label{diff:eq:vplus}\\
\left(y>\bar{y}_{st}\right). \nonumber
\end{gather}
The problem has 13 auxiliary conditions while the general solution has 12 constants of integration. We therefore need to apply 12 conditions in order to determine $V^{\pm}$ for a specified value of $a$ and use the 13th condition to determine $\tau$, from which we can find the marginal Rayleigh number using the formula
\begin{gather}
Ra=\tau^3 a^4 \frac{S}{1+S}. \label{diff:eq:calcrayleigh}
\end{gather}
To write the boundary conditions on $V$ in matrix form, we begin by noting, using (\ref{diff:eq:vminus})-(\ref{diff:eq:vplus}), that
\begin{gather*}
\left(D^2-a^2\right)^2V^-= \nonumber\\ S^{2/3}a^4 \tau^2\big[A_0\cosh{p_0 y}-\frac{1}{2}\left(i \sqrt{3}+1\right)A\cosh{p y} - \frac{1}{2}\left(-i \sqrt{3}+1\right)A^* \cosh{p^* y} \nonumber\\+B_0\sinh{p_0y}-\frac{1}{2}\left(i \sqrt{3}+1\right)B\sinh{py}- \frac{1}{2}\left(-i \sqrt{3}+1\right)B^*\sinh{p^*y}\big] ,
\end{gather*}
and
\begin{gather*}
(D^2-a^2)^2V^+= \nonumber\\ a^4 \tau^2\big[C_0\cos{q_0 y}+\frac{1}{2}\left(i \sqrt{3}-1\right)C\cosh{q y} - \frac{1}{2}\left(i \sqrt{3}+1\right)C^* \cosh{q^* y} \nonumber\\+D_0\sin{q_0y}+\frac{1}{2}\left(i \sqrt{3}-1\right)D\sinh{qy}- \frac{1}{2}\left(i \sqrt{3}+1\right)D^*\sinh{q^*y}\big] .
\end{gather*}
The conditions (\ref{diff:4bc:1})-(\ref{diff:4jc:2b}) can then be written in the matrix form $\mathbf{A}\mathbf{x}=\mathbf{b}$ for the vector of constants $\mathbf{x}$, which can be solved using
\begin{align}
\mathbf{x}=\mathbf{A}^{-1} \mathbf{b}. \label{diff:eq:matrix}
\end{align}
We then write the final condition (\ref{diff:4jc:3}), say, as
\begin{gather*}
\mathbf{c}^T\mathbf{x}=0,
\end{gather*}
where  $\mathbf{c}$ is a vector. A solution that satisfies all of the auxiliary conditions can therefore be found by solving the matrix equation (\ref{diff:eq:matrix}) for given values of $a$ and $S$ in Matlab to find $\mathbf{x}$ and then using Matlab's fzero function with $\mathbf{c}^T\mathbf{x}$ as the input. This will lead to a solution for $V^{\pm}$ and $\tau$, from which the marginal Rayleigh number can be found using equation (\ref{diff:eq:calcrayleigh}).
Note that the temperature perturbation profile can be recovered from the velocity perturbation profile using the equations
\begin{gather*}
\phi^\pm=\frac{ (1+S)}{S \tau^3 a^6}\left(D^2-a^2\right)V^\pm,
\end{gather*}
which are found by setting $\sigma=0$ in \eqref{diff:vtemperature} and using \eqref{diff:definitionoftau}.
\section{Discussion of results}\label{diff:sec:res}
In this section we present the results found by solving the problem in the marginal state using the method described at the end of the previous section. The aim is to calculate the critical Rayleigh number $Ra_c$, which characterises the conditions at the threshold of instability, and its dependence upon the stoichiometric coefficient $S$.

\begin{figure}
\subfigure[Marginal Rayleigh number versus wavenumber for $S=1$.]{
\includegraphics[scale=0.8]{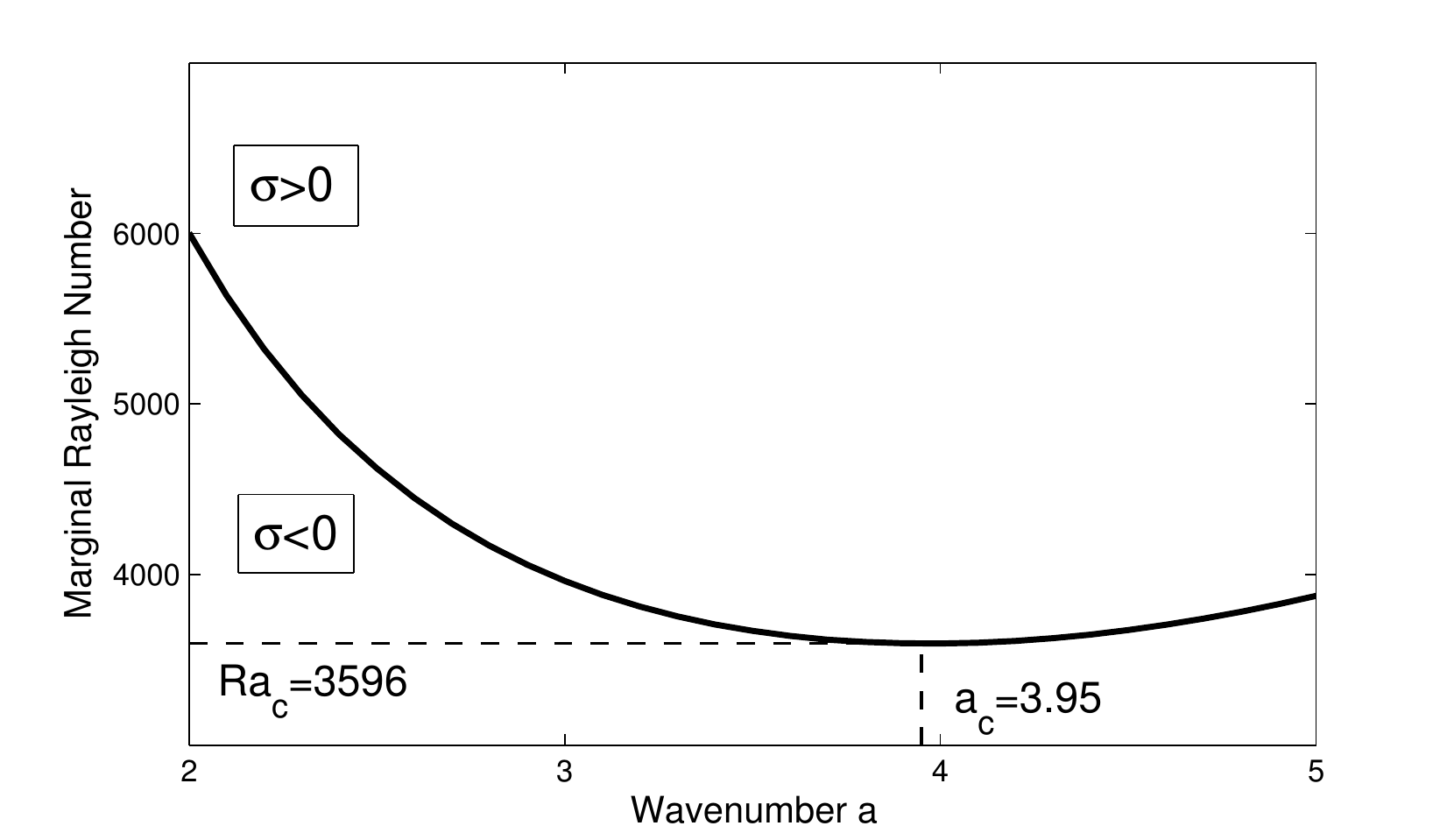}
}
\subfigure[Marginal Rayleigh number versus wavenumber for $S=5$.]{
\includegraphics[scale=0.8]{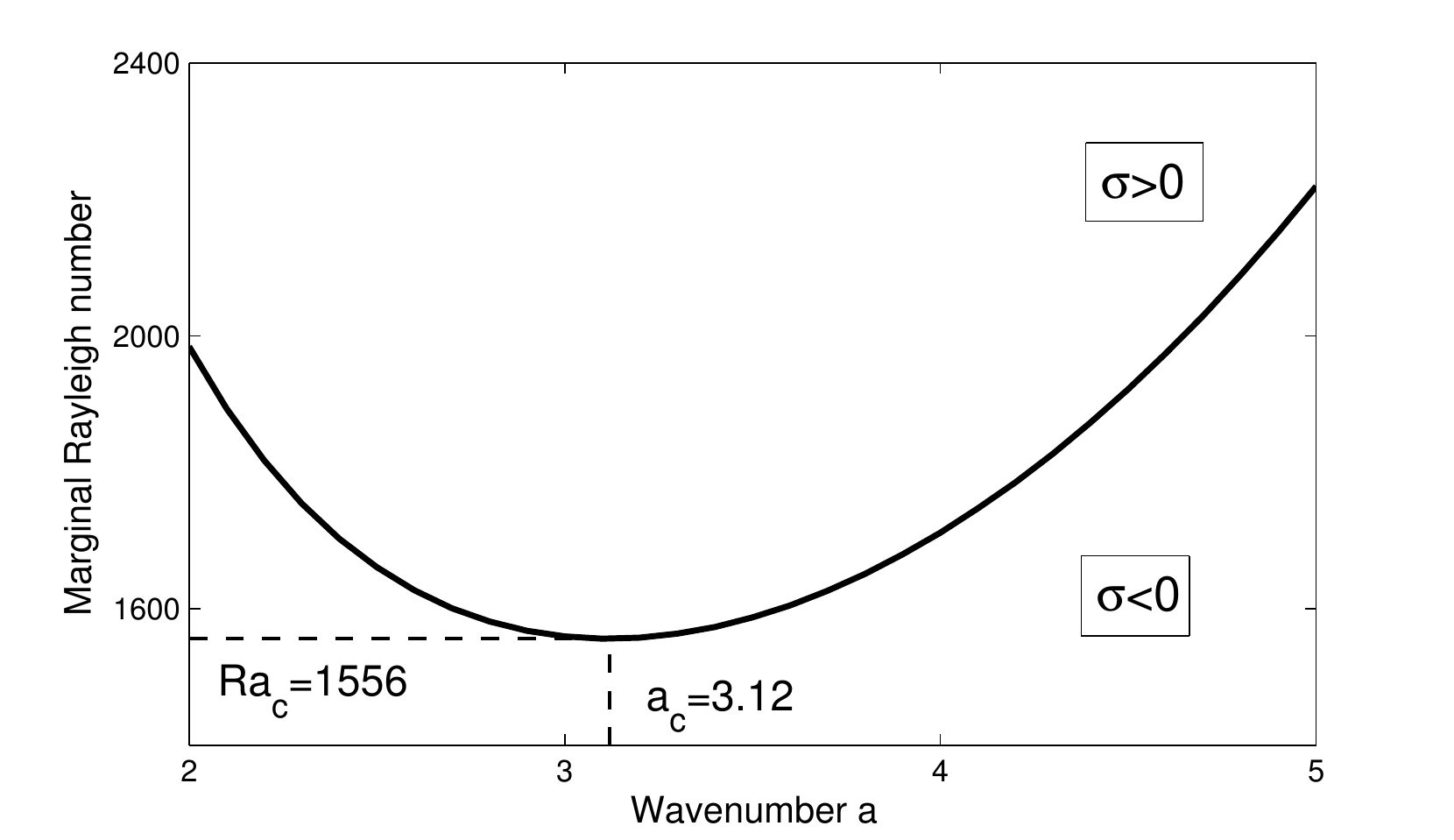}
}
\caption{Marginal Rayleigh number versus wavenumber for two selected values of the stoichiometric coefficient $S$. For each wavenumber the system is unstable if the Rayleigh number $Ra$ takes a value larger than the marginal Rayleigh number, and stable if $Ra$ is lower. For each value of $S$, the lowest marginal Rayleigh number is the critical Rayleigh number $Ra_c$ and the corresponding wavenumber is the critical wavenumber $a_c$.}
\label{diff:fig:ravswavenumber}
\end{figure}

\begin{figure}
\includegraphics[scale=0.8]{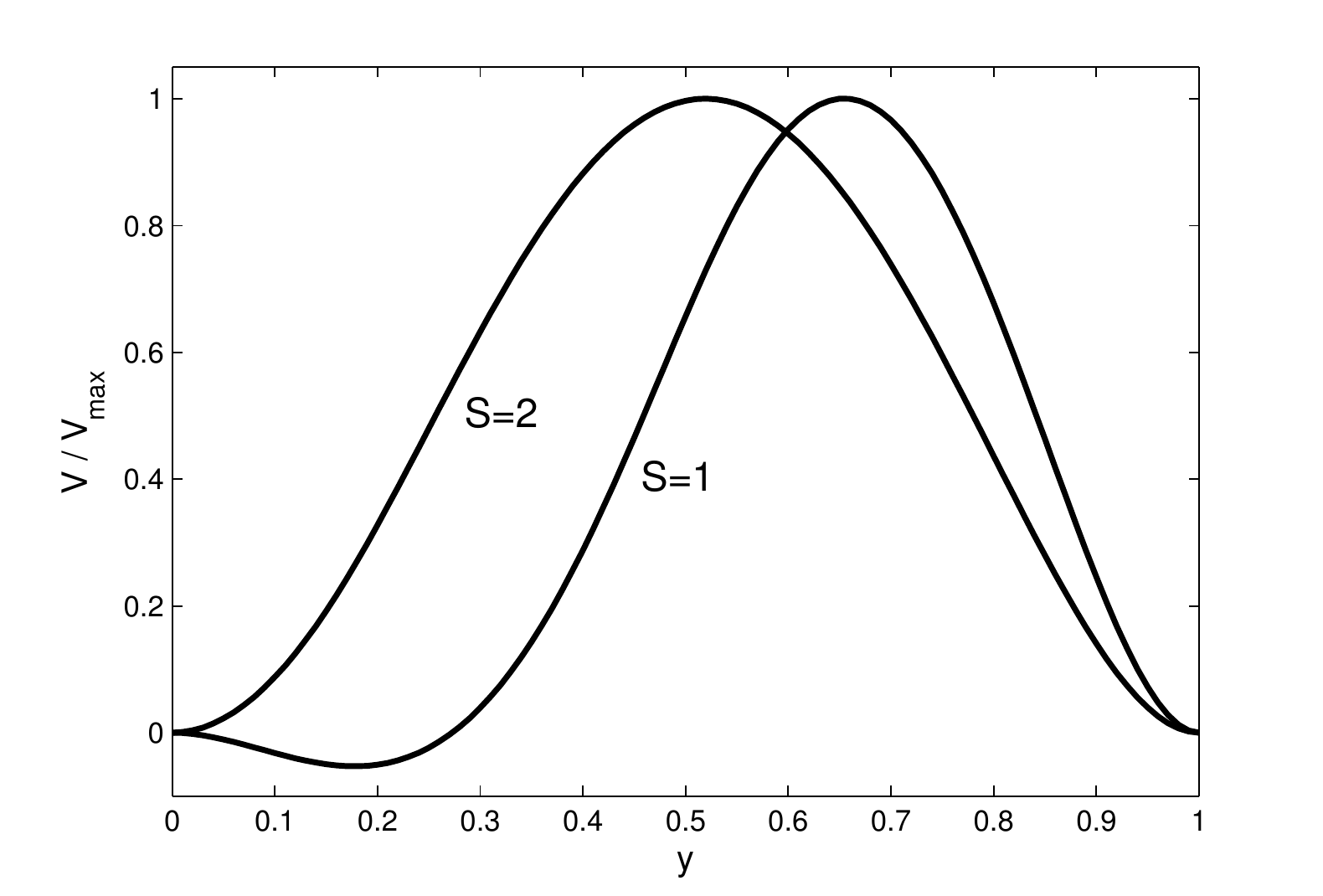}
\caption{Marginal velocity perturbation profile $V^{\pm}$ versus $y$ for two selected values of the stoichiometric coefficient $S$ (scaled by the maximum value for comparison). In each case the wavenumber $a$ is equal to its critical value $a_c$, given by $a_c=3.95$ for $S=1$ and $a_c=3.25$ for $S=2$.}
\label{diff:fig:veloc}
\end{figure}

We begin with plots of how the Rayleigh number in the marginal state varies with the wavenumber $a$ for two selected values of the stoichiometric coefficient $S$, provided in figure \ref{diff:fig:ravswavenumber}. Indicated in the figure are the calculated values of the critical Rayleigh number $Ra_c$  and the corresponding critical wavenumber $a_c$ for each value of $S$. It can be seen that, for a given value of $S$, if $Ra>Ra_c$ there is a band of wavenumbers for which the growth rate $\sigma>0$; thus if $Ra>Ra_c$ the perturbations made to the base state grow exponentially in time and the system is unstable. Velocity perturbation profiles versus $y$ at the onset of instability $Ra=Ra_c$ for selected values of $S$ are presented in figure \ref{diff:fig:veloc}.

\begin{figure}
\subfigure[Critical Rayleigh number versus flame position $y_{st}=1/(1+S)$.]{
\includegraphics[scale=0.8]{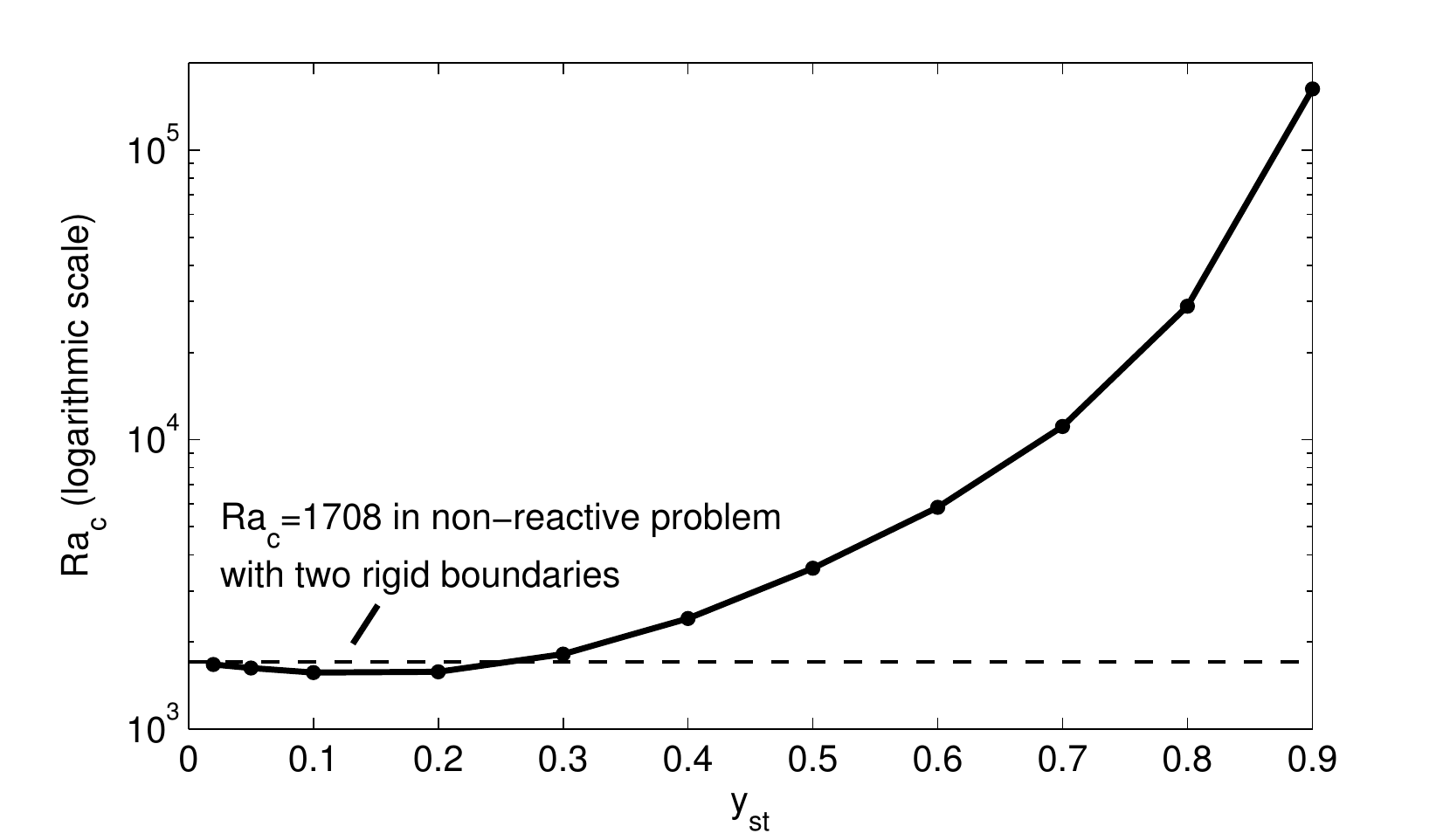}
}
\subfigure[Critical Rayleigh number versus stoichiometric coefficient $S$.]{
\includegraphics[scale=0.8]{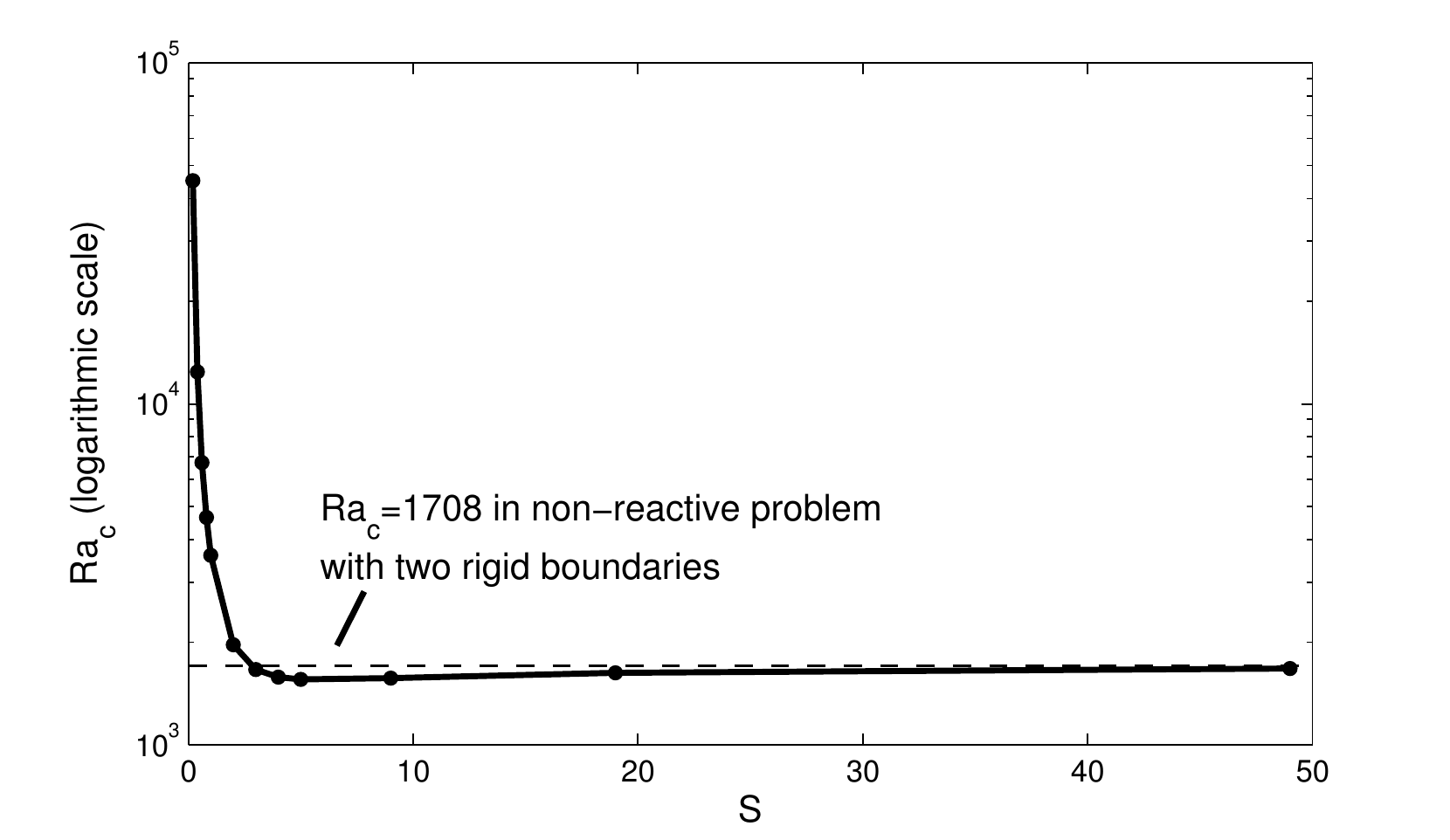}
}
\caption{The effect of the stoichiometric coefficient $S$ and flame position $y_{st}$ (which depends on $S$) on the critical Rayleigh number $Ra_c$. Indicated is the critical Rayleigh number in non-reactive Rayleigh--B\'{e}nard convection with two rigid boundaries.}
\label{diff:fig:crit_ra}
\end{figure}

Next we plot, in figure \ref{diff:fig:crit_ra}, the variation in $Ra_c$ with the stoichiometric coefficient $S$ and the flame position $y_{st}=1/(1+S)$. Indicated in each plot is the critical Rayleigh number given in \citet[][p. 39]{chandrasekhar1961hydrodynamic} for the non-reactive problem with two rigid boundaries. The fact that $Ra_c$ in our reactive case is found to be very close to the non-reactive critical Rayleigh number for large values of $S$, but not exactly equal in the limit $S \to \infty$, can be explained by considering the equations governing the marginal state. As $S \to \infty$, the problem in the upper half-space above the reaction sheet reduces to the non-reactive problem with two rigid boundaries, except for the conditions (\ref{diff:4jc:2a}) and (\ref{diff:4jc:2b}) at the reaction sheet. Thus, as $S\to\infty$ the critical Rayleigh number approaches a value slightly different to the non-reactive critical Rayleigh number, as can be seen in figure \ref{diff:fig:crit_ra}. It is found that as the flame moves away from the lower boundary, $Ra_c$ first decreases slightly then increases to very large values as the flame approaches the upper boundary.

It can be seen in figure \ref{diff:fig:crit_ra} that the order of magnitude of $Ra_c$ can significantly deviate from the order of magnitude of the critical Rayleigh number in the non-reactive case, especially as $y_{st} \to 1$. In order to facilitate comparison with the non-reactive case, we scale the Rayleigh number, by using as reference length the distance of the flame from the upper boundary $\left(1-y_{st}\right)L$ instead of $L$. This scaled Rayleigh number is more comparable to the Rayleigh number in the non-reactive case because both use as reference length the distance from the hot surface to the cold upper boundary. The scaled critical Rayleigh number is given by
\begin{gather*}
Ra_{c,\text{scaled}}=Ra_c\left (1-y_{st}\right)^3.
\end{gather*}

\begin{figure}
\subfigure[Scaled critical Rayleigh number versus the flame position $y_{st}=1/(1+S)$.]{
\includegraphics[scale=0.9]{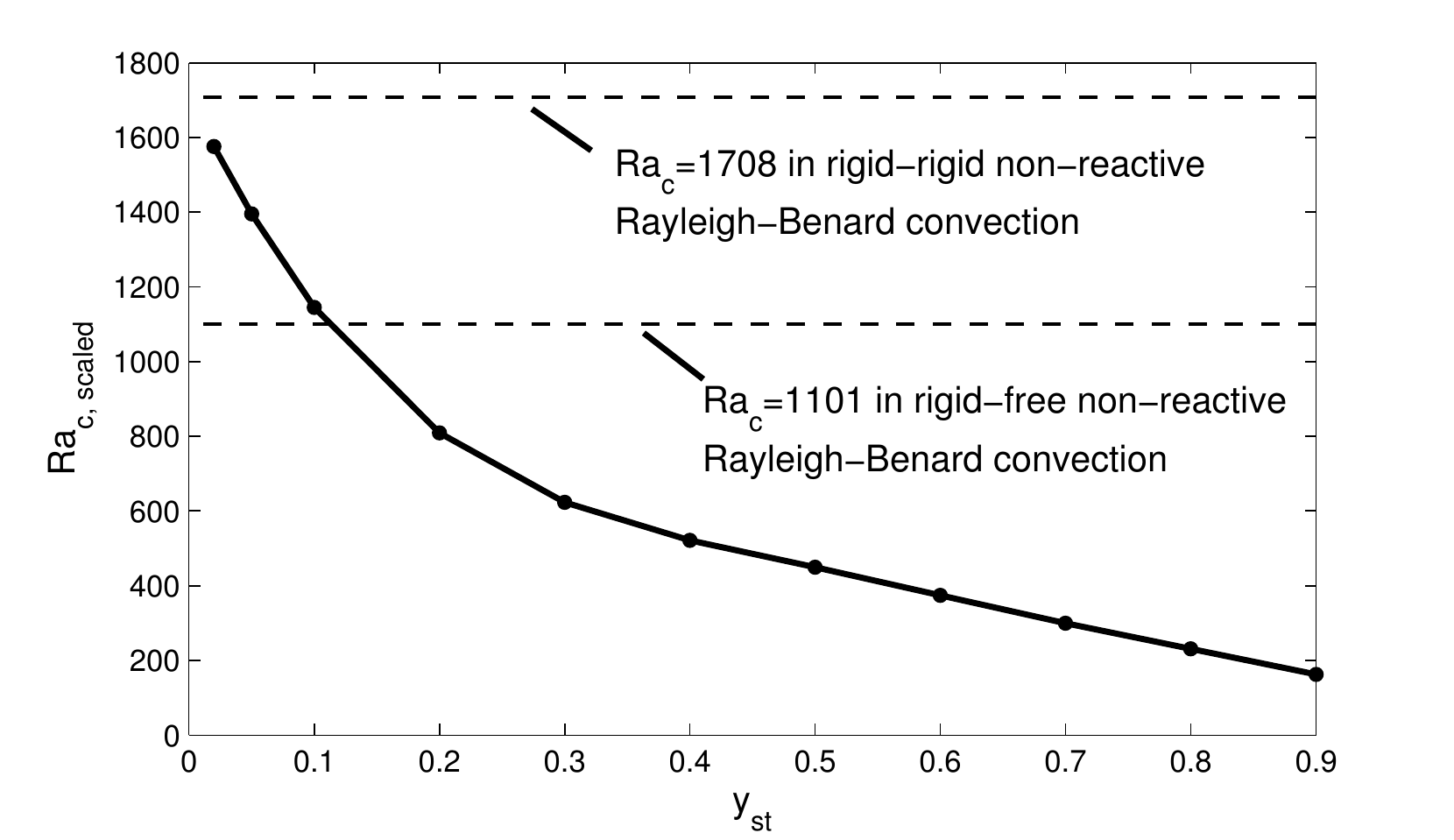}
}
\subfigure[Scaled critical Rayleigh number versus stoichiometric coefficient $S$.]{
\includegraphics[scale=0.9]{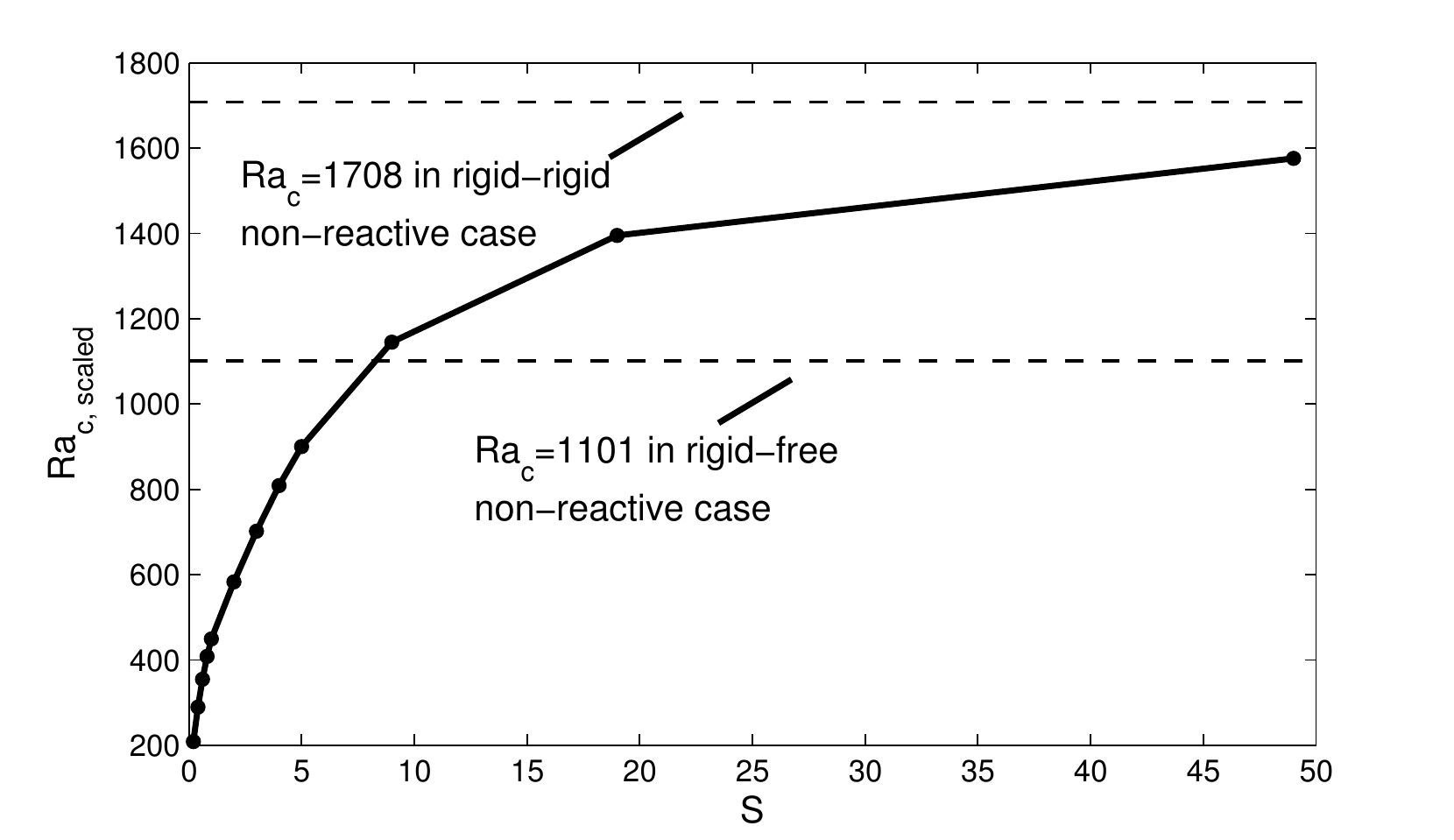}
}
\caption{The effect of the stoichiometric coefficient $S$ and flame position $y_{st}$ (which depends on $S$) on the scaled critical Rayleigh number $Ra_{c,{\text{scaled}}}$. Indicated are the critical Rayleigh numbers in non-reactive Rayleigh--B\'{e}nard convection with two rigid boundaries and with one rigid boundary and one free boundary.}
\label{diff:fig:scaled}
\end{figure}

In figure \ref{diff:fig:scaled} we plot $Ra_{c,\text{scaled}}$ against the flame position $y_{st}$ and the stoichiometric coefficient $S$ and compare with two non-reactive cases. The first is the case of two rigid boundaries; the second is the case of one rigid boundary and one free-surface boundary. Similarly to above it is found that, as $S \to \infty$, $Ra_{c,\text{scaled}}$ is of the order of magnitude of the critical Rayleigh number in the rigid-rigid non-reactive case. It is also found that for $S \approx 9$ (i.e. $y_{st} \approx 0.1$), $Ra_{c,\text{scaled}}$ is of the order of magnitude of the critical Rayleigh number in the rigid-free non-reactive case. As the flame approaches the upper boundary, in the limit $S \to 0$, $Ra_{c,\text{scaled}}$ decreases to lower values.

\section{Numerical study}\label{diff:section:numerical}
We now proceed to a full numerical treatment of the problem, using the finite element package Comsol Multiphysics to directly simulate the physical system. We solve the time-dependent equations (\ref{diff:eq:2machnondim1})-(\ref{diff:eq:mixfracreaction}) with boundary conditions (\ref{diff:eq:mixbc1})-(\ref{diff:eq:mixbc2}). As initial condition we use the planar strongly burning planar diffusion flame (which is found on the upper branch of the curve in figure \ref{diff:fig:scurve}). The aims in this section are: first, to investigate the nature of the instabilities that occur for values of the Rayleigh number slightly higher than $Ra_c$; second, to test the results of the linear stability analysis presented in the previous section against numerical results; finally, to test the effect of thermal expansion and finite chemistry on $Ra_c$.

All numerical calculations are performed for the parameter values $\beta=10$ and $Pr=1$, in a numerical domain of aspect ratio 10; this aspect ratio has been shown to be sufficiently large to approximate an infinite domain in the non-reactive case by \citet{gelfgat1999different}. The domain is discretised into a non-uniform grid of approximately 120,000 triangular elements, with local refinement around the reaction zone and the upper and lower boundaries; various tests have been performed to ensure the mesh-independence of the results. We use a value of $Da=2$x$10^4$ to approximate an infinite Damk\"{o}hler number in all calculations unless otherwise specified.

\begin{figure}
\subfigure[$S=1$; computed at $Ra=3700$.]{
\includegraphics[scale=0.8]{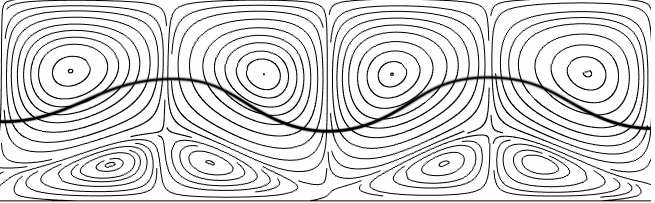}
}
\subfigure[$S=6$; computed at $Ra=1740$.]{
\includegraphics[scale=0.8]{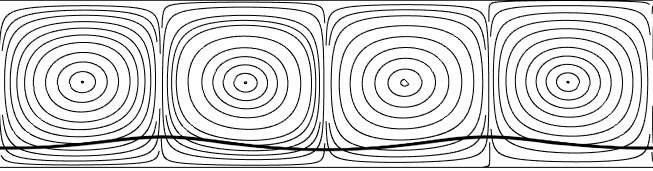}
}
\caption{Convection roll/cellular flame steady state streamlines for different values of the stoichiometric coefficient $S$ in the Boussinesq approximation. The black line represents the flame position. Calculations were performed the parameter values $Da=2\text{x}10^4$, $\beta=10$ and $Pr=1$ and for values of the Rayleigh number $Ra$ slightly higher than its critical value $Ra_c$ in each case.}
\label{diff:fig:cellular}
\end{figure}

We begin with an illustrative calculation, shown in figure \ref{diff:fig:cellular}, which displays the nature of the instability that occurs if $Ra$ takes a value slightly higher than $Ra_c$.  The figure shows the stationary, stable states that the system reaches for two selected values of $S$ when $Ra>Ra_c$. The mechanism of the fluid instability is similar to that of the non-reactive case, described in detail by \citet[pp.9--10]{chandrasekhar1961hydrodynamic} and \citet[p. 12]{getling1998rayleigh}, whereby fluid of higher density lies above the hotter, lower density fluid causing convection to occur. The induced flow enhances the transport of fuel and oxidiser to certain parts of the flame, generating a cellular structure, as shown in the figure.

\begin{figure}
\subfigure[Scaled critical Rayleigh numbers versus flame position $y_{st}=1/(1+S)$.]{
\includegraphics[scale=0.7]{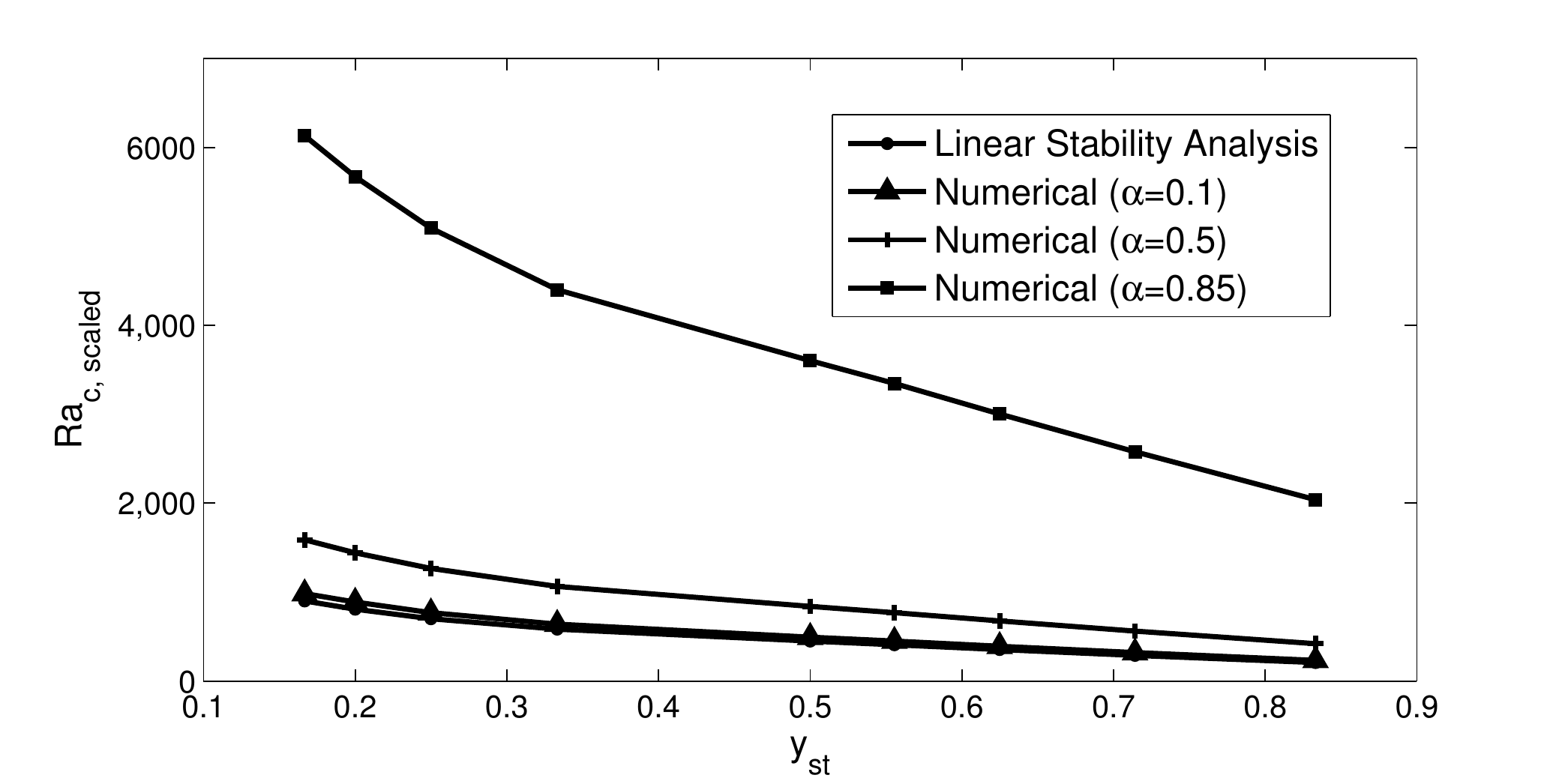}
}
\subfigure[Scaled critical Rayleigh numbers versus stoichiometric coefficient $S$.]{
\includegraphics[scale=0.7]{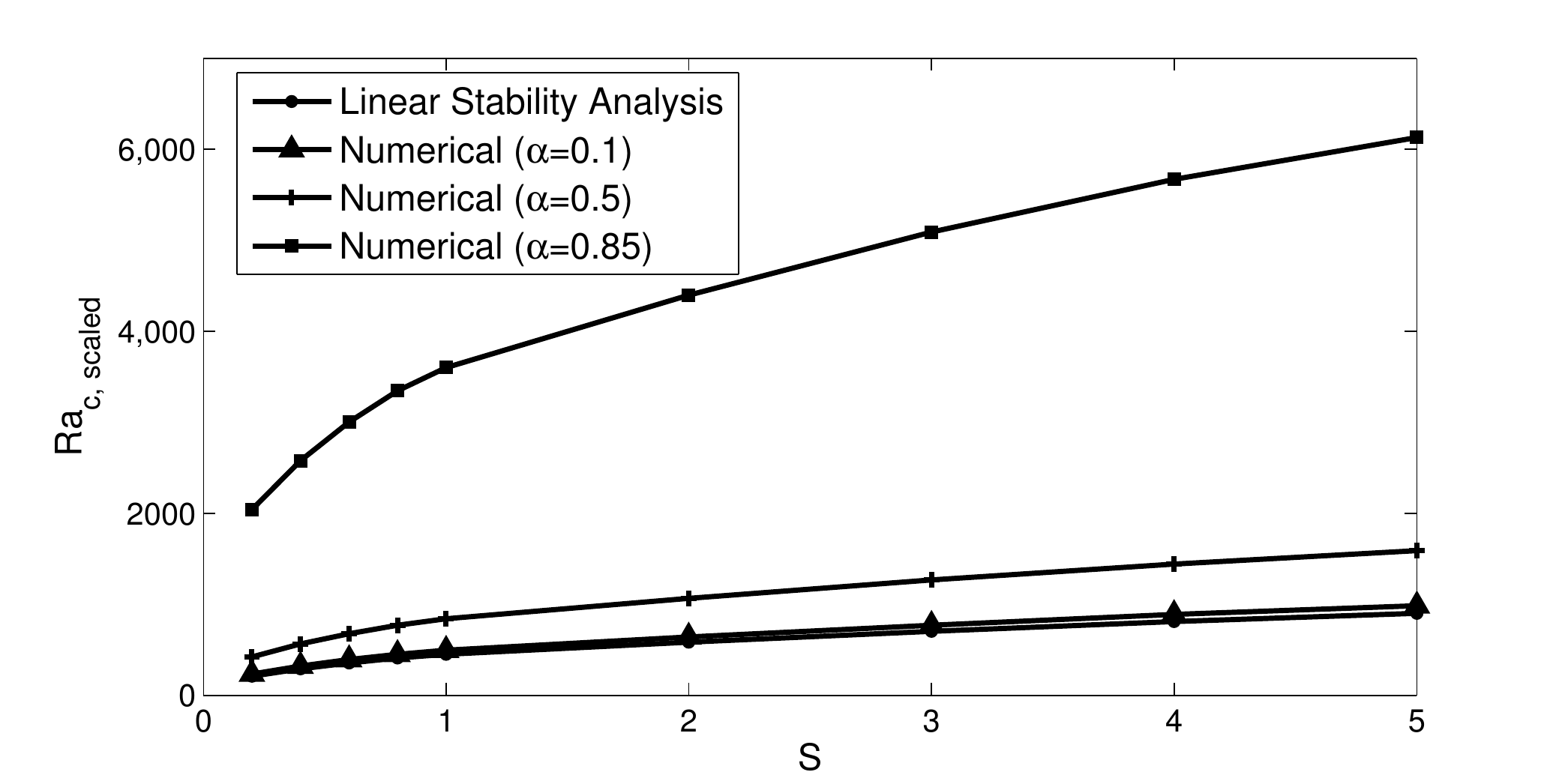}
}
\caption{Comparison of the behaviours of the scaled critical Rayleigh numbers $Ra_{c\text{, scaled}}$ versus (a) flame position $y_{st}$ and (b) stoichiometric coefficient $S$ as predicted by the linear stability analysis and the computations carried out for selected values of the thermal expansion coefficient $\alpha$. Numerical calculations were performed for the parameter values $Da=2\text{x}10^4$, $\beta=10$ and $Pr=1$.}
\label{diff:fig:comparison}
\end{figure}

Next, we test the validity of the linear stability results in the Boussinesq approximation and the effect of compressibility on the critical Rayleigh number. In figure \ref{diff:fig:comparison} we plot the relationship between (a) $Ra_{c\text{, scaled}}$ and $S$ and (b) $Ra_{c\text{, scaled}}$ and $y_{st}$ for several values of the thermal expansion coefficient $\alpha$. We also plot $Ra_{c\text{, scaled}}$ as predicted by the linear stability analysis, for comparison. As expected, the values of $Ra_{c\text{, scaled}}$ calculated for a low value of $\alpha$ are similar to the values calculated in the linear stability analysis. This is because the Boussinesq approximation is derived from an expansion in small $\alpha$ of the governing equations.

\begin{figure}
\subfigure[Strongly burning planar diffusion flame with no flow (quiescent state).]{
\includegraphics[scale=1,trim=60 80 0 70, clip=true]{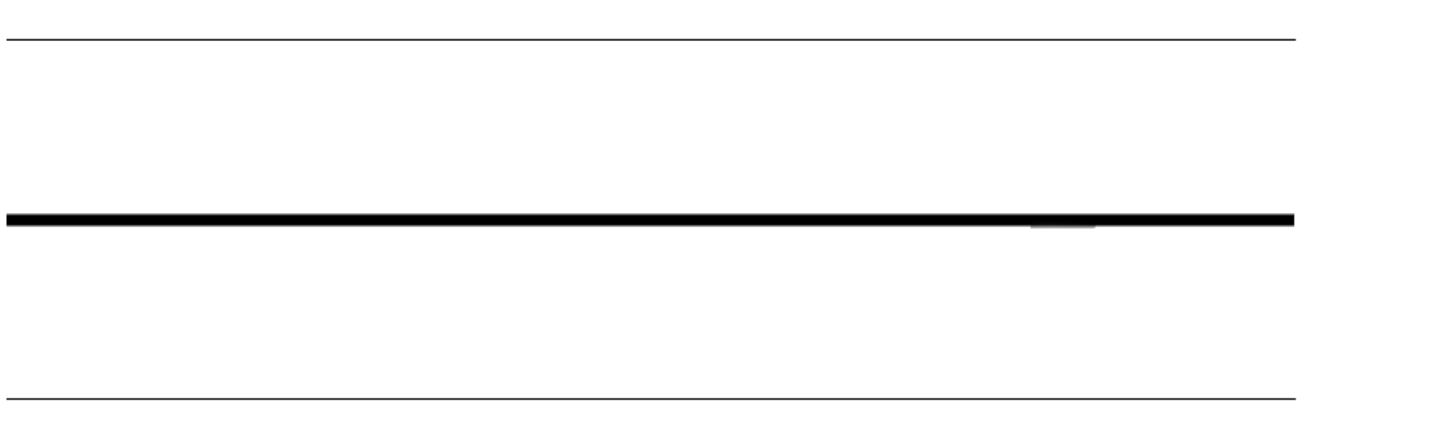}
}
\subfigure[Convection roll/cellular flame state with streamlines.]{
\includegraphics[scale=1,trim=60 80 0 70, clip=true]{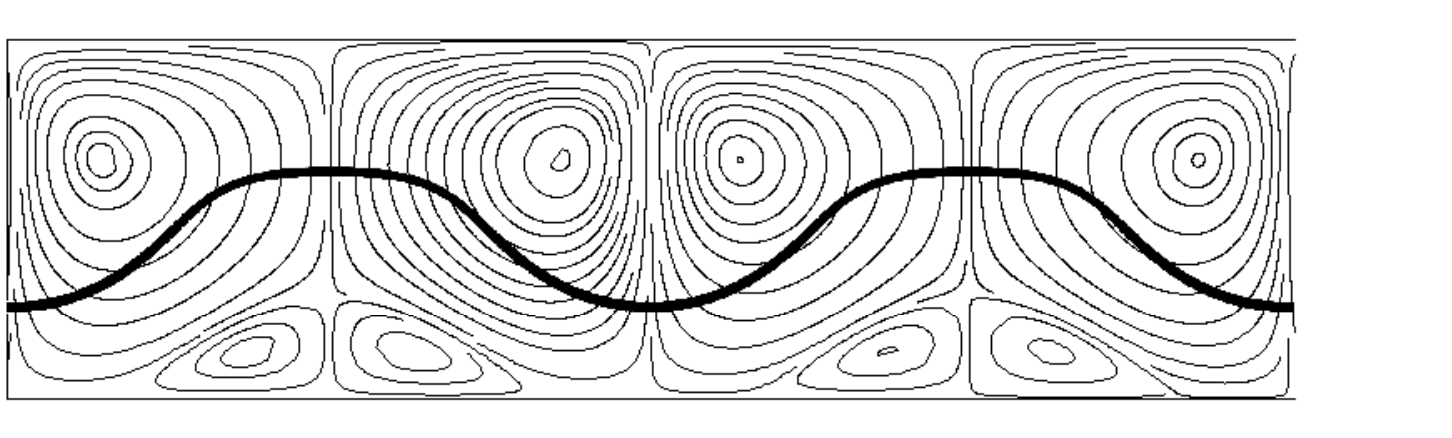}
}
\caption{Numerical calculations displaying an example of hysteresis in the non-Boussinesq system. The black line represents the flame position. Both calculations were performed for the parameter values $\alpha=0.85$, $Da=2\text{x}10^4$, $S=1$, $\beta=10$, $Pr=1$ and for $Ra=28000$, which is lower than $Ra_c$ for these parameter values. Both steady states are found to be stable, and since they both exist for $Ra<Ra_c$, this is an example of a finite-amplitude subcritical instability of the quiescent state.}
\label{diff:fig:hysterisis}
\end{figure}

The numerical results corresponding to larger, more realistic values of the thermal expansion coefficient show several discrepancies with those of the linear stability analysis. Firstly, the non-Boussinesq system is found to be more stable, a well known result in non-reactive Rayleigh--B\'{e}nard convection \citep[][]{bormann2001onset}. Secondly, the system is found to exhibit hysteresis at the onset of instability, whereby the system allows two different steady states for the same parameter values as shown in figure \ref{diff:fig:hysterisis}; again, it is well known in the literature for the non-reactive case that this is associated with departures from the Boussinesq approximation \citep[][]{frohlich1992large,getling1998rayleigh}. Despite these differences, figure \ref{diff:fig:comparison} shows that the relationship between $Ra_c$ and $S$ is found to be qualitatively similar for all of the selected values of $\alpha$.

\begin{figure}
\includegraphics[scale=0.75,trim=28 0 0 0, clip=true]{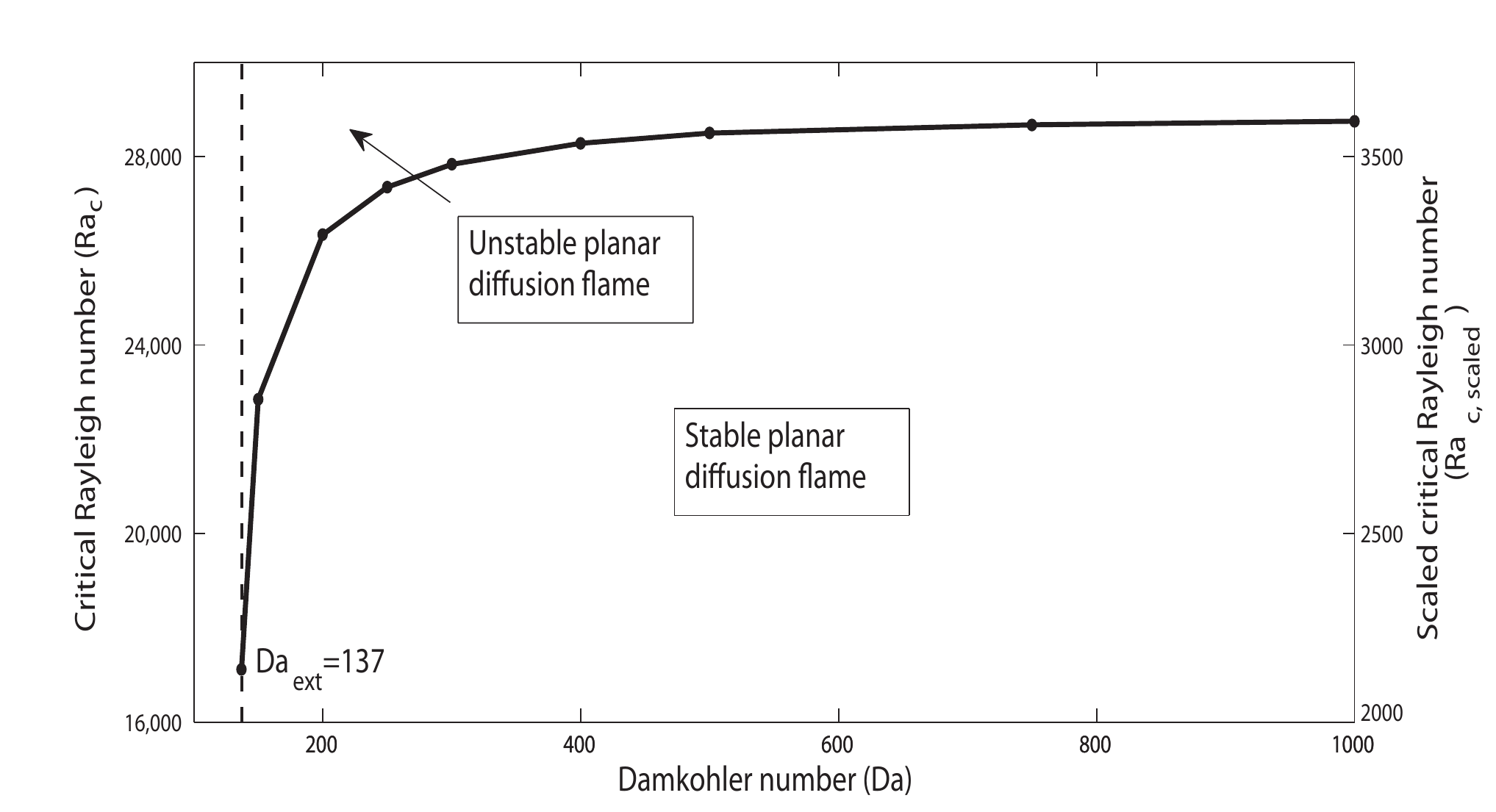}
\caption{The effect of the Damk\"{o}hler number $Da$ on the critical Rayleigh number $Ra_c$ and the scaled critical Rayleigh number $Ra_{c\text{, scaled}}$. The calculations were performed for the parameter values $\alpha=0.85$, $\beta=10$, $S=1$ and $Pr=1$. Indicated in the figure is the extinction value of $Da$, below which the strongly burning planar diffusion flame cannot exist.}
\label{diff:fig:finitechem}
\end{figure}

To close this section, we now test the effect of finite chemistry on the critical Rayleigh number by varying the Damk\"{o}hler number $Da$. The results, presented in figure \ref{diff:fig:finitechem}, show that $Ra_c$ decreases as $Da$ decreases. In the limit $Da \to \infty$ there is little change in $Ra_c$ as $Da$ varies; indeed, the curve in the figure can be seen to flatten out near $Da=1000$. Thus the value $Da=2\text{x}10^4$ used previously to approximate infinitely fast chemistry is sufficiently large. As $Da$ approaches its extinction value $Da_{\text{ext}}$ the system becomes considerably more unstable and $Ra_c$ is reduced significantly.

\section{Conclusion}
We have studied, using analytical and numerical methods, the stability of a planar diffusion flame in an infinitely long channel with porous walls, under gravitational effects. The conditions under which the diffusion flame becomes unstable have been determined by calculation of the critical Rayleigh number, which defines the threshold of instability. Such results do not seem to be available in the literature.

First, we have investigated the stability of the Burke--Schumann flame, using a linear stability analysis in the Boussinesq approximation. The relationship between the position of the flame in the channel (governed by the stoichiometric coefficient) and the critical Rayleigh number has been determined. The growth rate of the linear stability problem was first confirmed to be real using numerical methods, so that the system could be studied analytically in the marginal state using a similar method to that of the non-reactive problem.

Results have been presented, which show that as the flame approaches the lower boundary with increasing stoichiometric coefficient, the critical Rayleigh number is close to the well-known value it takes in the non-reactive case with two rigid boundaries. A rescaling of the Rayleigh number using the distance between the flame and the cold upper boundary as reference length was performed to aid in the comparison between the reactive and non-reactive problems. The scaled critical Rayleigh number was found to be (a) of the order of magnitude of the critical Rayleigh number in the non-reactive problem with two rigid boundaries, as the flame approaches the lower wall and (b) of the order of magnitude of the critical Rayleigh number in the non-reactive problem with one rigid and one free-surface boundary, when the flame is located a certain distance from the lower wall (which has been calculated).

Second, we have provided numerical solutions of the full set of governing equations for several values of the thermal expansion coefficient using a finite-element method with a high Damk\"{o}hler number. The results show that when the Rayleigh number is higher than its critical value, the fluid forms convection rolls as in the non-reactive case, which interact with the flame to generate cellular structures.

For weak values of the thermal expansion coefficient $\alpha$, which corresponds to conditions in which the Boussinesq approximation is expected to be valid, the numerical results show strong agreement with those of the linear stability analysis. For larger values of $\alpha$, there is qualitative agreement between the numerical and analytical results but several discrepancies caused by a higher thermal expansion coefficient. The non-Boussinesq system was found to be more stable than the system in the Boussinesq approximation and to exhibit hysteresis at the onset of instability, which are well known results in the non-reactive case.

Finally, we have investigated the effect of finite chemistry on the system. The results show that the system becomes less stable as the Damk\"{o}hler number $Da$ is decreased, especially as $Da$ approaches its extinction value $Da_{\text{ext}}$, below which the strongly burning diffusion flame cannot exist. For large values of $Da$ the decrease in the critical Rayleigh number with decreasing $Da$ was found to be small.

The results in this study are a crucial step in the understanding of gravitational effects on diffusion flames with important ramifications to combustion in non-uniform mixtures such as triple flames. Chapter \ref{chapter:triple} will concentrate on the effect of gravity on steadily propagating triple flames, using the characterisation of the instability of the trailing planar diffusion flame in this study as a basis.

\chapter{The effect of gravity and thermal expansion on the propagation of a triple flame in a horizontal channel}
\chaptermark{Effects of gravity on triple flames}
\label{chapter:triple}
\section{Introduction}
\label{triple:sec:intro}
The study of triple flames, which consist of two premixed branches and a trailing diffusion flame, has been extensive since their first experimental observation by Phillips \cite{phillips}. Early analytical studies were carried out by Ohki and Tsuge \cite{2}, followed by Dold and collaborators \cite{3,4}. These initial studies utilised the constant density approximation, decoupling the underlying hydrodynamics of the system from the equations of heat and mass. Most analytical studies since have used this approximation while investigating several practical aspects affecting triple flames. These aspects include preferential diffusion \cite{buckmaster1989anomalous,prefdif}, heat losses \cite{heatloss1,heatloss2,heatloss3}, reversibility of the chemical reaction \cite{reversability,daou2009asymptotic} and the presence of a parallel flow \cite{daou2010triple}. For further references see the review papers \cite{buckmaster2002edge} and \cite{chung2007stabilization}.

In this chapter, which is based on a paper by \citet{pearce2013effect}, we dispense with the constant density approximation in order to describe the coupled effect of thermal expansion and gravity on the propagation of a triple flame. To this end, it is imperative to first understand this effect on the ``strongly burning" diffusion flame, which forms one of the triple flame's branches. Steadily propagating triple flames are only expected for parameter values for which the diffusion flame exists and is stable. In Chapter \ref{chapter:diffusion}, we undertook a comprehensive study on the stability of a diffusion flame under the influence of gravity and thermal expansion in the channel configuration adopted in the present chapter (see also the paper by \citet{pearce2013rayleigh} on this topic). There is thus a clear understanding of the values that the parameters can take to ensure the existence and stability of the trailing diffusion flame.

The first attempt to understand the effects of variable density on triple flames was undertaken by Ruetsch \emph{et al.} \cite{5}, who investigated the problem numerically. They have also derived a scaling law describing the increase in the propagation speed of the triple flame above the planar premixed flame speed due to thermal expansion. The increase was attributed to the divergence of the flow field ahead of the flame. The result was confirmed in \cite{im1999structure} using numerical simulation; further related numerical studies were carried out in \cite{6,8,9,10}.

Triple flames propagating in a direction parallel to the direction of gravity have been investigated numerically and experimentally in \cite{11,12,chen2001numerical,echekki2004numerical}. It has been found that the propagation speed of a triple flame propagating downwards is decreased in comparison to that of a triple flame in the absence of gravity. The change in the propagation speed has been explained in \cite{chen2001numerical} as being due to an increase in the acceleration of the gas ahead of the triple flame leading edge, caused by buoyancy. Conversely, upward propagation leads to an increase in the propagation speed. To our knowledge no dedicated studies have been undertaken on triple flames propagating in a direction perpendicular to gravity.

In the present chapter we investigate the combined effect of thermal expansion and gravity on the propagation of a triple flame in a horizontal channel with rigid walls, through which fuel and oxidiser are injected. The main aim is to describe the behaviour of the triple flame in terms of three non-dimensional parameters; the flame-front thickness, the thermal expansion coefficient and the Rayleigh number.

The chapter is structured as follows. In \S \ref{triple:sec:formulation} we provide a non-dimensional formulation of the problem. In \S \ref{triple:sec:prelim} we present some important preliminary results used in later discussions, related to the planar premixed flame and to the existence and stability of the planar diffusion flame. In \S \ref{triple:sec:results}, we present the numerical results obtained. In particular, the effect of thermal expansion and gravity on a triple flame is described, with special emphasis on the relationship between the propagation speed $U$ and the flame-front thickness $\epsilon$ for various values of the thermal expansion coefficient $\alpha$ and the Rayleigh number $Ra$. We close the chapter with a summary of the main findings and recommendations for future studies.
\section{Formulation}
\label{triple:sec:formulation}
\begin{figure*}[t]
\includegraphics[scale=0.8]{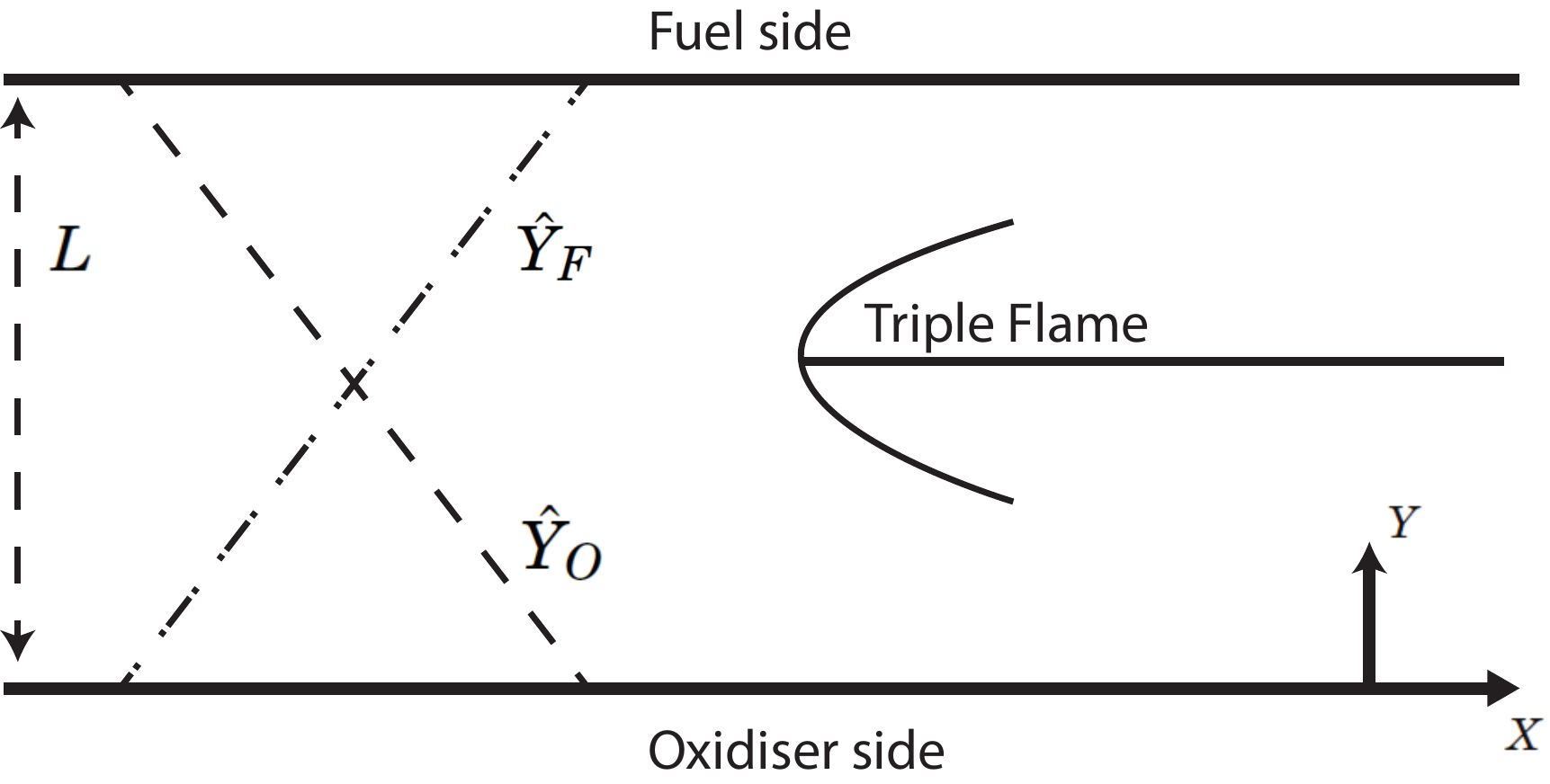}
\caption{An illustration of a triple flame in a channel of height $L$. The walls are assumed to be rigid and to have equal temperatures $\hat{T}=\hat{T}_u$. The mass fractions are prescribed by $\hat{Y}_F=\hat{Y}_{Fu},\text{ }\hat{Y}_O=0$ at the upper wall and $\hat{Y}_F=0,\text{ }\hat{Y}_O=\hat{Y}_{Ou}$ at the lower wall.}
\label{triple:fig:diagram}
\end{figure*}

We investigate the problem of a triple flame propagating through an infinitely long channel of height $L$, where fuel is provided at the upper wall and oxidiser at the lower wall, as shown in figure \ref{triple:fig:diagram}. For simplicity, the temperatures of the walls are assumed to be equal. The governing equations consist of the Navier--Stokes equations coupled to equations for temperature and mass fractions of fuel and oxidiser. The combustion is modelled as a
single irreversible one-step reaction of the form
\begin{align*}
\text{F}+s\text{O} \to (1+s)\text{Products}+q,
\end{align*}
where F denotes the fuel and O the oxidiser. The quantity $s$ denotes
the mass of oxidiser consumed and $q$ the heat released, both per unit mass of fuel.\\
The overall reaction rate $\hat{\omega}$ is taken to follow an Arrhenius law of the form
\begin{align*}
\hat{\omega}=\hat{\rho} B \hat{Y}_F \hat{Y}_O \exp{\left(-
E/R\hat{T}\right)}.
\end{align*}
Here $\hat{\rho}$, $\hat{Y}_F$, $\hat{Y}_O$, $R$, $\hat{T}$, $B$ and $E$ are the density, the fuel mass fraction, the oxidiser mass fraction, the
universal gas constant, the temperature, the pre-exponential factor and the activation
energy of the reaction, respectively.\\
\subsection{Governing Equations}
The governing equations in the low Mach number formulation will be written here using a co-ordinate system attached to the flame front. More precisely, if $\dot{x}_F\left(\hat{t}\right)$ denotes the propagation speed of the flame front relative to the laboratory (with $\dot{x}_F\left(\hat{t}\right)<0$ indicating a propagation to the left), we shall use the co-ordinates
\begin{gather*}
\left(\hat{t}',X',Y'\right) = \left(\hat{t}, X-x_F\left(\hat{t}\right),Y\right).
\end{gather*}
This leads to
\begin{gather*}
\frac{\partial}{\partial \hat{t}}=\frac{\partial}{\partial \hat{t}'}-\dot{x}_F\left(\hat{t}\right)\frac{\partial}{\partial X'}.
\end{gather*}
Dropping primes, we write the governing equations as the continuity equation
\begin{gather}
\frac{\partial \hat{\rho}}{\partial \hat{t}}+\frac{\partial}{\partial X}\left(\hat{\rho}\left(\hat{u}-\dot{x}_F\left(\hat{t}\right)\right)\right)+\frac{\partial}{\partial Y}\left( \hat{\rho} \hat{v}\right)=0, \label{triple:1}
\end{gather}
momentum equations
\begin{gather}
\hat{\rho}\frac{\partial \hat{u}}{\partial \hat{t}}+\hat{\rho}\left(\hat{u}-\dot{x}_F\left(\hat{t}\right)\right)\frac{\partial \hat{u}}{\partial X} + \hat{\rho} \hat{v} \frac{\partial \hat{u}}{\partial Y} +\frac{\partial \hat{p}}{\partial X}= \mu \left(\nabla^2\hat{u}+\frac{1}{3}\frac{\partial}{\partial X}\left(\nabla\cdot\mathbf{\hat{u}}\right)\right),\\
\hat{\rho}\frac{\partial \hat{v}}{\partial \hat{t}}+\hat{\rho}\left(\hat{u}-\dot{x}_F\left(\hat{t}\right)\right)\frac{\partial \hat{v}}{\partial X} + \hat{\rho} \hat{v} \frac{\partial \hat{v}}{\partial Y} +\frac{\partial \hat{p}}{\partial Y}= \mu \left(\nabla^2\hat{v}+\frac{1}{3}\frac{\partial}{\partial Y}\left(\nabla\cdot\mathbf{\hat{u}}\right)\right) + \left(\hat{\rho}-\hat{\rho}_0\right)g,
\end{gather}
temperature equation
\begin{gather}
\hat{\rho}\frac{\partial \hat{T}}{\partial \hat{t}}+\hat{\rho}\left(\hat{u}-\dot{x}_F\left(\hat{t}\right)\right)\frac{\partial \hat{T}}{\partial X} + \hat{\rho} \hat{v} \frac{\partial \hat{T}}{\partial Y}= \hat{\rho}D_T \nabla^2\hat{T}+\frac{q}{c_P}\hat{\omega},
\end{gather}
fuel and oxidiser mass fraction equations
\begin{gather}
\hat{\rho}\frac{\partial \hat{Y}_F}{\partial \hat{t}}+\hat{\rho}\left(\hat{u}-\dot{x}_F\left(\hat{t}\right)\right)\frac{\partial \hat{Y}_F}{\partial X} + \hat{\rho} \hat{v} \frac{\partial \hat{Y}_F}{\partial Y}= \hat{\rho}D_F \nabla^2\hat{Y}_F-\hat{\omega},\\
\hat{\rho}\frac{\partial \hat{Y}_O}{\partial \hat{t}}+\hat{\rho}\left(\hat{u}-\dot{x}_F\left(\hat{t}\right)\right)\frac{\partial \hat{Y}_O}{\partial X} + \hat{\rho} \hat{v} \frac{\partial \hat{Y}_O}{\partial Y}= \hat{\rho}D_O \nabla^2\hat{Y}_O-s\hat{\omega},
\end{gather}
and the equation of state
\begin{gather}
\hat{\rho}\hat{T}=\hat{\rho}_u \hat{T}_u \label{triple:state}.
\end{gather}
Here $\hat{p}$ is the hydrodynamic pressure and $D_T$, $D_F$, and $D_O$ denote the diffusion coefficients of heat, fuel and oxidiser, respectively. $\hat{T}_u$ refers to the temperature of the unburnt mixture which is also the temperature of both channel walls, while $\hat{\rho}_u$ is the density of the unburnt mixture. We assume that $\hat{\rho} D_T$, $\hat{\rho} D_F$ and $\hat{\rho} D_O$ are constant, as are the specific heat capacity $c_P$, thermal conductivity $\lambda$ and dynamic viscosity $\mu$. The flame speed $\dot{x}_F\left(\hat{t}\right)$ is an eigenvalue of the problem and must be determined as part of the solution.\\
The conditions as $X\to -\infty$ correspond to the frozen solution with no flow, which is independent of $X$ and is given by
\begin{gather}
\hat{T}=\hat{T}_u,\label{triple:frozen1}\\
\hat{Y}_F=\hat{Y}_{Fu}\frac{Y}{L},\label{triple:frozen2}\\
\hat{Y}_O=\hat{Y}_{Ou}\left(1-\frac{Y}{L}\right),\label{triple:frozen3}\\
\hat{u}=\hat{v}=0 \label{triple:frozen4},
\end{gather}
where $\hat{Y}_{Fu}$ and $\hat{Y}_{Ou}$ refer to the prescribed mass fractions at the fuel and oxidiser sides respectively. The channel walls are also considered rigid; thus the lateral boundary conditions are
\begin{gather}
\hat{T}=\hat{T}_u,\text{ }\hat{Y}_F=0,\text{ }\hat{Y}_O=\hat{Y}_{Ou},\text{ }\hat{u}=\hat{v}=0,\text{ at }Y=0,\label{triple:bc:dim1}\\
\hat{T}=\hat{T}_u,\text{ }\hat{Y}_F=\hat{Y}_{Fu},\text{ }\hat{Y}_O=0,\text{ }\hat{u}=\hat{v}=0,\text{ at }Y=L.\label{triple:bc:dim2}
\end{gather}
Downstream as $X \to \infty$ the solution corresponds to the one-dimensional strongly burning solution of the diffusion flame, which is again independent of $X$.\\
For large activation energies, the flame-front region is expected to be centred around the stoichiometric surface $Y=Y_{st}$ where $\hat{Y}_{O}=s\hat{Y}_F$. Upstream, the position of the stoichiometric surface can be determined from equations (\ref{triple:frozen1})-(\ref{triple:frozen3}) as
\begin{gather}
\frac{Y_{st}}{L}=\frac{1}{1+S},
\end{gather}
where $S \equiv s\hat{Y}_{Fu} / \hat{Y}_{Ou}$ is a normalised stoichiometric coefficient.\\
We now introduce the non-dimensional variables
\begin{align*}
x=\frac{X}{L},\text{ }y=\frac{Y}{L},\text{ }u=\frac{\hat{u}}{S_L^0},\text{ }v=\frac{\hat{v}}{S_L^0},\\
t=\frac{\hat{t}}{L \left/ S_L^0 \right.}, \text{ }
\theta=\frac{\hat{T}-\hat{T}_u}{\hat{T}_{ad}-\hat{T}_u},\text{ }y_F=\frac{\hat{Y}_F}{\hat{Y}_{F,st}},\text{ }\\ y_O=\frac{\hat{Y}_O}{\hat{Y}_{O,st}}, \text{ }p=\frac{\hat{p}}{\hat{\rho}_0 \left(S_L^0\right)^2},
\end{align*}
where the subscript 'st' denotes values at the upstream stoichiometric surface. Here $\hat{T}_{ad} \equiv\hat{T}_u+q\hat{Y}_{F,st}/ c_P$ is the adiabatic flame temperature, $\beta\equiv E\left(\hat{T}_{ad}-\hat{T}_u\right) / R\hat{T}_{ad}^2$ is the Zeldovich number or non-dimensional activation energy and $\alpha \equiv \left(\hat{\rho}_{u}-\hat{\rho}_{ad}\right)/\hat{\rho}_{u}$ is the thermal expansion coefficient. In non-dimensionalising we have used as unit speed
\begin{gather}
S_L^0=\left(4 Le_F Le_O \beta^{-3} Y_{O,st} \left(1-\alpha\right) D_T B \exp\left(-E/ R T_{ad}\right)\right)^{1/2}, \label{triple:eq:sl}
\end{gather}
which is the laminar burning speed of the stoichiometric planar flame to leading order for $\beta\gg 1$.\\
Inserting the scalings above into equations (\ref{triple:1})-(\ref{triple:state}) leads to the non-dimensional equations
\begin{gather}
\frac{\partial \rho}{\partial t}+\frac{\partial}{\partial x}\left(\rho(u+U(t))\right)+\frac{\partial}{\partial y}\left( \rho v\right)=0,\label{triple:nondim1}\\
\rho \frac{\partial u}{\partial t}+\rho(u+U(t))\frac{\partial u}{\partial x} + \rho v \frac{\partial u}{\partial y} +\frac{\partial P}{\partial x} = \epsilon Pr\nabla^2 u,\\
\rho \frac{\partial v}{\partial t}+\rho(u+U(t))\frac{\partial v}{\partial x} + \rho v \frac{\partial v}{\partial y} +\frac{\partial P}{\partial y} =  \epsilon Pr\nabla^2v+ \frac{\epsilon^2 Pr Ra}{\alpha }(1-\rho), \label{triple:nondim3}\\
\rho \frac{\partial \theta}{\partial t}+\rho(u+U(t))\frac{\partial \theta}{\partial x} + \rho v \frac{\partial \theta}{\partial y} = \epsilon \nabla^2 \theta + \frac{\epsilon^{-1} \omega}{1-\alpha},\\
\rho \frac{\partial y_F}{\partial t}+\rho(u+U(t))\frac{\partial y_F}{\partial x} + \rho v \frac{\partial y_F}{\partial y} = \frac{\epsilon}{ Le_F} \nabla^2 y_F - \frac{\epsilon^{-1} \omega}{1-\alpha},\\
\rho \frac{\partial y_O}{\partial t}+\rho(u+U(t))\frac{\partial y_O}{\partial x} + \rho v \frac{\partial y_O}{\partial y} = \frac{\epsilon}{Le_O} \nabla^2 y_O - \frac{\epsilon^{-1} \omega}{1-\alpha},\label{triple:nondim6}\\
\rho=\left(1+\frac{\alpha}{1-\alpha}\theta\right)^{-1},\label{triple:nondim7}
\end{gather}
where $P$ is a modified pressure given by $P=p-\frac{Pr}{3}\left(\nabla \cdot \bf{u}\right)$ and $U \equiv -\dot{x}_F\left/S_L^0\right.$ is the non-dimensional propagation speed relative to the laboratory. The non-dimensional parameters are defined as
\begin{gather*}
Ra=\frac{g\left(\hat{\rho}_{u}-\hat{\rho}_{ad}\right)L^3}{\nu \hat{\rho}_{u}D_T}, \quad \epsilon=\frac{l_{Fl}}{L}=\frac{D_T/S_L^0}{L},\\ \quad Le_F=\frac{D_T}{D_F}, \quad Le_O=\frac{D_T}{D_O},\quad \text{and}\quad Pr=\frac{\nu}{D_T},
\end{gather*}
which are the Rayleigh number, the flame-front thickness $l_{Fl}$ measured against the unit length $L$, the fuel and oxidiser Lewis numbers and the Prandtl number, respectively. Here $\nu$ is the kinematic viscosity $\nu=\mu \left/ \hat{\rho}_u \right.$. Note that $\epsilon$ is related to the Damk\"{o}hler number used in Chapter \ref{chapter:diffusion} by 
\begin{gather*}
Da=\frac{1}{\epsilon^{2}(1-\alpha)}.
\end{gather*}
The non-dimensional reaction rate is
\begin{gather}
\omega=\frac{\beta^3}{4Le_F Le_O} \rho y_F y_O \exp\left(\frac{\beta(\theta-1)}{1+\alpha_h(\theta-1)}\right),
\end{gather}
where $\alpha_h$ is a heat release parameter given by $\alpha_h=(\hat{T}_{ad}-\hat{T}_u)/\hat{T}_{ad}$. Note that in the low Mach number approximation the two parameters $\alpha \equiv \left(\hat{\rho}_{u}-\hat{\rho}_{ad}\right)/\hat{\rho}_{u}$ and $\alpha_h$ are in fact equal, which follows from equation (\ref{triple:state}). In this study, however, we leave the two distinct to aid comparison of our results with those previously obtained in the constant density approximation \cite{prefdif}, where $\alpha_h$ appears in the reaction term $\omega$ and only there. To assess the effect of thermal expansion we shall vary the coefficient $\alpha$, while maintaining $\alpha_h=0.85$ constant, as in \cite{prefdif,heatloss1}. Thus as $\alpha \to 0$ the equations in our study reduce to those of the constant density approximation\footnote{In asymptotic studies with $\beta \to \infty$, of course $\alpha_h$ is unimportant and can be set to zero. However, for finite values of $\beta$, say $\beta=10$ (which is commonly used in numerical studies), $\alpha_h$ has a noticeable effect.}.

Finally, (\ref{triple:bc:dim1})-(\ref{triple:bc:dim2}) imply that the boundary conditions are
\begin{align}
&\theta=0, \label{triple:bc1}\\
&y_F=(1+S)y,\\
&y_O=\frac{S+1}{S}(1-y),\\
&u=v=0 \quad \text{ as }\quad x \to -\infty,\text{ } y =0 \text{ or } y=1.\label{triple:bc3}\\
&\frac{\partial u}{\partial x}=\frac{\partial v}{\partial x}=\frac{\partial \theta}{\partial x}=\frac{\partial y_F}{\partial x}=\frac{\partial y_O}{\partial x}=0\quad \text{ as }\quad x \to \infty.\label{triple:bc4}
\end{align}
The non-dimensional problem is now fully formulated and is given by equations (\ref{triple:nondim1})-(\ref{triple:nondim7}), subject to the boundary conditions (\ref{triple:bc1})-(\ref{triple:bc4}). The non-dimensional parameters in this problem are $\alpha$, $\alpha_h$, $\beta$, $Pr$, $Ra$, $Da$, $S$, $Le_F$ and $Le_O$. Throughout this study we will make the assumption of unity Lewis numbers,
\begin{gather}
Le_F=Le_O=1.
\end{gather}
\begin{figure}[h]
\includegraphics[scale=0.8,trim=0 10 0 25,clip=true]{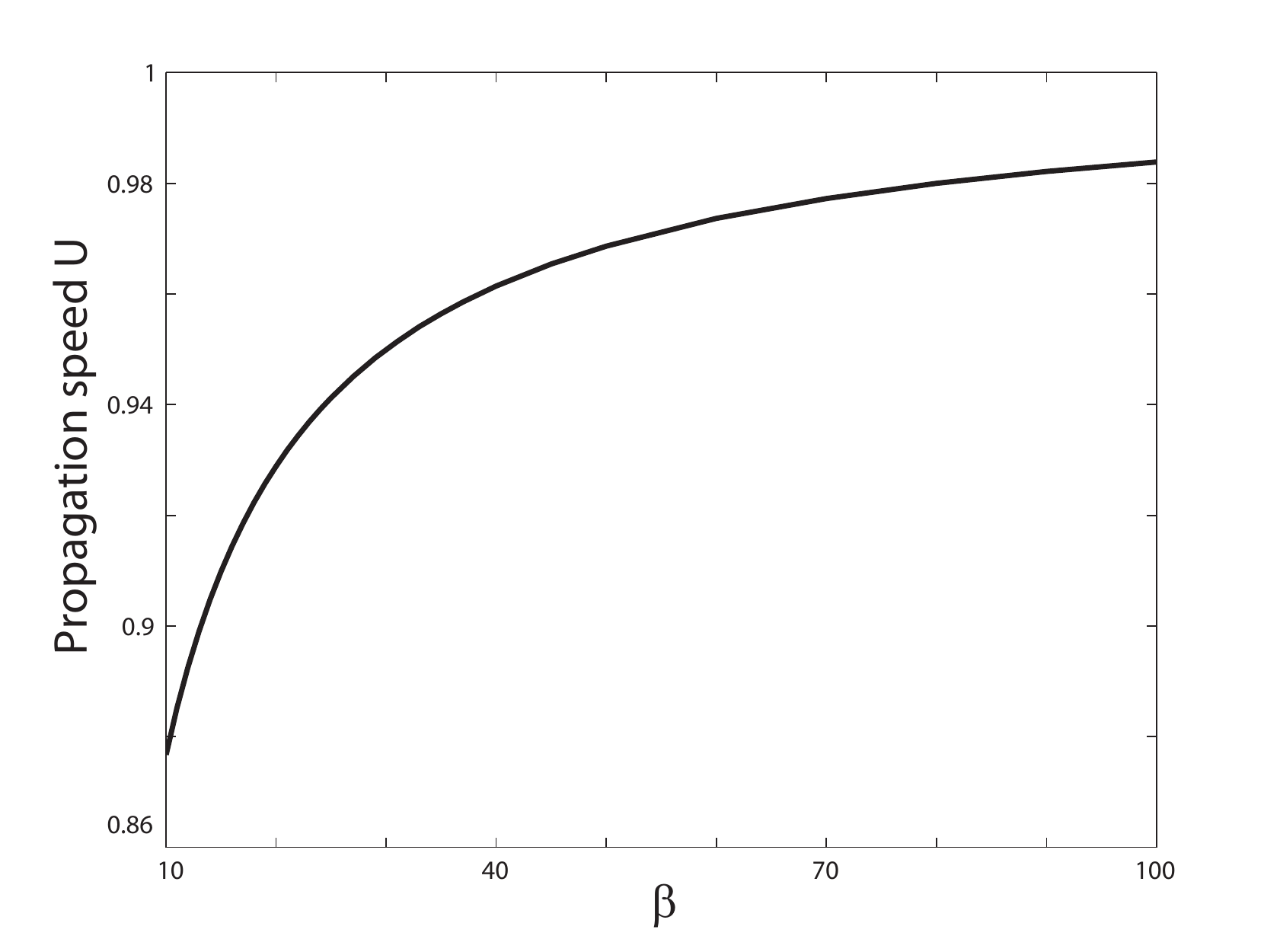}
\caption{The effect of $\beta$ on the numerically calculated value of the propagation speed, $U_{\text{planar}}$, for fixed values of the other parameters given by $\alpha=0.85$, $\epsilon=1$, $Le_F=Le_O=1$, $S=1$, $Pr=1$ and $Ra=0$.}
\label{triple:fig:planarbeta}
\end{figure}
\section[Preliminary studies]{Preliminary study: the planar premixed flame and the planar diffusion flame}
\label{triple:sec:prelim}
As a preliminary study whose results will be useful for subsequent discussions, in this section we investigate the planar premixed flame and the planar diffusion flame. We will begin with a discussion of how the propagation speed of the planar premixed flame is affected by the parameters $\alpha$ and $\beta$. We will then study the planar diffusion flame to determine the values of $\epsilon$, $\alpha$ and $Ra$ for which it is expected to exist in a stable state.
\subsection{Planar premixed flame}
\begin{figure}[h]
\includegraphics[scale=0.8,trim=20 10 0 20,clip=true]{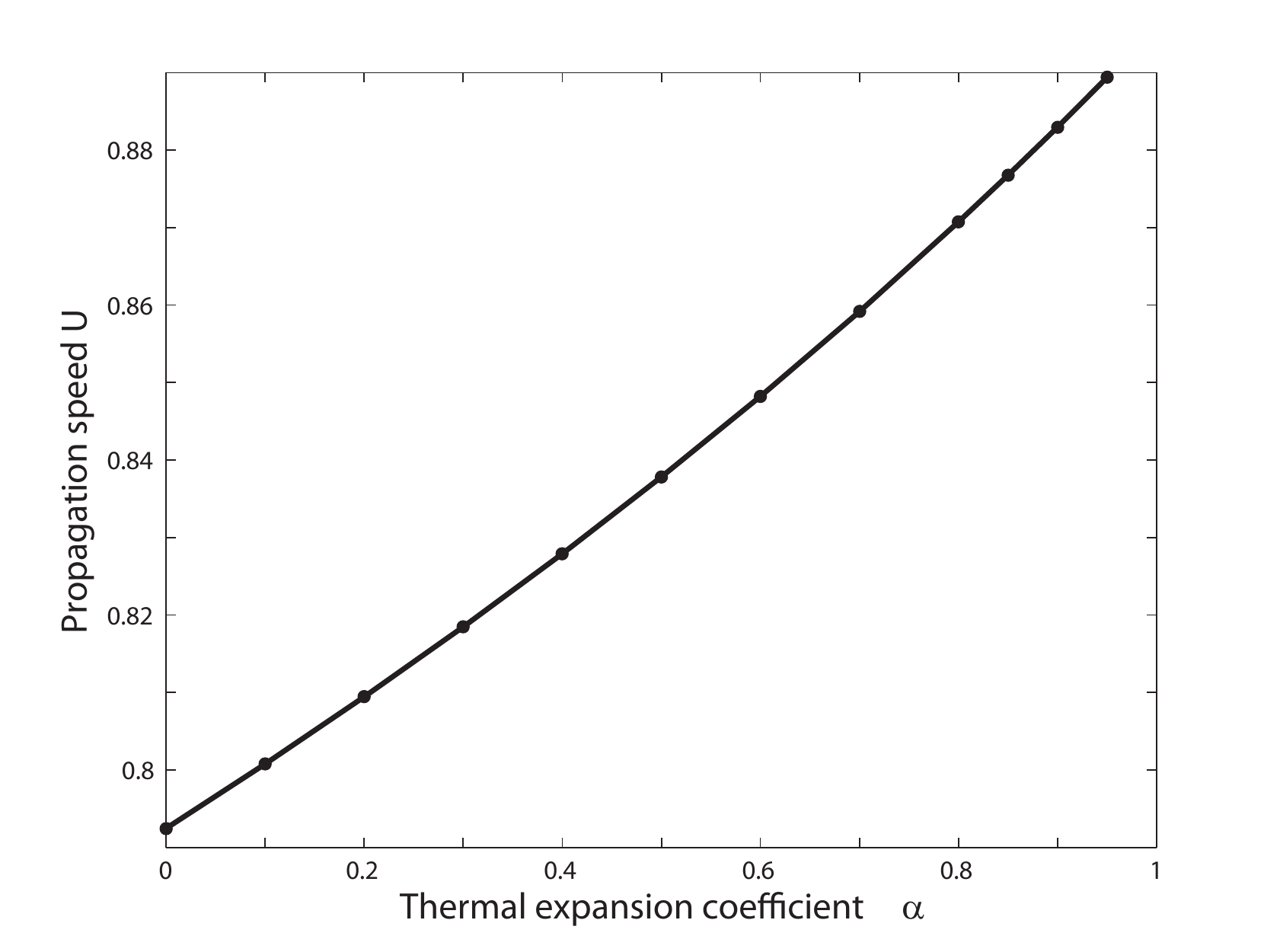}
\caption{The effect of the thermal expansion coefficient $\alpha$ on the numerically calculated value of the propagation speed, $U_{\text{planar}}$, for fixed values of the other parameters given by $\beta=10$, $\epsilon=1$, $Le_F=Le_O=1$, $S=1$, $Pr=1$ and $Ra=0$.}
\label{triple:fig:planaralpha}
\end{figure}
The equations governing the planar premixed flame are the stationary, $y$-independent form of equations (\ref{triple:nondim1})-(\ref{triple:nondim7}), subject to the boundary conditions
\begin{align}
&\theta=0, \quad y_F=1, \quad y_O=0,\quad u=v=0 \quad \text{ as }\quad x \to -\infty,\\
&\frac{\partial u}{\partial x}=\frac{\partial v}{\partial x}=\frac{\partial \theta}{\partial x}= \frac{\partial 
y_F}{\partial x}=\frac{\partial y_O}{\partial x}=0 \quad \text{ as }\quad x \to +\infty.
\end{align}

The stationary, one-dimensional problem is solved using the finite-element package Comsol Multiphysics. The aim is to investigate the effect of the parameters $\alpha$ and $\beta$ on the numerically calculated planar premixed flame propagation speed $U_{\text{planar}}$, which can be compared to its asymptotic value as $\beta \to \infty$. We give the other parameters fixed values of $\epsilon=1$, $Le_F=Le_O=1$, $S=1$, $Pr=1$, $\alpha_h=0.85$ and $Ra=0$.

In the limit $\beta \to \infty$ the (dimensional) propagation speed of the stoichiometric planar flame undergoing thermal expansion should approach the asymptotically derived value $S_L^0$, given by equation (\ref{triple:eq:sl}). Thus, since we have scaled the velocity by this value, the numerically calculated propagation speed $U_{\text{planar}}$ should approach unity for all values of $\alpha$ as $\beta \to \infty$. Figure \ref{triple:fig:planarbeta} shows that the numerically calculated value of $U_{\text{planar}}$ does indeed approach unity with increasing $\beta$ for a fixed value of $\alpha=0.85$, as expected.

For the remainder of this study we let $\beta$ take a typical value, namely $\beta=10$. From figure \ref{triple:fig:planarbeta} it can be seen that for this value of $\beta$, the propagation speed $U_{\text{planar}}$ deviates from its expected value of unity by about 12\%. However, to achieve a notable increase in accuracy would involve significant extra computational cost.

It will also be useful in later discussions to describe how $U_{\text{planar}}$ varies with the thermal expansion coefficient $\alpha$. Figure \ref{triple:fig:planaralpha} presents a plot of $U_{\text{planar}}$ versus $\alpha$ for a fixed value of $\beta=10$.
\begin{figure}[h]
\includegraphics[scale=0.8]{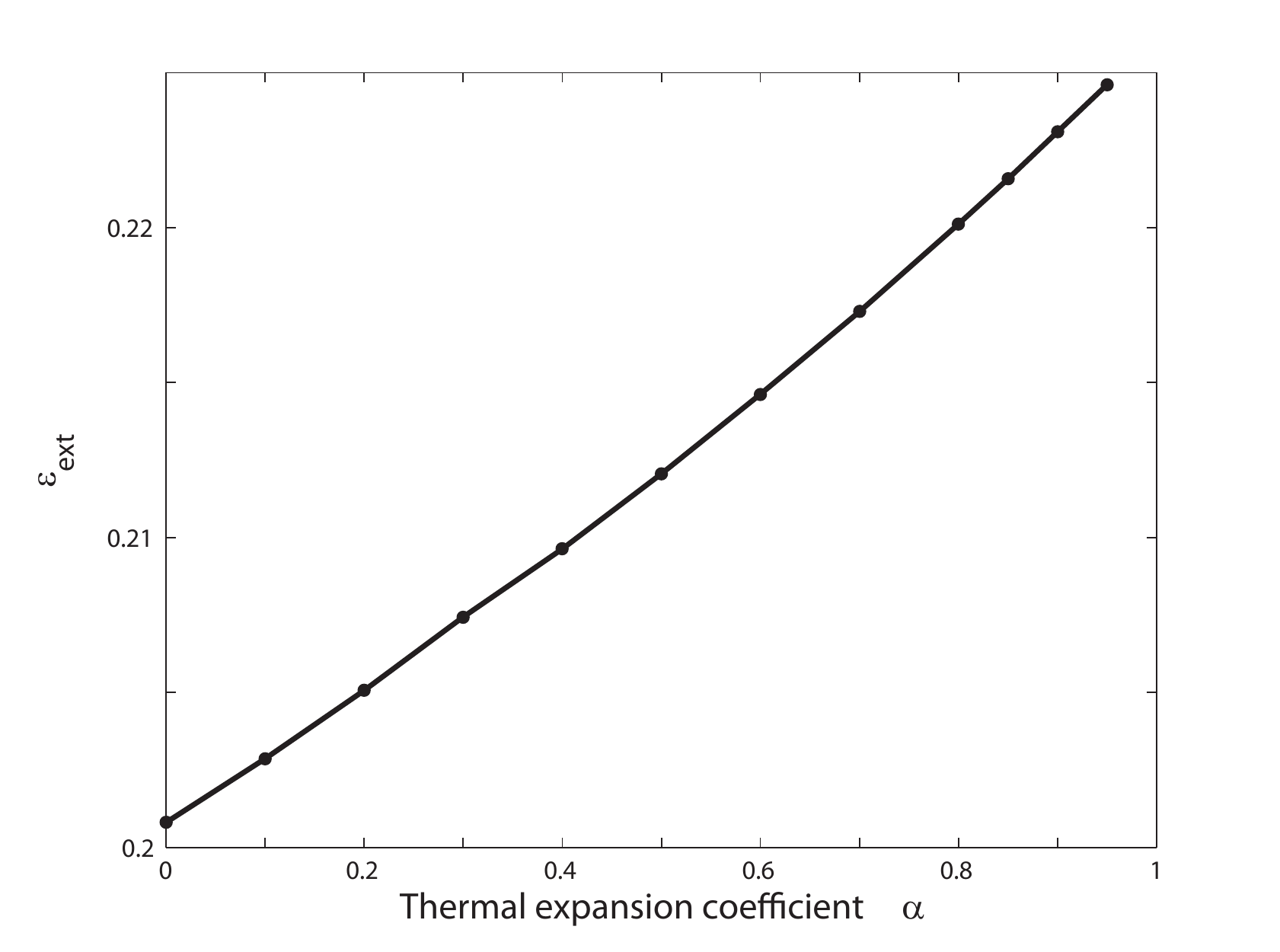}
\caption{The effect of the thermal expansion coefficient $\alpha$ on the value of $\epsilon$ corresponding to the extinction value of the planar diffusion flame, $\epsilon_{ext}$, for fixed values of the other parameters given by $\beta=10$, $Le_F=Le_O=1$, $S=1$, $Pr=1$ and $Ra=0$.}
\label{triple:fig:epsilonvsalpha}
\end{figure}
\subsection{Planar diffusion flame}
Steadily propagating triple flames are not expected if $\epsilon$ exceeds the extinction value $\epsilon_{\text{ext}}$ of the planar diffusion flame. Here we therefore numerically solve the underlying one-dimensional equations independent of $x$, to produce a plot of $\epsilon_{\text{ext}}$ versus $\alpha$, which is provided in figure \ref{triple:fig:epsilonvsalpha}.

\begin{figure*}[h]
\includegraphics[scale=0.8]{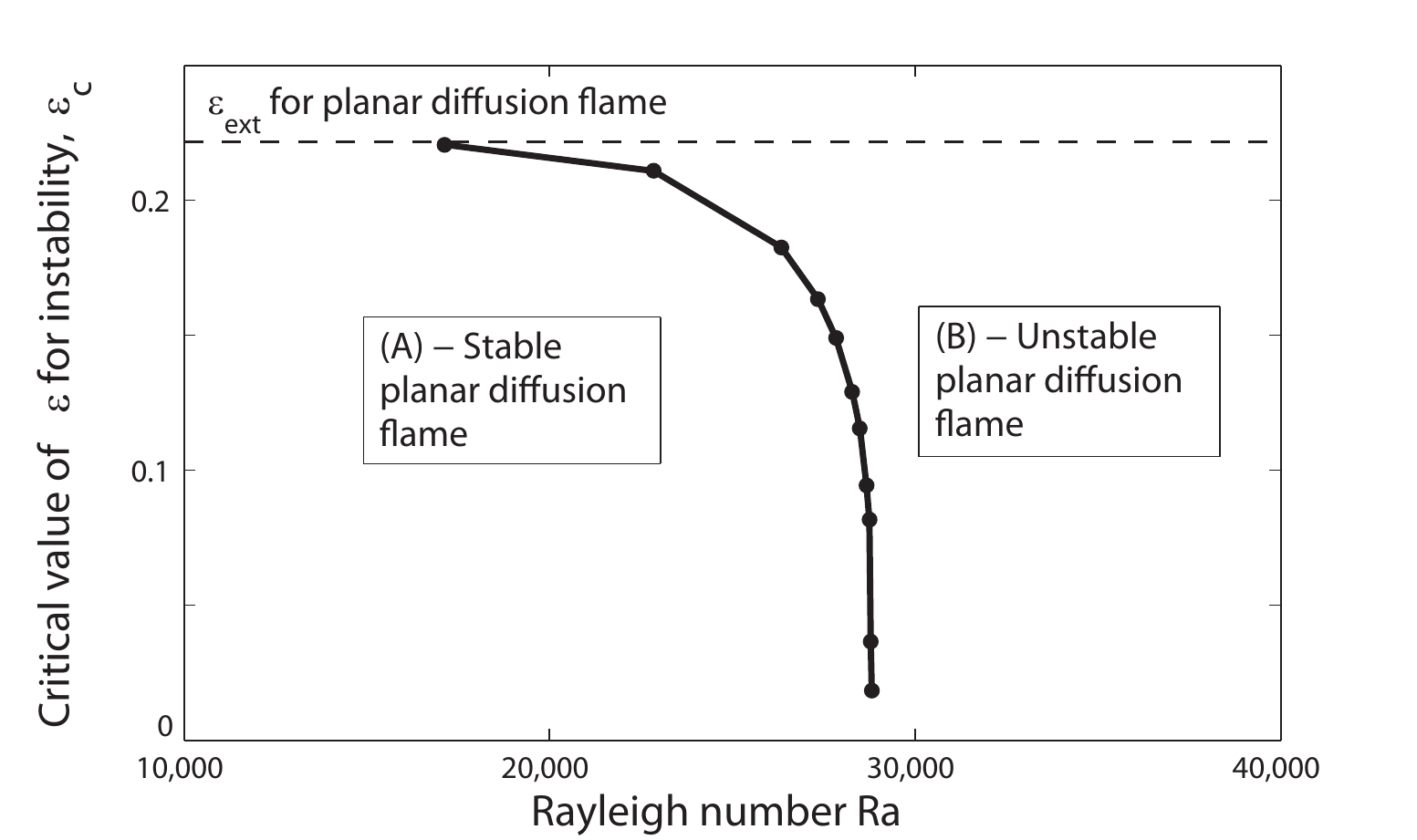}
\caption{The effect of the Rayleigh number $Ra$ on the value of $\epsilon$ for which the underlying planar diffusion flame becomes unstable, $\epsilon_c$, for fixed values of the other parameters given by $\beta=10$, $Le_F=Le_O=1$, $S=1$, $Pr=1$ and $\alpha=0.85$. The two regions in the diagram, labelled A and B, are the regions of stability and instability of the planar diffusion flame, respectively.}
\label{triple:fig:epsilonvsra}
\end{figure*}

A plot of the values of $\epsilon$ for which the trailing planar diffusion flame becomes unstable under gravity will also be useful in later discussions. This stability problem was investigated in Chapter \ref{chapter:diffusion} (see also the paper by \citet{pearce2013rayleigh}), from which important results relevant to our study are summarised in figure \ref{triple:fig:epsilonvsra}. Shown is an adaptation of figure \ref{diff:fig:finitechem}, plotting the critical value of $\epsilon$ versus the Rayleigh number $Ra$, for $\alpha=0.85$. This critical value defines the neutral stability curve which separates the two stability regions in the figure. These regions, labelled A and B, define the regions of stability and instability of the planar diffusion flame, respectively.
\section{Results for a triple flame propagating in a channel}
\label{triple:sec:results}
In this section we present the results obtained by solving the stationary form of equations (\ref{triple:nondim1})-(\ref{triple:nondim7}) with boundary conditions (\ref{triple:bc1})-(\ref{triple:bc4}), using the finite-element package Comsol Multiphysics. The main aim of the work is to calculate the propagation speed $U$ in terms of the parameters $\alpha$, $\epsilon$ and $Ra$, which represent thermal expansion, strain rate and gravity, respectively. The other parameters are assigned the values $\beta=10$, $\alpha_h=0.85$, $Le_F=Le_O=1$, $S=1$ and $Pr=1$ throughout this section. We begin with an investigation into the effect of thermal expansion on a triple flame in the absence of gravity. This is followed by a study on the combined effect of gas expansion and gravity.
\subsection{Effect of thermal expansion on a triple flame in the absence of gravity}
In this section we investigate the effect of thermal expansion on a triple flame in the absence of gravity. We therefore let $Ra=0$ throughout the section. Since the aim of the study is to calculate $U$, we will begin with a plot of $U$ versus $\epsilon$ for several values of the thermal expansion coefficient $\alpha$. In order to fully understand the effect of the two parameters $\epsilon$ and $\alpha$, we will end with a comparison of how the streamlines and reaction rate contours change with increasing $\epsilon$ for several fixed values of $\alpha$.
\subsubsection{Propagation speed of a triple flame}
\label{triple:sec:propagation}
\begin{figure*}[!]
\includegraphics[scale=0.8,trim=0 10 0 20,clip=true]{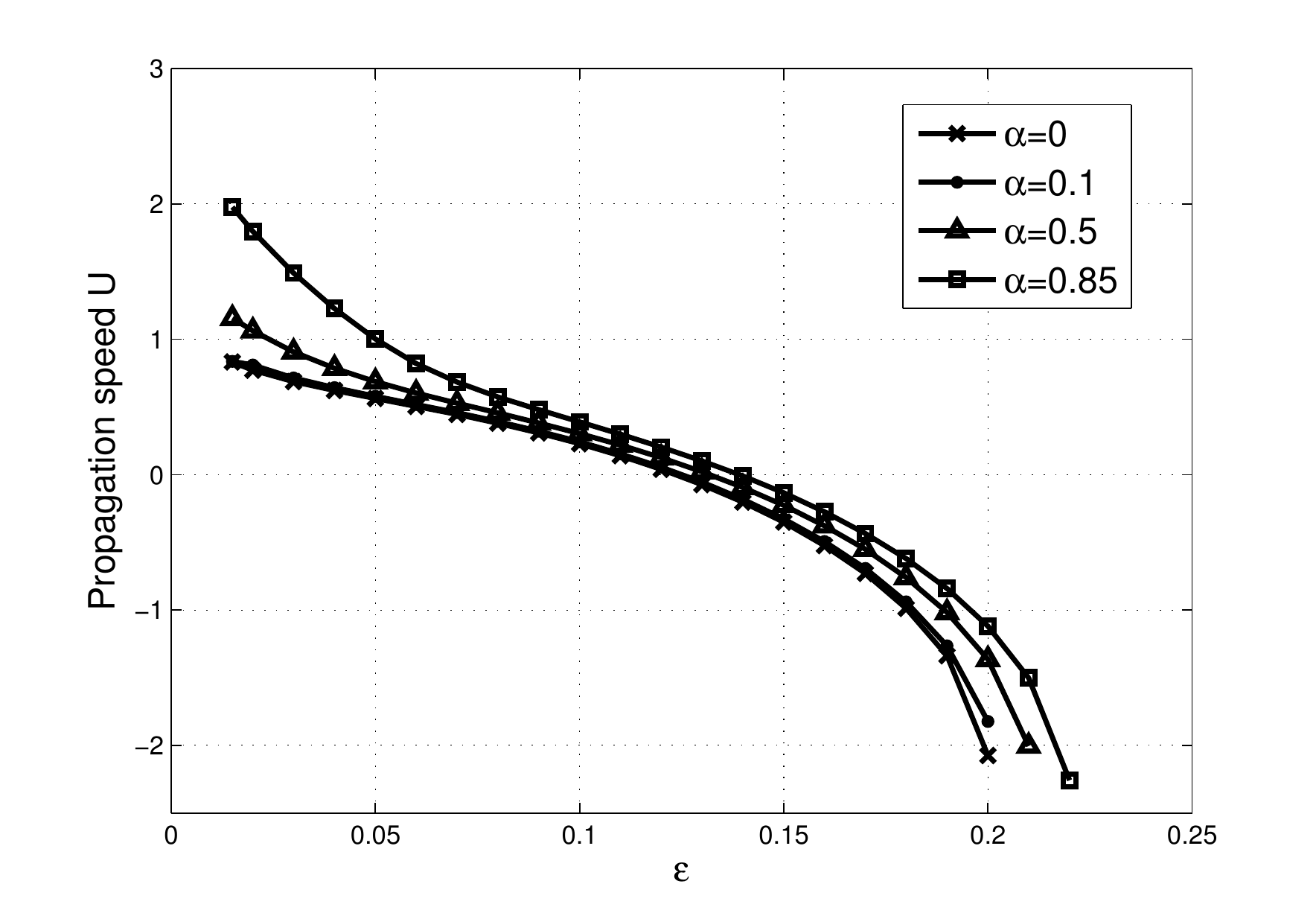}
\caption{The effect of $\epsilon$ on the propagation speed $U$ for several values of the thermal expansion coefficient $\alpha$ with fixed values of the other parameters given by $\beta=10$, $Le_F=Le_O=1$, $S=1$, $Pr=1$ and $Ra=0$. For each $\alpha$, $U$ is scaled by the numerical value calculated for the propagation speed of the planar premixed flame.}
\label{triple:fig:u_eps_alpha}
\end{figure*}

Figure \ref{triple:fig:u_eps_alpha} shows a plot of the propagation speed $U$ of the triple flame (scaled by the numerically calculated propagation speed $U_{\text{planar}}$ of the planar premixed flame, shown in figure \ref{triple:fig:planaralpha}) versus $\epsilon$ for several values of the thermal expansion coefficient $\alpha$. The maximum value $\epsilon$ can take for each $\alpha$ is the extinction value $\epsilon_{\text{ext}}$ of the trailing planar diffusion flame, which can be found in figure \ref{triple:fig:epsilonvsalpha}. The monotonic relationship between $U$ and $\epsilon$, with $U$ decreasing to negative values when $\epsilon$ is close to $\epsilon_{\text{ext}}$, is a well known property of constant density triple flames with Lewis numbers greater than or equal to unity \cite{prefdif}. It is found that this property remains valid for triple flames undergoing thermal expansion, for all values of $\alpha$, as shown in figure \ref{triple:fig:u_eps_alpha}.

It can also be seen in figure \ref{triple:fig:u_eps_alpha} that, if $\alpha>0$, $U$ approaches a value larger than that of the planar premixed flame as $\epsilon$ approaches zero (i.e. $U/U_{\text{planar}}>1$ as $\epsilon \to 0$). This result agrees with previous studies, which have found that thermal expansion causes an increase in the propagation speed of a triple flame above the speed of the planar premixed flame \cite{5,im1999structure}.

In fact, it can be derived (see \cite{5}) that under the influence of thermal expansion, and in the limit $\epsilon \to 0$, the ratio of the propagation speed of the triple flame to that of the planar premixed flame can be approximated by the formula
\begin{gather}
\frac{U}{U_\text{planar}}\mytilde\left(\frac{\rho_u}{\rho_b}\right)^{1/2}=\left(\frac{1}{1-\alpha}\right)^{1/2}\quad \text{ as }\epsilon \to 0.
\end{gather}
Thus in the case $\alpha=0.85$, say, the propagation speed of the triple flame (when scaled by the propagation speed of the planar premixed flame) can be expected to approach approximately the value 2.58 as $\epsilon \to 0$. Therefore our results are in good agreement with the predictions made in \cite{5}, as can be seen in figure \ref{triple:fig:u_eps_alpha}.

\subsubsection{Comparative cases for fixed $\alpha$}
\begin{figure*}[!]
\centering
\includegraphics[scale=1.24,trim=0 25 0 25,clip=true]{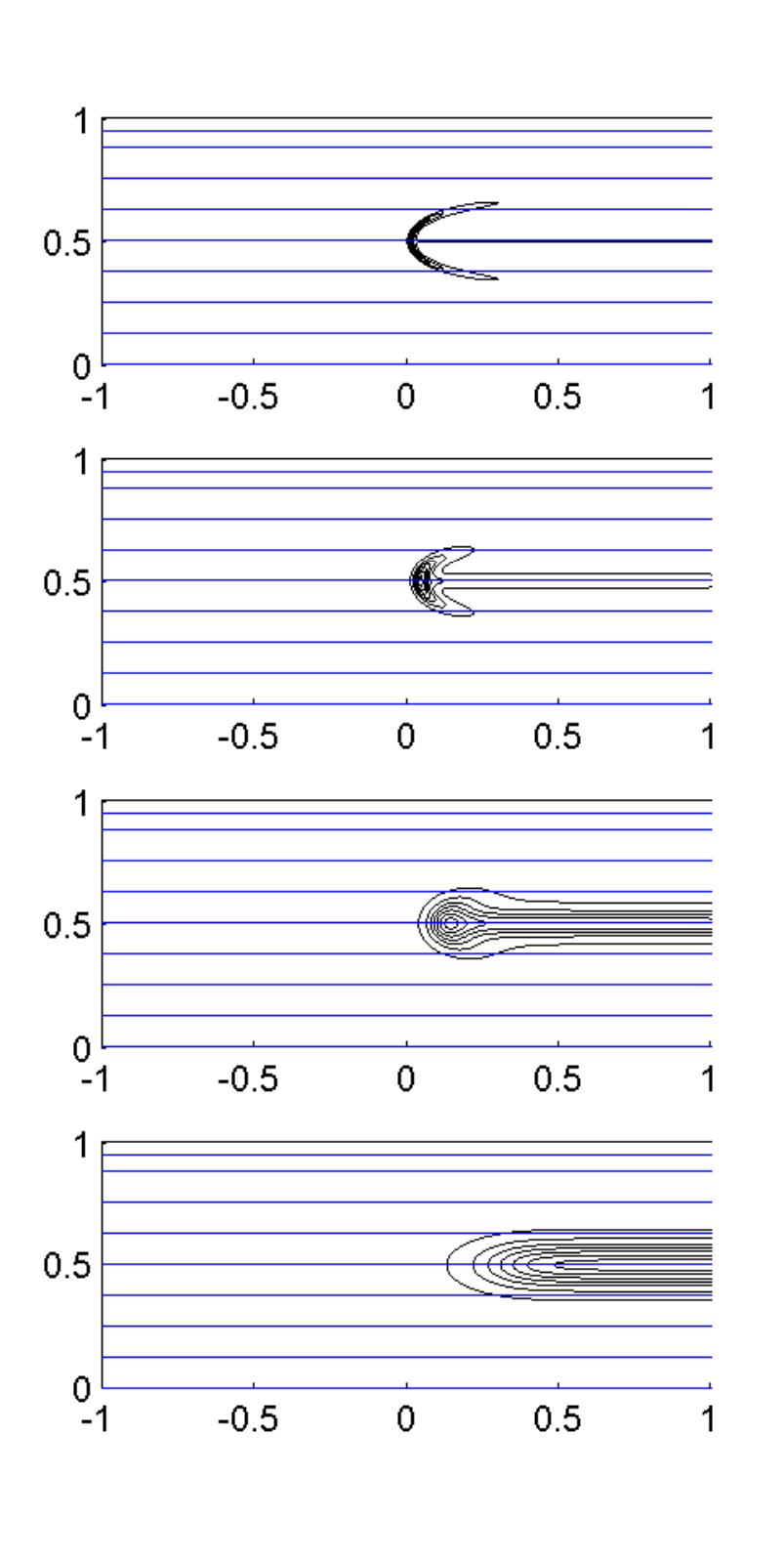}
\caption{Streamlines and reaction rate contours in a frame of reference attached to the flame-front for $\alpha=0$ and $\epsilon=0.015$, $\epsilon=0.05$, $\epsilon=0.12$ and $\epsilon=0.2$, respectively from top to bottom. The propagation speeds, when scaled by the numerically calculated propagation speed of a planar premixed flame, are given by $U=0.83$, $U=0.56$, $U=0.04$ and $U=-2.07$, respectively.}
\label{triple:fig:a_0}
\end{figure*}
\begin{figure*}[!]
\centering
\includegraphics[scale=1.24,trim=0 25 0 25,clip=true]{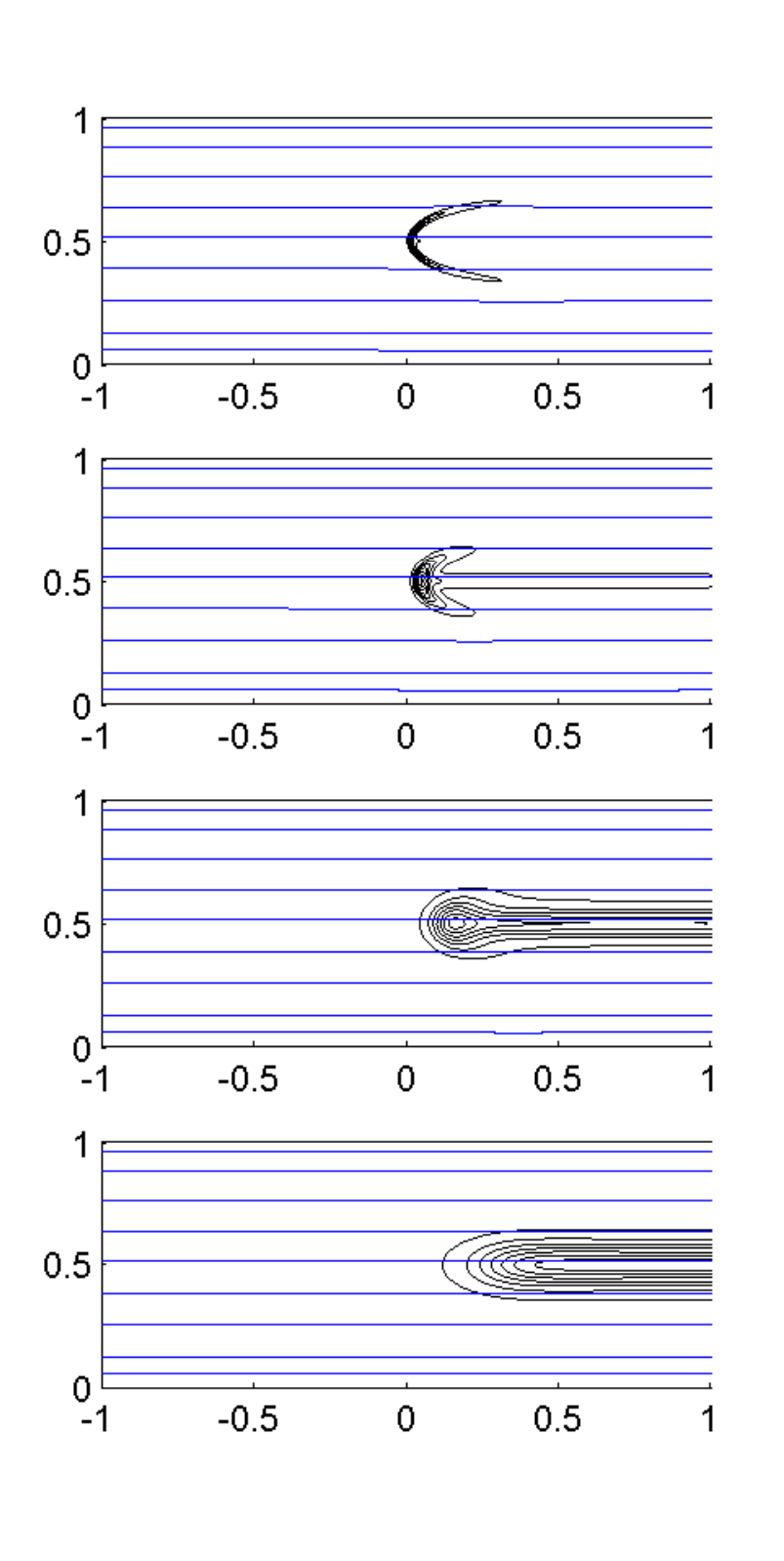}
\caption{Streamlines and reaction rate contours in a frame of reference attached to the flame-front for $\alpha=0.1$ and $\epsilon=0.015$, $\epsilon=0.05$, $\epsilon=0.13$ and $\epsilon=0.2$, respectively from top to bottom. The propagation speeds, when scaled by the numerically calculated propagation speed of a planar premixed flame, are given by $U=0.84$, $U=0.58$, $U=-0.06$ and $U=-1.82$, respectively.}
\label{triple:fig:a_10}
\end{figure*}
\begin{figure*}[!]
\centering
\includegraphics[scale=1.24,trim=0 25 0 25,clip=true]{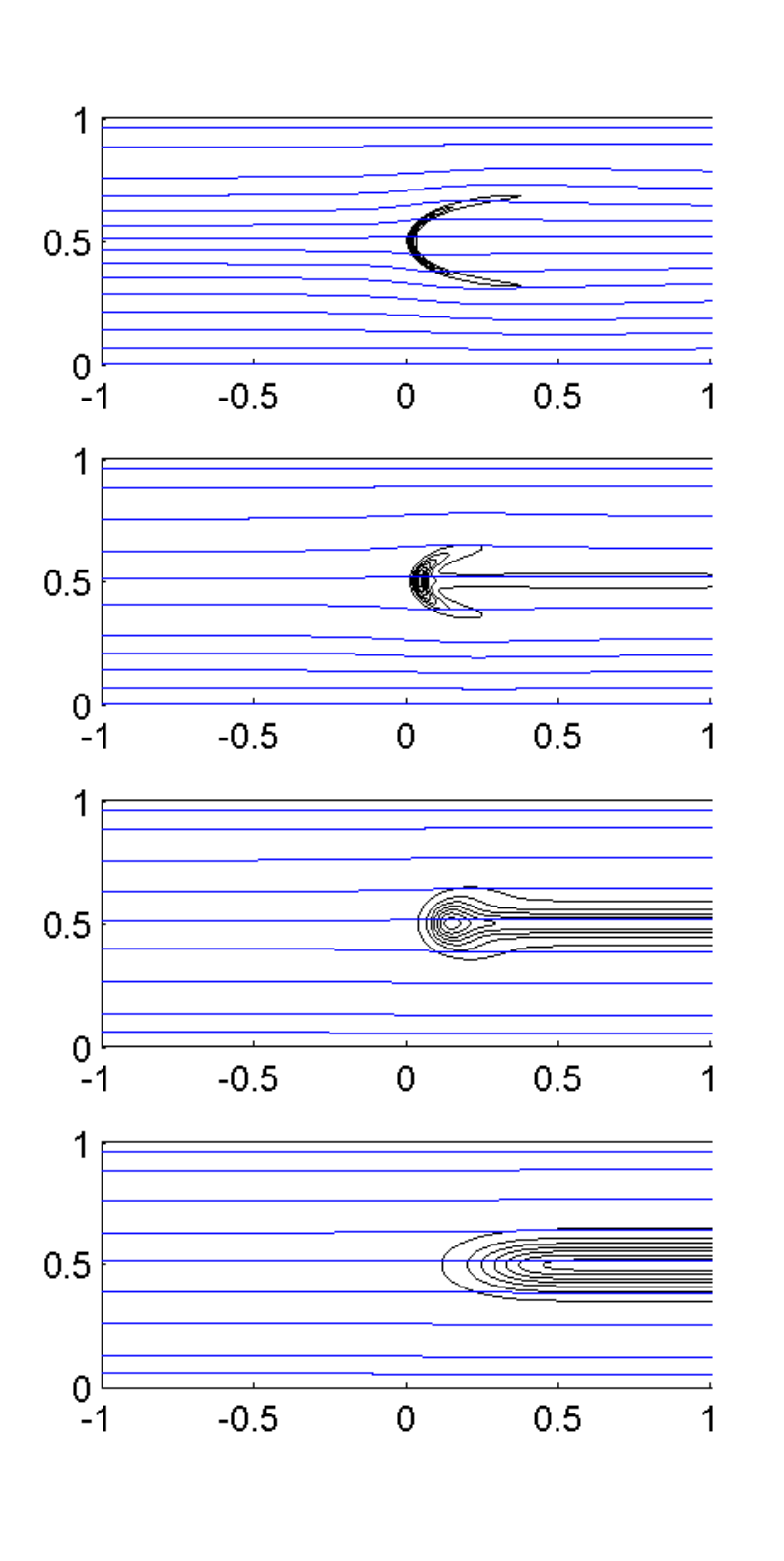}
\caption{Streamlines and reaction rate contours in a frame of reference attached to the flame-front for $\alpha=0.5$ and $\epsilon=0.015$, $\epsilon=0.05$, $\epsilon=0.13$ and $\epsilon=0.21$, respectively from top to bottom. The propagation speeds, when scaled by the numerically calculated propagation speed of a planar premixed flame, are given by $U=1.15$, $U=0.69$, $U=0.02$ and $U=-2.00$, respectively.}
\label{triple:fig:a_50}
\end{figure*}
\begin{figure*}[!]
\centering
\includegraphics[scale=1.24,trim=0 25 0 25,clip=true]{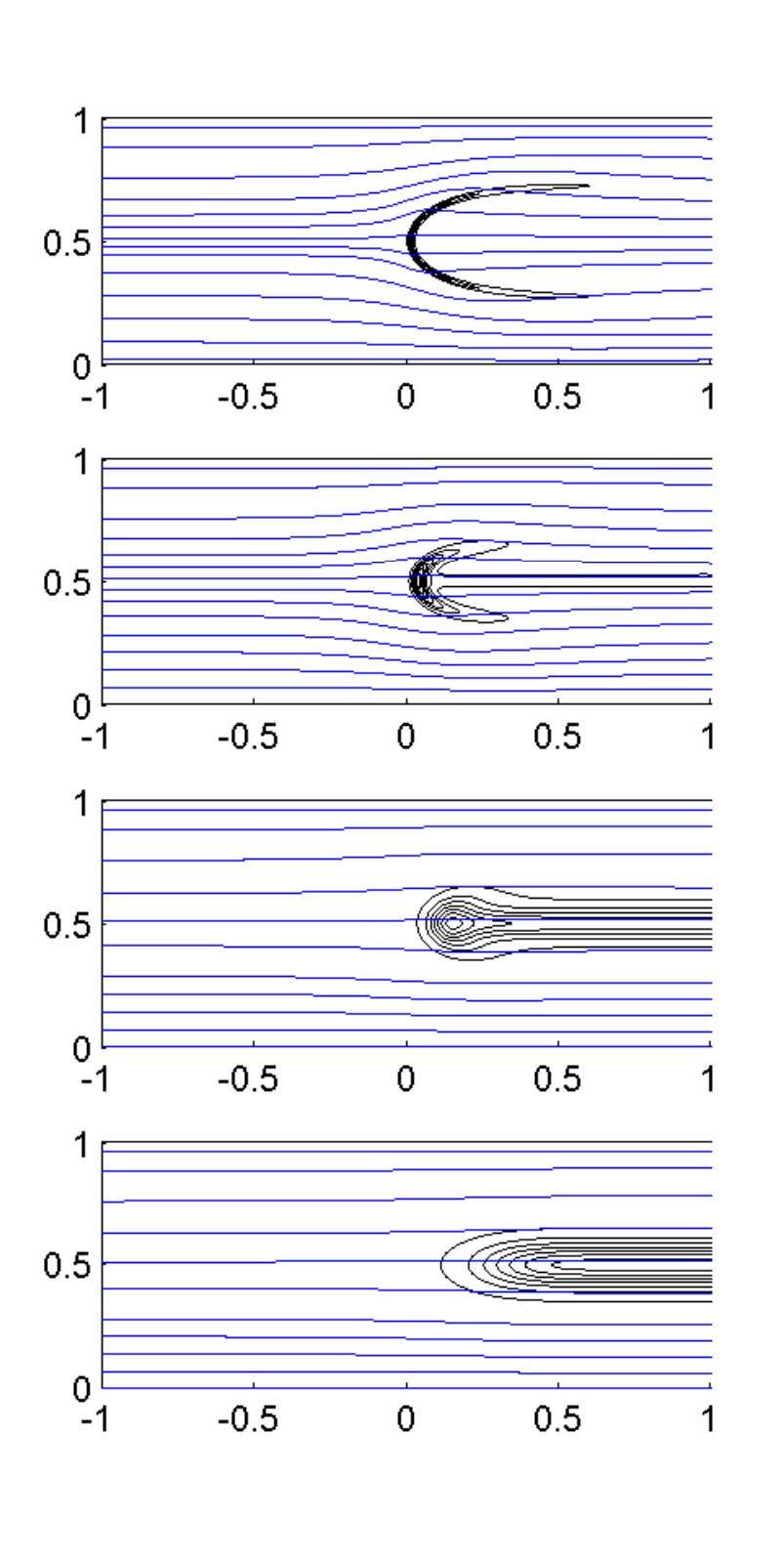}
\caption{Streamlines and reaction rate contours in a frame of reference attached to the flame-front for $\alpha=0.85$ and $\epsilon=0.015$, $\epsilon=0.05$, $\epsilon=0.14$ and $\epsilon=0.22$, respectively from top to bottom. The propagation speeds, when scaled by the numerically calculated propagation speed of a planar premixed flame, are given by $U=1.98$, $U=1.00$, $U=-0.01$ and $U=-2.26$, respectively.}
\label{triple:fig:a_85}
\end{figure*}
Figures \ref{triple:fig:a_0}--\ref{triple:fig:a_85} show the reaction rate contours and streamlines of the system for increasing values of $\epsilon$, with several fixed values of the thermal expansion coefficient $\alpha$. Examining, for example, figure \ref{triple:fig:a_85} shows the mechanism for the increase in triple flame speed above that of the planar premixed flame with thermal expansion as $\epsilon \to 0$, discussed in the previous section.

The figure shows that, for small $\epsilon$, ahead of the flame front the streamlines diverge. As explained in detail in \cite{5}, the divergence of the streamlines occurs because the fluid velocity normal to the flame front increases with thermal expansion, while the fluid velocity tangential to the flame front remains the same. The flow velocity vector is therefore bent towards the centreline (or stoichiometric surface) of the triple flame as it crosses the flame front. This must be accommodated by a divergence of the streamlines, which causes a drop in the horizontal component of the fluid velocity just ahead of the flame front. For small $\epsilon$, since the flame front is quasi-planar, the fluid velocity just ahead of the flame should be approximately equal to the propagation speed of the planar-premixed flame $S_L^0$. Thus, since the fluid velocity drops ahead of the flame front, the fluid velocity upstream of the flame must be above $S_L^0$.
\begin{figure*}
\includegraphics[scale=0.8,trim=27 10 0 0,clip=true]{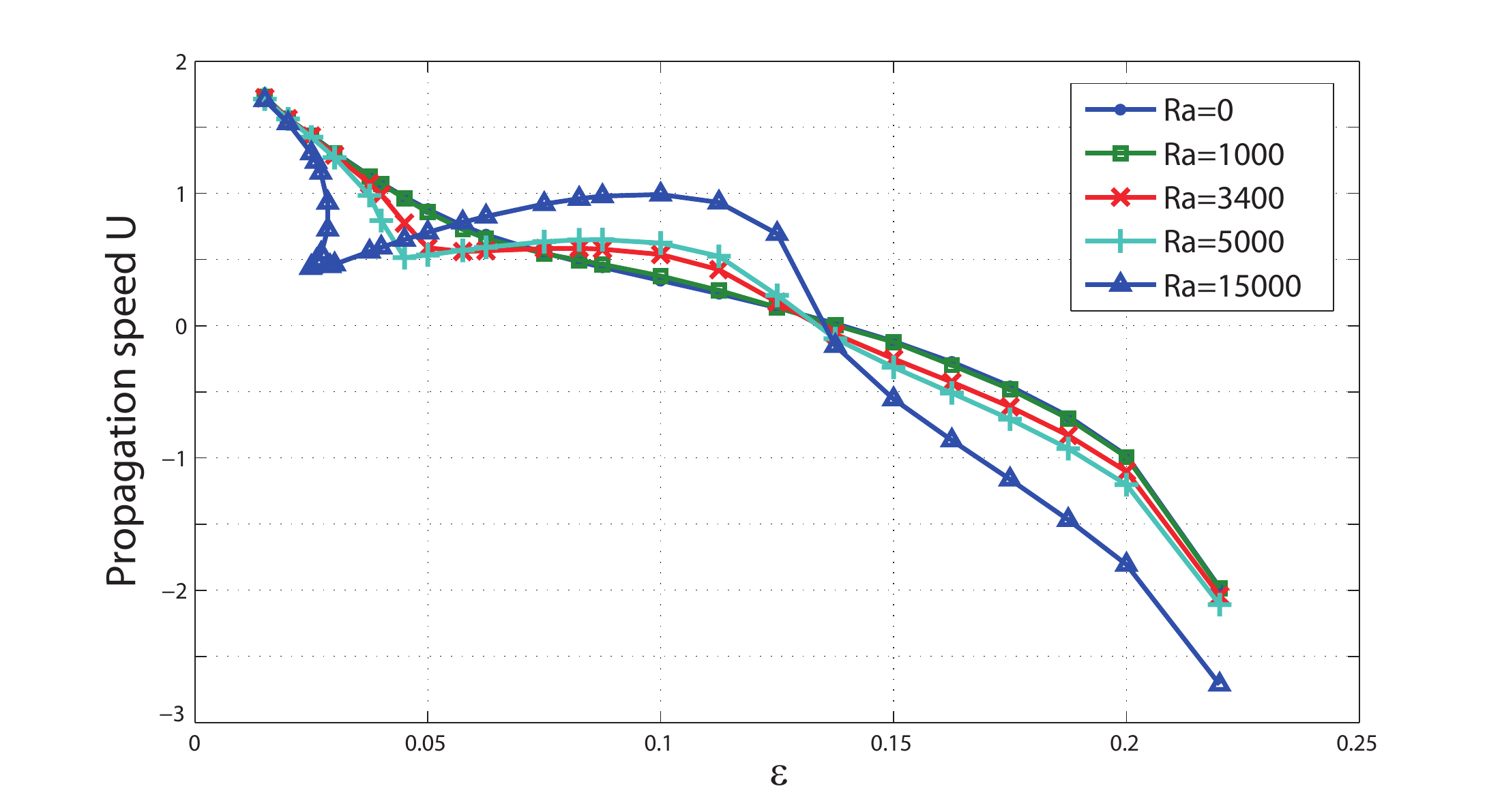}
\caption{The effect of $\epsilon$ on the propagation speed $U$ for selected values of the Rayleigh number with the values of the other parameters given by $\beta=10$, $Le_F=Le_O=1$, $S=1$, $Pr=1$ and $\alpha=0.85$.}
\label{triple:fig:grav_u}
\end{figure*}
\subsection{Effect of gravity on a triple flame}
\label{triple:sec:effectofgravity}
In this section we investigate the combined effect of thermal expansion and gravity on a triple flame. Throughout this section we let the thermal expansion coefficient take the typical value $\alpha=0.85$. Note that we only consider values in parameter space for which the underlying planar diffusion flame is stable, as shown in figure \ref{triple:fig:epsilonvsra}.  Since the aim of the study is to calculate the propagation speed $U$, we will begin with a plot of $U$ versus $\epsilon$ for several values of the Rayleigh number $Ra$. We will then plot graphs of $U$ versus $Ra$ for selected values of $\epsilon$ to further capture the complex relationships that are displayed between the physical parameters and the propagation speed. We will end with a comparison of how the streamlines and reaction rate contours change with increasing $\epsilon$ for several fixed values of $Ra$.
\begin{figure*}[!]
\subfigure{
\includegraphics[width=0.49\textwidth]{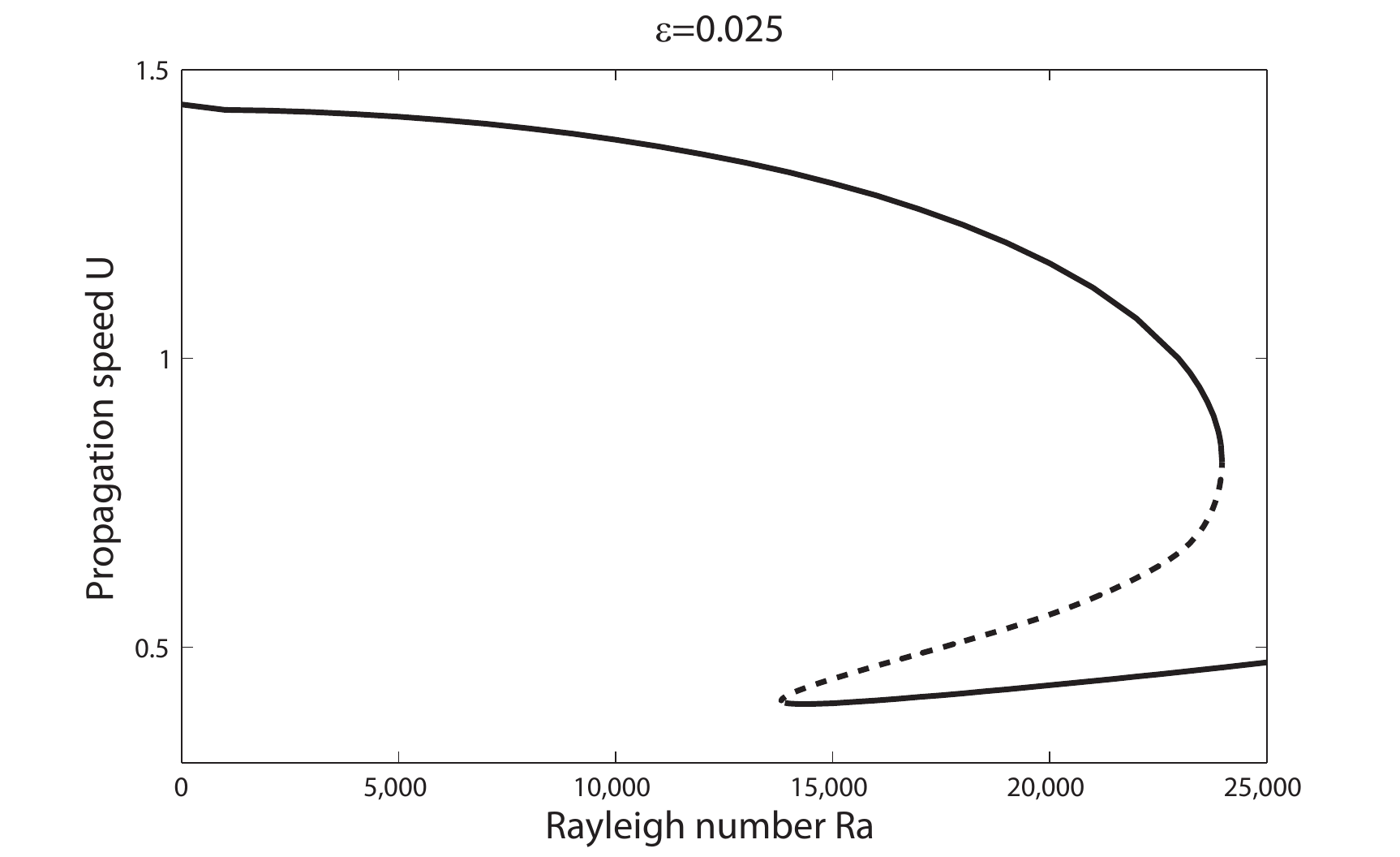} \label{triple:fig:u_ra_comp_a}
}
\subfigure{
\includegraphics[width=0.49\textwidth]{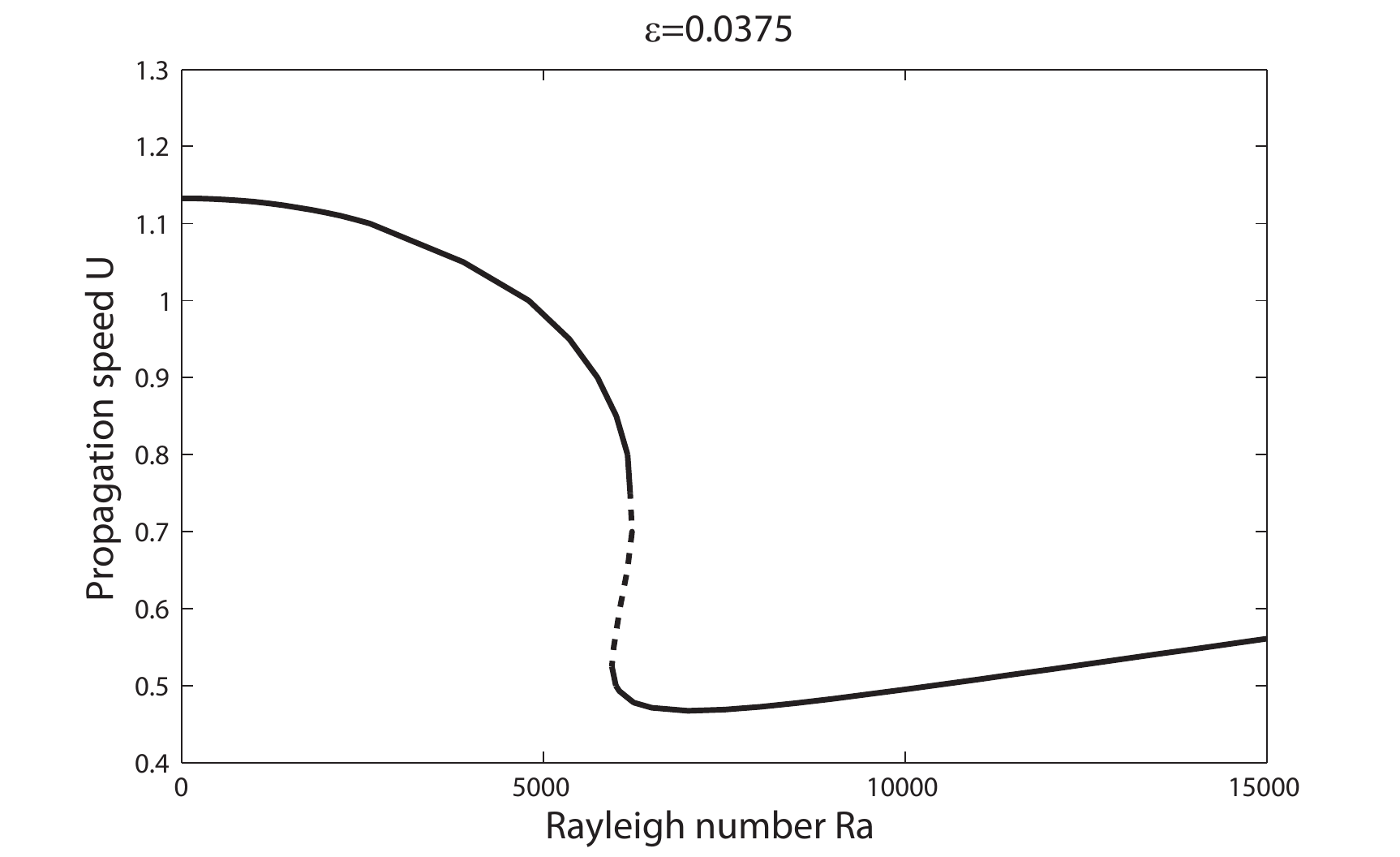} \label{triple:fig:u_ra_comp_b}
}
\subfigure{
\includegraphics[width=0.49\textwidth]{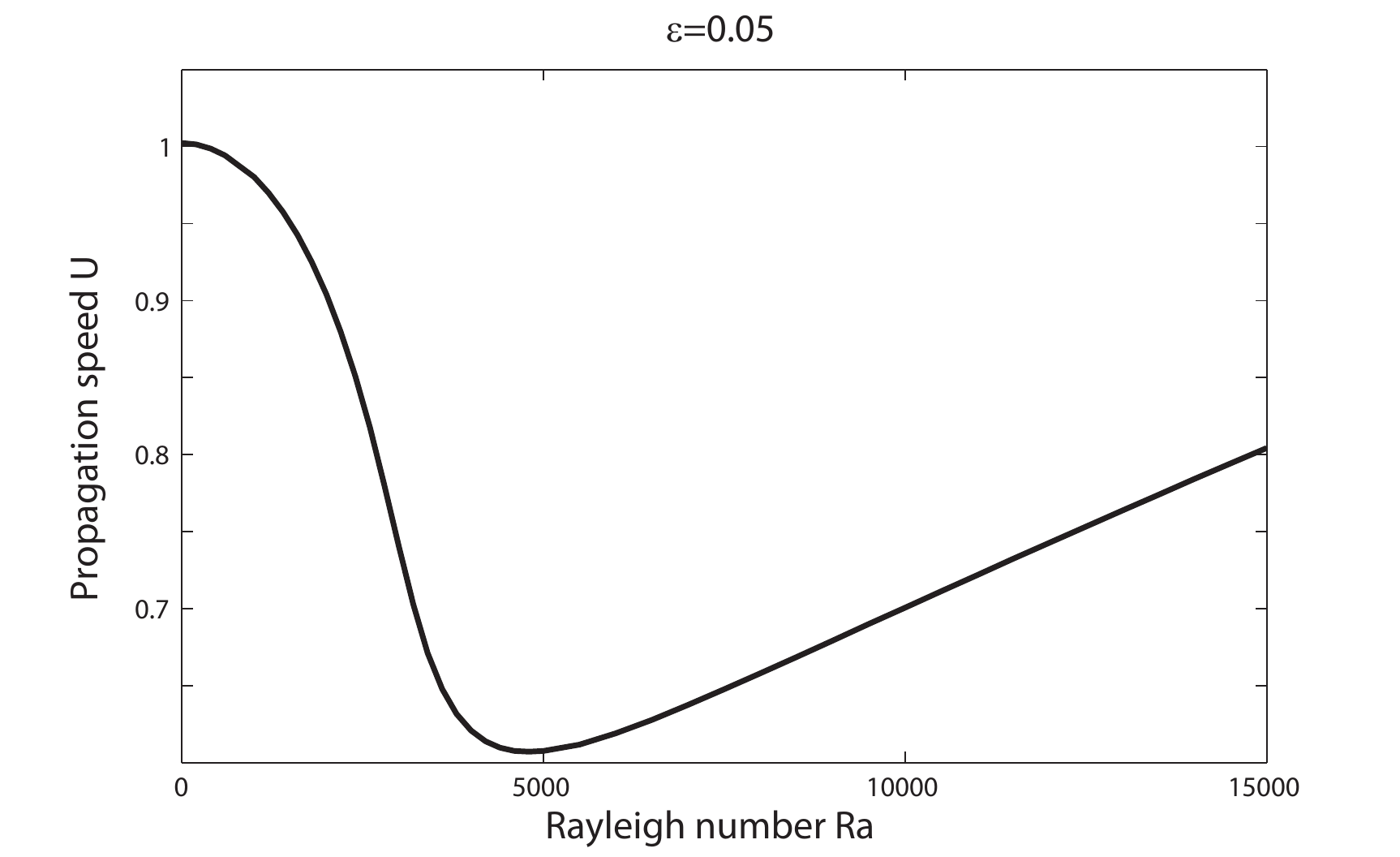} \label{triple:fig:u_ra_comp_c}
}
\subfigure{
\includegraphics[width=0.49\textwidth]{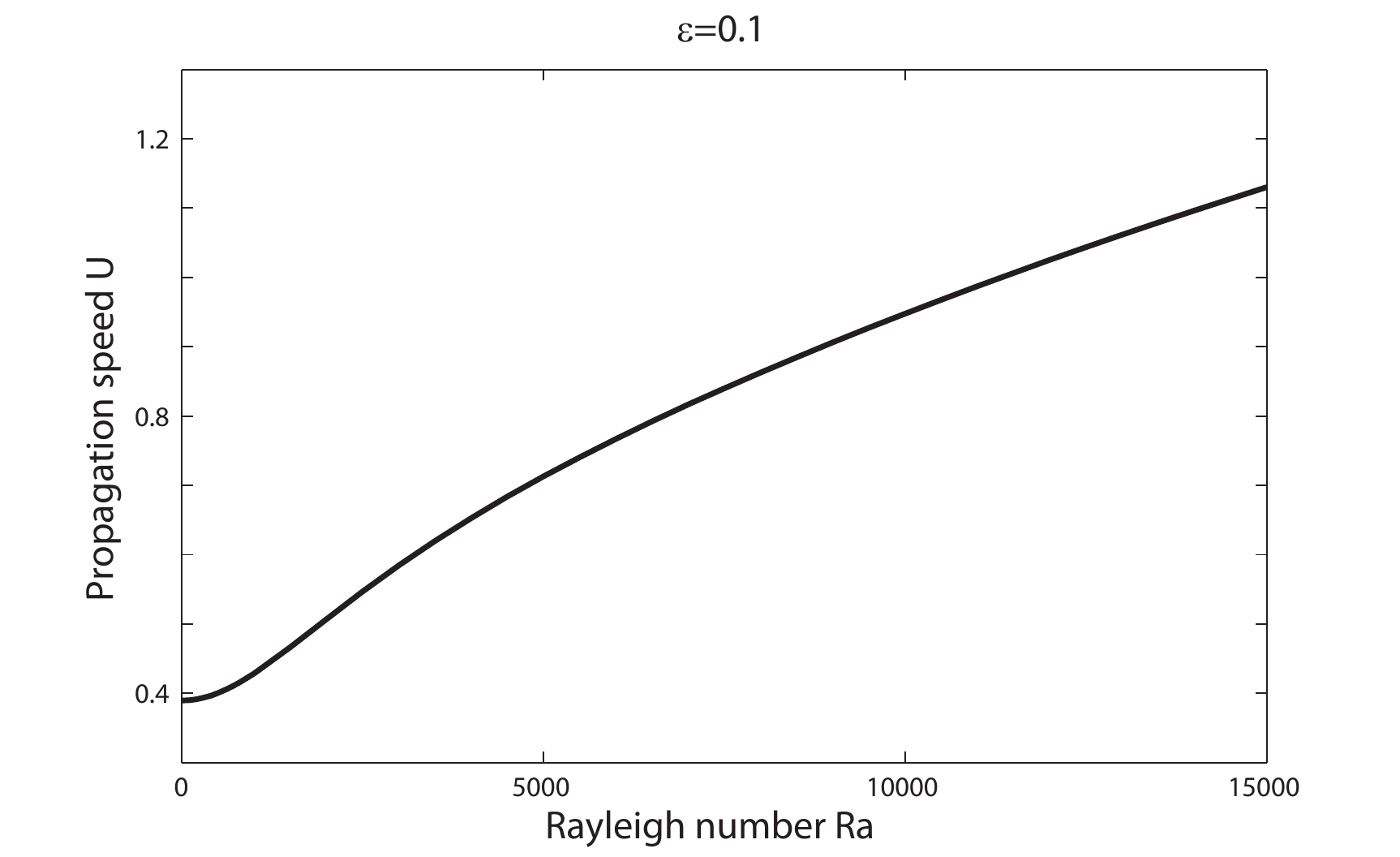} \label{triple:fig:u_ra_comp_d}
}
\subfigure{
\includegraphics[width=0.49\textwidth]{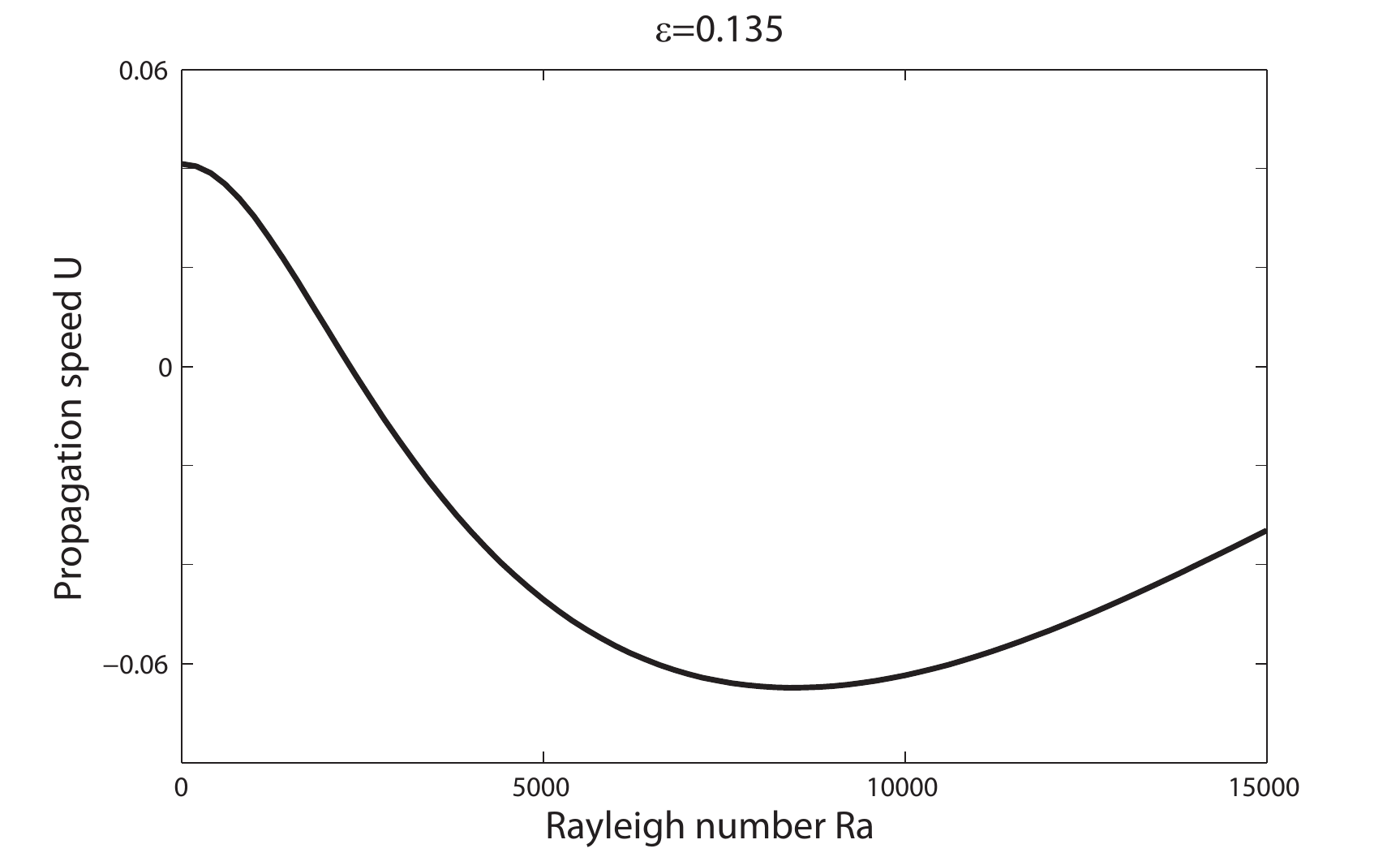} \label{triple:fig:u_ra_comp_e}
}
\subfigure{
\includegraphics[width=0.49\textwidth]{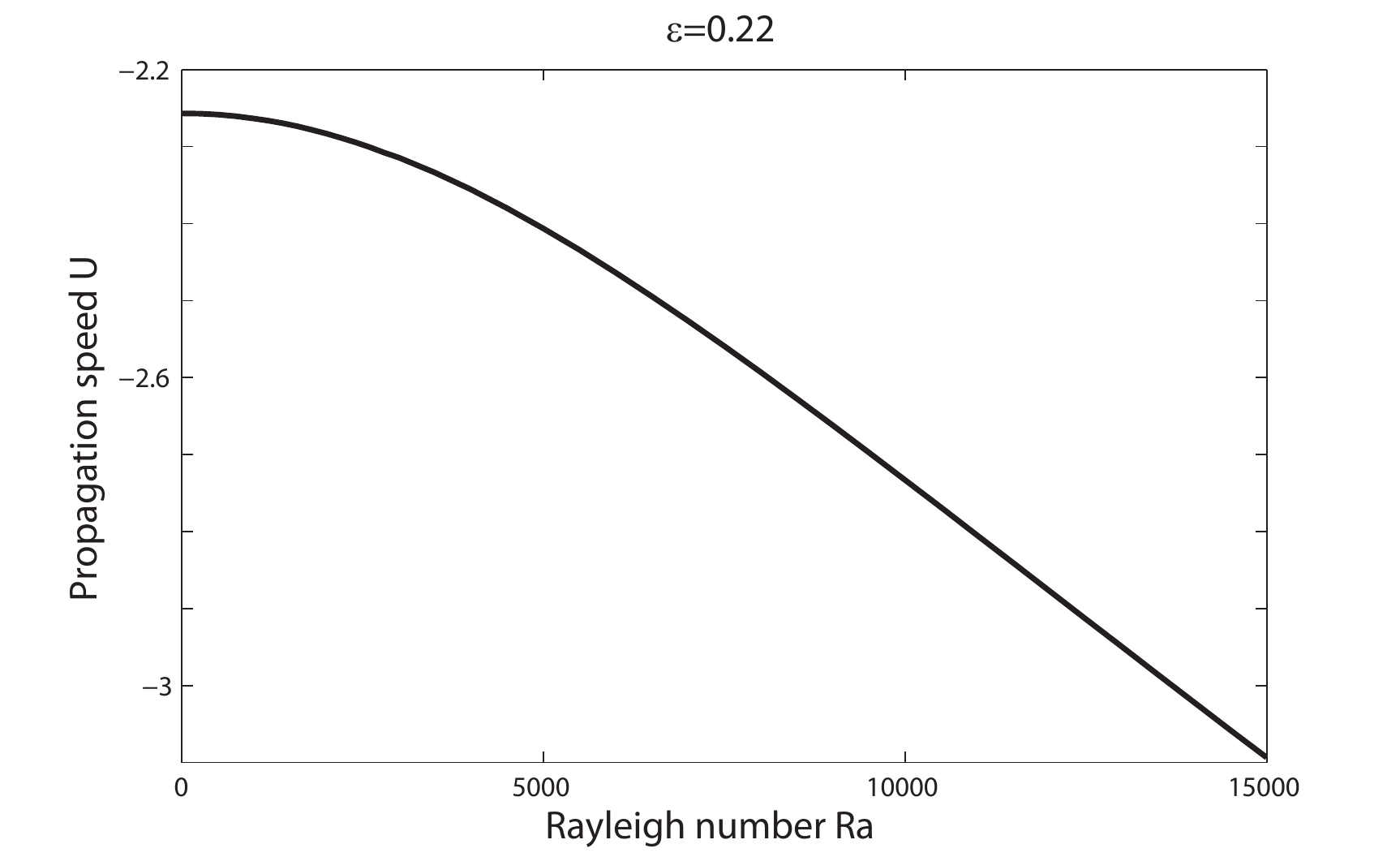} \label{triple:fig:u_ra_comp_f}
}
\caption{Comparison of the relationship between propagation speed $U$ and the Rayleigh number $Ra$ for several fixed values of $\epsilon$. The other parameters take fixed values given by $\beta=10$, $Le_F=Le_O=1$, $S=1$, $Pr=1$ and $\alpha=0.85$. A dashed line indicates that the steady solutions have been found to be unstable in time-dependent simulations; all other solutions have been found to be stable.}
\label{triple:fig:u_ra_comp}
\end{figure*}
\subsubsection{Propagation speed of a triple flame}
Figure \ref{triple:fig:grav_u} shows a plot of the propagation speed $U$ of the triple flame versus $\epsilon$, for selected values of the Rayleigh number $Ra$. The figure shows, firstly, that in the limit $\epsilon \to 0$ the Rayleigh number has very little effect on the propagation speed of a triple flame. This could be easily deduced by considering equation (\ref{triple:nondim3}) and noting that, as $\epsilon \to 0$, the buoyancy term does not enter the problem at $O\left(1\right)$ unless $Ra=O\left(\epsilon^{-2}\right)$. Secondly, the figure shows that there is a critical Rayleigh number, calculated as approximately $Ra=3400$, above which the graph of $U$ versus $\epsilon$ ceases to be monotonic. Finally, it shows the complex behaviour of the system for even higher values of $Ra$. It is found that there can exist three different steady solutions for some low values of $\epsilon$ (i.e. the system exhibits hysteresis). It is also found that at $\epsilon \approx 0.1$, there is a local maximum in the graph of $U$ versus $\epsilon$. As $\epsilon$ approaches its extinction value $\epsilon_{\text{ext}}$, $U$ is found to fall to negative values as in the case without gravity.

Some of the complex behaviour of the system can be captured by maintaining $\epsilon$ fixed and varying $Ra$. Graphs of $U$ versus $Ra$ for selected values of $\epsilon$ are plotted in figure \ref{triple:fig:u_ra_comp}. In figures \ref{triple:fig:u_ra_comp_a}--\ref{triple:fig:u_ra_comp_b}, in which $\epsilon$ takes a low value, the hysteresis displayed by the system can be clearly seen, whereby for certain values of $Ra$ there are three solutions for $U$. The middle branches of these hysteresis curves have been found to consist of unstable solutions.  For a slightly higher value of $\epsilon$ there is no longer found to be a multiplicity of solutions; the propagation speed has a local minimum at a certain value of $Ra$ before increasing as $Ra$ reaches higher values, as shown in figure \ref{triple:fig:u_ra_comp_c}. In figure \ref{triple:fig:u_ra_comp_d}, in which $\epsilon=0.1$, $U$ is seen to increase monotonically with $Ra$. Figure \ref{triple:fig:u_ra_comp_e} shows that if $U \approx 0$ for a triple flame without gravity, the propagation speed remains near zero as $Ra$ increases. Finally, figure \ref{triple:fig:u_ra_comp_f} shows that when $\epsilon$ is near its extinction value (with the propagation speed being negative), $U$ decreases monotonically with $Ra$.

For sufficiently small $\epsilon$, as in figures \ref{triple:fig:u_ra_comp_a}--\ref{triple:fig:u_ra_comp_c}, it can be seen that $U$ decreases with increasing $Ra$ for not too large $Ra$. This can be explained by considering the physical behaviour of the system, which we proceed to investigate next.
\begin{figure*}[!]
\centering
\includegraphics[scale=1.25,trim=0 25 0 25,clip=true]{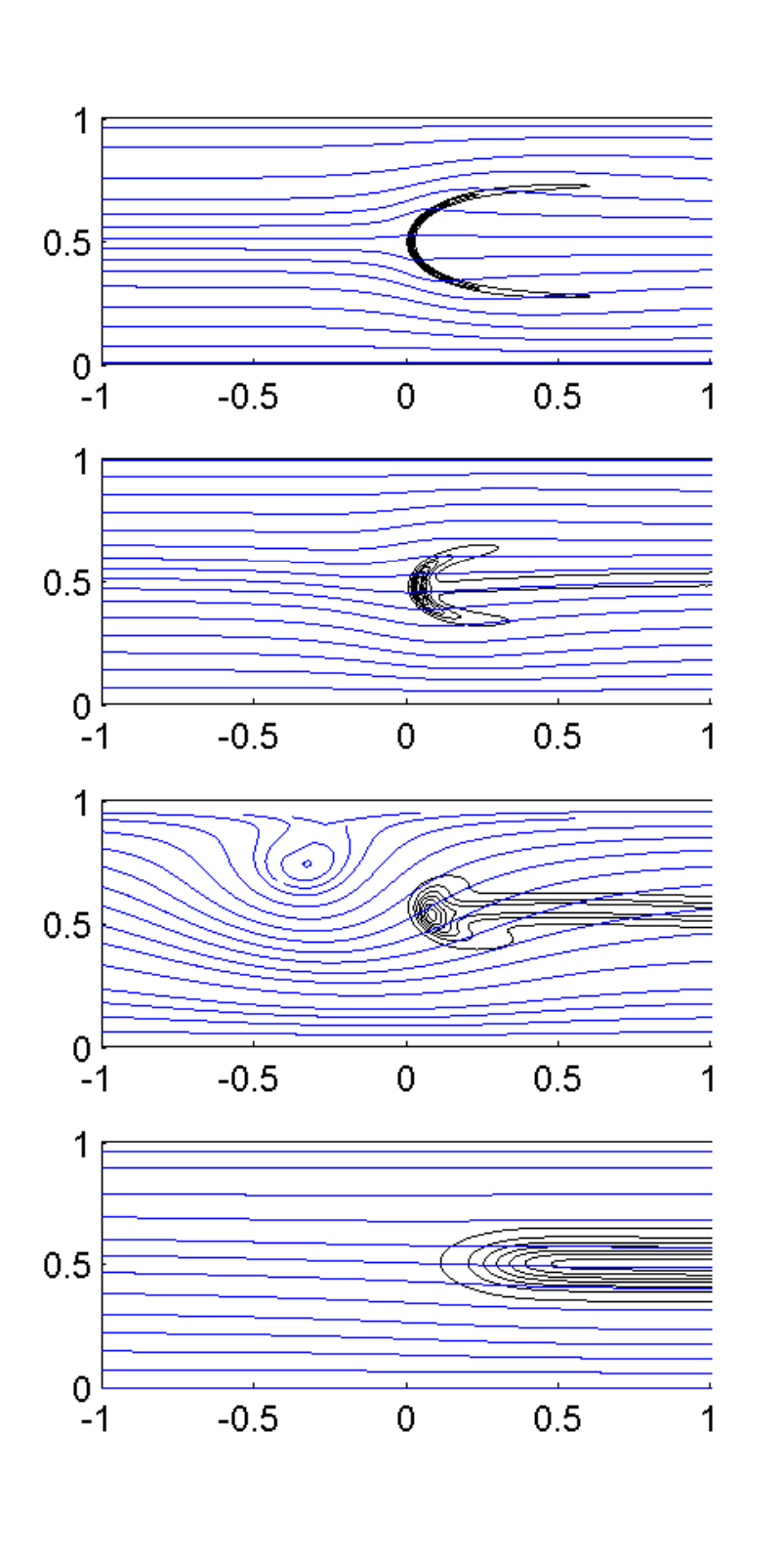}
\caption{Streamlines and reaction rate contours in a frame of reference attached to the flame-front for $Ra=1000$ and $\epsilon=0.015$, $\epsilon=0.05$, $\epsilon=0.1$ and $\epsilon=0.22$, respectively from top to bottom. The propagation speeds are given by $U=1.73$, $U=0.86$, $U=0.38$ and $U=-1.98$, respectively.}
\label{triple:fig:ra_1000}
\end{figure*}
\begin{figure*}[!]
\centering
\includegraphics[scale=1.25,trim=0 25 0 25,clip=true]{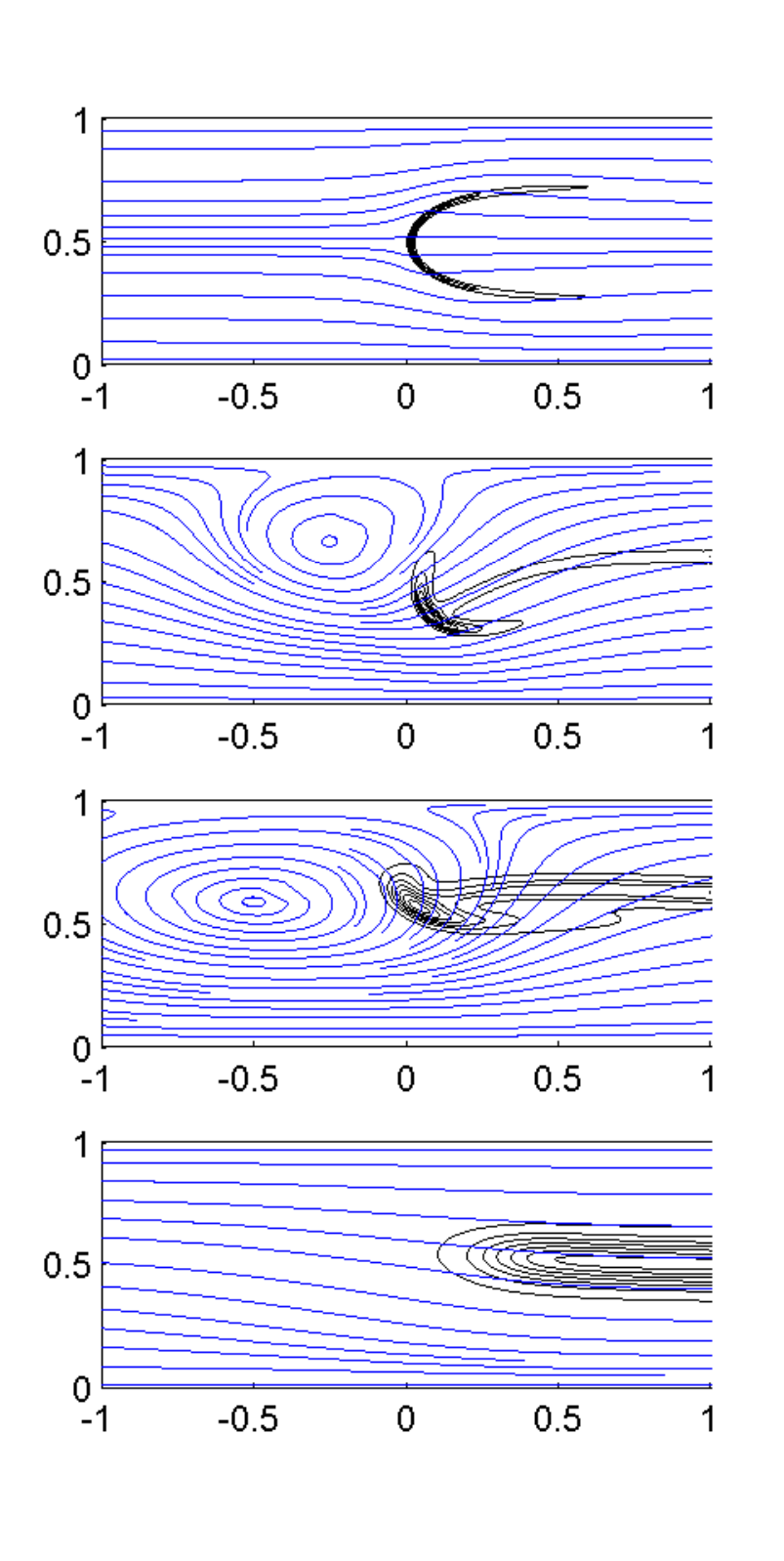}
\caption{Streamlines and reaction rate contours in a frame of reference attached to the flame-front for $Ra=5000$ and $\epsilon=0.015$, $\epsilon=0.05$, $\epsilon=0.1$ and $\epsilon=0.22$, respectively from top to bottom. The propagation speeds are given by $U=1.71$, $U=0.53$, $U=0.62$ and $U=-2.11$, respectively.}
\label{triple:fig:ra_5000}
\end{figure*}
\begin{figure*}[!]
\centering
\includegraphics[scale=1.25,trim=0 25 0 25,clip=true]{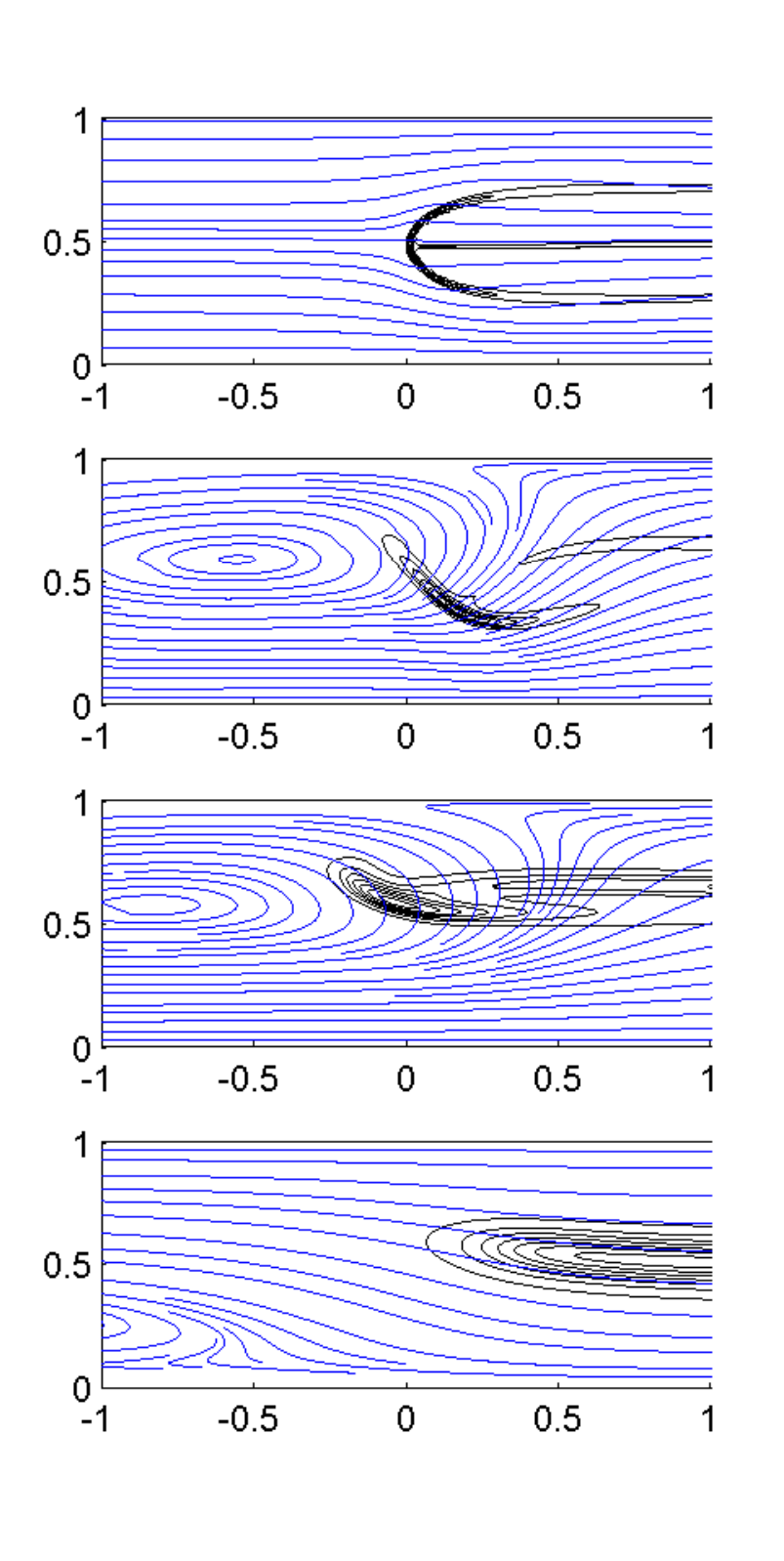}
\caption{Streamlines and reaction rate contours in a frame of reference attached to the flame-front for $Ra=15000$ and $\epsilon=0.015$, $\epsilon=0.05$, $\epsilon=0.1$ and $\epsilon=0.22$, respectively from top to bottom. The propagation speeds are given by $U=1.71$, $U=0.71$, $U=0.99$ and $U=-2.71$, respectively.}
\label{triple:fig:ra_15000}
\end{figure*}
\subsubsection{Comparative cases for fixed $Ra$}
Figures \ref{triple:fig:ra_1000}--\ref{triple:fig:ra_15000} show the reaction rate contours and streamlines of the system with increasing values of $\epsilon$, for several fixed values of $Ra$.\footnote{It is important to note that the solutions plotted for $\epsilon=0.015$ in figures \ref{triple:fig:ra_1000}--\ref{triple:fig:ra_15000} lie on the upper branch of the hysteresis curves described in \S \ref{triple:sec:propagation}, where applicable. All of the steady solutions plotted in figures \ref{triple:fig:ra_1000}--\ref{triple:fig:ra_15000} have been found to be stable by running time-dependent simulations.} From these diagrams it can be seen that buoyancy forces cause the formation of vortices upstream of the flame (or downstream of a negatively propagating triple flame). For higher values of $Ra$, the vortex formed is found to be larger in its size and the strength of its flow.

These vortices can be explained as being caused by the temperature gradient from cold to hot along the positive $x$-direction in the channel. It has been found that in a channel or pipe in the absence of a flame, differentially heated end walls cause the fluid in the channel or pipe to flow from hot to cold along the top of the domain, and from cold to hot along the bottom of the domain \cite{bejan1978fully}. The flow is explained in \cite{bejan1978fully} as being due to buoyancy forces caused by the change in density with temperature. A similar mechanism can explain the vortices caused by a triple flame in a channel and thus the reduction in the propagation speed of the triple flame for small values of $\epsilon$, when the Rayleigh number is increased above zero.
\begin{figure*}[!]
\centering
\includegraphics[scale=1,trim=0 50 0 0,clip=true]{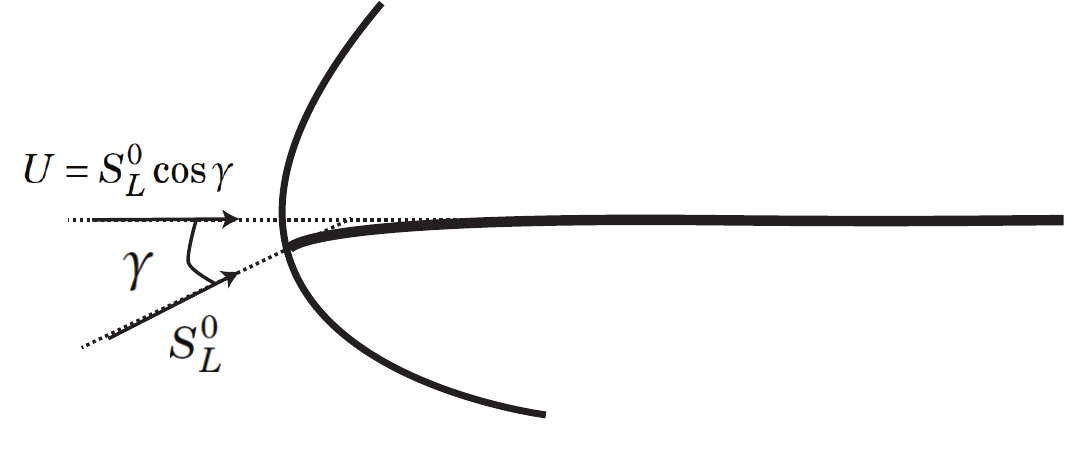}
\caption{An illustration of a triple flame for $\epsilon \to 0$ under small gravitational effects. In a frame of reference attached to the triple flame, the fluid flows across the (quasi-planar) flame front at an angle $\gamma$ to the horizontal at the planar premixed flame speed $S_L^0$; the fluid velocity along the centreline is therefore smaller than the planar premixed flame speed.}
\label{triple:fig:diagram_gravity}
\end{figure*}
The flow of fluid from hot to cold at the top of the domain and cold to hot at the bottom causes a downward flow in front of a positively propagating triple flame, bending the stoichiometric isosurface ahead of the flame downwards. Effectively this reduces the component of the propagation velocity in the horizontal direction.

More precisely, for small values of $\epsilon$ (for which the flame front is quasi-planar), the fluid velocity perpendicular to the flame front across the flame is approximately given by the planar premixed flame speed $S_L^0$. Thus, when the stoichiometric isosurface ahead of the flame is bent to an angle $\gamma$ to the horizontal by the downward flow ahead of the flame, the propagation speed can be expected to be approximately $U=S_L^0 \cos{\gamma}$, as shown in figure \ref{triple:fig:diagram_gravity}. Therefore for small $\epsilon$, an increase in the Rayleigh number (which causes the vortices ahead of the flame to be stronger, and hence bends the stoichiometric isosurface further downwards and increases $\gamma$) leads to a decrease in the propagation speed $U$, in the absence of other effects.

This physical behaviour is best illustrated by considering the triple flame under the influence of gravity but in the absence of thermal expansion, so that the acceleration of the flow field as it crosses the triple flame does not mask the effects of buoyancy. To do this, we must make use of the Boussinesq approximation, and expand the ideal gas equation of density, given by equation (\ref{triple:nondim7}), as 
\begin{gather}
\rho=1-\alpha\theta+O\left(\alpha^2\right).
\end{gather}
This gives $\rho=1$ to leading order in every term in equations (\ref{triple:nondim1})-(\ref{triple:nondim6}), except the buoyancy term in equation (\ref{triple:nondim3}), which becomes $\epsilon^2 Pr Ra \theta$.

\begin{figure*}[!]
\centering
\includegraphics[scale=0.85,trim=300 0 0 20,clip=true]{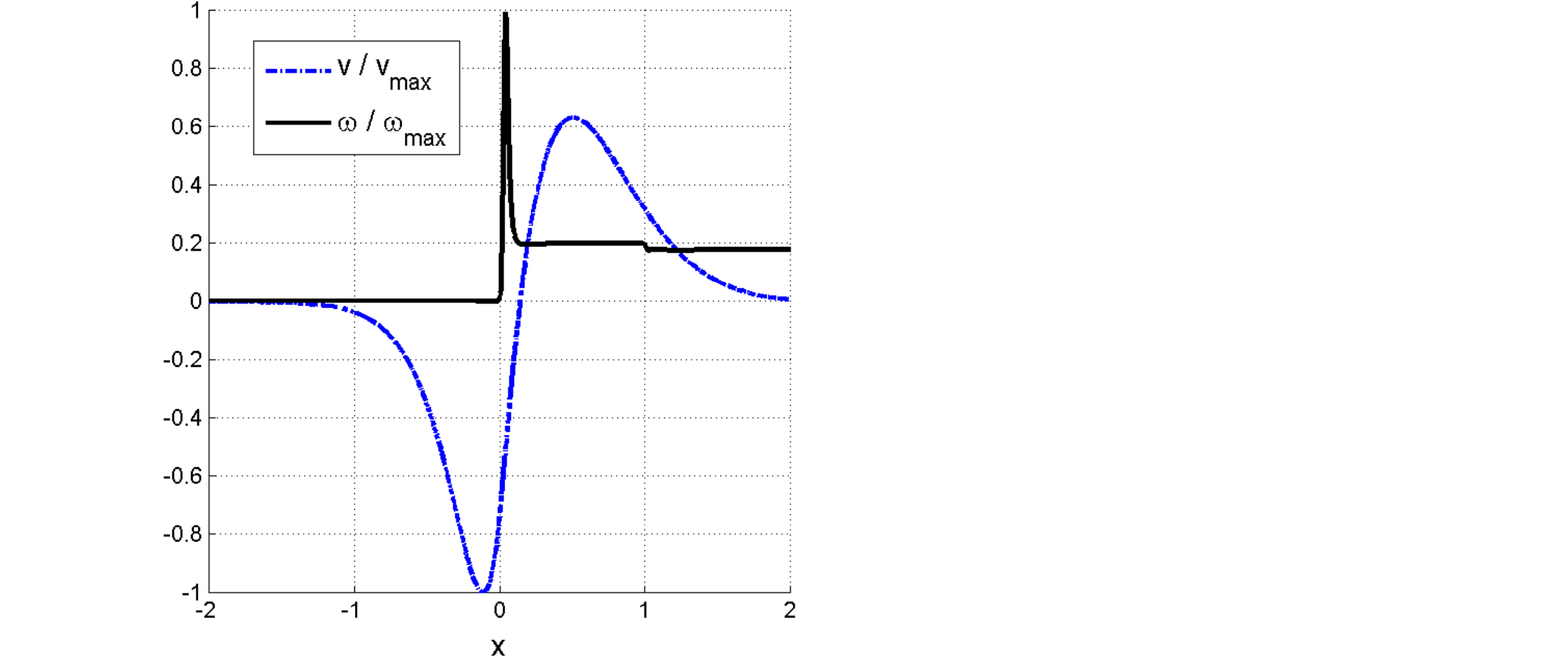}
\caption{The vertical component of the velocity vector and the reaction rate along the centreline of the triple flame located at $y=\frac{1}{2}$, both scaled by their maximum values, in the Boussinesq approximation ($\alpha\to 0$) for $\epsilon=0.05$ and $Ra=50$. The vertical velocity component is clearly negative just ahead of the flame-front as expected.}
\label{triple:fig:verticalvelocity}
\end{figure*}

Now we numerically solve equations (\ref{triple:nondim1})-(\ref{triple:nondim6}) with boundary conditions (\ref{triple:bc1})-(\ref{triple:bc4}) in the Boussinesq approximation, for a small value of the Rayleigh number, as $\epsilon \to 0$. Plotted in figure \ref{triple:fig:verticalvelocity} is the vertical component of the velocity along the centreline of the triple flame located at $y=\frac{1}{2}$. It can be seen that there is a downwards flow upstream of the point where the reaction rate reaches a maximum, and an upwards flow downstream of it. This clearly illustrates the vortex formed by the triple flame and the downwards flow ahead of the flame front.
\section{Conclusion}
\label{triple:sec:conc}
In this study, the effect of thermal expansion and gravity on a triple flame propagating in a horizontal channel where the fuel and oxidiser concentrations are prescribed at the walls has been investigated. This seems to be the first investigation dedicated to triple flame propagation in a direction perpendicular to gravity. The problem has been formulated in the low Mach number approximation and solved numerically. The effect of the flame-front thickness $\epsilon$ on the propagation speed $U$ has been described for several values of the thermal expansion coefficient $\alpha$ and the Rayleigh number $Ra$.

It has been found that the well-known monotonic relationship between $U$ and $\epsilon$ that is present in the constant density case (which arises in the limit $\alpha \to 0$) remains valid for $\alpha \neq 0$, when $Ra=0$ (i.e. in the absence of gravity). In fact, the influence of $\alpha$ on the triple flame for $Ra=0$ is found to agree with the conclusions of the study \cite{5}, where the physical mechanism for the increase in propagation speed has been explained.

Under the influence of gravity we have shown that the monotonic relationship between $U$ and $\epsilon$ is only present for values of $Ra$ below a critical value which has been determined. Further, it has been shown that, if $Ra$ takes a value higher than this critical value, there is a local maximum in the graph of $U$ versus $\epsilon$, as can be seen in figure \ref{triple:fig:grav_u}. The system has been shown to exhibit hysteresis for even higher values of the Rayleigh number.

The complex relationship between $U$ and $Ra$ has been further investigated by fixing $\epsilon$ and varying $Ra$. It has been found that the graph of $U$ versus $Ra$ (see figure \ref{triple:fig:u_ra_comp}) depends strongly on the value of $\epsilon$ chosen. Time-dependent simulations have shown that all of the steady solutions presented are stable, except for solutions on the middle branch of the hysteresis curves presented in figure \ref{triple:fig:u_ra_comp}. Finally, a physical argument has been provided, which explains the decrease of $U$ with increasing gravity for small values of $\epsilon$.

The results of this study provide valuable insight into the behaviour of a triple flame under gravitational effects and illustrate the complexity and variety of the scenarios that arise. A further aspect of the problem that has yet to be studied is the transient behaviour of a triple flame that is unstable due to gravitational effects. This will be addressed in Chapter \ref{chapter:ignition}, which involves the numerical solution of the time-dependent equations for values of $Ra$ higher than the critical Rayleigh number for the instability of a planar diffusion flame.

\chapter{Initiation and instability of triple flames subject to thermal expansion and gravity}
\chaptermark{Initiation and instability of triple flames}
\label{chapter:ignition}
\section{Introduction}
Understanding the transient behaviour of a flame from initiation to steady propagation, and in some cases instability, is a vital part of fundamental combustion research. In this chapter we study the problem of triple flame initiation in a mixing layer, taking the combined effect of thermal expansion and gravity into account. As well as investigating the effect of these phenomena on the energy required for initiation of triple flames from a two-dimensional ignition kernel, we study the transient behaviour of a triple flame with a trailing diffusion flame that is either stable or unstable due to gravitational effects.

The problem of ignition in homogeneous mixtures, which leads to the propagation of premixed flames, has been the focus of a large amount of research. In the particular case of spherically expanding premixed flames, a criterion for the energy required for ignition is provided by the thermal energy of a non-propagating, spherically symmetric solution of the governing equations known as a Zeldovich flame ball \cite{zeldovich1985mathematical}.  Flame balls have been found to be typically unstable under adiabatic conditions, and small perturbations can lead to either an outwardly propagating flame or an inwardly propagating flame and eventual extinction \cite{zeldovich1985mathematical,deshaies1984initiation,joulin1985point,champion1986spherical}. Further studies in the literature have investigated the transient behaviour of spherical premixed flames from an initial ignition kernel, depending on aspects such as heat loss and Lewis numbers (see e.g. \cite{tse2000computational,he2000critical,chen2007theoretical,chen2011critical}).

There has been considerably less focus on flame ignition in situations where the reactants are non-premixed. The work that has been done on both laminar and turbulent non-premixed ignition is summarised in the detailed review paper \cite{mastorakos2009ignition}. Most studies on non-premixed flame ignition have been concerned with autoignition, sometimes referred to as ``self-ignition". There are very few papers that have investigated the energy required for ``forced ignition" of non-premixed flames by an external heat source or spark. The transient behaviour of flames in inhomogeneous mixtures from forced ignition has been investigated using Direct Numerical Simulations (DNS) in the laminar case in \cite{im1999structure,ray2001ignition}, and more recently with the effects of turbulence included in \cite{chakraborty2007effects,chakraborty2008direct,chakraborty2010effects,patel2014localised}. These numerical studies do not, however, contain a detailed investigation of the energy required for ignition. To our knowledge there have been no dedicated investigations of the energy required for forced ignition of laminar two-dimensional triple flames in mixing layers, even without taking thermal expansion and gravity into account.

Some recent papers have extended the concept of Zeldovich flame balls to the inhomogeneous case, theoretically \cite{daou2014flame} and numerically \cite{daou2015flame} describing the existence and properties of flame balls in reactive mixing layers. Similarly to Zeldovich flame balls, these inhomogeneous flame balls may provide a corresponding criterion for the minimum energy for the successful flame ignition of axisymmetric flames in the mixing layer.  Here we extend the current understanding of flame ignition by providing a criterion for the minimum ignition energy for two-dimensional triple flames in the mixing layer, via a two-dimensional ignition kernel. This is achieved by first investigating steady, two-dimensional, non-propagating solutions of the governing equations which we refer to as ``flame tube" solutions. Such solutions have been observed in previous numerical simulations \cite{daou1999ignition,short2001edge,buckmaster2002edge,thatcher2002oscillatory} where the planar solutions are prone to cellular instabilities due to Lewis number effects \cite{short2001edge}, but these studies were not concerned with ignition. More relevant to the ignition problem are the papers \cite{buckmaster2000holes} and \cite{lu2004flame}. These studies include investigations of ``flame isolas" and ``flame disks", respectively, which are similar axisymmetric, stationary solutions of the governing equations. Although \cite{buckmaster2000holes} and \cite{lu2004flame} only partially address the ignition problem, in the paper \cite{lu2004flame}, non-propagating ``flame disk" solutions are argued to indicate that a minimum energy is required for ignition of axisymmetric flames in the mixing layer. To our understanding, no such study has yet been performed for two-dimensional triple flames in the mixing layer, as investigated in this chapter, with the effects of thermal expansion and gravity included.

In this chapter we also investigate the transient evolution of the propagating triple flames that arise when a perturbation is added to the unstable flame tube solutions. Steadily propagating triple flames are well studied. Aspects of these structures that have been investigated include preferential diffusion \cite{buckmaster1989anomalous,prefdif}, heat losses \cite{heatloss1,heatloss2,heatloss3}, reversibility of the chemical reaction \cite{reversability,daou2009asymptotic}, the presence of a parallel flow \cite{daou2010triple} and thermal expansion \cite{5,pearce2013effect}; for further references see the review papers \cite{buckmaster2002edge} and \cite{chung2007stabilization}. Here we are specifically interested in the combined effect of thermal expansion and gravity on both the ignition energy of triple flames from two-dimensional ignition kernels and the transient behaviour of the triple flames.

When investigating triple flames propagating perpendicular to the direction of gravity it is crucial to first understand whether or not the planar diffusion flame is stable, since this forms the trailing branch of the triple flame. In Chapter \ref{chapter:diffusion} we provided such a study; we have also investigated, in Chapter \ref{chapter:triple}, the combined effect of thermal expansion and gravity on steadily propagating triple flames (see also the papers by \citet{pearce2013effect, pearce2013rayleigh}). The current chapter completes the picture by investigating the transient behaviour of triple flames from their initiation in contexts where the underlying planar diffusion flame is either stable or unstable.

The chapter is structured as follows. In \S \ref{ignition:sec:form} we formulate the problem. In \S \ref{ignition:sec:res} we provide the results obtained from the numerical solution of the governing equations. \S \ref{ignition:sec:res} is split into three parts: the first part consists of important preliminary results; the second part provides the results of numerical simulations for values of the parameters where the underlying diffusion flame is stable; and the third part is concerned with areas of parameter space where the underlying diffusion flame is unstable. Finally, we end the chapter with conclusions in \S \ref{ignition:sec:conc}.

\section{Formulation}
\label{ignition:sec:form}
\begin{figure}
\includegraphics[scale=0.8]{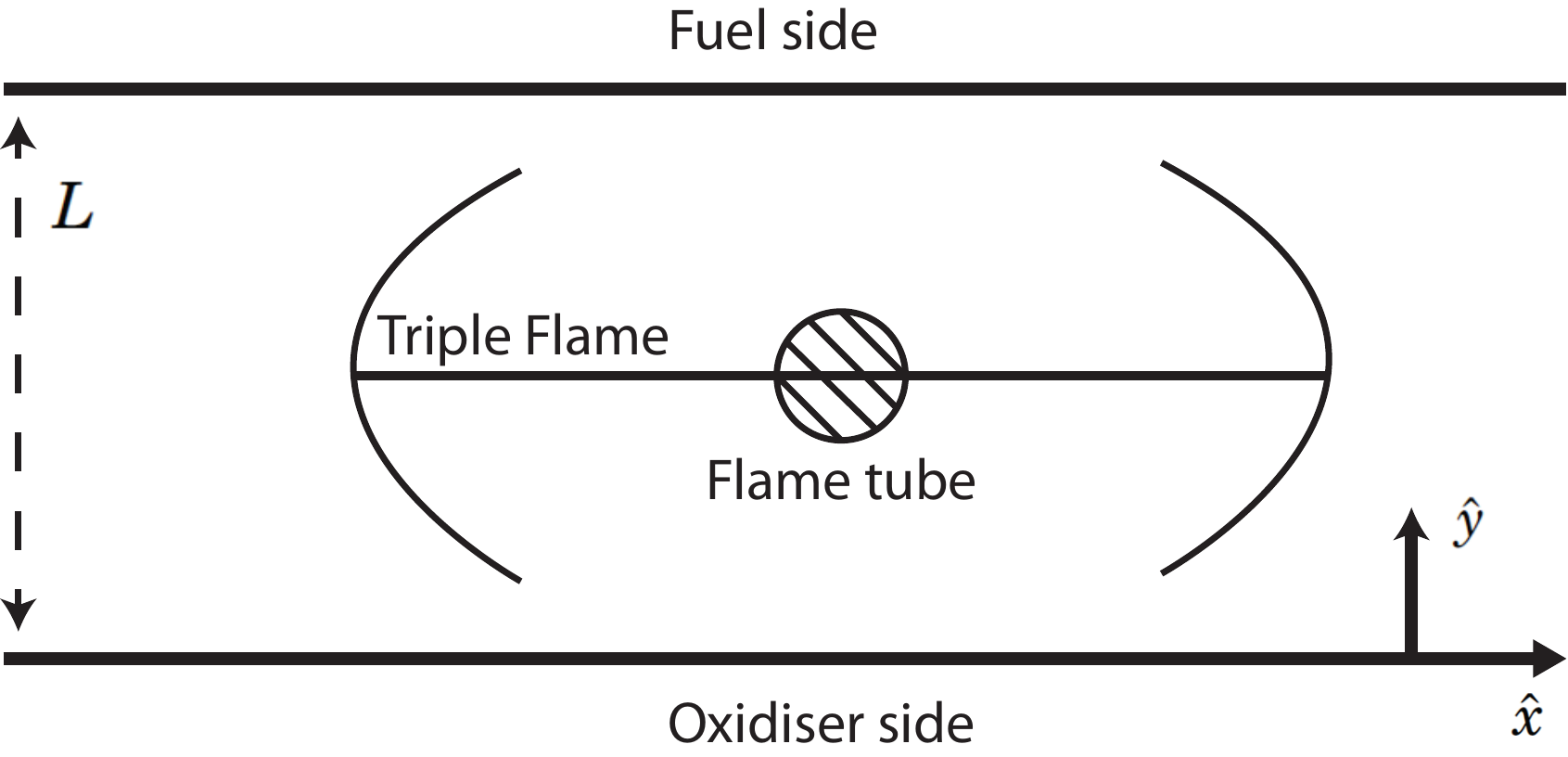}
\caption{An illustration of a pair of triple flames in a channel of height $L$. The walls are assumed to be rigid and to have equal temperatures $\hat{T}=\hat{T}_u$. The mass fractions are prescribed by $\hat{Y}_F=\hat{Y}_{Fu},\text{ }\hat{Y}_O=0$ at the upper wall and $\hat{Y}_F=0,\text{ }\hat{Y}_O=\hat{Y}_{Ou}$ at the lower wall. Also illustrated on the diagram is the stationary, non-propagating ``flame tube" solution that is used as initial condition in the numerical calculations throughout the chapter.}
\label{ignition:fig:diagram}
\end{figure}

We investigate the problem of triple flame initiation in an infinitely long channel of height $L$, where fuel is provided at the upper wall and oxidiser at the lower wall, as shown in figure \ref{ignition:fig:diagram}. The walls are taken to be rigid, porous, isothermal and of equal temperature. The governing equations for fuel and oxidiser are coupled to the Navier--Stokes equations for the fluid velocity vector $\hat{\mathbf{u}}$ in order to take thermal expansion and gravity into account. For simplicity, the combustion is modelled as a single irreversible one-step reaction of the form
\begin{align*}
\text{F}+s\text{O} \to (1+s)\text{Products}+q,
\end{align*}
where F denotes the fuel and O the oxidiser; $s$ denotes the mass of oxidiser consumed and $q$ the heat released, both per unit mass of fuel. The overall reaction rate $\hat{\omega}$ is taken to follow an Arrhenius law of the form
\begin{align*}
\hat{\omega}=\hat{\rho} B \hat{Y}_F \hat{Y}_O \exp{\left(-
E/R\hat{T}\right)}.
\end{align*}
Here $\hat{\rho}$, $\hat{Y}_F$, $\hat{Y}_O$, $R$, $\hat{T}$, $B$ and $E$ are the density, the fuel mass fraction, the oxidiser mass fraction, the
universal gas constant, the temperature, the pre-exponential factor and the activation
energy of the reaction, respectively.
\subsection{Governing equations}
We adopt the low Mach number formulation and assume that $\hat{\rho}D_T$, $\hat{\rho}D_F$ and $\hat{\rho}D_O$ are constant, where $D_T$, $D_F$ and $D_O$ are the diffusion coefficients of heat, fuel and oxidiser, respectively. We also assume that the specific heat capacity $c_P$, the thermal conductivity $\lambda$ and the dynamic viscosity $\mu$ are constant. These assumptions lead to the governing equations
\begin{gather}
\frac{\partial \hat{\rho}}{\partial \hat{t}}+\nabla \cdot \hat{\rho} \hat{\mathbf{u}}=0,\label{ignition:eq:1}\\
\hat{\rho}\frac{\partial \hat{\mathbf{u}}}{\partial \hat{t}}+\hat{\rho}\hat{\mathbf{u}}\cdot\nabla\hat{\mathbf{u}}+\nabla\hat{p}= \mu \left(\nabla^2\hat{\mathbf{u}}+\frac{1}{3}\nabla\left(\nabla\cdot\mathbf{\hat{u}}\right)\right)+\left(\hat{\rho}-\hat{\rho}_u\right)\hat{\mathbf{g}},\label{ignition:eq:2}\\
\hat{\rho}\frac{\partial \hat{T}}{\partial \hat{t}}+\hat{\rho}\hat{\mathbf{u}}\cdot\nabla\hat{T}= \hat{\rho}D_T \nabla^2\hat{T}+\frac{q}{c_P}\hat{\omega},\\
\hat{\rho}\frac{\partial \hat{Y}_F}{\partial \hat{t}}+\hat{\rho}\hat{\mathbf{u}}\cdot\nabla\hat{Y}_F= \hat{\rho}D_F \nabla^2\hat{Y}_F-\hat{\omega},\\
\hat{\rho}\frac{\partial \hat{Y}_O}{\partial \hat{t}}+\hat{\rho}\hat{\mathbf{u}}\cdot\nabla\hat{Y}_O= \hat{\rho}D_O \nabla^2\hat{Y}_O-s\hat{\omega},\\
\hat{\rho}\hat{T}=\hat{\rho}_u \hat{T}_u \label{ignition:eq:state}.
\end{gather}
Here $\hat{p}$ is the hydrodynamic pressure, $\hat{T}_u$ is the temperature of the unburnt mixture and the channel walls, and $\hat{\rho}_u$ is the density of the unburnt mixture. The mass fractions at the channel walls, which are assumed to be rigid, are also prescribed; the boundary conditions at the walls are therefore given by
\begin{gather}
\hat{T}=\hat{T}_u,\text{ }\hat{Y}_F=0,\text{ }\hat{Y}_O=\hat{Y}_{Ou},\text{ }\hat{u}=\hat{v}=0,\text{ at }\hat{y}=0,\label{ignition:bc:dim1}\\
\hat{T}=\hat{T}_u,\text{ }\hat{Y}_F=\hat{Y}_{Fu},\text{ }\hat{Y}_O=0,\text{ }\hat{u}=\hat{v}=0,\text{ at }\hat{y}=L.\label{ignition:bc:dim2}
\end{gather}
In this chapter we are concerned with the initiation of a pair of triple flames, which are expected to propagate in opposite directions \cite{im1999structure,ray2001ignition}; we therefore impose symmetry conditions at the centreline, which we take to be located at $\hat{x}=0$, and solve the problem for $\hat{x} \geq 0$. In the unburnt gas at $\hat{x}=\infty$ we assume that the induced flow is fully developed and the temperature and mass fractions are ``frozen". The boundary conditions at $\hat{x}=0$ and $\hat{x}=\infty$ are therefore given by
\begin{gather}
\hat{T}=\hat{T}_u,\quad
\hat{Y}_F=\hat{Y}_{Fu}\frac{\hat{y}}{L},\quad
\hat{Y}_O=\hat{Y}_{Ou}\left(1-\frac{\hat{y}}{L}\right),\quad
\pd{\hat{\mathbf{u}}}{\hat{x}}=0\text{ at }\hat{x}=\infty,\label{ignition:bc:frozen}\\
\pd{\hat{T}}{\hat{x}}=\pd{\hat{Y}_F}{\hat{x}}=\pd{\hat{Y}_O}{\hat{x}}=\hat{u}=\pd{\hat{v}}{\hat{x}}=0 \text{ at }\hat{x}=0.\label{ignition:bc:dimlast}
\end{gather}
For large activation energies, the flame-front region is expected to be centred around the stoichiometric surface $Y=Y_{st}$ where $\hat{Y}_{O}=s\hat{Y}_F$. In the unburnt gas at $\hat{x}=\infty$, the position of the stoichiometric surface can be determined from the frozen solution \eqref{ignition:bc:frozen} as
\begin{gather}
\frac{Y_{st}}{L}=\frac{1}{1+S},
\end{gather}
where $S \equiv s\hat{Y}_{Fu} / \hat{Y}_{Ou}$ is a normalised stoichiometric coefficient.\\
We now introduce the non-dimensional variables
\begin{equation}
\left.
\begin{aligned}
x=\frac{\hat{x}}{L},\text{ }y=\frac{\hat{y}}{L},\text{ }u=\frac{\hat{u}}{S_L^0},\text{ }v=\frac{\hat{v}}{S_L^0},&\\
t=\frac{\hat{t}}{L \left/ S_L^0 \right.}, \text{ }
\theta=\frac{\hat{T}-\hat{T}_u}{\hat{T}_{ad}-\hat{T}_u},\text{ }y_F=\frac{\hat{Y}_F}{\hat{Y}_{F,st}},\text{ }&\\ y_O=\frac{\hat{Y}_O}{\hat{Y}_{O,st}}, \text{ }p=\frac{\hat{p}}{\hat{\rho}_0 \left(S_L^0\right)^2},&
\end{aligned}
\quad \right\}
\label{ignition:eq:nondimensionalscalings}
\end{equation}
where the subscript 'st' denotes values at the stoichiometric surface. Here $\hat{T}_{ad} \equiv\hat{T}_u+q\hat{Y}_{F,st}/ c_P$ is the adiabatic flame temperature, $\beta\equiv E\left(\hat{T}_{ad}-\hat{T}_u\right) / R\hat{T}_{ad}^2$ is the Zeldovich number or non-dimensional activation energy and $\alpha \equiv \left(\hat{\rho}_{u}-\hat{\rho}_{ad}\right)/\hat{\rho}_{u}$ is the thermal expansion coefficient. In non-dimensionalising we have used as unit speed
\begin{gather}
S_L^0=\left(4 Le_F Le_O \beta^{-3} Y_{O,st} \left(1-\alpha\right) D_T B \exp\left(-E/ R T_{ad}\right)\right)^{1/2}, \label{ignition:eq:sl}
\end{gather}
which is the laminar burning speed of the stoichiometric planar flame to leading order for $\beta\gg 1$. The scalings \eqref{ignition:eq:nondimensionalscalings} are inserted into \eqref{ignition:eq:1}-\eqref{ignition:eq:state} to give the non-dimensional governing equations
\begin{gather}
\frac{\partial \rho}{\partial t}+\nabla\cdot\rho \mathbf{u}=0,\label{ignition:nondim1}\\
\rho \frac{\partial \mathbf{u}}{\partial t}+\rho\mathbf{u}\cdot\nabla\mathbf{u} + \nabla p = \epsilon Pr\left(\nabla^2 \mathbf{u}+\frac{1}{3}\nabla\left(\nabla\cdot\mathbf{u}\right)\right)+\frac{\epsilon^2 Pr Ra}{\alpha }\left(\rho-1\right)\frac{\hat{\mathbf{g}}}{|\hat{\mathbf{g}}|},\\
\rho \frac{\partial \theta}{\partial t}+ \rho\mathbf{u}\cdot\nabla\theta =\epsilon \nabla^2 \theta + \frac{\epsilon^{-1} \omega}{1-\alpha},\label{ignition:nondim4}\\
\rho \frac{\partial y_F}{\partial t}+\rho\mathbf{u}\cdot\nabla y_F = \frac{\epsilon}{ \Le} \nabla^2 y_F - \frac{\epsilon^{-1} \omega}{1-\alpha},\label{ignition:nondim6}\\
\rho \frac{\partial y_O}{\partial t}+\rho\mathbf{u}\cdot\nabla y_O = \frac{\epsilon}{ Le_O} \nabla^2 y_O - \frac{\epsilon^{-1} \omega}{1-\alpha},\label{ignition:nondim7}\\
\rho=\left(1+\frac{\alpha}{1-\alpha}\theta\right)^{-1}.\label{ignition:nondim8}
\end{gather}
The non-dimensional parameters are defined as
\begin{gather*}
Ra=\frac{g\left(\hat{\rho}_{u}-\hat{\rho}_{ad}\right)L^3}{\nu \hat{\rho}_{u}D_T}, \quad \epsilon=\frac{l_{Fl}}{L}=\frac{D_T/S_L^0}{L},\\ \quad Le_F=\frac{D_T}{D_F}, \quad Le_O=\frac{D_T}{D_O},\quad \text{and}\quad Pr=\frac{\nu}{D_T},
\end{gather*}
which are the Rayleigh number, the flame-front thickness $l_{Fl}$ measured against the unit length $L$, the fuel and oxidiser Lewis numbers and the Prandtl number, respectively. Here $\nu$ is the kinematic viscosity $\nu=\mu \left/ \tilde{\rho}_u \right.$. Note that $\epsilon$ is related to the Damk\"{o}hler number used in Chapter \ref{chapter:diffusion} by 
\begin{gather*}
Da=\frac{1}{\epsilon^{2}(1-\alpha)}.
\end{gather*}
The non-dimensional reaction rate is
\begin{gather}
\omega=\frac{\beta^3}{4Le_F Le_O} \rho y_F y_O \exp\left(\frac{\beta(\theta-1)}{1+\alpha_h(\theta-1)}\right),\label{ignition:eq:reaction}
\end{gather}
where $\alpha_h$ is a heat release parameter given by $\alpha_h=(\hat{T}_{ad}-\hat{T}_u)/\hat{T}_{ad}$. Although they are technically equal in the low Mach number approximation, here we have followed \cite{pearce2013effect} and kept the parameters $\alpha$ and $\alpha_h$ distinct; we set $\alpha_h=0.85$ throughout the chapter and vary $\alpha$ in order to assess the effect of thermal expansion without changing the Arrhenius reaction term \eqref{ignition:eq:reaction}. Thus when $\alpha=0$ the equations reduce to those of the constant density approximation.

Finally, with the scalings \eqref{ignition:eq:nondimensionalscalings} inserted into \eqref{ignition:bc:dim1}-\eqref{ignition:bc:dimlast}, the non-dimensional boundary conditions can be written
\begin{align}
\theta=0, \label{ignition:bc1}\quad
y_F=(1+S)y,\quad
y_O=\frac{S+1}{S}(1-y),\quad u=v=0\quad\text{ at }y =0 \text{, } y=1,&\\
\theta=0,\quad y_F=(1+S)y,\quad y_O=\frac{S+1}{S}(1-y),\quad \pd{u}{x}=\pd{v}{x}=0\quad\text{ at }x=\infty,&\\
\pd{\theta}{x}=\pd{y_F}{x}=\pd{y_O}{x}=u=\pd{v}{x}=0\quad\text{ at }x=0.&\label{ignition:bc4}
\end{align}
Suitable initial conditions must also be prescribed. The non-dimensional problem is now fully formulated and is given by equations (\ref{ignition:nondim1})-(\ref{ignition:eq:reaction}), subject to the boundary conditions (\ref{ignition:bc1})-(\ref{ignition:bc4}). The non-dimensional parameters in this problem are $\alpha$, $\alpha_h$, $\beta$, $Pr$, $Ra$, $\epsilon$, $S$, $Le_F$ and $Le_O$.
\subsection{Mixture fraction formulation}
In this section we simplify the problem in the special case of unity Lewis numbers
\begin{gather}
Le_F=Le_O=1.
\end{gather}
We set the Lewis numbers equal to unity in order to concentrate on the effects of thermal expansion and gravity on triple flames, without the added complication of thermo-diffusive instabilities. Following the method of \S \ref{diffusion:sec:mixfracform} in Chapter \ref{chapter:diffusion}, we note that for $Le_F=Le_O=1$ the \emph{mixture fraction} $Z$, defined as
\begin{gather}
Z=\frac{y_F+\theta}{1+S}=1-\frac{S}{1+S}\left(y_O+\theta\right),\label{ignition:eq:mixfracdef}
\end{gather}
satisfies
\begin{gather}
\rho \frac{\partial Z}{\partial t}+\rho\mathbf{u}\cdot\nabla Z = \epsilon \nabla^2 Z,\label{ignition:eq:mixfracform}\\
Z=0 \text{ at }y=0,\\
Z=1\text{ at }y=1.
\end{gather}
We can therefore solve the reaction-free equation \eqref{ignition:eq:mixfracform} and use equation \eqref{ignition:eq:mixfracdef} to find the fuel and oxidiser mass fractions if necessary. Thus the governing equations simplify to
\begin{gather}
\frac{\partial \rho}{\partial t}+\nabla\cdot\rho \mathbf{u}=0,\label{ignition:mix:nondim1}\\
\rho \frac{\partial \mathbf{u}}{\partial t}+\rho\mathbf{u}\cdot\nabla\mathbf{u} + \nabla p = \epsilon Pr\left(\nabla^2 \mathbf{u}+\frac{1}{3}\nabla\left(\nabla\cdot\mathbf{u}\right)\right)+\frac{\epsilon^2 Pr Ra}{\alpha }\left(\rho-1\right)\frac{\hat{\mathbf{g}}}{|\hat{\mathbf{g}}|},\\
\rho \frac{\partial \theta}{\partial t}+ \rho\mathbf{u}\cdot\nabla\theta =\epsilon \nabla^2 \theta + \frac{\epsilon^{-1} \omega}{1-\alpha},\label{ignition:mix:nondim4}\\
\rho \frac{\partial Z}{\partial t}+\rho\mathbf{u}\cdot\nabla Z = \epsilon \nabla^2 Z,\label{ignition:mix:nondim5}\\
\rho=\left(1+\frac{\alpha}{1-\alpha}\theta\right)^{-1}\label{ignition:mix:nondim8},
\end{gather}
which are subject to the boundary conditions
\begin{align}
\theta=0, \label{ignition:mix:bc1}\quad
Z=y,\quad u=v=0\quad\text{ at }y =0 \text{, } y=1,&\\
\theta=0,\quad Z=y,\quad \pd{u}{x}=\pd{v}{x}=0\quad\text{ at }x=\infty,&\\
\pd{\theta}{x}=\pd{Z}{x}=u=\pd{v}{x}=0\quad\text{ at }x=0,&\label{ignition:mix:bc4}
\end{align}
and suitable initial conditions. The non-dimensional problem is now formulated in terms of $\alpha$, $\alpha_h$, $\beta$, $Pr$, $Ra$, $\epsilon$ and $S$. In the next section we solve this problem numerically, with particular emphasis on the effect of $\alpha$, $\epsilon$ and $Ra$.

At this point it is instructive to briefly discuss the limitations of the model used. First, the model utilised here is two-dimensional. Although in Chapter \ref{chapter:diffusion} we showed that three-dimensional effects do not affect the critical Rayleigh number for the onset of gravitational instabilities of the planar diffusion flame, far beyond the threshold of instability these effects may become important. However, currently a transient simulation of the three-dimensional problem would involve prohibitively expensive computations. Understanding the two-dimensional problem is an important step towards a full description of the effect of gravity on triple flames. Second, we neglect the effects of heat-loss, differential diffusion and the stoichiometric coefficient $S$; these aspects of the problem would be interesting to investigate in future studies but are ignored here for the sake of simplicity and clarity. Finally, in simulations of the non-reactive problem of a fluid layer heated from below, it is well known that the problem can become dependent on the initial conditions at high values of the Rayleigh number \cite{foster1969effect}. In all time-dependent simulations in this chapter we take the initial conditions to be a steady flame tube solution (plus a small perturbation) and acknowledge that, far beyond the threshold of gravitational instability, different behaviour may arise if the initial conditions are taken to be different.

\section{Results and discussion}
\label{ignition:sec:res}
In this section we present the results obtained by numerically solving the problem \eqref{ignition:mix:nondim1}-\eqref{ignition:mix:bc4} in the finite element package Comsol Multiphysics. This has been extensively tested in combustion applications, including our previous publications on diffusion flames \cite{pearce2013rayleigh} and triple flames \cite{pearce2013effect}. The domain is covered by a grid of approximately 200,000 triangular elements, with local refinement around the reaction zone. Solutions have been tested to be independent of the mesh and the size of the domain.

The section is split into three parts. First, we provide some important preliminary results. Second, we investigate the transient initiation of triple flames in situations where the underlying diffusion flame is stable, including a study on the energy required for ignition. Third, we investigate the ignition energy and transient behaviour of a triple flame with a trailing diffusion flame that is unstable due to gravitational effects.

In order to provide a criterion for the energy required for initiation of triple flames, we investigate steady, non-propagating solutions of equations \eqref{ignition:mix:nondim1}-\eqref{ignition:mix:bc4}, which we refer to as \emph{flame tubes}. The energy of these flame tube solutions can be the required criterion for the \emph{ignition energy} of a triple flame. We define the energy $E$ of a flame (per unit depth) by
\begin{gather}
E=\int_{x=-\infty}^{\infty}\int_{y=0}^1 \theta~ \mathrm{d}y~\mathrm{d}x=2\int_{x=0}^{\infty}\int_{y=0}^1 \theta~ \mathrm{d}y~\mathrm{d}x.
\end{gather}
It is found that if a flame tube with a small ``energy-increasing" perturbation is used as an initial condition for time-dependent simulations, a triple flame forms, which propagates into the unburnt gas. A flame tube with a small ``energy-decreasing" perturbation leads to extinction. The perturbations are added to the temperature field as a kernel of finite size, surrounding the flame tube, and of order of magnitude $10^{-3}$. Throughout the chapter we provide illustrative examples of the two outcomes for various values of the parameters.
\begin{figure}
\centering
\subfigure[]{
\includegraphics[scale=0.9, trim=0 0 0 0,clip=true]{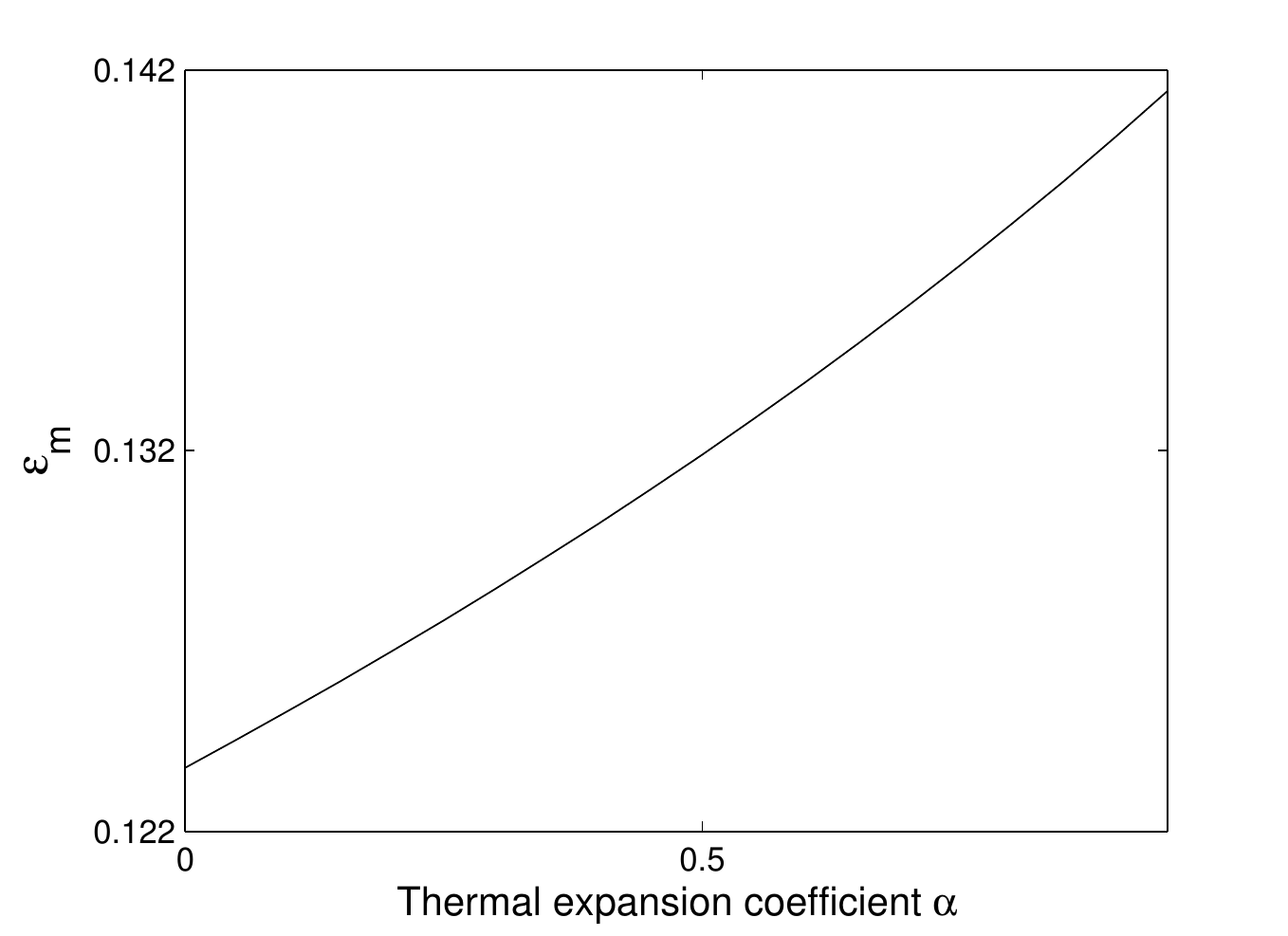}\label{ignition:fig:ext_a}}
\subfigure[]{\includegraphics[scale=0.9, trim=0 0 0 0,clip=true]{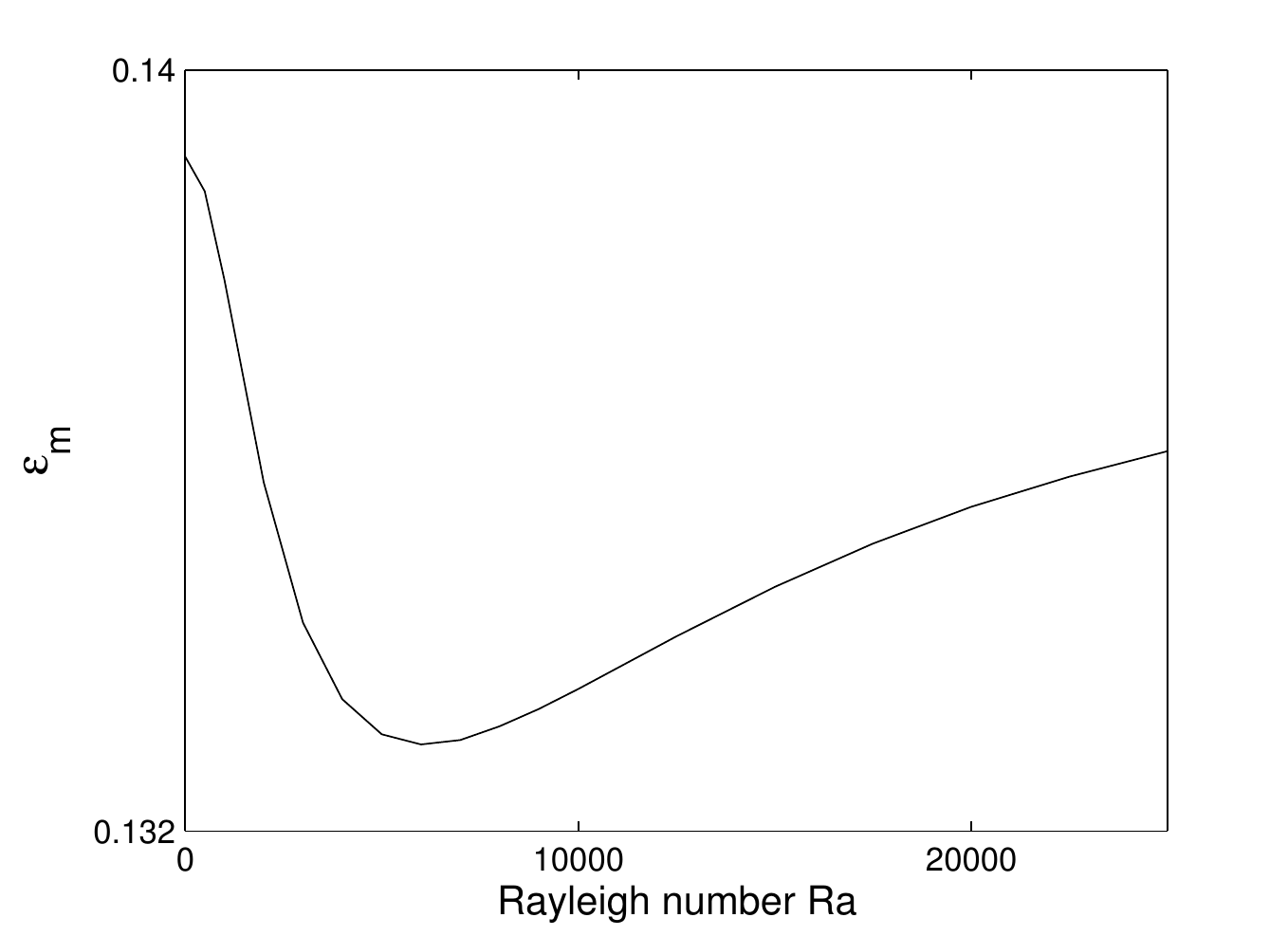}\label{ignition:fig:ext_ra}}
\caption{The value of $\epsilon$ at which a triple flame has zero propagation speed, denoted $\epsilon_m$, versus a) the thermal expansion coefficient $\alpha$, for $Ra=0$ and b) the Rayleigh number $Ra$, for $\alpha=0.85$. The other parameters are given the fixed values $\alpha_h=0.85$, $\beta=10$, $Pr=1$ and $S=1$.}
\label{ignition:fig:ext}
\end{figure}
\subsection{Preliminary results}
\label{ignition:sec:preliminary}
In this section we provide some preliminary results that will be useful in the following sections. We begin by investigating the effects of $\alpha$ and $Ra$ on $\epsilon_{m}$, \emph{the value of} $\epsilon$ \emph{below which positively propagating triple flames cannot exist}. This preliminary study is important because flame tubes are not expected to exist for $\epsilon >\epsilon_{m}$. Figure \ref{ignition:fig:ext} shows a) the effect of $\alpha$ on $\epsilon_m$ for $Ra=0$ and b) the effect of $Ra$ on $\epsilon_m$ for $\alpha=0.85$.

It should be noted that technically flame tube solutions also do not exist in the asymptotic limit $\epsilon \to 0$. This can be seen by considering the problem \eqref{ignition:mix:nondim1}-\eqref{ignition:mix:bc4} in the limit of infinite activation energy and taking $\epsilon \to 0$. The problem reduces to finding stationary, two dimensional tubes in a homogeneous mixture; such a problem is known to have no solution. This is because the leading order term for the temperature is governed by the cylindrically symmetrical Laplace equation in the unburnt gas, whose only solution that satisfies the boundary conditions in the far field is $\theta=0$. This, of course, is a contradiction since the temperature should be given by $\theta=1$ at the reaction zone.

Next, we mention results obtained in Chapter \ref{chapter:diffusion}, on the instability of a planar diffusion flame in an infinitely long channel. The results are summarised in figure \ref{triple:fig:epsilonvsra}, which plots the critical Rayleigh number $Ra_c$ against $\epsilon$ in the case $\alpha=0.85$. For values of the Rayleigh number below $Ra_c$, a stable planar diffusion flame exists and a stable triple flame is expected to propagate along the channel. If $Ra>Ra_c$, the planar diffusion flame is unstable.
\begin{figure}
\centering
\includegraphics[width=\textwidth, trim=0 0 0 0,clip=true]{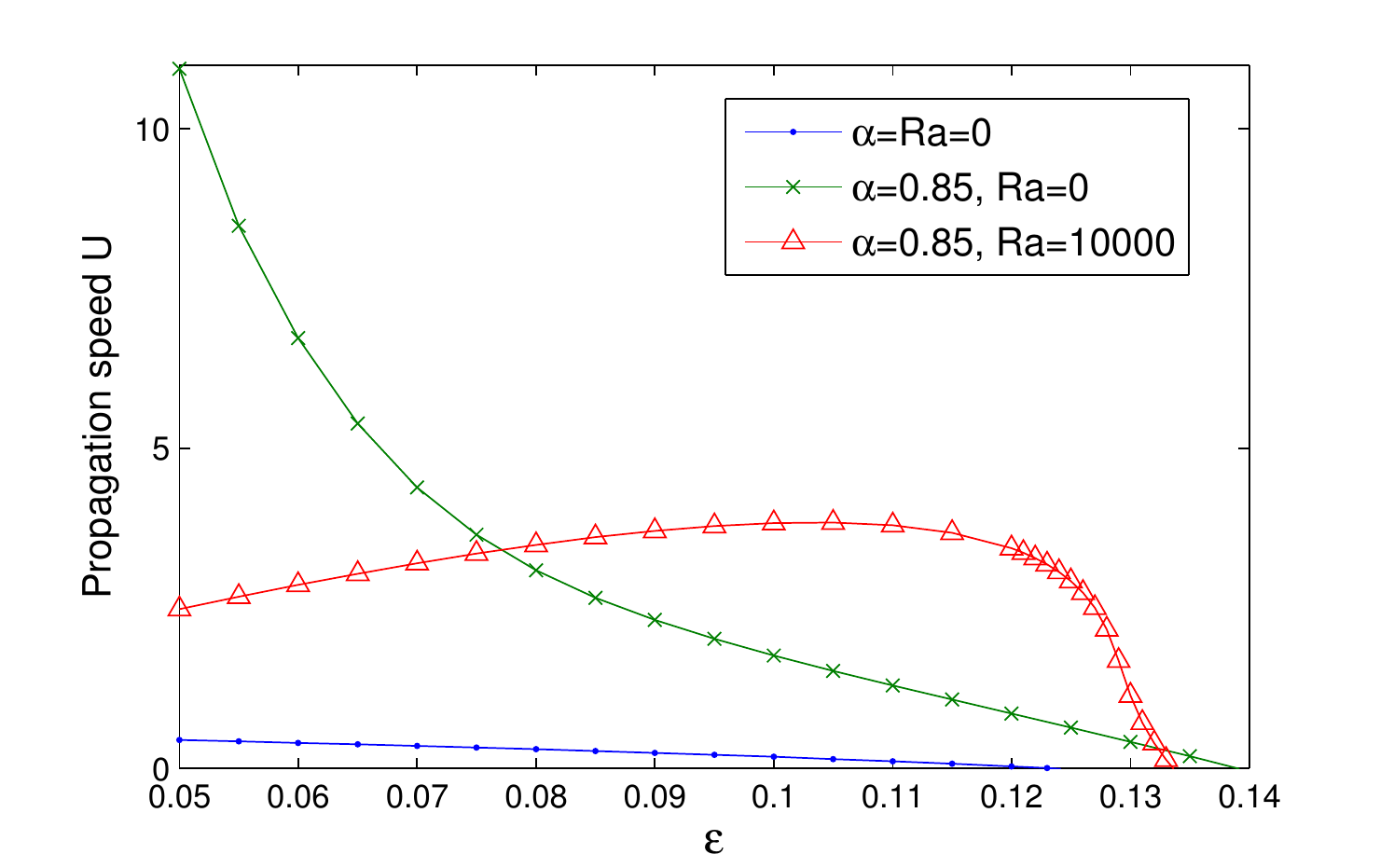}
\caption{Propagation speed $U$ of a triple flame versus $\epsilon$ for selected values of $\alpha$ and $Ra$. The propagation speed rises to very large values for small $\epsilon$ in the case $\alpha=0.85$, $Ra=0$. This can be attributed to the shear flow induced by the channel walls, since if the calculations are repeated for ``free-slip" walls with $u_y=0$, the propagation speed is much smaller. In this figure $\beta=10$, $S=1$ and $\alpha_h=0.85$.}
\label{ignition:fig:steadytriple}
\end{figure}

Finally, figure \ref{ignition:fig:steadytriple} shows the propagation speed of steadily propagating triple flames versus $\epsilon$, for selected values of the parameters, corresponding to the range of values that will be used for time-dependent simulations in later sections. These results were obtained by solving \eqref{ignition:mix:nondim1}-\eqref{ignition:mix:bc4} in a coordinate system attached to the flame front, which travels at speed $U$ relative to the laboratory, i.e. $\left(t,x,y\right)\to\left(t,x+Ut,y\right)$. Figure \ref{ignition:fig:steadytriple} is important because it is expected that if initiation is successful, a steadily propagating triple flame with propagation speed given in the figure will be the result.

Figure \ref{ignition:fig:steadytriple} provides similar results to those found in Chapter \ref{chapter:triple}. The difference between the results in figure \ref{ignition:fig:steadytriple} and the results in figures \ref{triple:fig:u_eps_alpha} and \ref{triple:fig:grav_u} is in the boundary conditions imposed; in the current chapter we consider a pair of triple flames propagating away from each other towards open boundaries, with symmetric conditions imposed at $x=0$. Chapter \ref{chapter:triple} was concerned with a triple flame propagating into an undisturbed fluid, with zero gradients imposed downstream.
\subsection{Initiation of triple flames}
We proceed with an investigation of the energy required for initiation of a triple flame in an infinitely long channel and the transient behaviour of a triple flame after initiation. In this section we focus on values of the parameters for which the underlying diffusion flame is \emph{stable}, as summarised in figure \ref{triple:fig:epsilonvsra}. The effects of the flame-front thickness $\epsilon$, the thermal expansion coefficient $\alpha$ and the Rayleigh number $Ra$ will all be investigated in separate sections.

\subsubsection{Effect of $\epsilon$}
\begin{figure}
\centering
\includegraphics[scale=0.9, trim=0 0 0 0,clip=true]{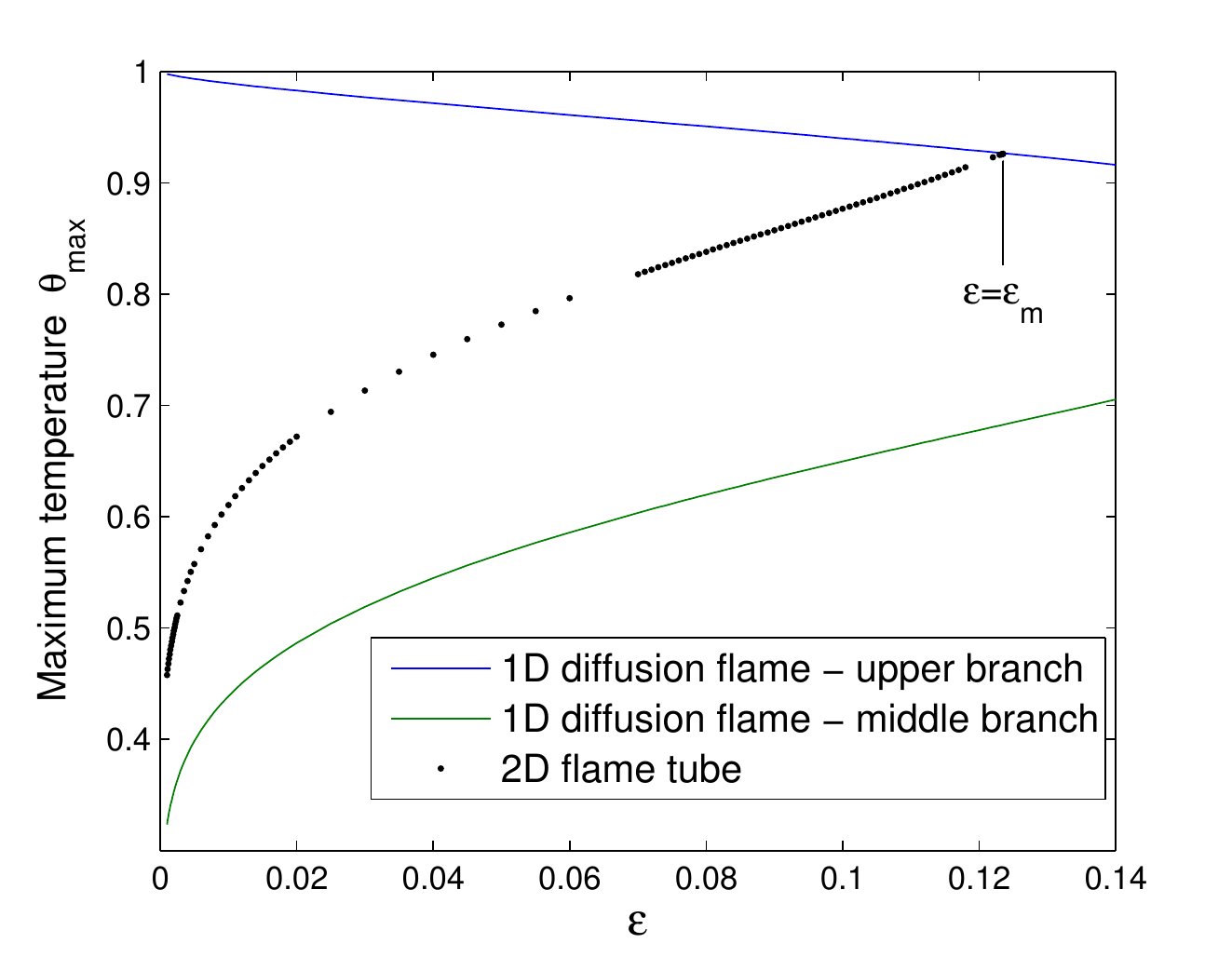}
\caption{Maximum temperature $\theta_{\text{max}}$ versus $\epsilon$ for one-dimensional diffusion flame solutions, which generate an S-shaped curve, and two-dimensional flame tube solutions. The lower branch of the S-shaped curve is not shown. Indicated on the figure is $\epsilon_m$, the value where a triple flame has zero propagation speed (see figure \ref{ignition:fig:ext}); above this value no flame tube solutions are found. Computations are performed for $\alpha=0$, $\beta=10$ and $\alpha_h=0.85$.}
\label{ignition:fig:Scurve}
\end{figure}
In this section we investigate the effect of the flame-front thickness $\epsilon$ on the initiation of a triple flame. The ignition energy $E$ is given by the energy of steady flame tube solutions of equations \eqref{ignition:mix:nondim1}-\eqref{ignition:mix:bc4}. It is instructive to begin by showing where the flame tube solutions lie in parameter space and in particular in relation to the well-known \emph{S-shaped curve} of one-dimensional diffusion flame solutions.

The S-shaped curve, generated by plotting the maximum temperature of one-dimensional, steady diffusion flame solutions of \eqref{ignition:mix:nondim1}-\eqref{ignition:mix:bc4} (with $u=v=0$) in terms of $\epsilon$, is given in figure \ref{ignition:fig:Scurve}. The curve consists of an upper (stable) ``strongly burning" branch, a middle (unstable) branch and a lower (stable) ``weakly burning" branch (not shown). As expected, the maximum value of $\epsilon$ for which flame tubes can exist is given by $\epsilon_m$, the value at which a triple flame has zero propagation speed. The curve of flame tube temperature versus $\epsilon$ terminates on the upper branch of the S-shaped curve at $\epsilon=\epsilon_m$. The figure shows that, although flame tube solutions do not exist in the limit $\epsilon \to 0$ (as discussed in \S \ref{ignition:sec:preliminary}), solutions have been calculated numerically for very small values of $\epsilon$.
\begin{figure}
\centering
\subfigure[]{
\includegraphics[scale=0.9, trim=0 0 0 0,clip=true]{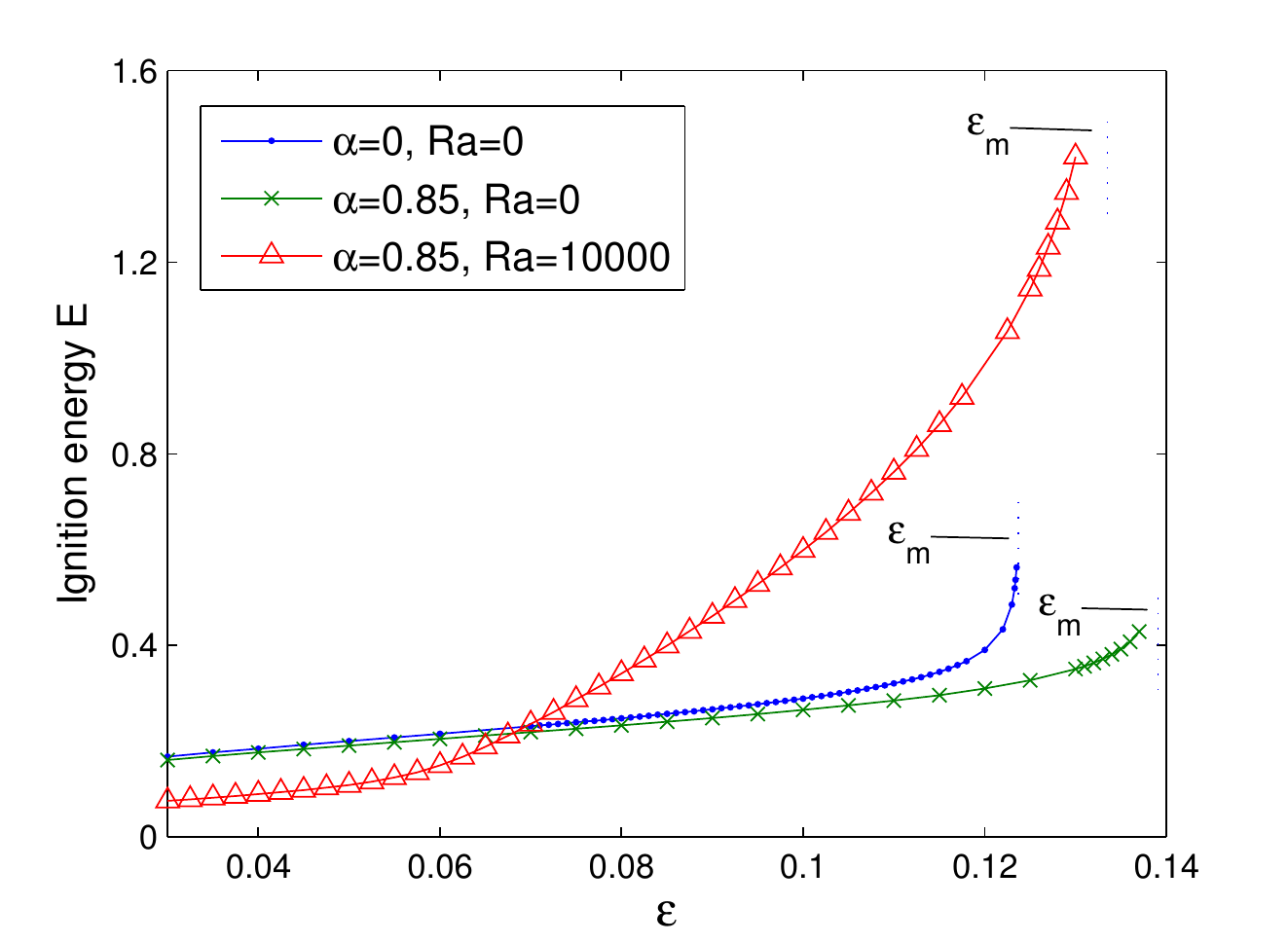}\label{ignition:fig:energy_epsilon}}
\subfigure[]{
\includegraphics[scale=0.9, trim=0 0 0 0,clip=true]{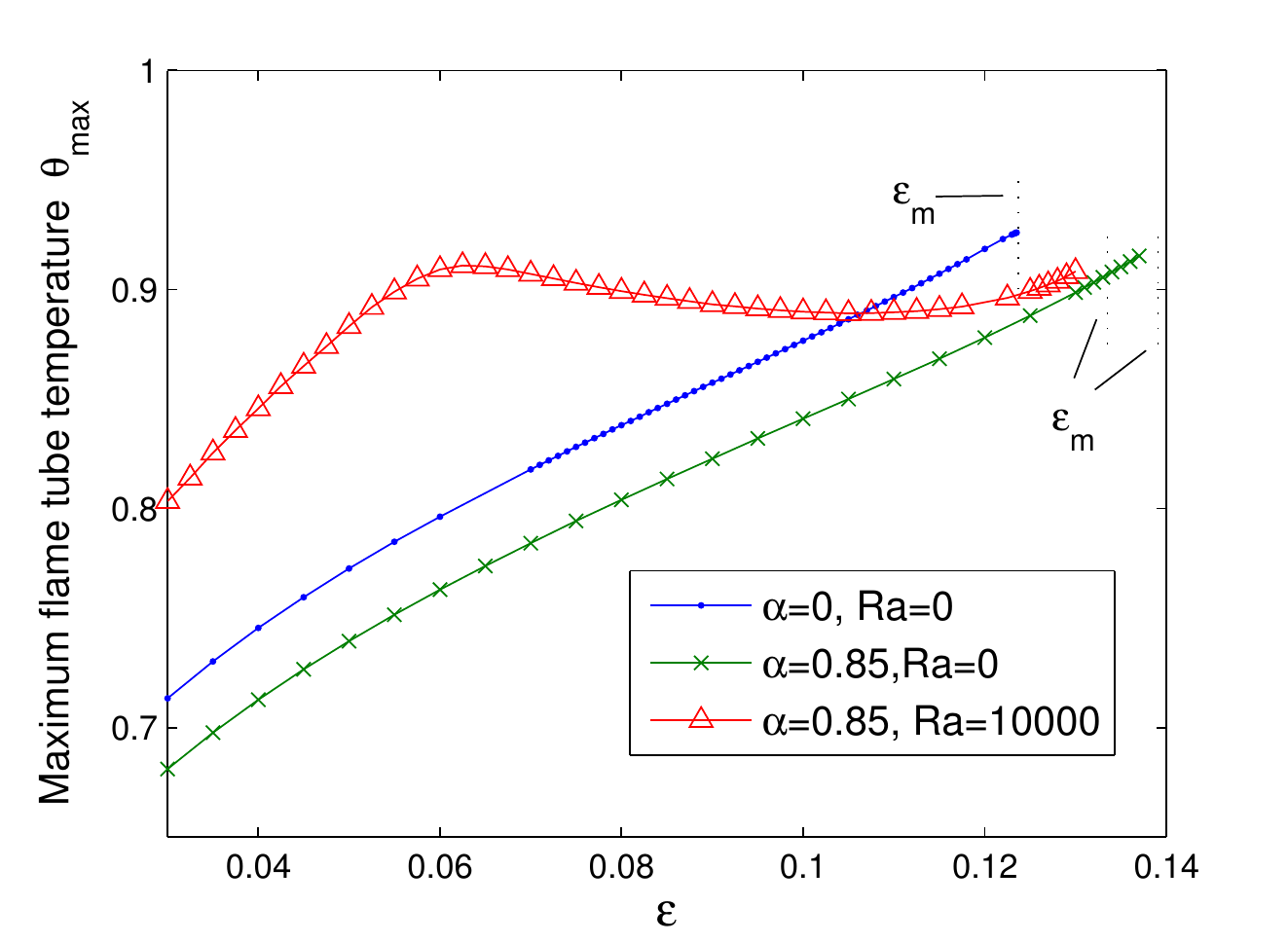}\label{ignition:fig:temp_epsilon}}
\caption{(a) Ignition energy $E$ and (b) maximum flame tube temperature $\theta_{\text{max}}$ versus $\epsilon$ for selected values of $\alpha$ and $Ra$. Also indicated on the figure for each case is the value of $\epsilon$, denoted $\epsilon_m$, at which the triple flame propagation speed is zero, which is taken from figure \ref{ignition:fig:ext}. Flame tube solutions are not found for $\epsilon>\epsilon_m$. The other parameters are given the fixed values $\alpha_h=0.85$, $\beta=10$, $Pr=1$ and $S=1$.}
\label{ignition:fig:energy_temp_epsilon}
\end{figure}

We plot the flame tube energy $E$ and the maximum flame tube temperature $\theta_{\text{max}}$ versus $\epsilon$ for selected values of $\alpha$ and $Ra$ in figure \ref{ignition:fig:energy_temp_epsilon}. It can be seen that $E$ monotonically increases with increasing $\epsilon$ in all cases. When thermal expansion is present but gravity is not taken into account, the ignition energy for each value of $\epsilon$ is lower than in the constant density case $\alpha=0$, $Ra=0$. It can be seen that the ignition energy increases sharply as $\epsilon$ approaches $\epsilon_m$, which is the value of $\epsilon$ at which the triple flame speed is zero (see figure \ref{ignition:fig:ext}) and above which no flame tube solutions are found. The effects of thermal expansion and gravity will be described in more detail in the following sections.

\begin{figure}
\centering
\subfigure[]{
\includegraphics[scale=0.9, trim=0 0 0 0,clip=true]{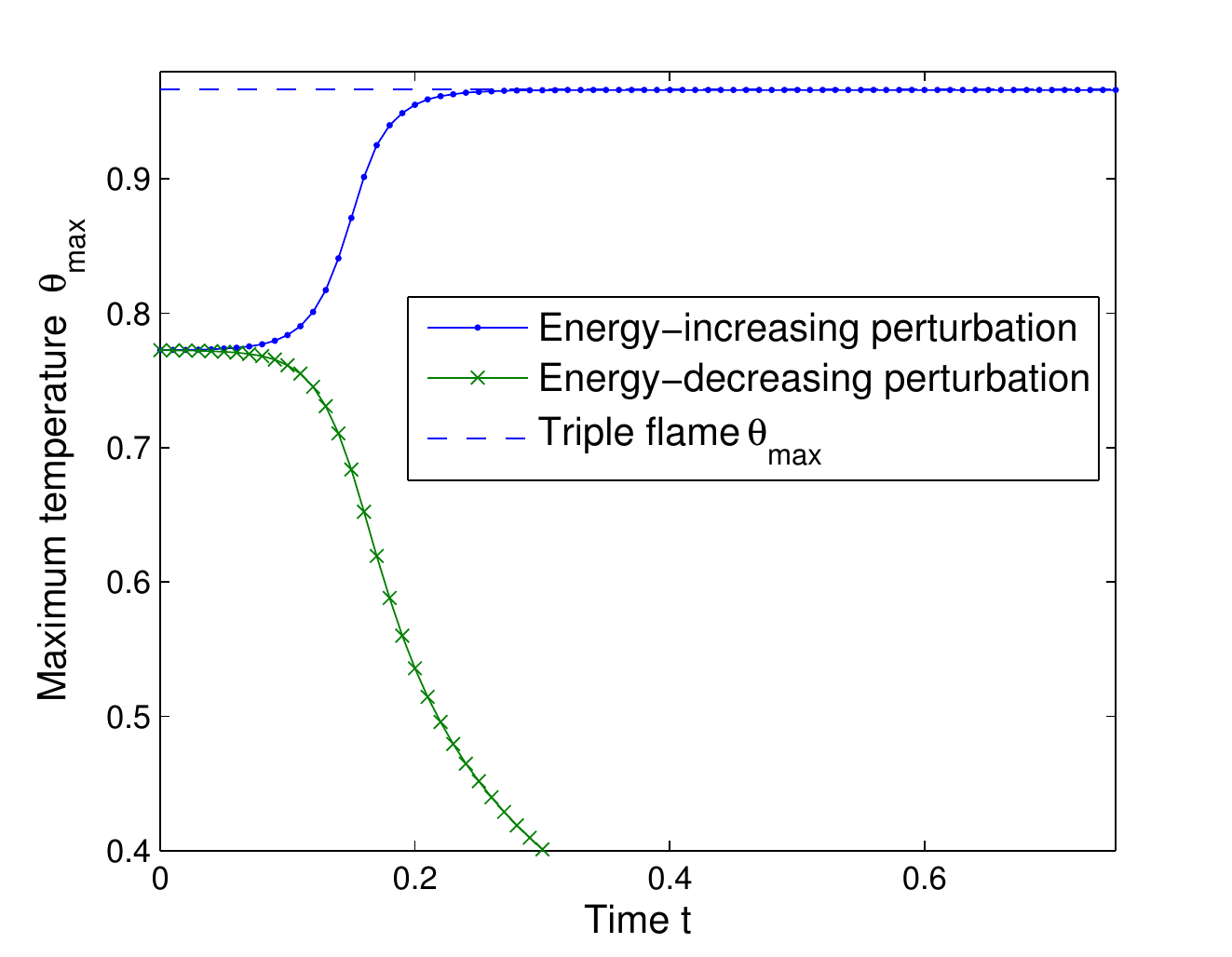}\label{ignition:fig:temp_e_05_a_0}}
\subfigure[]{
\includegraphics[scale=0.9, trim=0 0 0 0,clip=true]{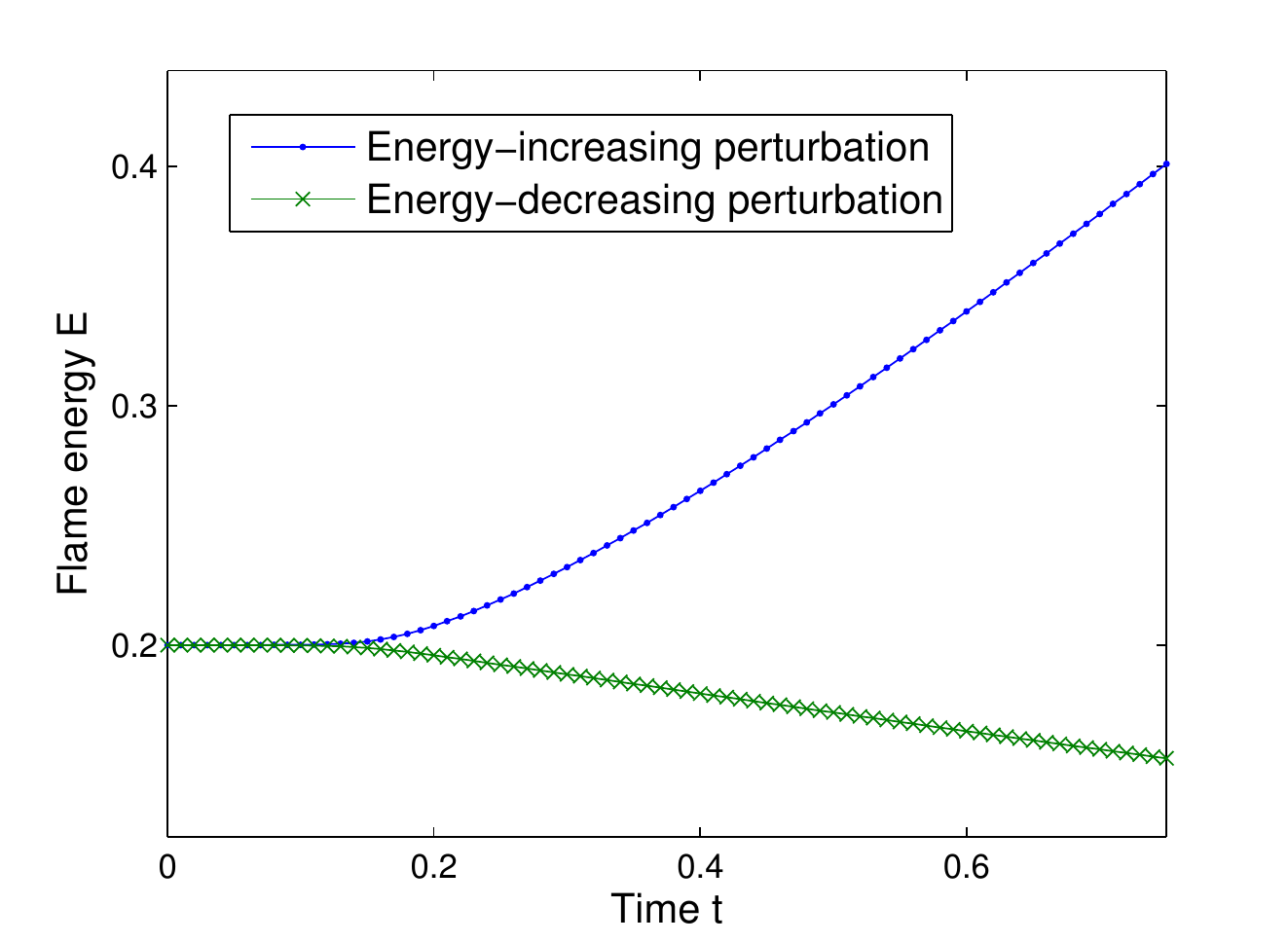}\label{ignition:fig:temp_e_05_a_0}}
\caption{Transient behaviour of flames with a small energy-decreasing or energy-increasing perturbation added to the unstable flame tube solution for $\epsilon=0.05$, showing a) maximum temperature $\theta_{max}$ and b) flame energy $E$, for $\alpha=0$, $Ra=0$. Also included in a) is the maximum temperature of a triple flame, which the ignited flame tends towards. The other parameters are given the fixed values $\alpha_h=0.85$, $\beta=10$, $Pr=1$ and $S=1$.}
\label{ignition:fig:energy_temp_e_05_a_0}
\end{figure}

\begin{figure}
\centering
\subfigure[]{
\includegraphics[scale=0.9, trim=0 0 0 0,clip=true]{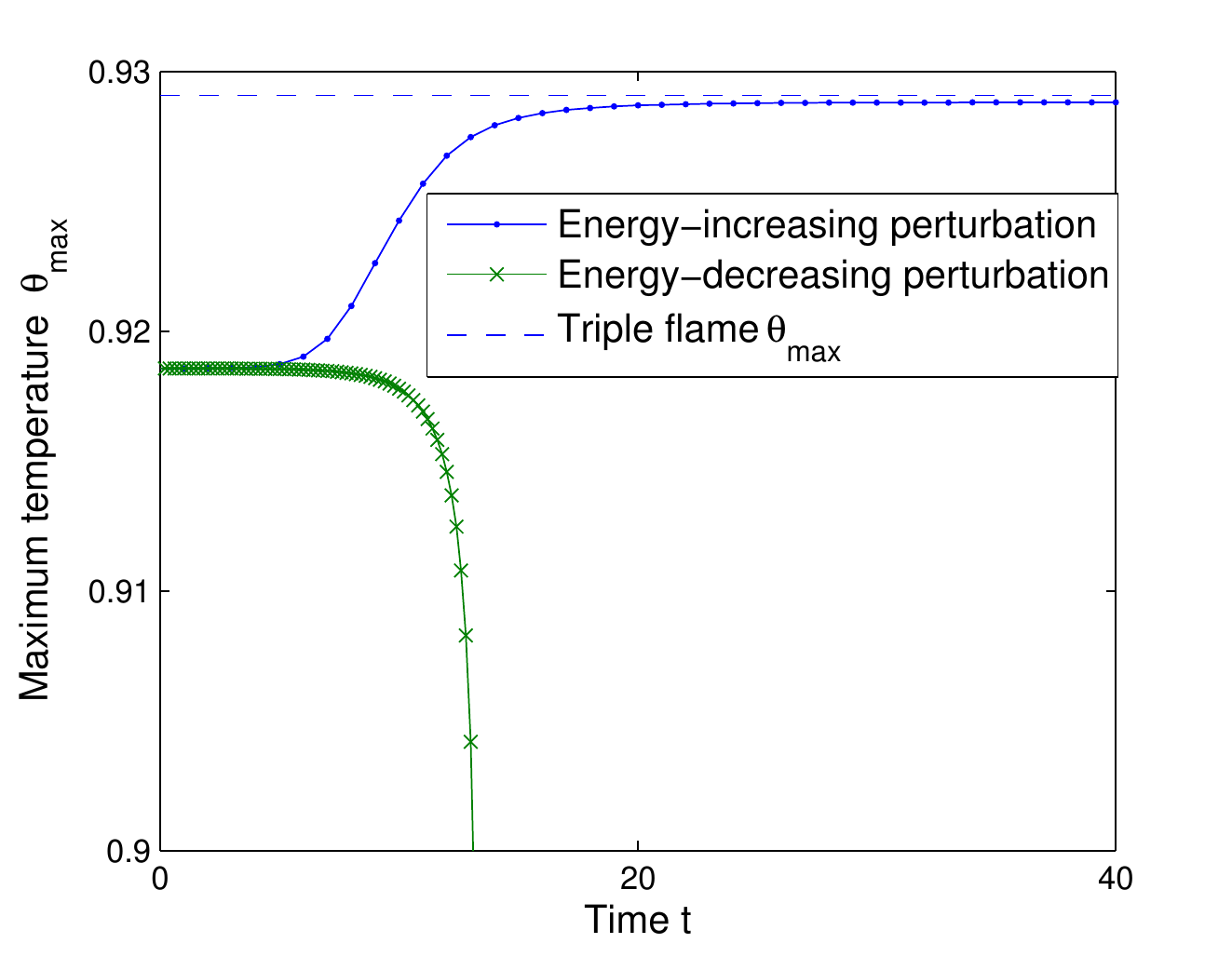}\label{ignition:fig:temp_e_12_a_0}}
\subfigure[]{
\includegraphics[scale=0.9, trim=0 0 0 0,clip=true]{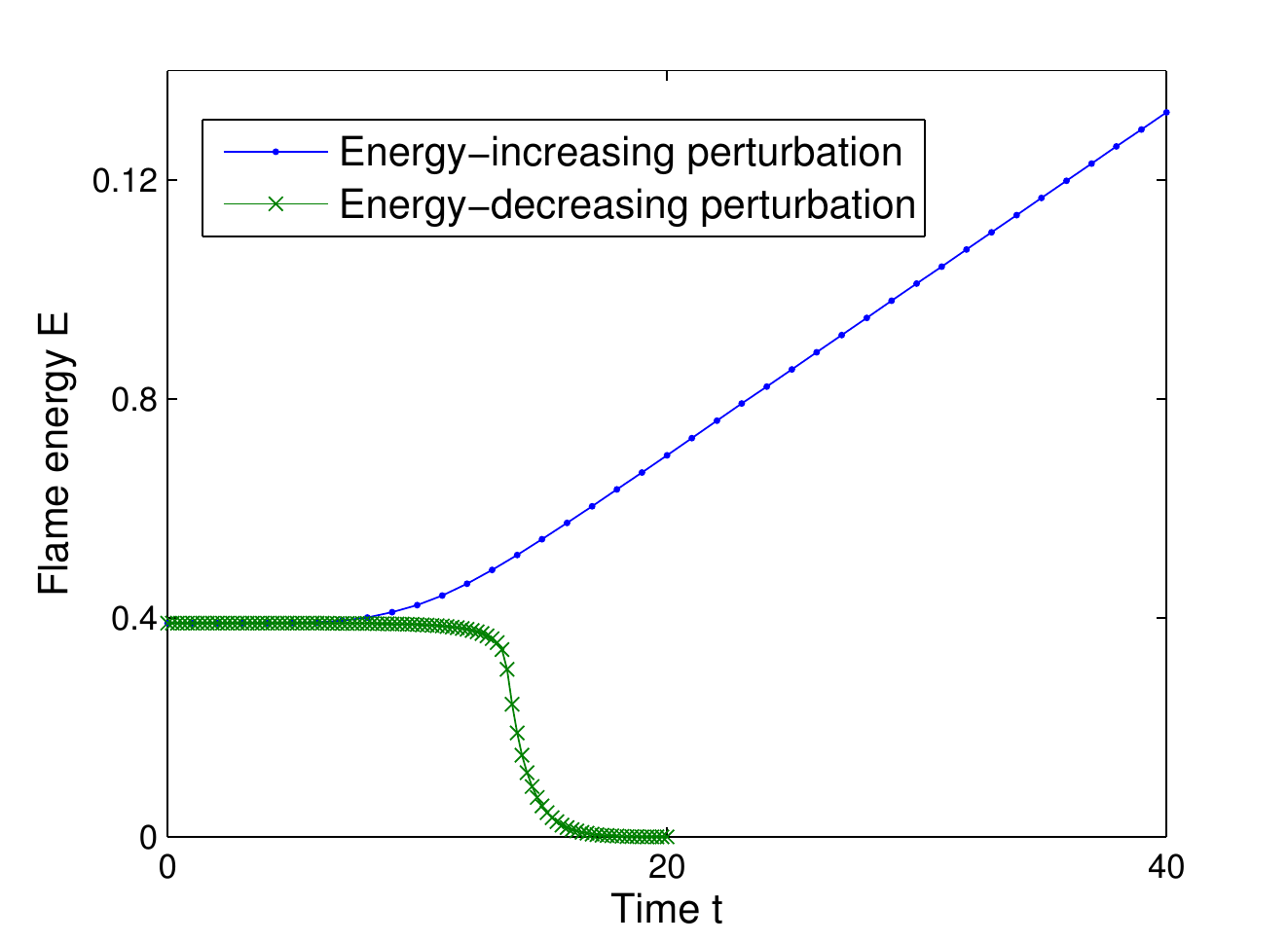}\label{ignition:fig:temp_e_12_a_0}}
\caption{Transient behaviour of flames with a small energy-decreasing or energy-increasing perturbation added to the unstable flame tube solution for $\epsilon=0.12$, showing a) maximum temperature $\theta_{max}$ and b) flame energy $E$, for $\alpha=0$, $Ra=0$. Also included in a) is the maximum temperature of a triple flame, which the ignited flame tends towards. The other parameters are given the fixed values $\alpha_h=0.85$, $\beta=10$, $Pr=1$ and $S=1$.}
\label{ignition:fig:energy_temp_e_12_a_0}
\end{figure}

Now we investigate the transient behaviour of the unstable flame tube solutions for $\alpha=0$, $Ra=0$ when they are subject to small perturbations. Plotted in figures \ref{ignition:fig:energy_temp_e_05_a_0} and \ref{ignition:fig:energy_temp_e_12_a_0}, in which $\epsilon=0.05$ and $\epsilon=0.12$ respectively, are the maximum temperature $\theta_{\text{max}}$ and flame energy $E$ for flame tubes subject to either an energy-increasing or an energy-decreasing perturbation. These figures show that, if an energy-increasing perturbation is added, the flame tube solution will evolve in time into a steadily propagating triple flame, with speed given in figure \ref{ignition:fig:steadytriple}. If an energy-decreasing perturbation is added, the flame extinguishes and the maximum temperature and flame energy both decay to zero.

\begin{figure}
\subfigure[]{
\includegraphics[scale=0.8, trim=0 0 0 0,clip=true]{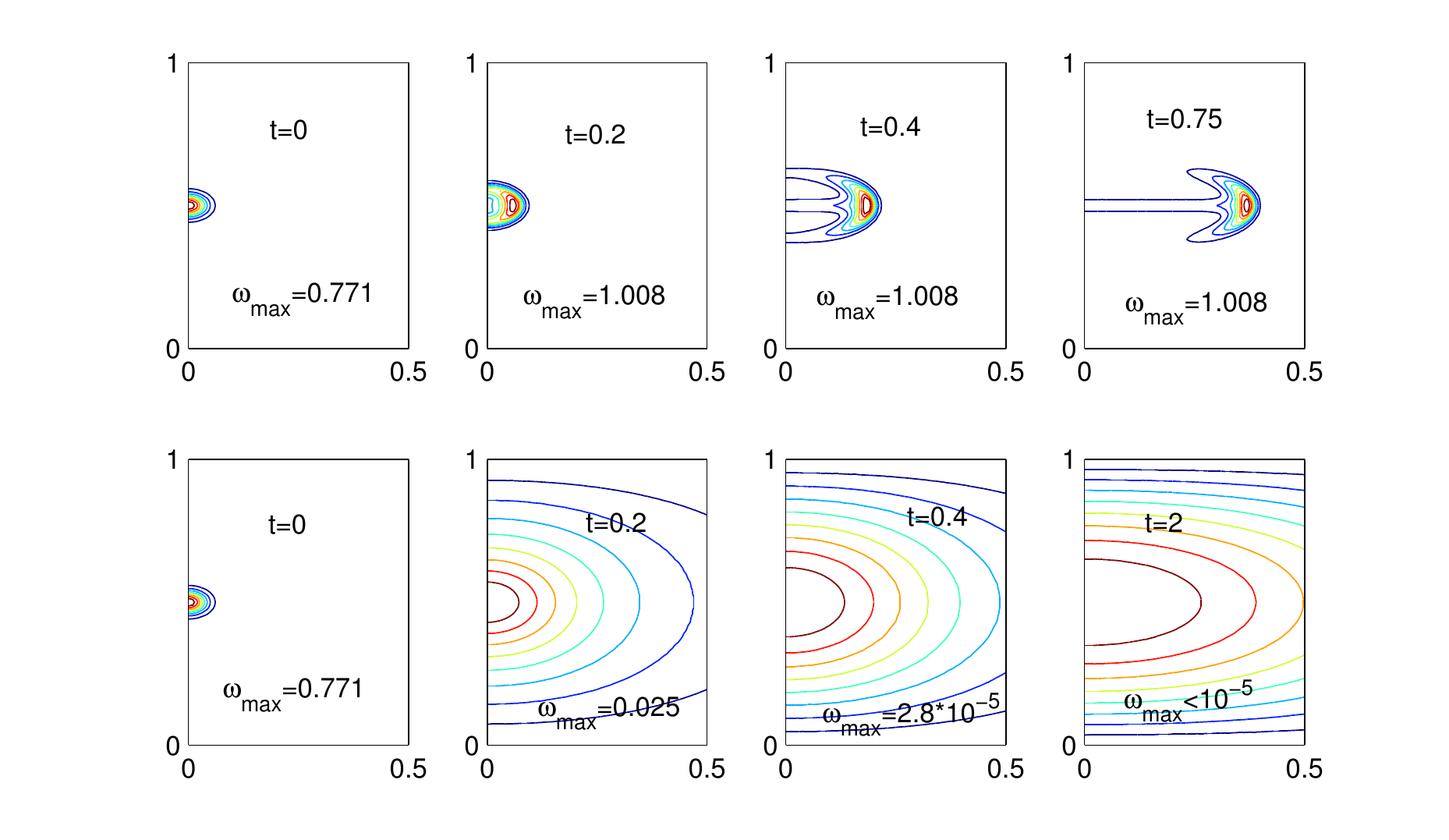}\label{ignition:fig:reaction_eps_05_a_0}}
\subfigure[]{
\includegraphics[scale=0.8, trim=0 0 0 0,clip=true]{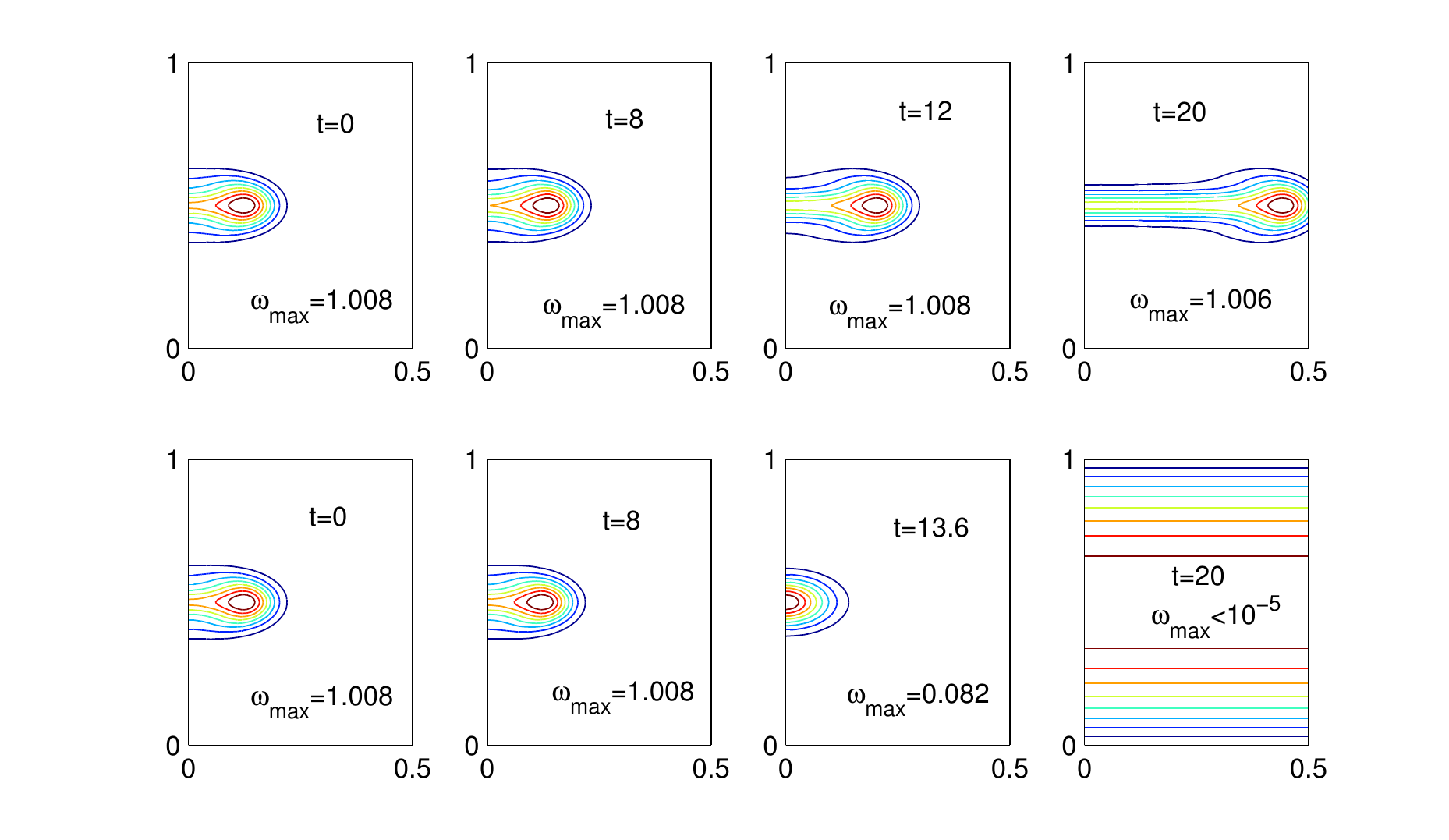}\label{ignition:fig:reaction_eps_012_a_0}}
\caption{Transient behaviour of a flame tube solution (shown at $t=0$) for a) $\epsilon=0.05$, b) $\epsilon=0.12$, with either an energy-increasing (upper figures) or energy-decreasing perturbation (lower figures), for  $\alpha=0$, $Ra=0$. Shown are 8 reaction rate contours equally spaced up to the maximum value $\omega_{\text{max}}$, which is labelled on each figure. The other parameters are given the fixed values $\alpha_h=0.85$, $\beta=10$, $Pr=1$ and $S=1$.}
\label{ignition:fig:reaction_eps_05_012_a_0}
\end{figure}

Illustrative examples of this behaviour for two selected values of $\epsilon$ are shown in figure \ref{ignition:fig:reaction_eps_05_012_a_0}. The upper rows in figures \ref{ignition:fig:reaction_eps_05_a_0} and \ref{ignition:fig:reaction_eps_012_a_0} show, for $\epsilon=0.05$ and $\epsilon=0.12$ respectively, reaction rate contours for a flame tube subject to an energy-increasing perturbation, which leads to a triple flame. The lower rows in figures \ref{ignition:fig:reaction_eps_05_a_0} and \ref{ignition:fig:reaction_eps_012_a_0} show a flame tube decaying to extinction due to an energy-decreasing perturbation.
\subsubsection{Effect of $\alpha$}
\begin{figure}
\centering
\subfigure[]{
\includegraphics[scale=0.9, trim=0 0 0 0,clip=true]{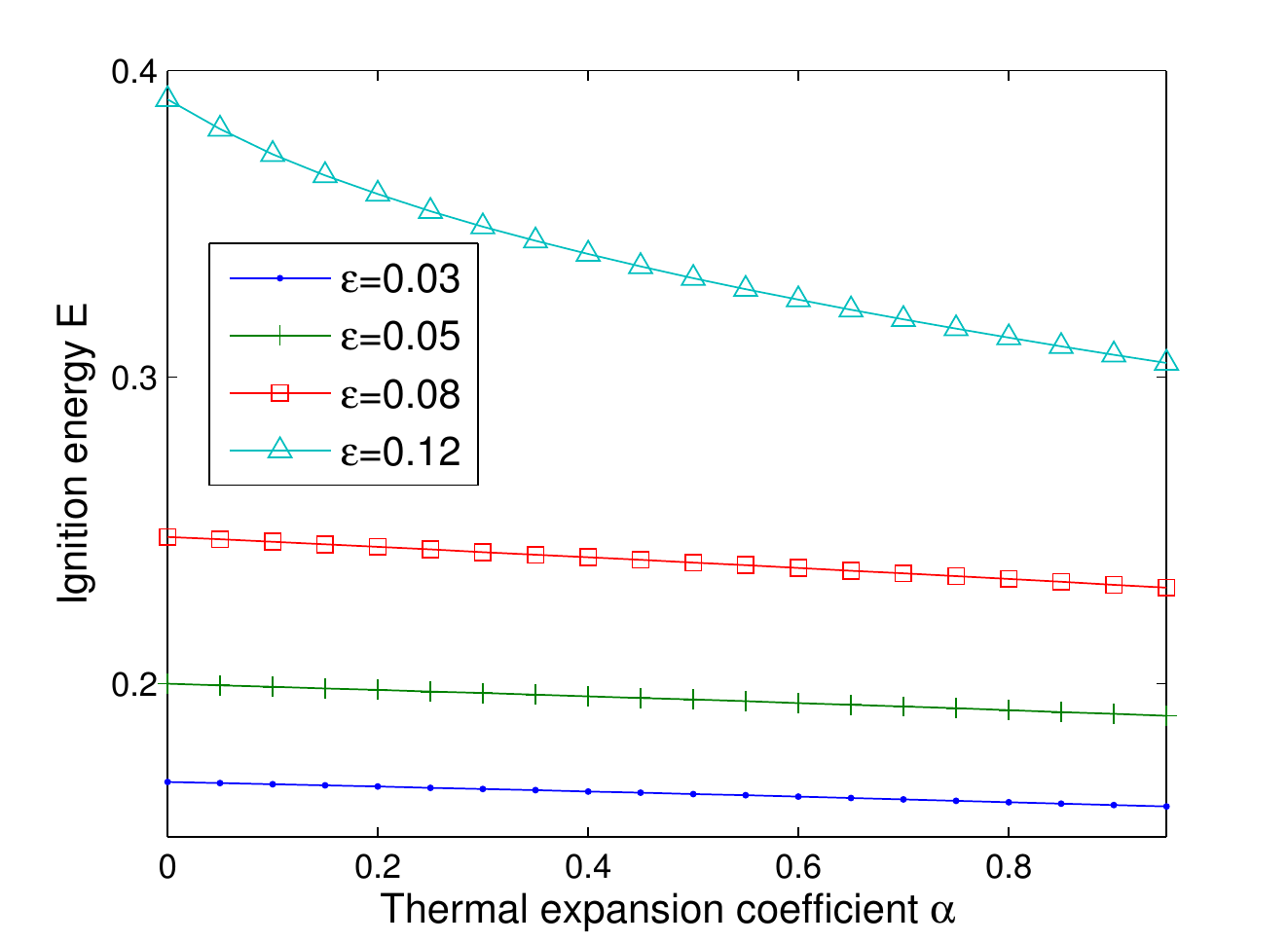}\label{ignition:fig:energy_alpha}}
\subfigure[]{
\includegraphics[scale=0.9, trim=0 0 0 0,clip=true]{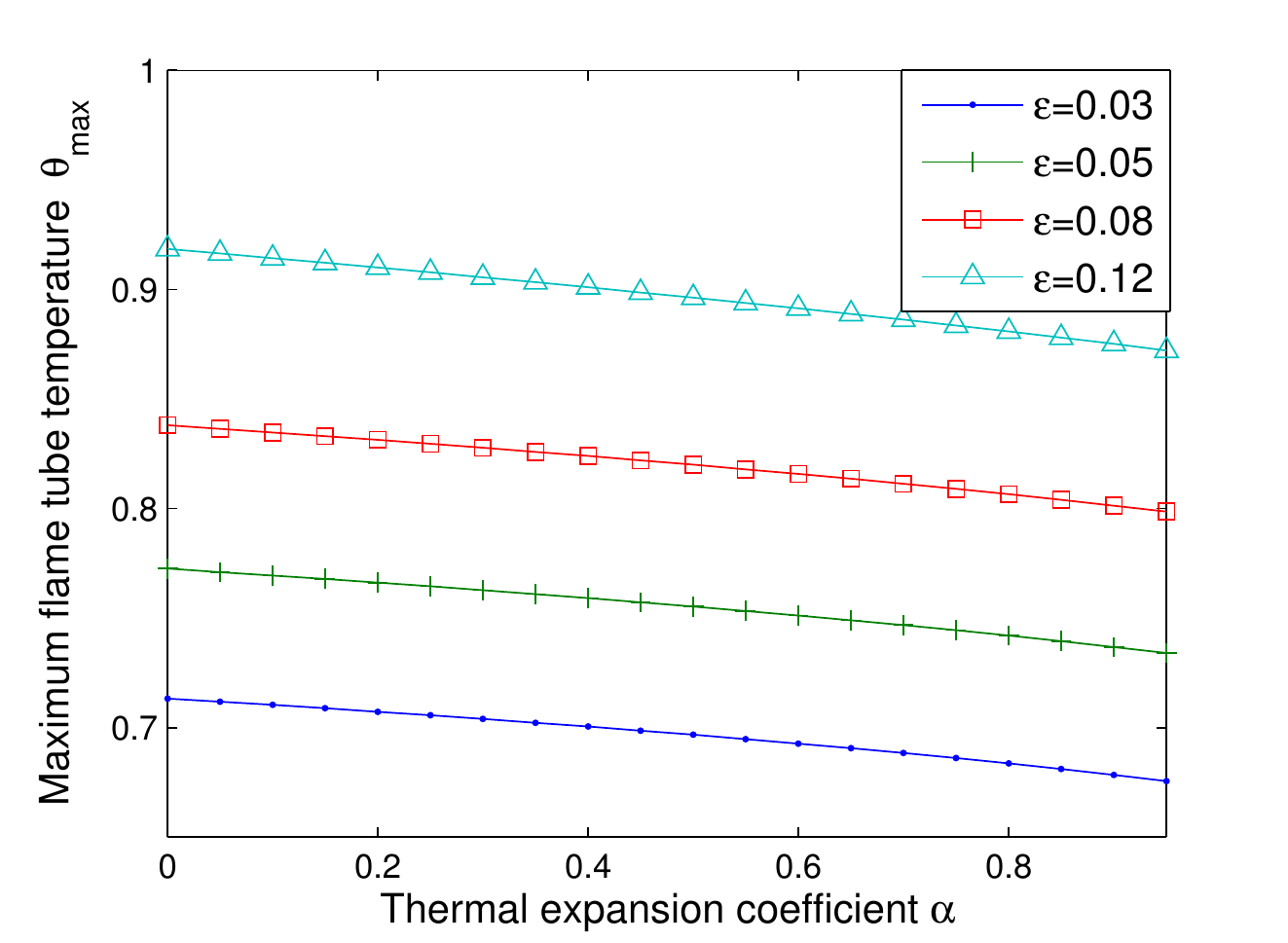}\label{ignition:fig:temp_alpha}}
\caption{(a) Ignition energy $E$ and (b) maximum flame tube temperature $\theta_{\text{max}}$ versus the thermal expansion coefficient $\alpha$ for selected values of $\epsilon$. The other parameters are given the fixed values $Ra=0$, $\alpha_h=0.85$, $\beta=10$, $Pr=1$ and $S=1$.}
\label{ignition:fig:energy_temp_alpha}
\end{figure}
\begin{figure}
\centering
\subfigure[]{
\includegraphics[scale=0.9, trim=0 0 0 0,clip=true]{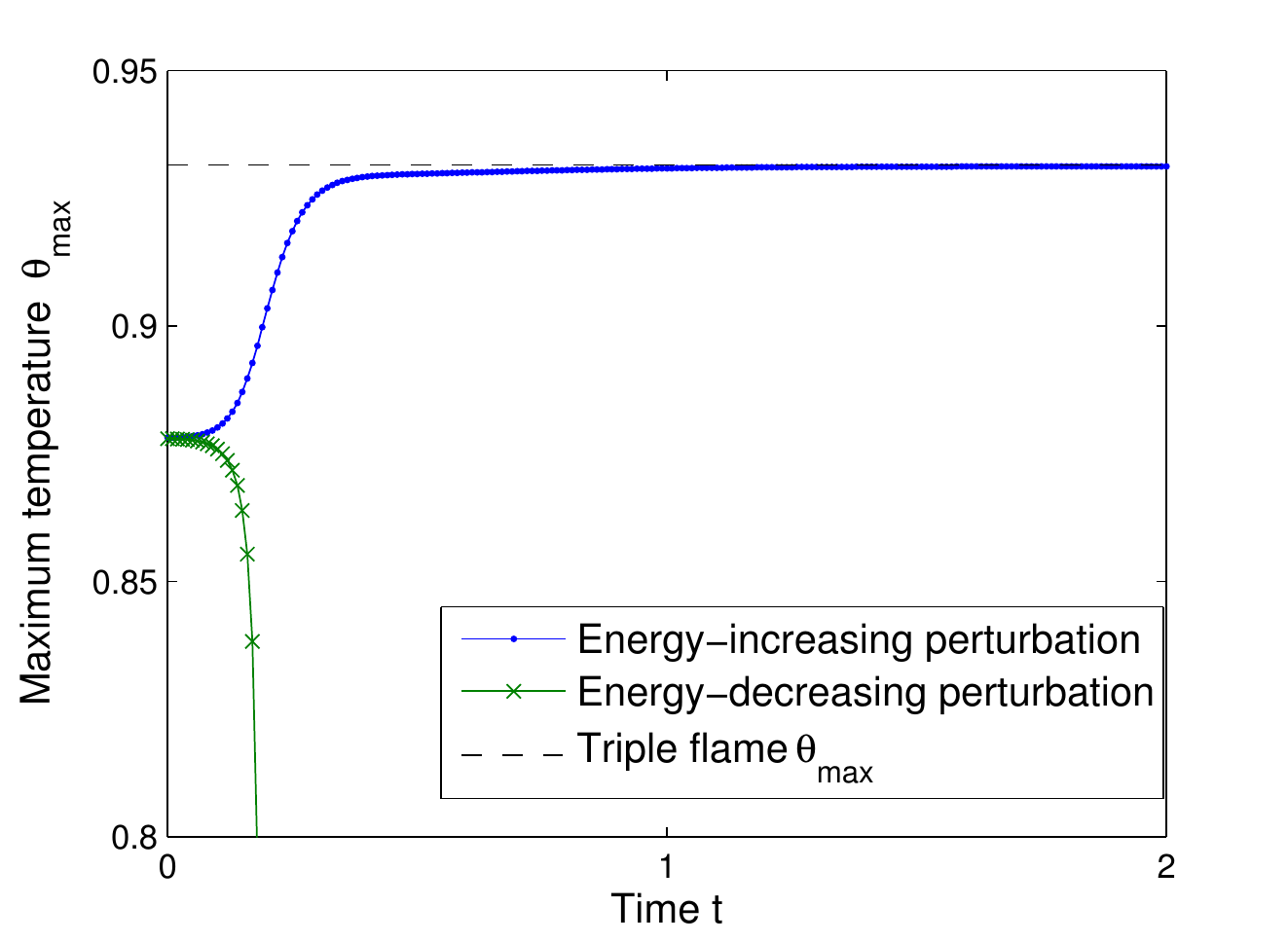}\label{ignition:fig:temp_e_12_a_085}}
\subfigure[]{
\includegraphics[scale=0.9, trim=0 0 0 0,clip=true]{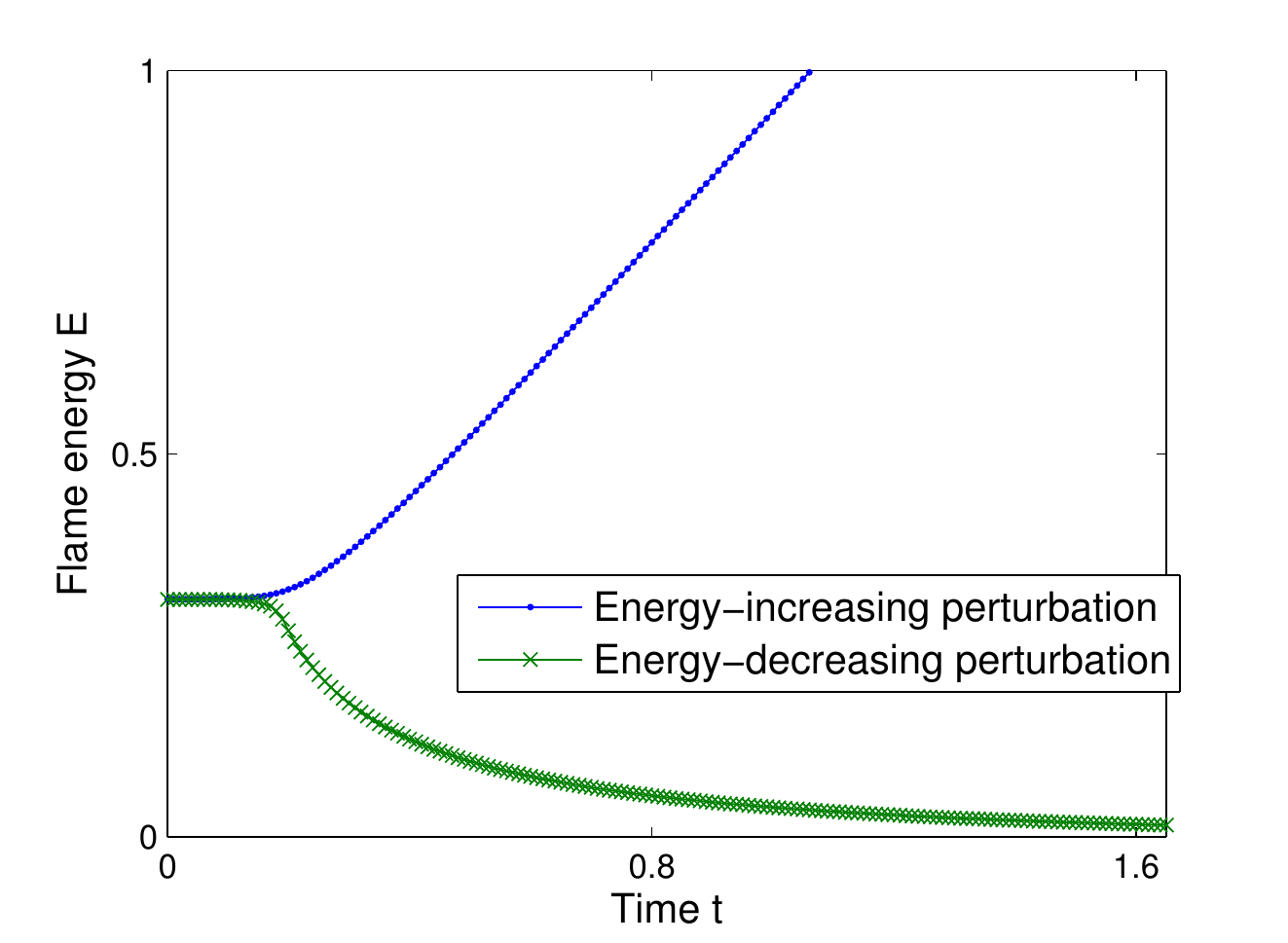}\label{ignition:fig:temp_e_12_a_085}}
\caption{Transient behaviour of flames with a small energy-decreasing or energy-increasing perturbation added to the unstable flame tube solution for $\epsilon=0.12$, $\alpha=0.85$ and $Ra=0$, showing a) maximum temperature $\theta_{max}$ and b) flame energy $E$. Also included in a) is the maximum temperature of a triple flame, which the ignited flame tends towards. The other parameters are given the fixed values $\alpha_h=0.85$, $\beta=10$, $Pr=1$ and $S=1$.}
\label{ignition:fig:energy_temp_e_12_a_085}
\end{figure}
\begin{figure}
\includegraphics[scale=0.8, trim=70 0 0 0,clip=true]{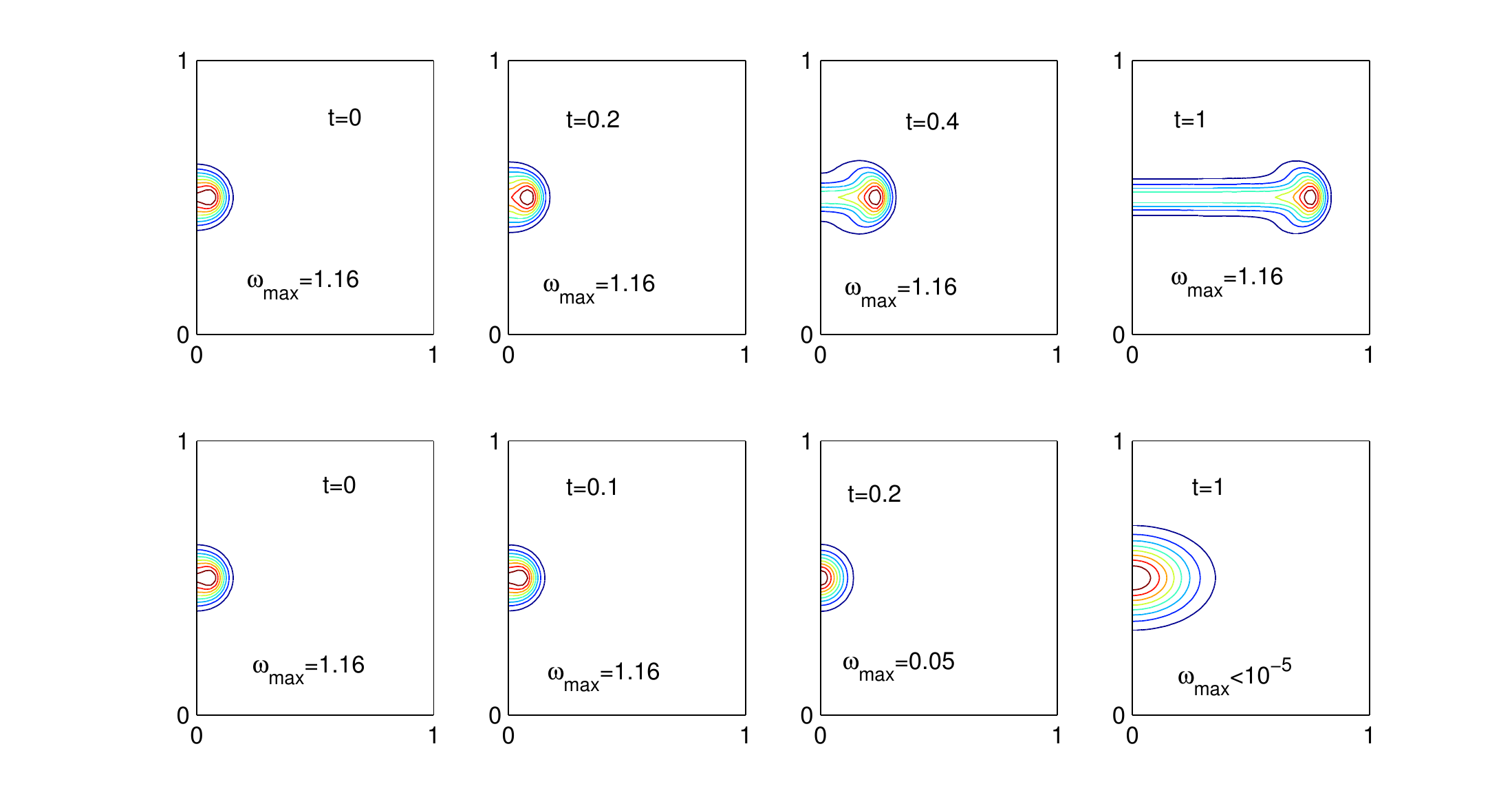}\label{ignition:fig:reaction_eps_012_a_85}
\caption{Transient behaviour of a flame tube solution (shown at $t=0$) for $\epsilon=0.12$, with either an energy-increasing (upper row) or energy-decreasing perturbation (lower row), for  $\alpha=0.85$, $Ra=0$. Shown are 8 reaction rate contours equally spaced up to the maximum value $\omega_{\text{max}}$, which is labelled on each figure. The other parameters are given the fixed values $\alpha_h=0.85$, $\beta=10$, $Pr=1$ and $S=1$.}
\label{ignition:fig:reaction_eps_012_a_85}
\end{figure}
In this section we investigate the effect of thermal expansion on the ignition of a triple flame in the absence of gravity. Figure \ref{ignition:fig:energy_temp_alpha} shows the ignition energy $E$ and the maximum flame tube temperature $\theta_{max}$ versus the thermal expansion coefficient $\alpha$ for selected values of $\epsilon$. It can be seen that, for a fixed value of $\epsilon$, increasing $\alpha$ leads to a monotonic decrease in the ignition energy $E$. A key result of these calculations is that for all values of $\alpha$, the flame tube solutions have $u=v=0$. This is physically plausible since $u=v=p=0$ solves the stationary form of the governing equations \eqref{ignition:mix:nondim1}-\eqref{ignition:mix:nondim8} if $Ra=0$, as well as the boundary conditions \eqref{ignition:mix:bc1}-\eqref{ignition:mix:bc4}. This is a significant simplification of the governing equations and represents an efficient way to numerically calculate flame tube solutions, if gravity is not accounted for.

Now we investigate the transient behaviour of the unstable flame tube solutions when thermal expansion is accounted for, by setting $\alpha=0.85$. Figure \ref{ignition:fig:energy_temp_e_12_a_085}, with $\epsilon=0.12$, plots a) maximum temperature $\theta_{\text{max}}$ and b) flame energy $E$ against time after an energy-increasing and an energy-decreasing perturbation.  As can be seen from the figure, after an energy-increasing perturbation the maximum temperature increases monotonically from that of the flame tube to that of the steadily propagating triple flame, which has propagation speed given in figure \ref{ignition:fig:steadytriple}. An energy-decreasing pertubation again leads to extinction.

An illustrative example of this behaviour for $\epsilon=0.12$ is shown in figure \ref{ignition:fig:reaction_eps_012_a_85}. The upper row shows the reaction rate for a flame tube subject to an energy-increasing perturbation, which leads to a triple flame. The lower row in figure \ref{ignition:fig:reaction_eps_012_a_85} shows a flame tube decaying to extinction due to an energy-decreasing perturbation.
\subsubsection{Effect of $Ra$}
\begin{figure}
\centering
\subfigure[]{
\includegraphics[scale=0.9, trim=0 0 0 0,clip=true]{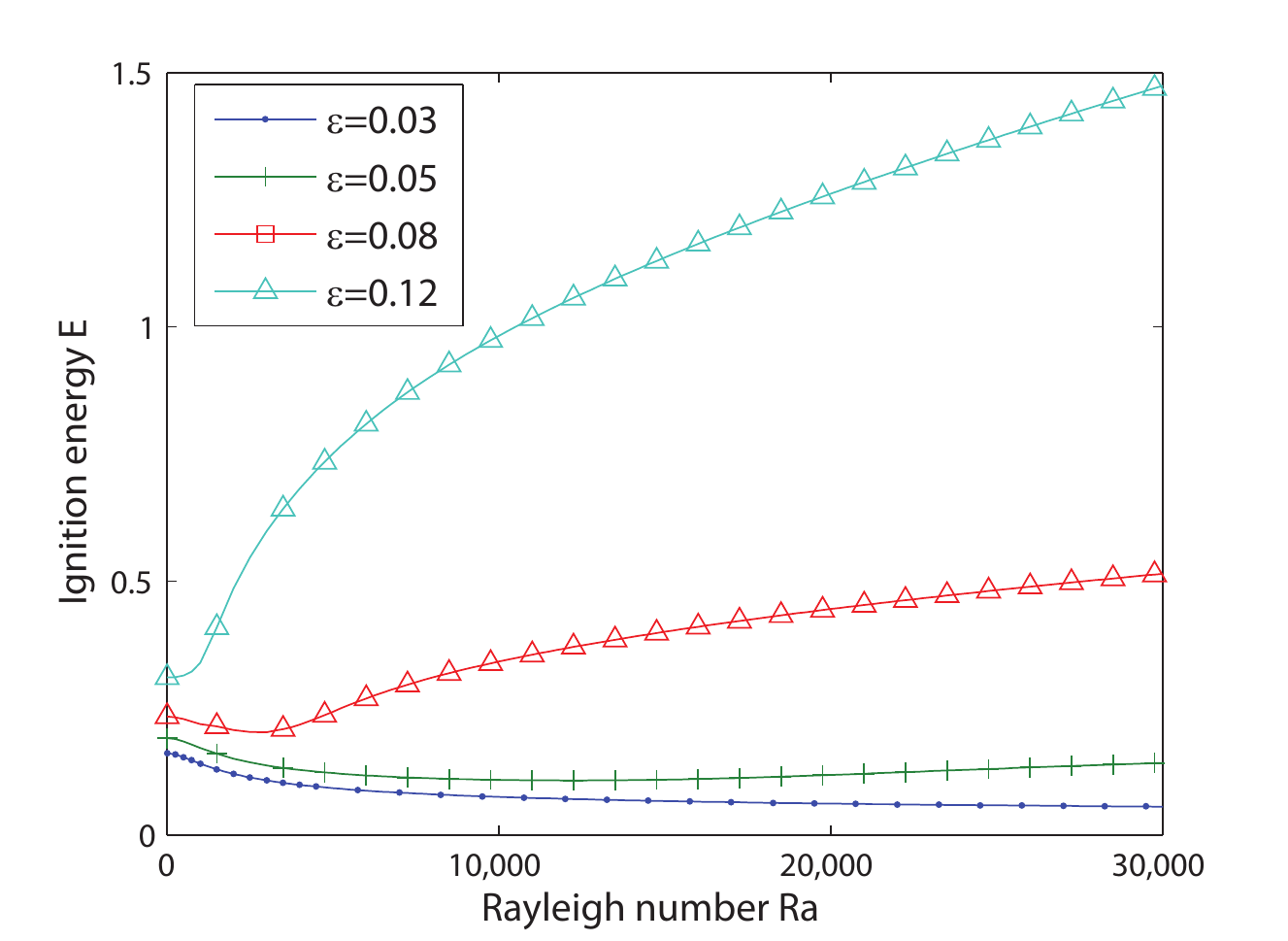}\label{ignition:fig:energy_ra}}
\subfigure[]{
\includegraphics[scale=0.9, trim=0 0 0 0,clip=true]{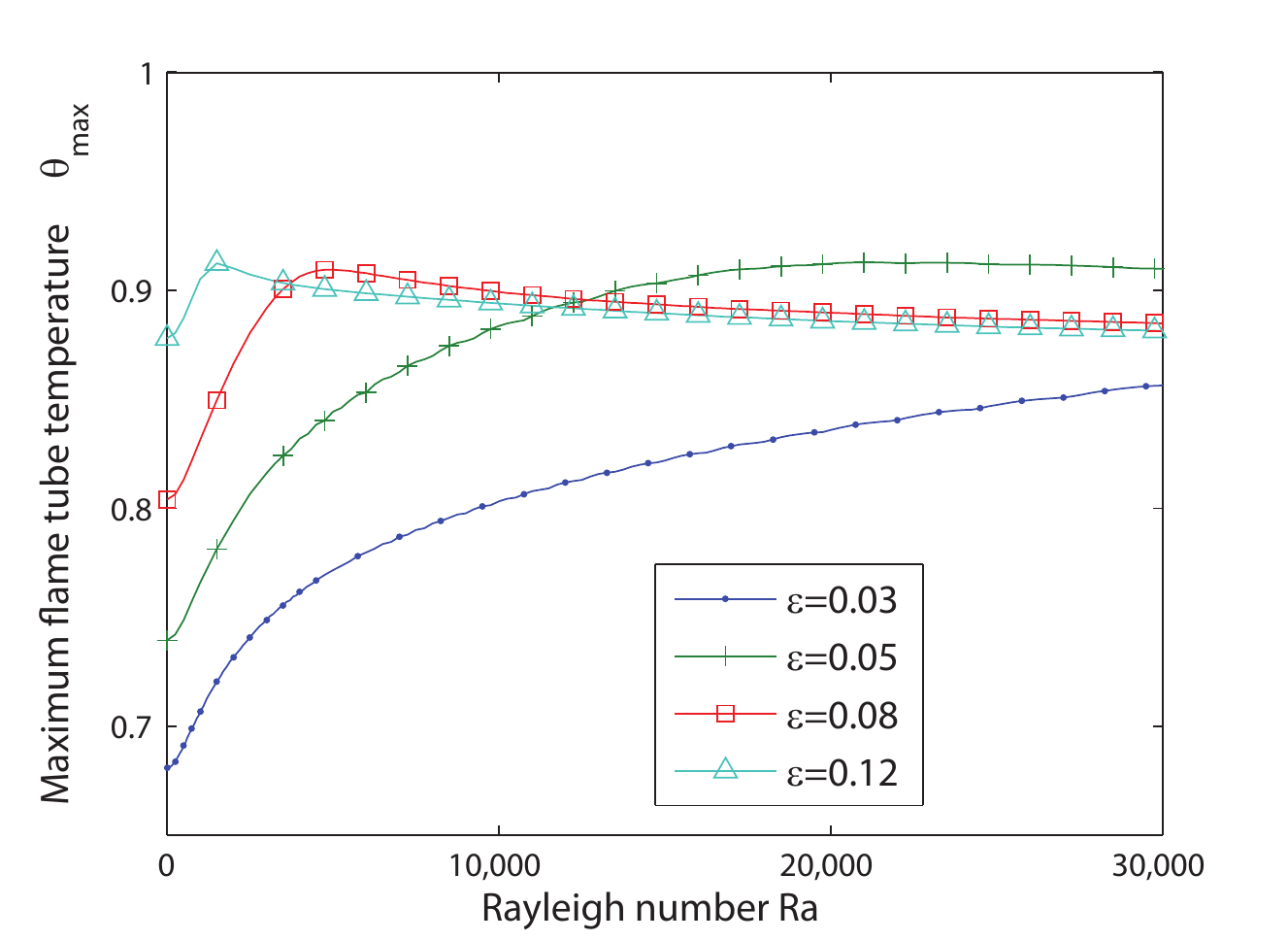}\label{ignition:fig:temp_ra}}
\caption{(a) Ignition energy $E$ and (b) maximum flame tube temperature $\theta_{\text{max}}$ versus the Rayleigh number $Ra$ for selected values of $\epsilon$. The other parameters are given the fixed values $\alpha=0.85$, $\alpha_h=0.85$, $\beta=10$, $Pr=1$ and $S=1$. Note that the line markers are for illustration purposes and calculations were also performed for values of $Ra$ in between the values labelled with markers.}
\label{ignition:fig:energy_temp_ra}
\end{figure}
In this section we investigate the combined effect of thermal expansion and gravity on the initiation of a triple flame.  Figure \ref{ignition:fig:energy_temp_ra} shows the ignition energy $E$ and the maximum flame tube temperature $\theta_{max}$ versus the Rayleigh number $Ra$ for selected values of $\epsilon$. It can be seen that the relationship between $E$ and $Ra$ is dependent upon the value of $\epsilon$, however for all values of $Ra$ an increase in $\epsilon$ leads to an increase in the ignition energy $E$. For large values of $Ra$, an increase in $\epsilon$ leads to a larger increase in $E$ than for smaller values of $Ra$.

Unlike in the previous sections, when gravity is present the flame tube solutions do induce a flow due to the temperature gradient from the flame tube at $x=0$ to the cold gas at $x=\pm\infty$, as was found in the context of triple flames in Chapter \ref{chapter:triple}. An important result that can be seen in figure \ref{ignition:fig:energy_temp_ra} is that flame tube solutions still exist for values of $Ra$ where the planar diffusion flame is unstable - this will be discussed further in the following section.

\begin{figure}
\centering
\subfigure[]{
\includegraphics[scale=0.9, trim=0 0 0 0,clip=true]{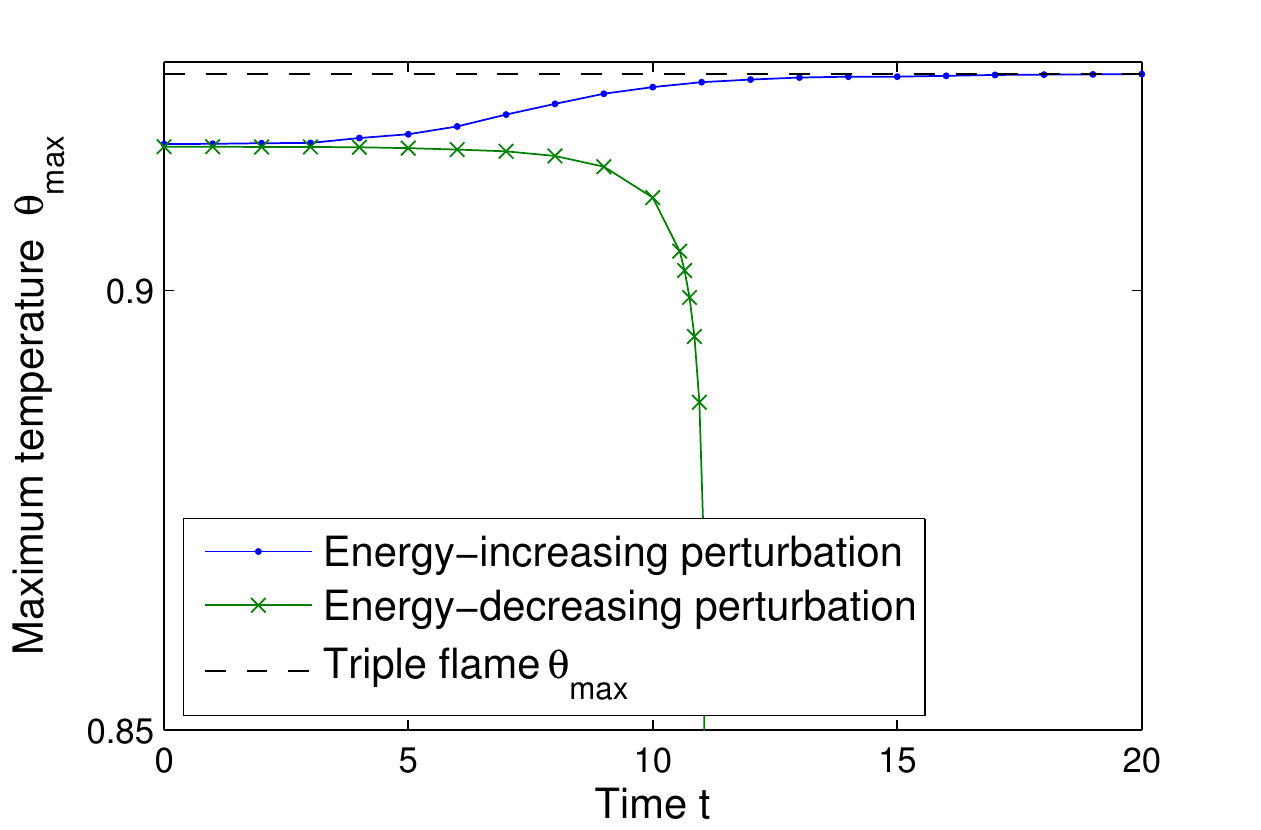}\label{ignition:fig:temp_e_133_ra_10000}}
\subfigure[]{
\includegraphics[scale=0.9, trim=0 0 0 0,clip=true]{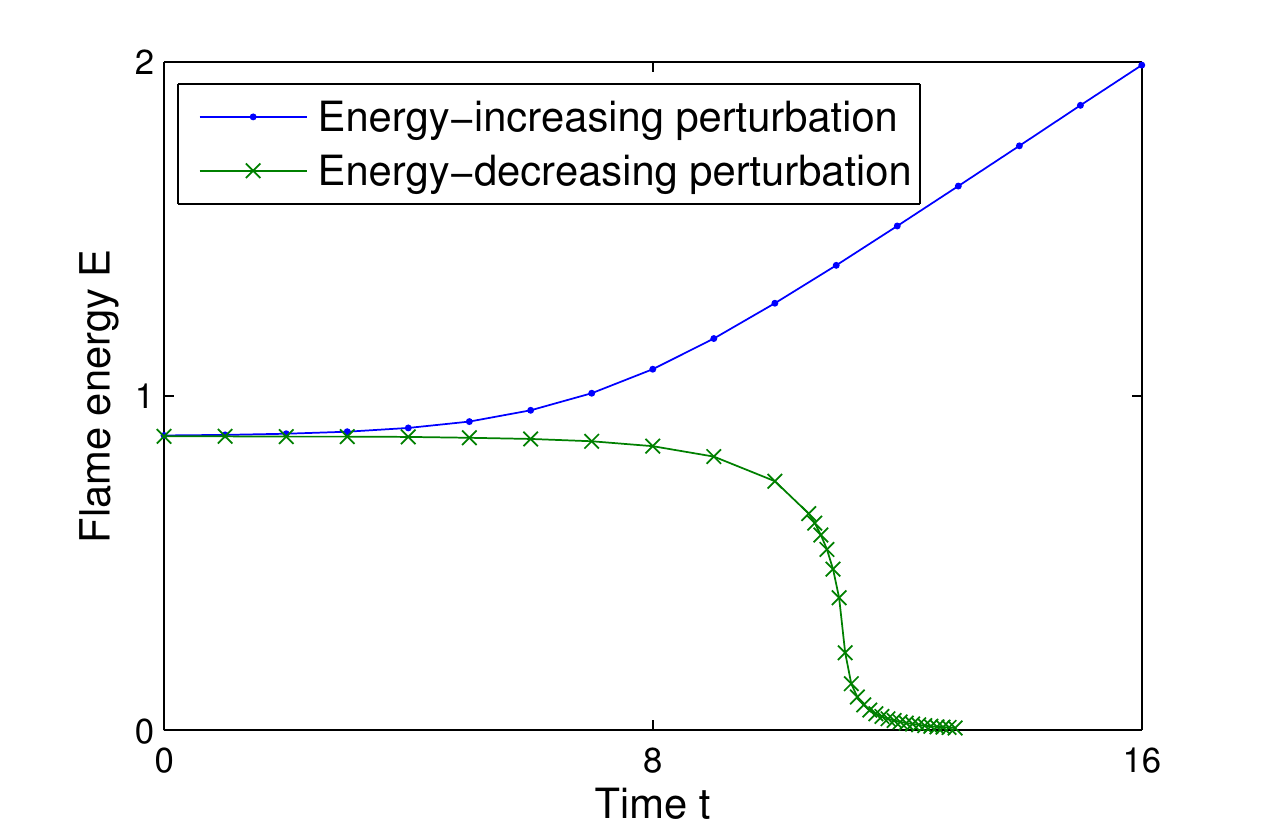}\label{ignition:fig:energy_e_133_ra_10000}}
\caption{Transient behaviour of flames with a small energy-decreasing or energy-increasing perturbation added to the unstable flame tube solution for $\epsilon=0.1325$, showing a) maximum temperature $\theta_{max}$ and b) flame energy $E$, for $\alpha=0.85$ and $Ra=10000$. Also included in a) is the maximum temperature of a triple flame, which the ignited flame tends towards. The other parameters are given the fixed values $\alpha_h=0.85$, $\beta=10$, $Pr=1$ and $S=1$.}
\label{ignition:fig:energy_temp_e_133_ra_10000}
\end{figure}
\begin{figure}
\includegraphics[scale=0.6, trim=50 0 0 0,clip=true]{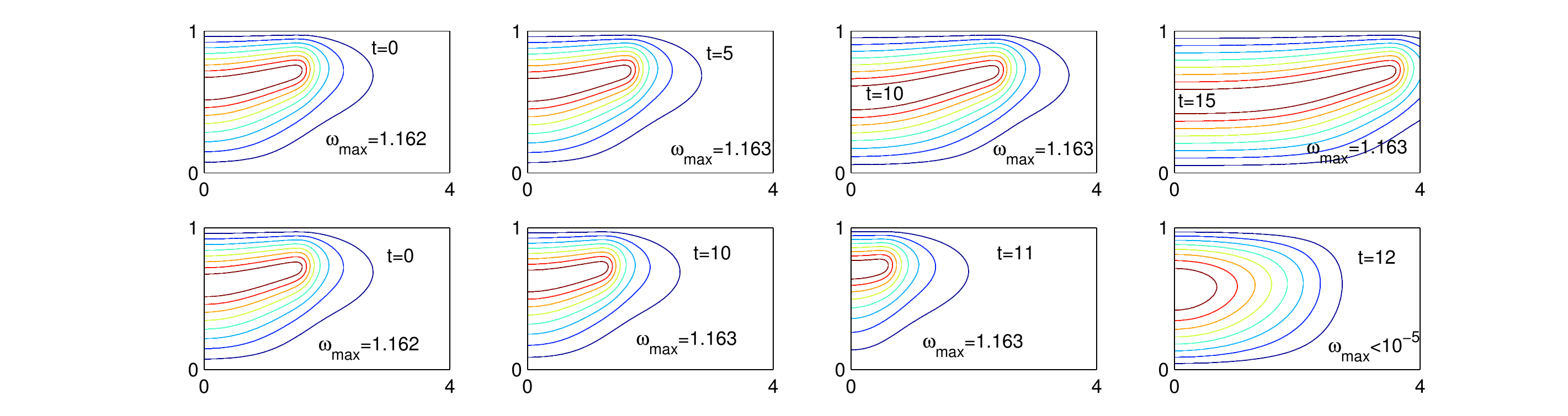}
\caption{Transient behaviour of a flame tube solution (shown at $t=0$) for $\epsilon=0.1325$, with either an energy-increasing (top row) or energy-decreasing perturbation (bottom row), for $\alpha=0.85$ and $Ra=10000$. Shown are 8 reaction rate contours equally spaced up to the maximum value $\omega_{\text{max}}$, which is labelled on each figure. The other parameters are given the fixed values $\alpha_h=0.85$, $\beta=10$, $Pr=1$ and $S=1$.}\label{ignition:fig:reaction_eps_0133_ra_10000}
\end{figure}

Now we investigate the transient behaviour of the unstable flame tube solutions when thermal expansion and gravity are accounted for, by setting $\alpha=0.85$ and $Ra=10000$. Figure \ref{ignition:fig:energy_temp_e_133_ra_10000} plots a) maximum temperature $\theta_{\text{max}}$ and b) flame energy $E$ against time after an energy-increasing and an energy-decreasing perturbation, for $\epsilon=0.1325$.  As can be seen from the figure, after an energy-increasing perturbation the maximum temperature increases monotonically from that of the flame tube to that of the steadily propagating triple flame, which has propagation speed given in figure \ref{ignition:fig:steadytriple}. An energy-decreasing pertubation again leads to extinction. An illustrative example of this behaviour for $\epsilon=0.1325$ is shown in figure \ref{ignition:fig:reaction_eps_0133_ra_10000}.
\subsection{Gravitational instability of triple flames}
\begin{figure}
\includegraphics[scale=0.9, trim=0 0 0 0,clip=true]{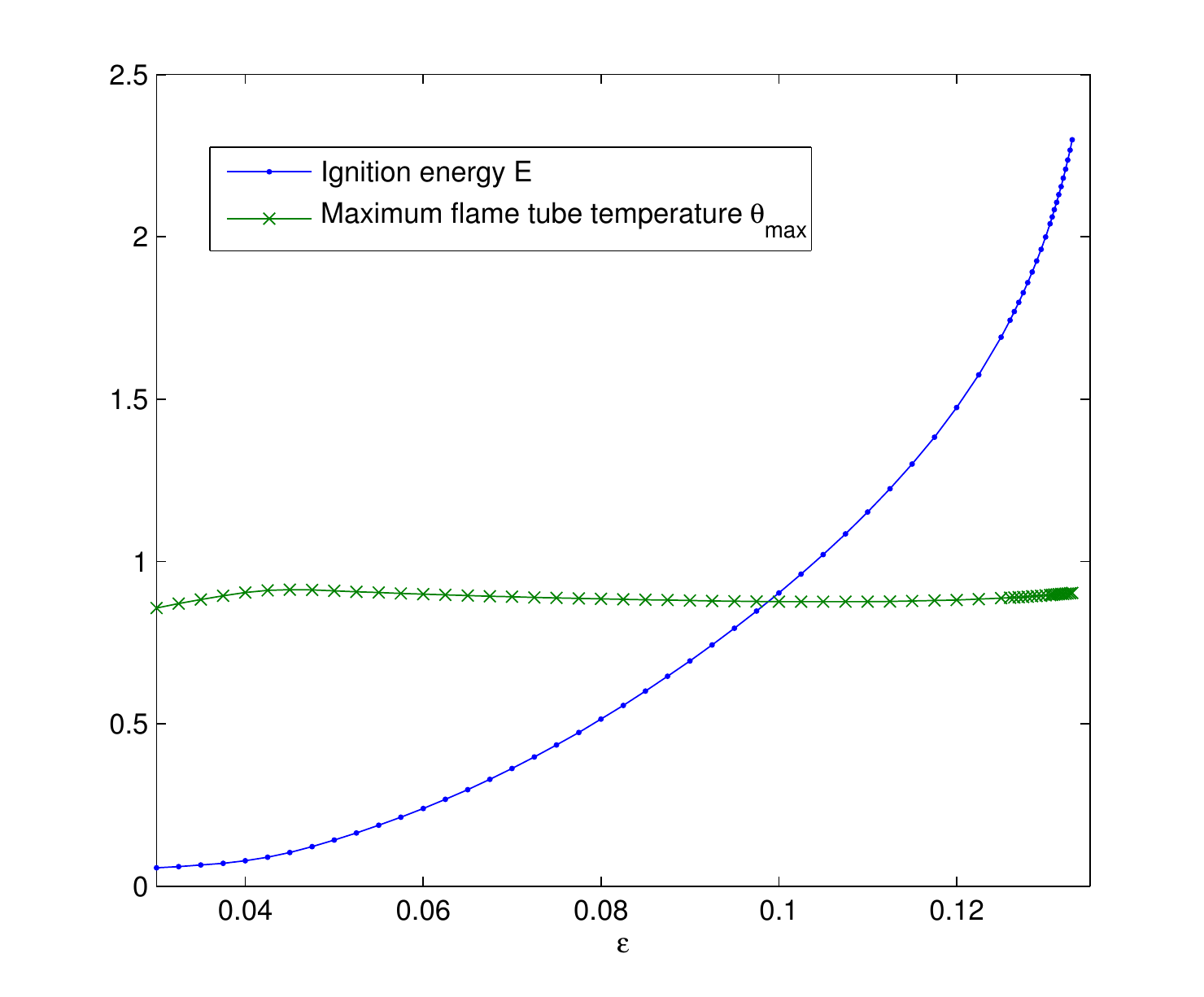}
\caption{Ignition energy $E$ and maximum flame tube temperature $\theta_{\text{max}}$ versus the $\epsilon$ for $Ra=30000$, which is larger than the critical Rayleigh number for the gravitational instability of the planar diffusion flame. The other parameters are given the fixed values $\alpha=0.85$, $\alpha_h=0.85$, $\beta=10$, $Pr=1$ and $S=1$.}\label{ignition:fig:energy_temp_epsilon_ra_30000}
\label{ignition:fig:energy_temp_ra_30000}
\end{figure}
In this section we investigate the initiation of triple flames in situations where the underlying diffusion flame is unstable. As was demonstrated in figure \ref{ignition:fig:energy_temp_ra} of the previous section, unstable flame tube solutions still exist for values of the Rayleigh number where the underlying planar diffusion flame is unstable. Therefore, we can still provide a criterion for the ignition energy of a triple flame in this case. Plots of the ignition energy $E$ and maximum flame tube temperature $\theta_{\text{max}}$ versus $\epsilon$ for $Ra=30000$ are provided in figure \ref{ignition:fig:energy_temp_epsilon_ra_30000}. For this value of the Rayleigh number, the underlying diffusion flame is unstable for all values of $\epsilon$, as can be seen in figure \ref{triple:fig:epsilonvsra}.
\begin{figure}
\includegraphics[width=\textwidth, trim=53 0 25 0,clip=true]{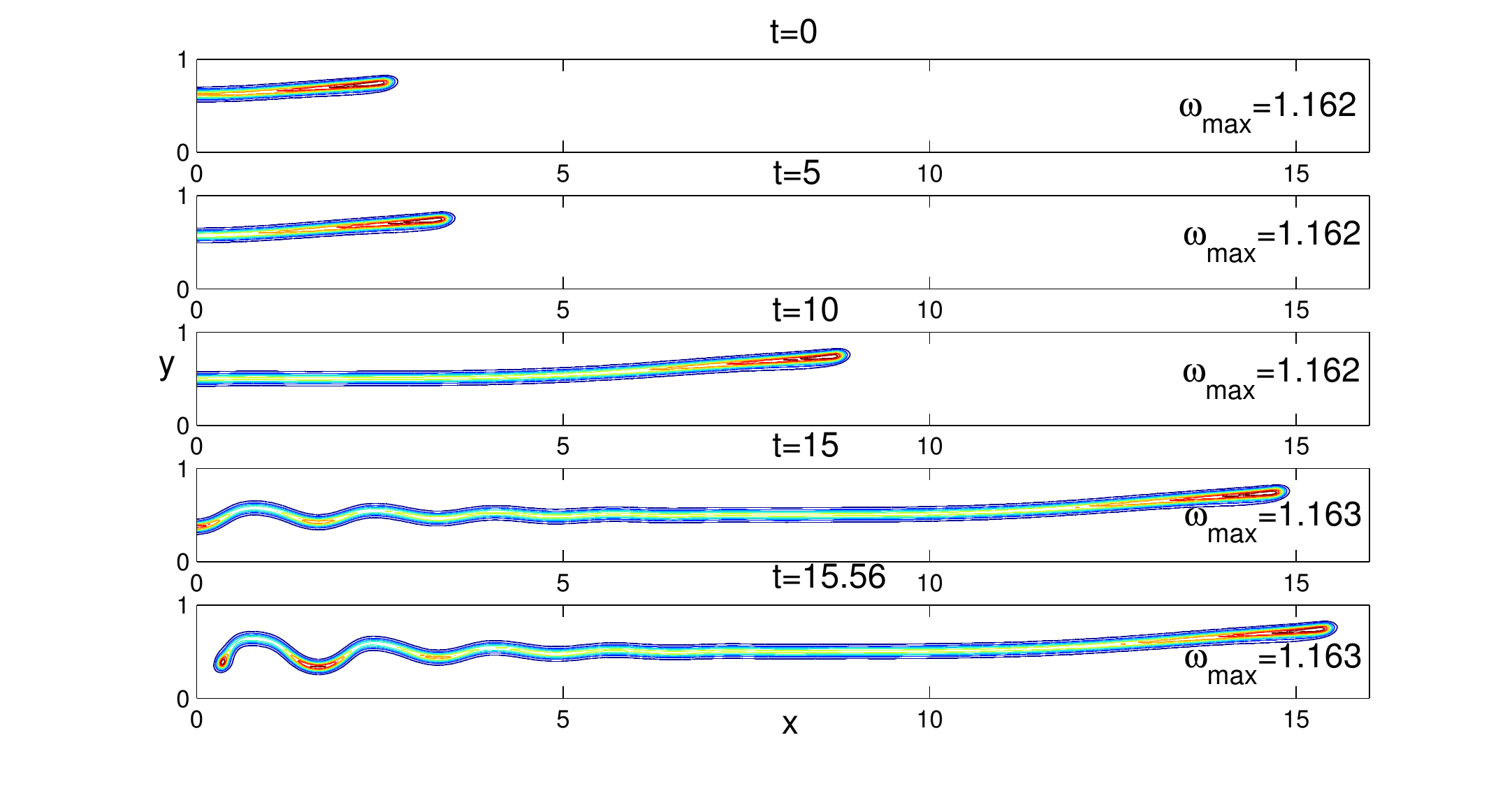}
\caption{Transient behaviour of a flame tube solution (shown at $t=0$) for $\epsilon=0.135$, with an energy-increasing perturbation, for $\alpha=0.85$ and $Ra=30000$. Shown are 8 reaction rate contours equally spaced up to the maximum value $\omega_{\text{max}}$, which is labelled on each figure. The other parameters are given the fixed values $\alpha_h=0.85$, $\beta=10$, $Pr=1$ and $S=1$.}\label{ignition:fig:reaction_eps_0135_ra_30000}
\end{figure}
\begin{figure}
\includegraphics[scale=1, trim=0 0 0 0,clip=true]{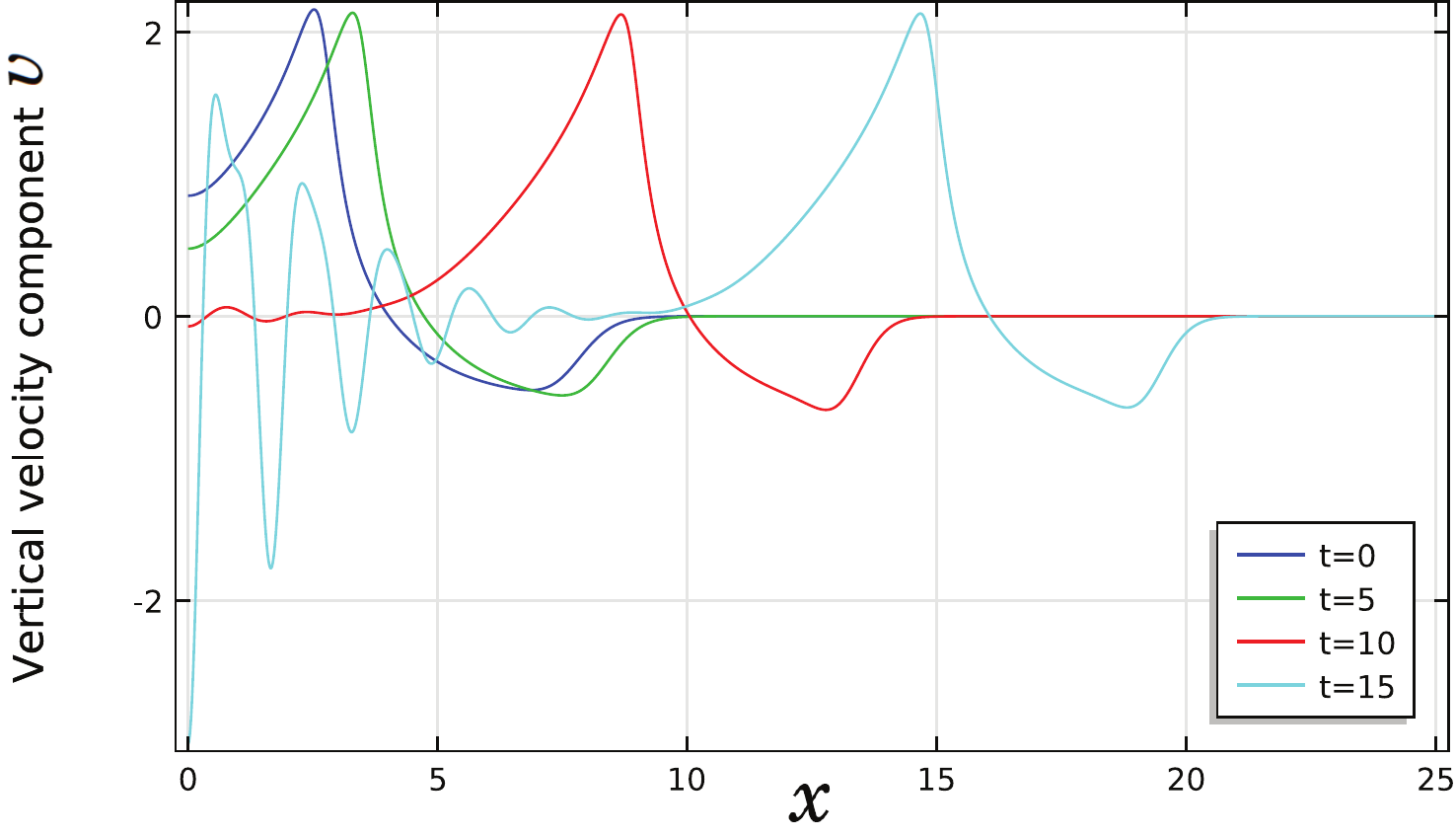}
\caption{Vertical component of the velocity $v$ along the centreline of the channel located at $y=0.5$, at several values of time $t$, for $\alpha=0.85$ and $Ra=30000$. The other parameters are given the fixed values $\alpha_h=0.85$, $\beta=10$, $Pr=1$ and $S=1$.}\label{ignition:fig:vertical_instability}
\end{figure}

We can therefore use flame tubes (with energy shown in figure \ref{ignition:fig:energy_temp_epsilon_ra_30000}) as an initial condition for time-dependent calculations to investigate the evolution of triple flames with a trailing diffusion flame which is unstable due to gravitational effects. The transient behaviour of the triple flames that arise after an energy-increasing perturbation to the flame tube solutions for $Ra=30000$ is shown in figure \ref{ignition:fig:reaction_eps_0135_ra_30000}, which plots reaction rate contours at several points in time. It can be seen that initially, a triple flame propagates, similarly to the flames that were shown in the previous section. After a certain amount of time, however, the trailing diffusion flame becomes unstable due to gravity, causing convection in the fluid, which leads to a cellular trailing diffusion flame. The induced convection can be strong enough to cause local extinction, which occurs first at the axis of symmetry located at $x=0$, as can be seen from figure \ref{ignition:fig:reaction_eps_0135_ra_30000}.

The induced convection can be seen in figure \ref{ignition:fig:vertical_instability}, which plots the vertical component of the velocity $v$ along the centreline of the channel located at $y=0.5$, at several values of the time $t$. The figure shows the vortex induced by the temperature gradient from the hot triple flame to the cold unburnt gas, as discussed in detail in \S \ref{triple:sec:effectofgravity} of Chapter \ref{chapter:triple}. This vortex propagates along the channel with the tip of the triple flame. The triple flame initially leaves behind a trailing diffusion flame. As can be seen from figure \ref{ignition:fig:vertical_instability}, at approximately $t=10$ the diffusion flame starts to induce convection. The magnitude of the induced convection increases, with the largest magnitude of $v$ occurring at the axis of symmetry $x=0$, which causes the local extinction that was seen in figure \ref{ignition:fig:reaction_eps_0135_ra_30000}.
\section{Conclusion}
\label{ignition:sec:conc}
In this chapter we have investigated the combined effect of thermal expansion and gravity on the ignition energy of triple flames by studying stationary, non-propagating solutions of the governing equations, which have been referred to as ``flame tubes". We have also described the transient initiation of two-dimensional triple flames in the mixing layer, using the (unstable) flame tube solutions as initial conditions. Finally, we have investigated the time-dependent behaviour, from initiation to instability, of triple flames in situations where the underlying diffusion flame is unstable due to gravitational effects.

The chapter provides several new contributions. It is the first work to provide a detailed parametric study of ``flame tube" solutions in two-dimensional mixing layers, providing an equivalent criterion for the ignition energy of triple flames to the one provided by Zeldovich flame balls for premixed flames and corresponding flame balls in axisymmetric mixing layers. It has been found that the flame tube solutions exist only for $\epsilon<\epsilon_m$, where $\epsilon_m$ is the value of the flame-front thickness $\epsilon$ for which a triple flame has zero propagation speed. Flame tubes are not expected to exist in the asymptotic limit $\epsilon \to 0$, but have been found to exist numerically for small values of $\epsilon$. The flame tubes are unstable and it has been found that an ``energy-increasing" perturbation leads to the propagation of a triple flame and an ``energy-decreasing" perturbation leads to extinction.

This study is also the first to describe the initiation of triple flames in a horizontal channel with the combined effect of thermal expansion and gravity taken into account. In cases where gravity is not present, it is found that the flow velocity $u=v=0$, so that the Navier--Stokes equations do not need to be solved. However, when gravity is present, a flow is induced by the temperature gradient from the flame tube to the cold gas. The relationship between $Ra$ and the energy $E$ of the flame tubes is found to depend on $\epsilon$. A key result from these calculations is that flame tube solutions are still found to exist for values of the Rayleigh number above its critical value for the gravitational instability of the planar diffusion flame. Therefore a criterion for the ignition energy of a triple flame can still be given even though its trailing diffusion flame is expected to be unstable.

The final contribution of the chapter is the investigation of the transient behaviour of a triple flame from its initiation (using a flame tube as initial condition) to the eventual instability of its trailing diffusion flame for large enough values of $Ra$. It has been found that initially a triple flame forms and propagates into the unburnt gas. After a certain amount of time, the trailing diffusion flame becomes unstable and forms convection rolls, as predicted in Chapter \ref{chapter:diffusion}. These can lead to local extinction of the trailing diffusion flame, which occurs first at the axis of symmetry located at $x=0$.

\chapter{Conclusion}
\label{chapter:conc}
The main aim of this thesis has been to assess the combined effect of thermal expansion and gravity on the propagation and stability of flames in inhomogeneous mixtures. Flame propagation has been investigated in situations where the inhomogeneity is prescribed in the unburned gas, either a) as a non-uniform flow field against which a premixed flame propagates or b) as a stratification of the fuel and oxidiser, leading to the propagation of a triple flame.

In \textbf{Chapter \ref{chapter:premixed1}}, we have investigated the propagation of a premixed flame through a narrow channel against a Poiseuille flow of large amplitude, taking the effect of the flame on the flow into account through the action of thermal expansion. The problem has been studied analytically to determine the effective propagation speed $U_T$ for a Peclet number $\Pe=O(1)$, in the thick flame asymptotic lmit $\epsilon \to \infty$ with both finite and asymptotically infinite values of the activation energy $\beta$.  The limit $\epsilon \to \infty$, with $\Pe=O(1)$ is equivalent to taking the amplitude $A$ of the Poiseuille flow $A\to\infty$, with $\Pe=O(1)$. It has been found that, in the distinguished limit considered, a two-dimensional premixed flame propagating through a rectangular channel against a Poiseuille flow can be described by a problem which corresponds to the propagation of a one-dimensional premixed flame with an effective diffusion coefficient, given by
\begin{gather*}
D_{\text{eff}}=D_T\left(1+\frac{8}{945} \Pe^2 \right),
\end{gather*}
in the constant density case and
\begin{gather*}
D_{\text{eff}}=D_T\left(1+\frac{8}{945} \Pe^2 \frac{\tilde{\rho}^2}{\tilde{\rho}^2_u} \right),
\end{gather*}
in the variable density case. These effective values correspond to those found in studies of enhanced dispersion due to a Poiseuille flow in non-reactive fluids, known as Taylor dispersion. A premixed flame propagating through a channel in the limit $\epsilon \to \infty$, with $\Pe=O(1)$ can therefore be considered to be in the Taylor regime. The derived one-dimensional problem has been solved numerically for a full range of values of the Peclet number and activation energy. An analytical solution to the one-dimensional problem has also been obtained in the limit $\beta \to \infty$ in both the constant density and variable density cases. Further results have been obtained by solving the full system of governing equations for a wide range of values of the parameters. The asymptotic results have been found to show strong agreement with the numerical results.

The analytical result obtained in Chapter \ref{chapter:premixed1}, given by (\ref{result:infinitebeta}), predicts that in the thick flame limit considered, the effective propagation speed has a value that depends only on the Peclet number. This result may be used to provide a possible explanation of the so-called bending effect of the turbulent premixed flame speed when plotted in terms of the turbulence intensity for fixed values of the Reynolds number (see e.g. \cite[][]{bradley1992fast,ronney1995some}). Our distinguished limit, namely $\epsilon \to \infty$ with $\Pe$ fixed (note that the Reynolds number and Peclet number are equal if $Pr=1$), mimics the experimental conditions of \citet{bradley1992fast} and can be used to shed some light on the experimental findings. A more thorough discussion of the relevance of the asymptotic results obtained in the thick flame limit to the bending effect is provided in \textbf{Chapter \ref{chapter:premixed2}}. In this chapter, further asymptotic results have been provided, in order to provide a full asymptotic picture of the relationship between the effective propagation speed and the parallel flow amplitude $A$. A numerical study for a full range of values of $\Pe$ and $A$ has also been undertaken.  The numerical results show that as the prescribed inflow amplitude $A$ increases to large values, the effective propagation speed $U_T$ approaches a constant value, which depends on the Peclet number. This result is in line with both the asymptotic results \eqref{result:infinitebeta} and \eqref{premixed2:res:2} and with the available experimental results on turbulent combustion. The asymptotic and numerical results both show that for small values of $A$, the relationship between $U_T$ and $A$ is linear. Finally, we have investigated the effect of thermal expansion on the problem; it has been found that for variable density premixed flames the effective propagation speed still approaches a constant value which depends on the Peclet number for large values of $A$, which again mimics the behaviour of the turbulent premixed flame experimental findings.

In \textbf{Chapter \ref{chapter:diffusion}}, we have investigated the combined effect of gravity and thermal expansion on the stability of a planar diffusion flame in an infinitely long channel with rigid walls. The conditions under which the diffusion flame becomes unstable have been determined by calculation of the critical Rayleigh number, which defines the threshold of instability. First, we have investigated the stability of the Burke--Schumann flame, using a linear stability analysis in the Boussinesq approximation. The growth rate of the linear stability problem has been confirmed to be real using numerical methods, so that it is possible to study the system analytically in the marginal state using a similar method to that of the non-reactive problem. Results have been presented, which show that as the flame approaches the lower boundary of the channel with increasing stoichiometric coefficient, the critical Rayleigh number is close to the well-known value it takes in the non-reactive case with two rigid boundaries. Second, we have performed numerical calculations of the full system of governing equations for several values of the thermal expansion coefficient, for large values of the Damk\"{o}hler number. The results show that when the Rayleigh number is higher than its critical value, the fluid forms convection rolls as in the non-reactive case, which interact with the flame to generate cellular structures. The numerical results show strong agreement with those of the linear stability analysis. Finally, we have investigated the effect of finite chemistry on the system. The results show that the system becomes less stable as the Damk\"{o}hler number $Da$ is decreased.

Next, in \textbf{Chapter \ref{chapter:triple}}, the combined effect of thermal expansion and gravity on a triple flame propagating steadily in a horizontal channel has been investigated. The results obtained in Chapter \ref{chapter:diffusion} have been used to identify areas of parameter space where a triple flame is expected to propagate steadily. The effect of the flame-front thickness $\epsilon$ on the propagation speed $U$ of a triple flame has been described for several values of the thermal expansion coefficient $\alpha$ and the Rayleigh number $Ra$. It has been found that the well-known monotonic relationship between $U$ and $\epsilon$ that is present in the constant density case (which arises in the limit $\alpha \to 0$) remains valid for $\alpha \neq 0$, when $Ra=0$ (i.e. in the absence of gravity). However, we have shown that under the influence of gravity the monotonic relationship between $U$ and $\epsilon$ is only present for values of $Ra$ below a critical value, which has been determined. Further, it has been shown that, if $Ra$ takes a value higher than this critical value, there is a local maximum in the graph of $U$ versus $\epsilon$. The system has been shown to exhibit hysteresis for even higher values of the Rayleigh number. The results in this chapter provide valuable insight into the behaviour of a triple flame under gravitational effects and illustrate the complexity and variety of the scenarios that arise.

Finally, in \textbf{Chapter \ref{chapter:ignition}}, we have studied the combined effects of thermal expansion and gravity on the transient evolution of triple flames from their initiation via a hot ignition kernel, as well as the energy required for ignition of triple flames. We have identified two-dimensional, non-propagating solutions of the governing equations, which have been referred to as ``flame tubes" and which can provide a criterion for the minimum energy for forced ignition of two-dimensional triple flames in the mixing layer. We have provided a detailed parametric study of flame tubes, including the combined effect of thermal expansion and gravity. We have shown that ``energy-increasing" perturbations to flame tubes (which are unstable) lead to the propagation of a triple flame and ``energy-decreasing" perturbations lead to extinction. It has been found that flame tubes still exist for values of the Rayleigh number above the critical value for gravitational instability of the planar diffusion flame, so that a criterion for the ignition energy is still available for triple flames in this case. Finally, the transient behaviour of a triple flame with a trailing diffusion flame that is unstable due to gravitational effects has been investigated. It has been found that, after initiation, a triple flame forms initially before the trailing diffusion flame becomes unstable and forms a cellular flame due to convection rolls in the surrounding fluid. The induced convection can be strong enough to cause local extinction.

There are many opportunities for future work in the area of flame propagation in inhomogeneous mixtures. In the area of premixed flames, studied in Chapters \ref{chapter:premixed1} and \ref{chapter:premixed2}, the effect of more complicated flows on premixed flame propagation would help to further understand the interaction between laminar premixed flames and flow. This would provide a platform for a thorough understanding of the behaviour of turbulent premixed flames, which is currently lacking, as can be seen by the lack of a convincing explanation for the bending effect of turbulent flames, discussed in Chapter \ref{chapter:premixed2}. Further work in this area could also investigate the Taylor dispersion regime of premixed flames discussed in Chapter \ref{chapter:premixed1} for more complex flows.

In the area of diffusion flames, discussed in Chapter \ref{chapter:diffusion}, work could be done to further understand the behaviour of diffusion flames that are unstable due to gravitational effects, including a detailed parametric study of their behaviour beyond the threshold of gravitational instability. It would also be interesting to further investigate the behaviour of the triple flames studied in Chapters \ref{chapter:triple} and \ref{chapter:ignition} in situations where the planar diffusion flame is unstable due to gravitational effects. A future study could also repeat the work of Chapters \ref{chapter:diffusion}-\ref{chapter:ignition} in the case of triple flames in a counterflow configuration. Such a study could build on the theoretical and numerical investigations in this thesis, which were the first to describe in detail the combined effect of thermal expansion and gravity on the propagation and stability of triple flames propagating in a direction perpendicular to the direction of gravity.

\bibliography{bibliog}
\bibliographystyle{unsrtnat}
\appendix
\chapter[Cylindrical channel asymptotic analysis]{Premixed flames in a cylindrical channel: asymptotic analysis in the limit $\epsilon \to \infty$}
\label{appendix:premixed1}
In this section we derive the results provided in \S \ref{premixed1:sec:cylindrical}, namely the one-dimensional boundary problem satisfied by a premixed flame propagating against a prescribed Poiseuille inflow in a narrow cylindrical channel. As was done in \S \ref{premixed1:sec:asymptotics} in the case of a rectangular channel, we consider the steady form of equations \eqref{nondim1}-\eqref{nondim8} with $\Le=1$, so that only the temperature equation needs to be considered, since $y_F=1-\theta$. We consider the cylindrical coordinate system $(r,z)$, with fluid velocity $(u_r,u_z)$. In the limit $\epsilon \to \infty$, we introduce a rescaled coordinate
\begin{gather}
\xi=\frac{z}{\epsilon},\label{aeq:scaling}
\end{gather}
so that the governing equations (\ref{nondim1})-(\ref{nondim7}) can be written
\begin{gather}
\frac{\partial}{\partial \xi}\left(\rho(u_z+U)\right)+\epsilon\frac{1}{r}\frac{\partial}{\partial r}\left( \rho r u_r\right)=0,\label{a:nondim1}\\
\rho(u_z+U)\frac{\partial u_z}{\partial \xi} +\epsilon \rho u_r \frac{\partial u_z}{\partial r} +\frac{\partial p}{\partial \xi} = \nonumber\\Pr\left( \frac{4}{3}\sd{u_z}{\xi}+\sd{u_z}{r}+\frac{\epsilon^2}{r}\frac{\partial}{\partial r}\left(r\pd{u_z}{r}\right)+\frac{\epsilon}{3}\left(\frac{\partial^2 u_r}{\partial\xi\partial r}+\frac{1}{r}\pd{u_r}{\xi}\right)\right),\\
\rho(u_z+U)\frac{\partial u_r}{\partial \xi} +\epsilon \rho u_r \frac{\partial u_r}{\partial r} +\epsilon\frac{\partial p}{\partial r} = Pr\left( \sd{u_r}{\xi}+\frac{4\epsilon^2}{3r}\frac{\partial}{\partial r}\left(r\pd{u_r}{r}\right)+\frac{\epsilon}{3r}\pd{u_z}{\xi}\right),\\
\rho(u_z+U)\frac{\partial \theta}{\partial \xi} + \epsilon\rho u_r \frac{\partial \theta}{\partial r} = \sd{\theta}{\xi}+\frac{\epsilon^2}{r}\frac{\partial}{\partial r}\left(r\pd{\theta}{r}\right) + \frac{ \omega}{1-\alpha},\label{a:nondim4}
\end{gather}
where the density $\rho$ is given by \eqref{nondim8} and the reaction term $\omega$ is given by \eqref{eq:reaction2}.
These equations are subject to the cylindrical form of boundary conditions \eqref{bc:1}-\eqref{bc:last}, which can be written
\begin{gather}
\pd{\theta}{r}=\pd{u_z}{r}=u_r=\pd{p}{r}=0 \quad \text{at } r=0, \label{abc:1}\\
\pd{\theta}{r}=u_z=u_r=0 \quad \text{at } r=1,\label{abc:2}\\
\theta=0,\quad u_z=A\left(1-r^2\right)=\epsilon \Pe \left(1-r^2\right), \quad u_r=0\quad \text{at } \xi=-\infty,\label{abc:3}\\
\pd{\theta}{\xi}=\pd{u_z}{\xi}=\pd{u_r}{\xi}=p=0 \quad \text{at } \xi=+\infty. \label{abc:last}
\end{gather}
As in \S \ref{premixed1:sec:asymptotics}, we now introduce expansions for $\epsilon \to \infty$ in the form
\begin{equation}
\left.
\begin{aligned}
&U=-\frac{1}{2}\epsilon \Pe+U_0+\epsilon^{-1}U_1+...\\  &u_z=\epsilon u_0 +u_1+...\quad u_r=v_0+\epsilon^{-1}v_1+...\\&\theta=\theta_0+\epsilon^{-1}\theta_1+...\quad p=\epsilon^3 p_0+\epsilon^2 p_1+...
\end{aligned}
\right\}
\label{aeq:expansions}
\end{equation}
where $U_0$ is the leading order approximation to the effective flame speed $U_T$ defined in (\ref{effective}), which in this case is given by
\begin{gather}
U_T\equiv U+\frac{1}{2}\epsilon \Pe.
\end{gather}
Substituting (\ref{aeq:expansions}) into equations (\ref{a:nondim1})-(\ref{a:nondim4}), to leading order we obtain
\begin{gather}
\pd{}{\xi}\left(\rho_0\left(u_0-\frac{1}{2}\Pe\right)\right)+\frac{1}{r}\frac{\partial}{\partial r}\left(\rho_0 r v_0 \right)=0\label{aleading1},\\
\pd{p_0}{\xi}=\frac{Pr}{r}\frac{\partial}{\partial r}\left( r \pd{u_0}{r} \right),\label{aleading2} \\
\pd{p_0}{r}=0,\label{aleading3}\\
\frac{1}{r}\frac{\partial}{\partial r}\left( r \pd{\theta_0}{r} \right)=0.\label{aleading4}
\end{gather}
Equations (\ref{aleading3}) and (\ref{aleading4}) can be integrated with respect to $r$ to give $p_0=p_0(\xi)$ and $\theta_0=\theta_0(\xi)$, after considering the boundary conditions (\ref{abc:1})-(\ref{abc:2}) on $\theta_0$, so that $\rho_0=\rho_0(\xi)$ from \eqref{nondim8}.\\
Now we look for a separable solution for $u_0(\xi,y)$ in the form
\begin{gather}
u_0(\xi,r)=\hat{u}_0(r)\check{u}_0(\xi).\label{asplitveloc}
\end{gather}
Substituting (\ref{asplitveloc}) into equation (\ref{aleading2}) gives
\begin{gather}
\frac{1}{r}\frac{\partial}{\partial r}\left( r \pd{\hat{u}_0}{r} \right)=\frac{1}{\check{u}_0 Pr}\pd{p_0}{\xi},\label{asplitveloc2}
\end{gather}
where $C$ is a constant. Equation (\ref{asplitveloc2}) can be integrated twice with respect to $r$, using the boundary conditions (\ref{abc:1}) and (\ref{abc:2}), to yield
\begin{gather*}
\hat{u}_0(r)=C(1-r^2),
\end{gather*}
so that
\begin{gather}
u_0(\xi,r)=\check{u}_0(\xi)(1-r^2), \label{appendix:splitveloc3}
\end{gather}
where $C$ has been absorbed into $\check{u}_0(\xi)$. Integrating equation (\ref{aleading1}) with respect to $r$ from $r=0$ to $r=1$, we obtain
\begin{gather}
\pd{}{\xi}\left(\rho_0(\xi)\left(\frac{1}{4}\check{u}(\xi)-\frac{1}{4}\Pe\right)\right)=0,\label{aleadcont}
\end{gather}
after using boundary conditions (\ref{abc:1})-(\ref{abc:2}) on $v_0$. Equation (\ref{aleadcont}) implies that
\begin{gather*}
\frac{1}{4}\rho_0(\xi)\left(\check{u}_0(\xi)-\Pe\right)=\frac{1}{4}\left(\check{u}_0\left(\xi \to -\infty\right)-\Pe\right)=0,
\end{gather*}
using the fact that $\rho_0\left(\xi \to -\infty\right)=1$ from equation \eqref{nondim8} and boundary condition (\ref{abc:3}). Thus $\check{u}_0(\xi)=\Pe$, so that, from \eqref{appendix:splitveloc3}
\begin{gather}
u_0=\Pe(1-r^2).\label{aleadingu}
\end{gather}
Equation (\ref{aleading1}) can then integrated with respect to $r$, using (\ref{aleadingu}) and condition (\ref{abc:1}), to yield
\begin{gather}
v_0=-\frac{1}{r\rho_0}\pd{\rho_0}{\xi}\Pe\left(\frac{r^2}{4}-\frac{r^4}{4}\right).\label{aleadingv}
\end{gather}
Now, at $O\left(\epsilon\right)$ in equation (\ref{a:nondim4}) we have
\begin{gather*}
\rho_0\left(u_0-\frac{1}{2}\Pe\right)\pd{\theta_0}{\xi}=\frac{1}{r}\frac{\partial}{\partial r}\left( r \pd{\theta_1}{r} \right),
\end{gather*}
which, after using (\ref{aleadingu}) and condition (\ref{abc:1}), can be integrated twice with respect to $r$ to give
\begin{gather}
\theta_1=\rho_0\pd{\theta_0}{\xi}\Pe\left(\frac{r^2}{8}-\frac{r^4}{16}\right)+\check{\theta}_1(\xi).\label{atheta_1}
\end{gather}
Next we look to $O\left(1\right)$ in equation (\ref{a:nondim1}) to find
\begin{gather}
\pd{}{\xi}\left(\rho_1\left(u_0-\frac{1}{2}\Pe\right)\right)+\pd{}{\xi}\left(\rho_0\left(u_1+U_0\right)\right)+\frac{1}{r}\pd{}{r}\left(\rho_0v_1\right)+\frac{1}{r}\pd{}{r}\left(\rho_1v_0\right)=0.\label{aorder1}
\end{gather}
Equation (\ref{aorder1}) can be integrated first with respect to $r$ from $r=0$ to $r=1$, utilising the boundary conditions (\ref{abc:1})-(\ref{abc:2}) on $v_0$, and then with respect to $\xi$ to give
\begin{gather}
\int_0^1 r\left(\rho_1\left(u_0-\frac{1}{2}\Pe\right)\right)\mathrm{d}r+\int_0^1 r\left(\rho_0\left(u_1+U_0\right)\right)\mathrm{d}r=K.\label{acont1}
\end{gather}
To evaluate $K$, we use boundary conditions (\ref{abc:3}) to obtain
\begin{gather}
K=\int_0^1 r\left(\rho_1(\xi \to -\infty)\left(u_0(\xi \to -\infty)-\frac{1}{2}\Pe\right)\right)\mathrm{d}r\nonumber\\+\int_0^1 r\left(\rho_0(\xi \to -\infty)\left(u_1(\xi \to -\infty)+U_0\right)\right)\mathrm{d}r=\frac{U_0}{2}.\label{acont2}
\end{gather}
Finally, at $O\left(1\right)$ of equation (\ref{a:nondim4}) we have
\begin{gather}
\rho_0\left(u_1+U_0\right)\pd{\theta_0}{\xi}+\rho_1\left(u_0-\frac{1}{2}\Pe\right)\pd{\theta_0}{\xi}+\rho_0\left(u_0-\frac{1}{2}\Pe\right)\pd{\theta_1}{\xi}+\rho_0v_0\pd{\theta_1}{r}=\nonumber\\\sd{\theta_0}{\xi}+\frac{1}{r}\frac{\partial}{\partial r}\left(r \pd{\theta_2}{r}\right)+\frac{\omega_0}{1-\alpha},\label{atheta_order1}
\end{gather}
where $\omega_0(\xi)=\omega\left(\theta_0,\rho_0\right)$. Integrating (\ref{atheta_order1}) with respect to $r$ from $r=0$ to $r=1$, using the boundary conditions (\ref{abc:1})-(\ref{abc:2}) on $\theta$ and substituting (\ref{aleadingu}), (\ref{aleadingv}), (\ref{atheta_1}), (\ref{acont1}) and (\ref{acont2}) leads to
\begin{gather}
U_0\pd{\theta_0}{\xi}-\pd{}{\xi}\left(\left(1+\frac{1}{192}\Pe^2\rho_0^2\right)\pd{\theta_0}{\xi}\right)=\frac{\omega_0}{1-\alpha} \label{result:afinitebeta},
\end{gather}
with
\begin{equation}
\left.
\begin{aligned}
&\rho_0=\left(1+\frac{\alpha}{1-\alpha}\theta_0\right)^{-1}, \\ 
&\omega_0=\frac{\beta^2}{2}\rho_0\left(1-\theta_0\right)\exp\left(\frac{\beta \left(\theta_0-1\right)}{1+\alpha\left(\theta_0-1\right)}\right),
\end{aligned}
\right\}
\label{ares}
\end{equation}
subject to the boundary conditions
\begin{gather}
\theta_0(-\infty)=0, \quad \theta_{0\xi}(+\infty)=0. \label{ares:bc}
\end{gather}
The problem \eqref{ares}-\eqref{ares:bc} is the required one-dimensional boundary value problem, with an effective diffusion coefficient, as given in \S \ref{premixed1:sec:cylindrical}.

\chapter[Constant density solution for $A\to 0$]{Premixed flames in a narrow channel: constant density solution for $A\to 0$, $Pe=O(1)$}
\label{appendix:premixed2}
In this section we provide an asymptotic result used in \S \ref{premixed2:sec:results}. We consider the constant density form of the infinite activation energy governing equations \eqref{b:1}-\eqref{jump:last}. For an imposed Poiseuille flow of amplitude $A=\epsilon \Pe$, this is achieved by considering equation \eqref{b:4} with $u=\epsilon \Pe\left(1-\eta^2\right)$, $v=0$ and $\rho=1$, which gives
\begin{gather}
\left(\epsilon \Pe\left(1-\eta^2\right)+U+\epsilon f''\left(\eta\right)\right)\pd{\theta}{\xi}=\epsilon \left(1+f'\left(\eta\right)^2\right)\sd{\theta}{\xi}+\epsilon\sd{\theta}{\eta}-2\epsilon f'\left(\eta\right)\frac{\partial^2\theta}{\partial\eta\partial\xi}. \label{appendix:constdensityinfinitebeta}
\end{gather}
Using \eqref{b:bc:1}-\eqref{jump:last}, equation \eqref{appendix:constdensityinfinitebeta} is subject to boundary conditions and jump conditions given by
\begin{gather}
\pd{\theta}{\eta}=f'\left(\eta\right)=0 \quad \text{at } \eta=0,\text{ }\eta=1, \label{appendix:const:bc:1}\\
\theta=0,\quad \text{at } \xi=-\infty, \label{appendix:const:bc:minusinfty}\\
\theta=1, \quad \theta_\xi=\epsilon^{-1}\left(1+f'\left(\eta\right)^2\right)^{-1\left/ 2\right.}, \quad \text{at}\quad \xi=0\label{appendix:const:jump:last}.
\end{gather}

The asymptotic limit considered in this section is $A\to0$, with $\Pe=O(1)$. This limit is equivalent to $\epsilon \to 0$, with $\Pe=O(1)$. The aim is to find an asymptotic formula for the effective propagation speed $U_T$, which is defined in \eqref{effective}. The method is similar to the one used in \cite{daou2001flame}.

We begin by introducing the stretched variable
\begin{gather}
\hat{\xi}=\frac{\xi}{\epsilon},
\end{gather}
so that \eqref{appendix:constdensityinfinitebeta}-\eqref{appendix:const:jump:last} become, using the definition \eqref{effective} for $U_T$,
\begin{gather}
\left(\epsilon \Pe\left(\frac{1}{3}-\eta^2\right)+U_T+\epsilon f''\left(\eta\right)\right)\pd{\theta}{\hat{\xi}}= \left(1+f'\left(\eta\right)^2\right)\sd{\theta}{\hat{\xi}}+\epsilon^2 \sd{\theta}{\eta}-2\epsilon f'\left(\eta\right)\frac{\partial^2\theta}{\partial\eta\partial\hat{\xi}},\label{constsmalleps:1}\\
\pd{\theta}{\eta}=f'\left(\eta\right)=0 \quad \text{at } \eta=0,\text{ }\eta=1, \label{appendix:const2:bc:1}\\
\theta=0,\quad \text{at } \hat{\xi}=-\infty, \label{appendix:const2:bc:minusinfty}\\
\theta=1, \quad \theta_{\hat{\xi}}=\left(1+f'\left(\eta\right)^2\right)^{-1\left/ 2\right.}, \quad \text{at}\quad \hat{\xi}=0\label{appendix:const2:jump:last}.
\end{gather}

Letting $\epsilon \to 0$ and expanding variables in succesive powers of $\epsilon$ as
\begin{gather}
U_T=U_0+\epsilon U_1+...,\quad \theta=\theta_0+\epsilon \theta_1+...,\quad f=f_0+\epsilon^{1/2} f_1+...
\end{gather}
we find that to leading order in equation \eqref{constsmalleps:1}, the solution is the planar flame solution
\begin{gather}
U_0=1,\quad \theta_0=\exp\hat{\xi},\quad f_0'=0. \label{pe1:leading}
\end{gather}
Then, at $O\left(\epsilon\right)$ in \eqref{constsmalleps:1}, we have, using \eqref{pe1:leading},
\begin{gather}
\left(\Pe\left(\frac{1}{3}-\eta^2\right)+U_1\right)\exp\hat{\xi}+\pd{\theta_1}{\hat{\xi}}=\sd{\theta_1}{\hat{\xi}}+f_1'^2\exp\hat{\xi}. \label{pe1:secondorder}
\end{gather}
Now, in this case \eqref{appendix:const2:bc:minusinfty}-\eqref{appendix:const2:jump:last} lead to
\begin{align*}
\theta_1=0\quad\text{at } \hat{\xi}=-\infty,&\quad\\
\theta_1=0,\quad\theta_{1\hat{\xi}}=-\frac{f_1'^2}{2}\quad\text{at }\hat{\xi}=0,&
\end{align*}
so equation \eqref{pe1:secondorder} can be integrated from $\hat{\xi}=-\infty$ to $\hat{\xi}=0$ to give
\begin{gather}
\Pe\left(\frac{1}{3}-\eta^2\right)+U_1=\frac{f_1'^2}{2}.
\end{gather}
Hence, applying condition \eqref{appendix:const2:bc:1} at $\eta=1$ we obtain
\begin{gather}
U_1=\frac{2}{3}\Pe,\label{appendix:1stordereffective}
\end{gather}
and so
\begin{gather}
f_1'=-\sqrt{2\Pe\left(1-\eta^2\right)}. \label{pe1:shape2}
\end{gather}
Now, equation \eqref{pe1:shape2} leads to $f_1''\left(\eta=1\right)\to \infty$, so we must consider the boundary layer near $\eta=1$ in order to find the next order solution to the effective propagation speed $U_T$. In the boundary layer the necessary scalings are found to be
\begin{gather}
\eta=1-2\epsilon^{1/3}\hat{\eta}, \quad f'=\epsilon^{2/3}\psi,
\end{gather}
and we expand the effective propagation speed, using \eqref{pe1:leading} and \eqref{appendix:1stordereffective}, as
\begin{gather}
U_T=1+\frac{2}{3}\epsilon\Pe+\epsilon^{4/3} U_2. \label{appendix:effectiveexpansion1}
\end{gather}
Then, to $O\left(\epsilon^{4/3}\right)$ in equation \eqref{constsmalleps:1} we find, after integrating from $\hat{\xi}=-\infty$ to $\hat{\xi}=0$ and applying the boundary/jump conditions \eqref{appendix:const2:bc:minusinfty}-\eqref{appendix:const2:jump:last},
\begin{gather}
\psi'+\psi^2=8A\hat{\eta}+2U_2,\label{bl:ricatti}
\end{gather}
which is subject to the boundary conditions
\begin{gather}
\psi_0\left(0\right)=0,\quad \psi\left(\eta\to\infty\right)\sim -2 A^{1/2}\sqrt{2\eta}.\label{bl:ricattibc}
\end{gather}
 Equation \eqref{bl:ricatti} is a Ricatti equation; the problem \eqref{bl:ricatti}-\eqref{bl:ricattibc} has been solved in \cite{daou2001flame} to give
\begin{gather}
U_2=-2K A^{2\left/ 3\right.},\label{appendix:effectiveexpansion2}
\end{gather}
where $K\approx1.02$ (note that here the flow amplitude is denoted $A$, whereas $u_0$ is used in \cite{daou2001flame}). Then, using \eqref{appendix:effectiveexpansion1} and \eqref{appendix:effectiveexpansion2} with the fact that $A=\epsilon \Pe$, we have
\begin{gather}
U_T=1+\frac{2}{3}A-2KA^{4/3}\Pe^{-2/3},\label{bl:res1}
\end{gather}
in the limit $\epsilon \to 0$ with $\Pe=O(1)$, which is equivalent to the limit $A\to 0$ with $\Pe=O(1)$. This is the result discussed in \S \ref{premixed2:sec:results}.


\end{document}